%% file: AmpMod_final.tex
\documentclass[preprint2]{aastex}
\usepackage{amsmath}
\usepackage{graphicx,multicol} 
\usepackage{times}
\def\etal{\it et~al.}

\makeatletter
\newcommand\notsotiny{\@setfontsize\notsotiny{7.5}{7.5}}
\makeatother

\shorttitle{Periodic Modulations in Pulsars}
\shortauthors{Basu, Mitra \& Melikidze}

\begin{document}

\title{Periodic Modulation : Newly emergent emission behaviour in Pulsars}

\author{Rahul Basu\altaffilmark{1,2}, Dipanjan Mitra\altaffilmark{3,2}, Giorgi I. Melikidze\altaffilmark{2,4}}

\altaffiltext{1}{Inter-University Centre for Astronomy and Astrophysics, Pune, 411007, India, rahulbasu.astro@gmail.com.}
\altaffiltext{2}{Janusz Gil Institute of Astronomy, University of Zielona G\'ora, ul. Szafrana 2, 65-516 Zielona G\'ora, Poland.}
\altaffiltext{3}{National Centre for Radio Astrophysics, Tata Institute of Fundamental Research, Pune 411007, India.}
\altaffiltext{4}{Evgeni Kharadze Georgian National Astrophysical Observatory, Abastumani, Georgia.}

%\email{rahulbasu.astro@gmail.com}

%\date{Accepted\ldots Received\ldots ; in original form\ldots}

%\pagerange{\pageref{firstpage}--\pageref{lastpage}} \pubyear{2017}

%\maketitle

%\label{firstpage}

\begin{abstract}
Periodic modulations are seen in normal period pulsars ($P >$ 0.1 sec) over 
timescales ranging from a few seconds to several minutes. Such modulations have
usually been associated with the phenomenon of subpulse drifting. A number of 
recent studies have shown subpulse drifting to exhibit very specific physical 
characteristics : i) drifting is seen only in conal components of the pulse 
profile and is absent in central core emission; ii) drifting pulsars are 
distributed over a narrow range of spin-down energy loss ($\dot{E}$), where 
pulsars with $\dot{E} <$ 2$\times$10$^{32}$ erg~s$^{-1}$ show this behaviour, 
iii) drifting periodicity ($P_3$) is anti-correlated with $\dot{E}$, such that 
pulsars with lower values of $\dot{E}$ tend to have longer $P_3$. These 
detailed characterisations of drifting behaviour on the other hand also 
revealed the presence of other distinct periodic modulations, which can be 
broadly categorised into two types, periodic nulling and periodic amplitude 
modulation. In contrast to drifting these periodic phenomena are seen across 
the entire profile in both the core and conal components simultaneously and are 
not restricted to any specific $\dot{E}$ range. In this work we have assembled 
an exhaustive list of around 70 pulsars which show such periodic modulations, 
22 of which were newly detected using observations from the Giant Meterwave 
Radio Telescope and the remaining compiled from past publications. The presence
of such a significant group in the pulsar population suggests that periodic 
nulling and periodic amplitude modulations to be newly emergent phenomena in 
pulsars with their physical origin distinct from subpulse drifting.
\end{abstract}


\section{Introduction}
\noindent
Radio emission from normal period pulsars ($P >$ 0.1 seconds) shows variations 
in their single pulses which exhibit different periodicities. Periodic 
behaviour over timescales of several hundred pulses is studied using longitude 
resolved fluctuation spectra \citep[LRFS,][]{bac70a,bac73}. LRFS requires 
single pulse sequence to be arranged in the form of a pulse stack, which is a 
two dimensional representation with pulse longitude along the abscissa and 
every subsequent period placed along the ordinate. Fourier transforms are 
carried out along each longitude range of the pulse stack to form LRFS. Single 
pulses are composed of one or more components which are known as subpulses. 
Peak frequency in the fluctuation spectra represents periodicity of subpulse 
repetition at any given pulse longitude. Phase variations corresponding to the 
peak amplitude indicate the time delay with which subpulse appears with respect
to a specific reference longitude. 

Periodic modulations have been studied in great detail in the literature, with 
\citet{rud75} suggesting subpulse drifting as primary mechanism for such 
variations. However, detailed phenomenological studies in recent past have 
revealed other likely sources for periodic behaviour in single pulses. We 
present a careful distinction between these possibilities in the discussion 
below.
 
\subsection{Subpulse Drifting}
\noindent
The phenomenon of subpulse drifting is associated with systematic 
variations of subpulses within pulse window \citep{dra68}. Detailed studies
of subpulse drifting in a large number of pulsars \citep{ran86,gil00,des01,
wel06,wel07,bas16,bas18a,bas19a} reveal two important characteristics: 
evolution of drift pattern with line of sight (LOS) geometry, where the 
drifting is different for each component in profile, and dependence of drifting
periodicity ($P_3$) on spin-down energy loss ($\dot{E}$). 

Classification of profile types can be used to understand evolution of drifting
with LOS geometry. Average radio emission beam is expected to consist of a 
central core component, surrounded by two concentric conal rings 
\citep{ran90,ran93}. The observed profile diversity is believed to be related 
to different LOS traverse across emission beam. When LOS cuts across edge of 
beam, conal Single (S$_d$) profiles are seen. As LOS traverses progressively
more interior regions of beam, conal Double (D), conal Triple ($_c$T) and conal
Quadruple ($_c$Q) profile classes are observed. Core dominated profiles, viz., 
core Single (S$_t$), core-cone Triple (T) and core-double cone Multiple (M) are
expected to arise due to central LOS traverse of emission beam. In some pulsars
with core emission, one of the conal pair is too weak to be detected and they 
are classified as T$_{1/2}$. 

Subpulse drifting is a strictly conal phenomenon, with no drifting seen in core
component \citep{ran86}. In addition, drifting behaviour also evolves from 
outer edge to central regions of emission beam. Systematic large scale phase 
variations are seen across entire profile in S$_d$ and D profile classes. On 
the other hand phase variations for each conal component in $_c$T and $_c$Q 
profiles are usually different and show jumps and reversals from one component 
to another. In most M profiles drifting is largely phase stationary in outer 
conal components, but shows significant phase changes in inner cones 
\citep{bas19a,bas19b}. 

Subpulse drifting shows a clear dependence on $\dot{E}$, which is different 
from other periodic behaviour. Drifting is seen in pulsars with $\dot{E} <$ 
5$\times$10$^{32}$ erg~s$^{-1}$. In addition, a correlation can also be 
inferred between drifting periodicity ($P_3$) and $\dot{E}$. If the periodicity
measured in LRFS is not considered to be aliased, pulsars with low $\dot{E}$ 
generally has longer $P_3$ than pulsars with higher $\dot{E}$. This 
anti-correlation becomes stronger under the assumption that subpulse motion is 
lagging behind co-rotation speed. In this scenario negative drifting with 
subpulse motion from trailing to leading edge of profile has $P_3 >2P$, and 
positive drifting with subpulse motion towards trailing edge has $P < P_3 < 
2P$. The estimated dependence is given as $P_3\propto\dot{E}^{-0.6\pm0.1}$ 
\citep{bas16}. 

Subpulse drifting is associated with non-stationary plasma flow in Inner 
Acceleration Region (IAR) of pulsars, which is best explained using a partially
screened gap model \citep[PSG,][]{gil03,sza15}. PSG model considers a steady 
flow of ions from stellar surface which screens the electric field in IAR by a 
screening factor ($\eta$) :
\begin{equation}
\eta = 1 - \rho_i/\rho_{GJ}; ~~~~~~~~\rho_{GJ} = \vec{\Omega}\cdot\vec{B}/2c.
\label{eq_eta}
\end{equation}
Here, $\rho_i$ is density of ions and $\rho_{GJ}$ the Goldreich-Julien density 
\citep{gol69}. The plasma responsible for radio emission is generated in IAR in
the form of sparking discharges \citep{rud75}. Sparks do not co-rotate with the
star, but lag behind co-rotation speed which results in subpulse drifting. 

In presence of PSG the drift speed of sparks ($v_{sp}$) can be expressed as 
\citep{bas16} :
\begin{equation}
v_{sp} = \eta (E/B)c,
\label{eq_drift}
\end{equation} 
where $B$ is magnetic field and $E$ is co-rotation electric field. The 
subpulses can be associated with spark motion in IAR. It is speculated that IAR
is packed with circular sparks which are also equidistant, represented by size 
$h$ \citep{gil00}. The time of repetition of sparks at any longitude, which is 
also drifting periodicity ($P_3$), is estimated as 
\begin{equation}
P_3 = 2 h/v_{sp}. 
\label{eq_P3_1}
\end{equation}
When $\eta$ is small ($\eta$ $\sim$0.1) it can be shown that $P_3$ in PSG is 
estimated as \citep{sza13} :
\begin{equation}
P_3 = \frac{1}{2\pi\eta {\rm cos}\alpha},
\label{eq_P3}
\end{equation}
where $\alpha$ is inclination angle between rotation and magnetic axes. The 
anti-correlation between $P_3$ and $\dot{E}$ can be derived from 
eq.(\ref{eq_P3_1}) and eq.(\ref{eq_P3}) using basic physical approximations. 
When the full energy outflow from polar cap is associated with $\dot{E}$ using 
factor $\xi$ we obtain \citep{bas16} :
\begin{equation}
P_3 = 2\times10^{-9}\left(\frac{\gamma_0}{\xi}\right)\left(\frac{\dot{E}}{\dot{E}_1}\right)^{-0.5},
\label{eq_P3edot}
\end{equation}
where $\gamma_0\sim10^6$ is Lorentz factor of primary particles in IAR, 
$\xi\sim10^{-3}-10^{-4}$ is scaling factor obtained from the fraction of 
non-thermal X-ray emission \citep{bec09}, and $\dot{E}_1$ = 4$\times$10$^{31}$
erg~s$^{-1}$.

\subsection{Other Periodic Modulations}
\noindent
Recent studies have revealed presence of additional periodic behaviours seen in
fluctuation spectra, which include periodic nulling \citep{her07} and periodic 
amplitude modulation. Periodic nulling is seen in conal as well as core-cone 
profiles, where the core components also vanishes along with cones in a 
periodic manner. This prompted \citet{bas17} to identify periodic nulling to be
a different phenomenon. In addition certain pulsars with core emission show low
frequency modulation in intensity, where the core component also participates 
\citep{bas16,mit17}. This behaviour is known as periodic amplitude modulation. 

There are clear differences between physical parameters of subpulse drifting 
and other periodic modulations. A significant number of pulsars showing this 
behaviour have high $\dot{E}$ in excess of 5$\times$10$^{32}$ erg~s$^{-1}$. 
Periodic nulling and periodic amplitude modulation are usually seen as low
frequency features in fluctuation spectra. The corresponding longer 
periodicities are not correlated with $\dot{E}$ unlike subpulse drifting. 

Contrasting behaviour of drifting and other periodic modulations are 
clearly illustrated in pulsars where both effects are seen in pulse sequence 
\citep{bas17,bas19a}. In all such cases the low frequency feature in 
fluctuation spectra can be identified as periodic modulation. One example is 
the pulsar B2003$-$08, with a M type profile, which has subpulse drifting and 
periodic nulling. Subpulse drifting is seen only in conal pairs, where the 
outer pairs have phase stationary behaviour and the inner pairs exhibit large 
scale bi-drifting behaviour. Periodic nulling is seen across entire profile as 
a phase stationary behaviour \citep{bas19b}. Another pulsar B1737+13, with M 
type profile, shows presence of both subpulse drifting and periodic amplitude 
modulation. Drifting only affects the conal components, but amplitude 
modulation is seen across all components \citep{for10}.

Physical mechanism of periodic modulations is still unknown, but is expected to
be different from subpulse drifting. Subpulse drifting can be explained using 
standard physics of non-stationary flow in the PSG as explained above. On the 
other hand origin of a number of other physical phenomena in pulsar radio
emission cannot be explained using this standard model. In certain pulsars 
subpulses show presence of quasi-periodic structures which are also called 
microstructures. They likely originate due to temporal modulations of 
non-stationary plasma flow resulting in alternating radial emitting regions
interspersed with relatively less bright parts \citep{mit15}. However, presence
of such modulations require variations in the plasma generation process in PSG,
which are currently unknown.
 
Pulsar radio emission also shows the presence of nulling \citep{bac70b}, 
and mode changing \citep{bac70c}, where emission switches from one steady state
to another. These phenomena are not periodic, with rapid transition between 
different states (usually within a period), which switch back to the initial 
state after irregular intervals. A notable exception is the periodic swooshing 
events in pulsars B0919+06 and B1859+07 \citep{ran06,wah16}. Their physical 
origin, unlike drifting, also cannot be explained using the steady state 
conditions in IAR, and requires changes in plasma generation process which are 
still unexplored. 

A detailed classification of subpulse drifting from a complete list of 
pulsars has been reported in \citet{bas19a}. However, no equivalent study 
exists for periodic nulling and periodic amplitude modulation, which we plan to
address in this work. We have carried out measurements of nulling and 
periodicities in a large number of pulsars, observed using the Giant Meterwave 
Radio Telescope (GMRT), as well as conducted an exhaustive literature survey. 
In section \ref{sec:obs} we describe details of observations from GMRT as well 
as analysis schemes used to determine nulling and periodic behaviour. Section 
\ref{sec:intmod} carries out a collective study of all known pulsars showing 
periodic modulations, including those previously reported in literature. A 
detailed discussion comparing physical properties of different periodic 
behaviours in pulsars is presented in section \ref{sec:disc}. Finally, section 
\ref{sec:sum} summarizes primary results and conclusions of our studies.

\section{Observations and Analysis}\label{sec:obs}
\noindent
A large sample of pulsars was assembled by \citet{bas19a} to study drifting
behaviour in the population. This list included more than thirty pulsars 
observed with GMRT and another fifty from archival observations \citep{mit11}. 
Many of these sources exhibited nulling and periodic modulations, whose 
properties we have explored in this work. We have carried out detailed nulling 
and fluctuation spectral analysis of 62 pulsars as shown in Table \ref{tabobs}.

Pulsars were observed in `Phased-Array' mode at 325 MHz frequency band. We 
analysed sources without any previous studies of nulling or periodic 
modulation. Details of observations as well as instrumental setup are available
in \cite{bas19a}. Initially, recorded signals from each pulsar were converted 
into well calibrated, baseline corrected, single pulse sequence \citep{mit16,
bas16}. Subsequently, number of different analyses were carried out, which 
are briefly summarized below\citep{bas17}. 

\begin{deluxetable}{cccccccccccc}
\tabletypesize{\notsotiny}
\tablecaption{Estimations of Nulling and Modulation Periodicity \label{tabobs}}
\tablehead{
 \colhead{PSR} & \colhead{P} & \colhead{N$_p$} & \colhead{NF} & \colhead{N$_T$} & \colhead{$\langle BL\rangle$} & \colhead{$\langle NL\rangle$} & \colhead{$f_p$} & \colhead{FWHM} & \colhead{S$_M$} & \colhead{$P_M$} & \colhead{Remarks}\\
 \colhead{} & \colhead{(s)} & \colhead{} & \colhead{(\%)} &\colhead{} & \colhead{($P$)} & \colhead{($P$)} & \colhead{(cy/$P$)} & \colhead{(cy/$P$)} & \colhead{($P$/cy)} & \colhead{($P$)} & \colhead{}
 }
\startdata
  B0105+65  & 1.284 & ~3020 & ... & ... & ... & ... & ... & ... & ... & ... & No Nulls \\
  B0138+59  & 1.223 & ~1960 & ~7.6$\pm$1.0 & ~89 & ~19.0 & ~2.8 & 0.023$\pm$0.017 & 0.040 & ~21.2 & 44$\pm$30 & Periodic Nulls \\
  B0320+39  & 3.032 & ~2073 & 0.7$\pm$0.2 & ~14 & 140.4 & ~1.0 & ... & ... & ... & ... & Short Nulls \\
  B0355+54  & 0.156 & 13113 & ... & ... & ... & ... & ... & ... & ... & ... & Low emission \\
  B0402+61  & 0.595 & ~3041 & ... & ... & ... & ... & ... & ... & ... & ... & No Nulls \\
 J0421$-$0345 & 2.161 & ~1226 & ... & ... & ... & ... & ... & ... & ... & ... & Low emission \\
 B0447$-$12 & 0.438 & ~2742 & ... & ... & ... & ... & ... & ... & ... & ... & No Nulls \\
 B0450$-$18 & 0.549 & ~2738 & ... & ... & ... & ... & 0.064$\pm$0.023 & 0.055 & ~18.1 & 15.6$\pm$5.7 & Amp. Mod. \\
  B0450+55  & 0.341 & ~2658 & ... & ... & ... & ... & 0.111$\pm$0.016 & 0.037 & ~27.4 & 9.0$\pm$1.3 & Amp. Mod. \\
  B0523+11  & 0.354 & ~3436 & ... & ... & ... & ... & ... & ... & ... & ... & No Nulls \\
 B0559$-$05 & 0.396 & ~3900 & ... & ... & ... & ... & ... & ... & ... & ... & No Nulls \\
  B0609+37  & 0.444 & ~2049 & ... & ... & ... & ... & ... & ... & ... & ... & Low emission \\
 B0621$-$04 & 1.039 & ~1357 & ... & ... & ... & ... & 0.0137$\pm$0.0003 & 0.0007 & 313.5 & 73.1$\pm$1.6 & Low emission \\
 B0727$-$18 & 0.510 & ~3537 & ... & ... & ... & ... & ... & ... & ... & ... & Low emission \\
 B0740$-$28 & 0.167 & ~3629 & ... & ... & ... & ... & ... & ... & ... & ... & No Nulls \\
  B0809+74  & 1.292 & ~~890 & 1.2$\pm$0.4 & ~~6 & 145.2 & ~~2.0 & ... & ... & ... & ... & Short Nulls \\
 B0818$-$13 & 1.238 & ~2425 & 0.9$\pm$0.1 & ~16 & 130.6 & ~~1.3 & ... & ... & ... & ... & Short Nulls \\
  B0820+02  & 0.865 & ~1376 & 0.5$\pm$0.2 & ~~7 & 181.1 & ~~1.0 & ... & ... & ... & ... & Short Nulls \\
 B0905$-$51 & 0.254 & ~2358 & ... & ... & ... & ... & ... & ... & ... & ... & No Nulls \\
 B0906$-$17 & 0.402 & ~2244 & 12.4$\pm$1.1 & ... & ... & ... & ... & ... & ... & ... & Low emission \\
  B0919+06  & 0.431 & ~4100 & ... & ... & ... & ... & ... & ... & ... & ... & No Nulls \\
 B0932$-$52 & 1.445 & ~1614 & 1.9$\pm$0.1 & ~30 & ~48.7 & ~1.5 & 0.029$\pm$0.016 & 0.038 & ~~7.6 & 35$\pm$19 & Periodic Nulls \\
  B1112+50  & 1.656 & ~1977 & 34.8$\pm$1.4 & 215 & ~~3.4 & ~5.8 & ... & ... & ... & ... & Med. Nulls \\
  B1237+25  & 1.382 & ~~946 & 8.9$\pm$1.0 & ~40 & ~21.2 & ~~2.1 & 0.039$\pm$0.007 & 0.018 & ~39.4 & 25.7$\pm$4.9 & Periodic Nulls \\
  B1322+83  & 0.670 & ~2672 & ... & ... & ... & ... & ... & ... & ... & ... & Low emission \\
  B1508+55  & 0.740 & ~1784 & 5.2$\pm$0.4 & ~95 & ~16.7 & ~~2.0 & 0.066$\pm$0.032 & 0.076 & ~~2.6 & 15.1$\pm$7.3 & Periodic Nulls \\
 B1510$-$48 & 0.455 & ~2104 & ... & ... & ... & ... & 0.027$\pm$0.003 & 0.007 & ~57.9 & 36.6$\pm$4.1 & Low emission \\
 B1540$-$06 & 0.709 & ~5660 & ... & ... & ... & ... & ... & ... & ... & ... & No Nulls \\
  B1541+09  & 0.748 & ~3035 & ... & ... & ... & ... & 0.066$\pm$0.021 & 0.050 & ~~6.2 & 15.1$\pm$4.8 & Amp. Mod. \\
 B1556$-$44 & 0.257 & ~3550 & ... & ... & ... & ... & ... & ... & ... & ... & No Nulls \\
 B1601$-$52 & 0.658 & ~4551 & ... & ... & ... & ... & 0.047$\pm$0.015 &  0.036 & ~~1.7 & 21.4$\pm$7.0 & Low emission \\
 B1604$-$00 & 0.422 & ~3248 & 0.15$\pm$0.07 & ... & ... & ... & 0.029$\pm$0.011 & 0.026 & ~11.9 & ~34$\pm$13 & Amp. Mod. \\
  B1612+07  & 1.207 & ~~914 & ... & ... & ... & ... & ... & ... & ... & ... & Low emission \\
 B1642$-$03 & 0.388 & ~1502 & ... & ... & ... & ... & 0.078$\pm$0.032 & 0.075 & ~~5.9 & 12.8$\pm$5.2 & Amp. Mod. \\
 J1650$-$1654 & 1.750 & ~2162 & ... & ... & ... & ... & 0.016$\pm$0.011 & 0.026 & ~~5.1 & 64$\pm$45 & Low emission \\
 B1700$-$18 & 0.804 & ~1849 & ... & ... & ... & ... & 0.023$\pm$0.009 & 0.021 & ~~8.9 & 43$\pm$15 & Low emission \\
 B1717$-$29 & 0.620 & ~1850 & ... & ... & ... & ... & ... & ... & ... & ... & No Nulls \\
 B1718$-$02 & 0.478 & ~3075 & ... & ... & ... & ... & ... & ... & ... & ... & Low emission \\
 B1742$-$30 & 0.367 & ~1900 & ... & ... & ... & ... & ... & ... & ... & ... & Low emission \\
 B1804$-$08 & 0.164 & ~1700 & ... & ... & ... & ... & ... & ... & ... & ... & Low emission \\
 B1822$-$09 & 0.769 & ~2328 & ... & ... & ... & ... & ... & ... & ... & ... & No Nulls \\
 B1831$-$04 & 0.290 & ~2000 & ... & ... & ... & ... & ... & ... & ... & ... & Low emission \\
 B1845$-$19 & 4.308 & ~2223 & 27.2$\pm$1.7 & 410 & ~~3.4 & ~~2.0 & ... & ... & ... & ... & Short Nulls \\
 B1851$-$14 & 1.147 & ~1024 & ... & ... & ... & ... & ... & ... & ... & ... & Low emission \\
 J1857$-$1027 & 3.687 & ~1000 & $>$30 & ... & ...  & ... & 0.056$\pm$0.033 & 0.077 & ~11.2 & 18$\pm$10 & Periodic Nulls \\
 B1857$-$26 & 0.612 & ~1955 & 4.3$\pm$0.5 & 198 & ~~8.6 & ~~1.2 & ... & ... & ... & ... & Short Nulls \\
  B1905+39  & 1.236 & ~1124 & 10.0$\pm$0.5 & 80 & ~11.5 & ~~2.3 & 0.021$\pm$0.011 & 0.026 & ~26.9 & 47$\pm$24 & Periodic Nulls \\
 B1907$-$03 & 0.505 & ~1279 & ... & ... & ... & ... & ... & ... & ... & ... & No Nulls \\
  B1929+10  & 0.227 & ~1824 & ... & ... & ... & ... & 0.088$\pm$0.006 & 0.014 & ~33.0 & 11.4$\pm$0.8 & Amp. Mod. \\
 B1937$-$26 & 0.403 & ~1904 & ... & ... & ... & ... & ... & ... & ... & ... & Low emission \\
  B1952+29  & 0.427 & ~3512 & ... & ... & ... & ... & 0.039$\pm$0.018 & 0.042 & ~~1.9 & 26$\pm$12 & Low emission \\
  B2016+28  & 0.558 & ~5358 & ... & ... & ... & ... & ... & ... & ... & ... & No Nulls \\
  B2021+51  & 0.529 & ~5598 & ... & ... & ... & ... & 0.045$\pm$0.030 & 0.070 & ~~5.5 & 22$\pm$15 & Low emission \\
 B2043$-$04 & 1.547 & ~1980 & ... & ... & ... & ... & ... & ... & ... & ... & No Nulls \\
 B2045$-$16 & 1.962 & ~1199 & 9.2$\pm$0.7 & ~69 & ~14.4 & ~~2.8 & 0.023$\pm$0.013 & 0.031 & ~25.5 & 43$\pm$25 & Periodic Nulls \\
  B2053+21  & 0.815 & ~2198 & ... & ... & ... & ... & ... & ... & ... & ... & No Nulls \\
  B2111+46  & 1.015 & ~2000 & 8.7$\pm$0.5 & 130 & ~11.1 & ~~4.1 & 0.021$\pm$0.005 & 0.011 & ~70.4 & 48$\pm$10 & Periodic Nulls \\
  B2217+47  & 0.538 & ~2179 & ... & ... & ... & ... & ... & ... & ... & ... & No Nulls \\
  B2224+65  & 0.683 & ~2049 & ... & ... & ... & ... & ... & ... & ... & ... & Low emission \\
  B2310+42  & 0.349 & ~3583 & 5.1$\pm$0.8 & ~54 & ~36.6 & ~~2.9 & 0.045$\pm$0.015 & 0.034 & ~11.4 & 22.3$\pm$7.2 & Periodic Nulls \\
  B2319+60  & 2.256 & ~2121 & 15.2$\pm$1.1 & 100 & ~15.0 & ~~6.2 & 0.017$\pm$0.005 & 0.012 & ~39.6 & 58$\pm$17 & Periodic Nulls \\
 B2327$-$20 & 1.644 & ~1854 & 10.7$\pm$1.0 & 130 & ~11.9 & ~~2.3 & 0.048$\pm$0.023 & 0.054 & ~13.3 & 20.8$\pm$9.9 & Periodic Nulls \\
%  &  &  &  &  &  &  &  &  &  &  &  &  \\
\enddata
\end{deluxetable}

\begin{figure*}
\begin{center}
\begin{tabular}{@{}cr@{}}
\mbox{\includegraphics[angle=0,scale=0.62]{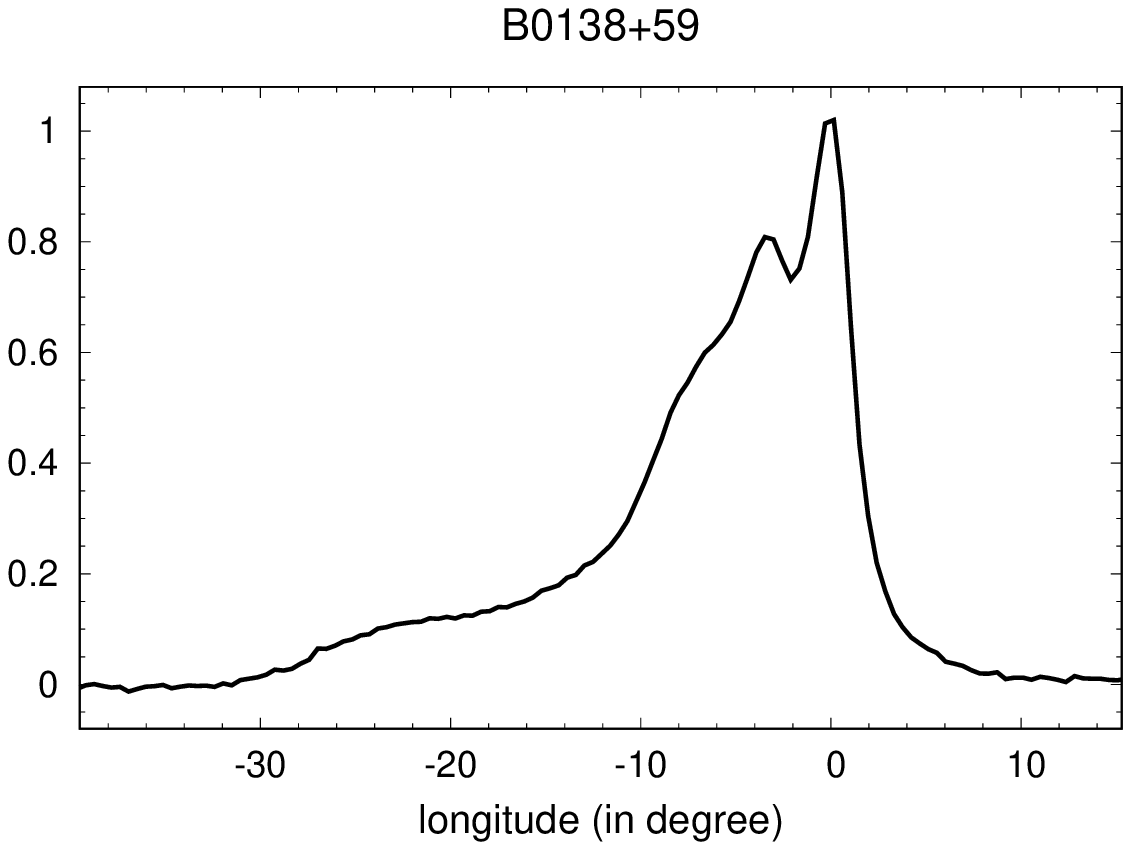}} &
\hspace{20px}
\mbox{\includegraphics[angle=0,scale=0.62]{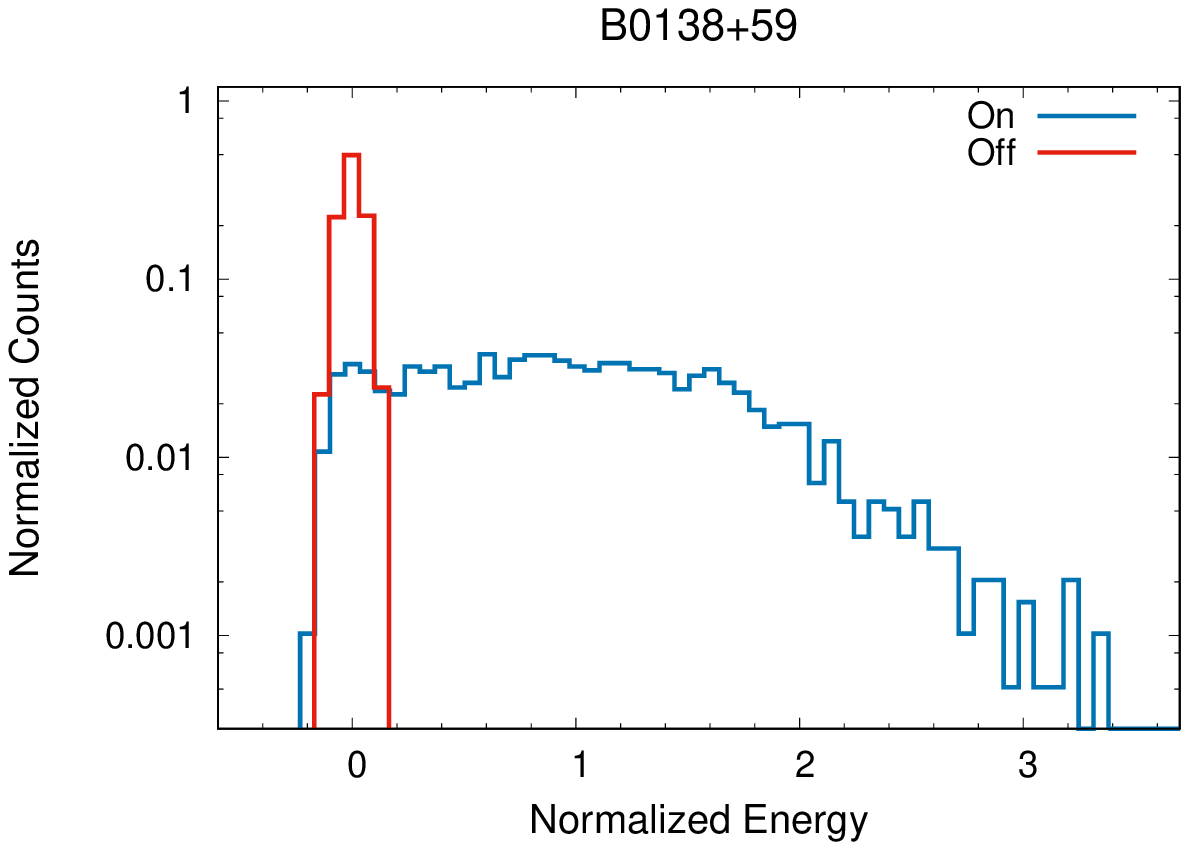}} \\
\end{tabular}
\caption{The figure shows average profile (left panel) and pulse energy 
distribution (right panel) of the pulsar B0138+59. (The complete figure set (62
images) is available.)}
\label{fig_enerdist}
\end{center}
\end{figure*}

\begin{figure*}
\begin{center}
\begin{tabular}{@{}cr@{}}
\mbox{\includegraphics[angle=0,scale=0.62]{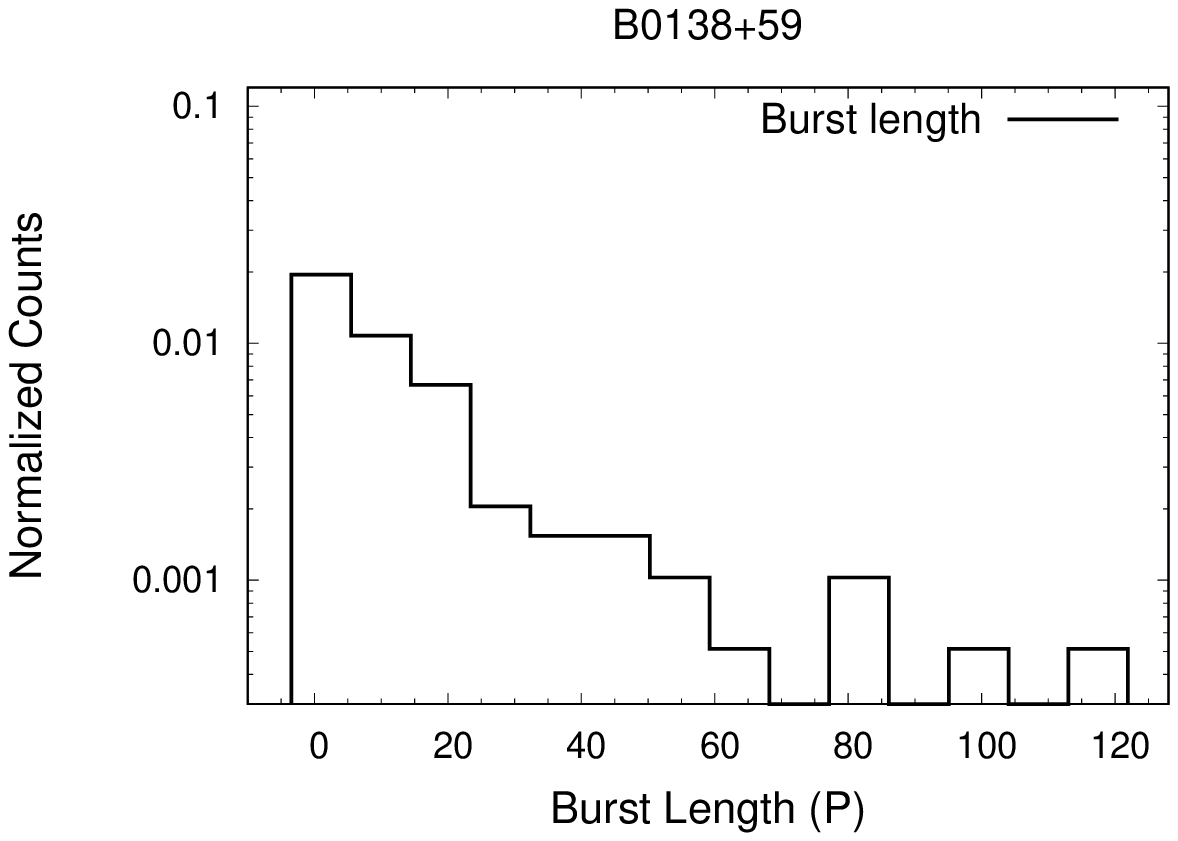}} &
\hspace{20px}
\mbox{\includegraphics[angle=0,scale=0.62]{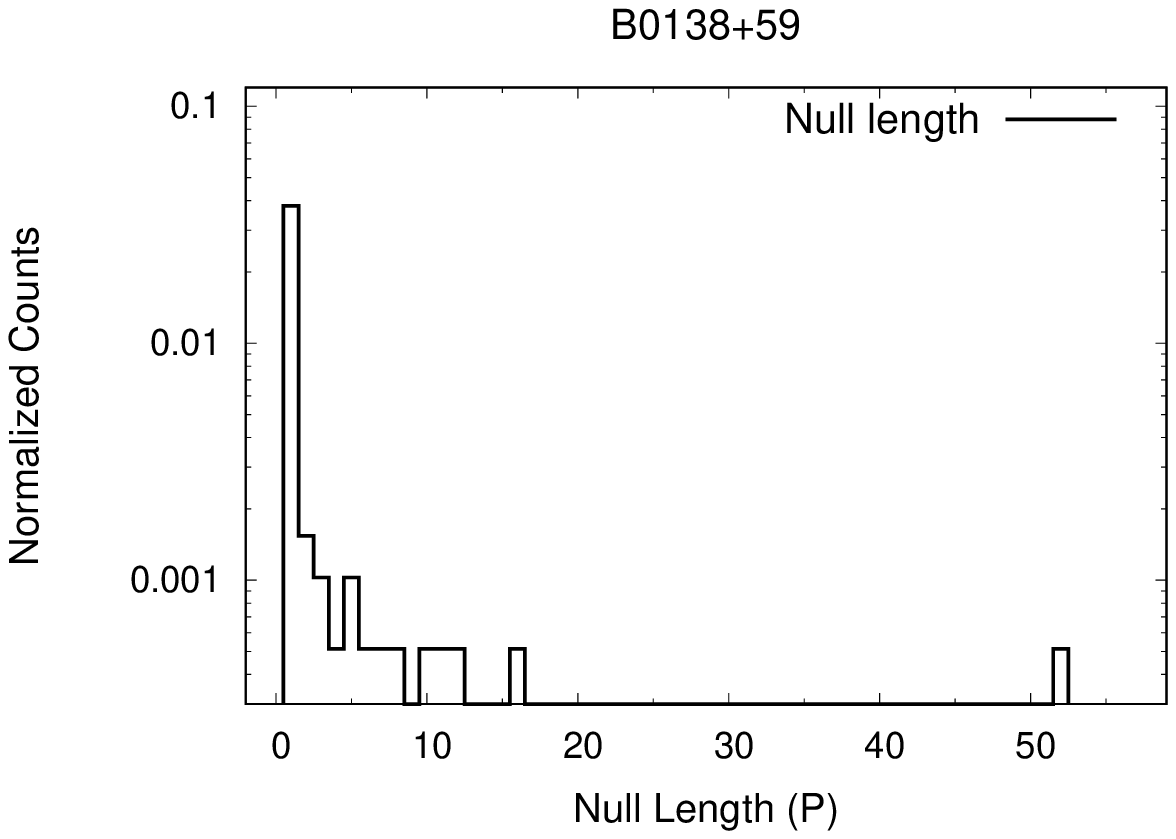}} \\
\end{tabular}
\caption{Burst length (left panel) and Null length (right panel) distributions 
of the pulsar B0138+59. (The complete figure set (17 images) is available.)}
\label{fig_lendist}
\end{center}
\end{figure*}

Nulling was studied by identifying suitable on and off-pulse windows in 
profile. Average energies in these windows were calculated for each pulse. 
Pulse energy distribution for the entire sequence was estimated to search for 
presence of nulling. Figure \ref{fig_enerdist} shows an example of on and 
off-pulse energy distributions for the nulling pulsar B0138+59. On-pulse 
energies show bimodal behaviour where null pulses are coincident with off-pulse
distribution. Nulling fraction (NF) was estimated by fitting Gaussian functions
to null and the off-pulse distributions and finding suitable scaling relation 
between them \citep{rit76}. Error in NF was calculated from statistical errors 
of functional fits. In a few pulsars there were short duration (a few periods),
infrequent, but clear nulls, which were too few to resemble fully formed 
Gaussian functions. In such cases the number of null pulses were individually 
counted to estimate NF. Error in NF was estimated as $\delta$NF = 
$\sqrt{n_p}$/N$_p$, where $n_p$ corresponded to all null pulses and N$_p$ is 
total number of pulses. 

In Table \ref{tabobs} we report 20 pulsars where nulling was present. 19 
pulsars in the list show clear separation between on and off-pulse 
distributions, indicating absence of nulling. Last column in the Table 
identifies 17 pulsars without any nulling. In addition, two pulsars B0450$-$18 
and B1642$-$03 also had no detectable nulls. Detection sensitivities of single 
pulses in remaining 23 pulsars were not sufficient to rule out presence of 
nulling. Figure \ref{fig_lendist} shows an example of null length and burst 
length distributions for the pulsar B0138+59. There were 17 pulsars where such 
distributions could be estimated. The Table also lists total number of 
transitions from null to burst sequence ($N_T$) as well as average duration of 
nulls ($\langle NL\rangle$) and bursts ($\langle BL\rangle$) in these 17 
pulsars.

\begin{figure*}
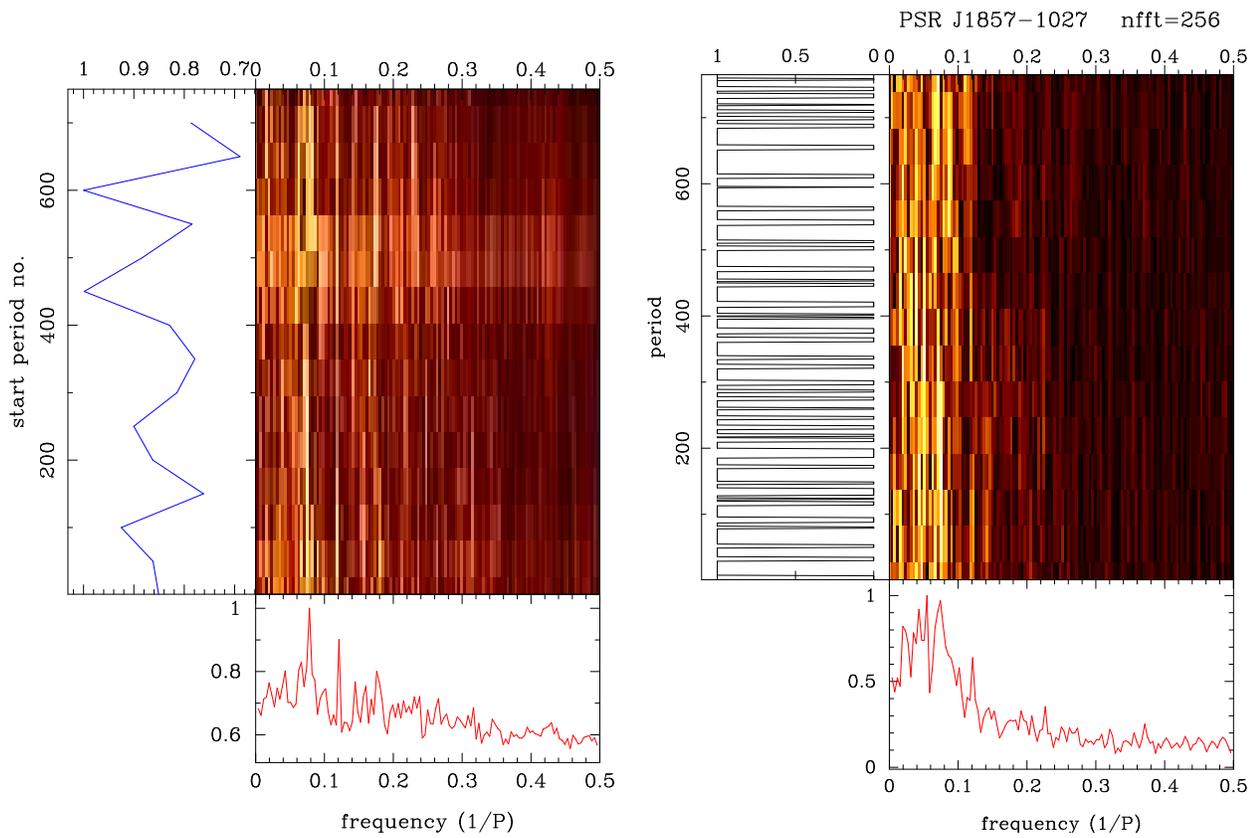

\begin{center}
\begin{tabular}{@{}cr@{}}
\mbox{\includegraphics[angle=0,scale=0.42]{J1857-1027_LRFSavg_256.ps}} &
%\hspace{20px}
\mbox{\includegraphics[angle=0,scale=0.42]{J1857-1027_nullfft_256.ps}} \\
\end{tabular}
\caption{Time evolution of LRFS (left panel) and Null-Burst time series FFT 
(right panel) in the pulsar J1857$-$1027. (The complete figure set (24 images) 
is available.)}
\label{fig_ampmod}
\end{center}
\end{figure*}

We have also investigated the presence of periodic modulations in these 
pulsars. This involved determining time variations in LRFS as well as harmonic 
resolved fluctuation spectra \citep[HRFS,][]{des01}, as described in 
\citet{bas16}. In nulling pulsars, with well defined null and burst pulses, 
presence of periodic nulling was also explored. Nulls were identified as `0' 
and bursts as `1' and Fourier transforms of this binary series was 
determined \citep[see][for details]{bas17}. Figure \ref{fig_ampmod} shows time 
varying LRFS and nulling FFT in the pulsar J1857$-$1027 with periodic nulling. 

Periodic modulations were detected in 24 pulsars, where 11 pulsars exhibited
periodic nulling and another 6 showed periodic amplitude modulations. Exact 
nature of periodic behaviour in remaining 7 pulsars was indeterminate due to 
lower sensitivity of detections. Only 10 out of 17 pulsars with clearly 
separated null and burst pulses had periodic nulling. Additionally, the pulsar 
J1857$-$1027 showed presence of clear nulls of 5-10$P$ durations which were 
also periodic in nature (see figure \ref{fig_ampmod}), but the emission was too
weak for other nulling analysis. Table \ref{tabobs} reports peak modulation 
frequency ($f_p$), width of peak feature (FWHM), strength of the feature 
($S_M$), defined as relative height of feature from baseline divided by FWHM, 
and periodicity ($P_M$). Error in estimating peak frequency is given as $\delta
f_p$ = FWHM/2$\sqrt{2ln(2)}$ \citep{bas16}.

\section{Pulsars with Periodic Modulations}\label{sec:intmod}
\noindent
We have investigated all known pulsars with periodic modulations which are 
different from subpulse drifting. There are around 70 pulsars exhibiting 
periodic behaviour either in the form of periodic nulling, periodic amplitude 
modulations or indeterminate cases. Individual pulsars belonging to each group 
are described below along with their basic physical properties.

\subsection{Periodic Nulling}\label{sec:null}

\begin{table*}
\begin{center}
\caption{Periodic Nulling
\label{tabnull}}
{\begin{tabular}{cccccc}
\tableline
  PSR & $\dot{E}$ & $P_N$ & Profile &  & REF \\
\tableline
  & (2$\times$10$^{32}$~erg/s) & ($P$) &  &  & \\ 
\tableline
  B0031$-$07  & 0.096 & ~75$\pm$14 & S$_d$ & Drift, Mode & [1] \\
   B0138+59   & 0.042 & ~44$\pm$30 & M & ... & [2] \\
   B0301+19   & 0.096 & 103$\pm$34 & D & Drift & [3] \\
   B0525+21   & 0.151 & ~46$\pm$4~ & D & ... & [3] \\
   B0751+32   & 0.071 & ~73$\pm$10 & D & ... & [3] \\
   B0823+26   & 2.260 & ~14$\pm$3~ & S$_t$ & Mode & [4] \\
   B0834+06   & 0.650 & ~16$\pm$4~ & D & Drift & [3]\\
  B0932$-$52  & 0.305 & ~35$\pm$19 & S$_d$ & Drift & [2] \\
   B1133+16   & 0.440 & ~29$\pm$2~ & D & ... & [5] \\
   B1237+25   & 0.072 & ~26$\pm$5~ & M & Drift, Mode & [2] \\
   B1508+55   & 2.440 & ~15$\pm$7~ & T & ... & [2] \\
  J1649+2533  & 0.106 & ~27$\pm$2~ & ... & ... & [3] \\
  B1706$-$16  & 4.470 & 130$\pm$70 & S$_t$ & ... & [1] \\
 J1727$-$2739 & 0.101 & 206$\pm$33 & D & Drift & [1] \\
  B1738$-$08  & 0.053 & ~34$\pm$8~ & $_c$Q? & Drift & [1] \\
  J1819+1305  & 0.060 & ~64$\pm$8~ & $_c$Q & ... & [6] \\
  B1819$-$22  & 0.041 & 134$\pm$33 & D & Drift, Mode & [7] \\
%   B1839+09   & 3.880 & ~37$\pm$3~ & S$_t$ & ... & [3] \\
 J1857$-$1027 & 0.042 & ~18$\pm$10 & T & ... & [2] \\
   B1905+39   & 0.057 & ~47$\pm$24 & M & ... & [2] \\
   B1918+19   & 0.320 & ~85$\pm$14 & $_c$T & Drift, Mode & [3] \\
   B1944+17   & 0.056 & 600$\pm$52 & $_c$T & Drift, Mode & [1] \\
  B2003$-$08  & 0.046 & ~41$\pm$4~ & M & Drift, Mode & [8] \\
   B2034+19   & 0.045 & ~57$\pm$6~ & T & ... & [3] \\
  B2045$-$16  & 0.287 & ~51$\pm$20 & T & Drift & [1] \\
   B2111+46   & 0.135 & ~48$\pm$10 & T & ... & [2] \\
   B2303+30   & 0.146 & ~43$\pm$8~ & S$_d$ & Drift, Mode & [3] \\
   B2310+42   & 0.520 & ~32$\pm$11 & $_c$T & Drift & [1] \\
   B2319+60   & 0.121 & ~58$\pm$17 & $_c$Q & Drift, Mode & [2] \\
   B2327$-$20 & 0.206 & ~19$\pm$1~ & T & ... & [1] \\
%  &  &  &  &  &  \\
\tableline
\end{tabular}}
\tablenotetext{~}{1-\cite{bas17}; 2-This paper; 3-\cite{her09}; 4-\cite{bas19c}; 5-\cite{her07}; 6-\cite{ran08}; 7-\cite{bas18b}; 8-\cite{bas19b}.}
\end{center}
\end{table*}

\noindent
Table \ref{tabnull} presents twenty nine pulsars which show presence of 
periodic nulling. The Table also describes various physical characteristics of 
these pulsars including their $\dot{E}$ values, nulling periodicity ($P_N$), 
profile classification, presence of subpulse drifting and/or mode changing in 
pulse sequence, and references for initial detection of periodic nulling in 
each pulsar. Periodic nulling was first detected in the pulsar B1133+16 by 
\citet{her07} and subsequently its presence was reported in PSR J1819+1305 
by \citet{ran08}. Detailed work of \cite{her09} identified additional 9 pulsars
to show this behaviour, which was further increased to a total of 19 pulsars by
\citet[][see Table 3 of this paper]{bas17}. Subsequently, periodic nulling was 
also detected in PSR B0823+26 \citep{bas19c}, PSR B1819$-$22 \citep{bas18b} and 
PSR B2003$-$08 \citep{bas19b}. We have identified an additional 8 pulsars which
exhibit periodic nulling in Table \ref{tabobs}. 

Physical characteristics of these pulsars indicate that periodic nulling 
coexists with subpulse drifting and mode changing. In this list there are 15 
pulsars which show subpulse drifting and 9 pulsars also has presence of 
multiple emission modes. In 8 pulsars all three phenomena are seen in the same 
pulse sequence. Profile classification shows that periodic nulling is not 
restricted to any particular profile type but seen across all classes; 
S$_d$ = 3; S$_t$ = 2; D = 7; $_c$T = 3; $_c$Q = 3; T = 6; M = 4; not classified
= 1.

\subsection{Periodic Amplitude Modulation}\label{sec:ampmod}

\begin{table*}
%\footnotesize
\begin{center}
\caption{Periodic Amplitude Modulation
\label{tabamp}}
{\begin{tabular}{ccccc}
\tableline
 PSR & $\dot{E}$ & $P_A$ & Profile &  \\
\tableline
  & (2$\times$10$^{32}$~erg/s) & ($P$) &  &  \\
\tableline
  B0450$-$18  & ~~6.850 & ~16$\pm$6~ & T & ... \\
   B0450+55   & ~11.850 & ~~9$\pm$1~ & T & ... \\
   B0823+26   & ~~2.260 & ~~6$\pm$1~ & S$_t$ & Mode \\
  B1055$-$52  & 150.500 & ~21$\pm$2~ & ... & ... \\
   B1541+09   & ~~0.204 & ~15$\pm$5~ & T & ... \\
  B1600$-$49  & ~~5.750 & ~50$\pm$26 & T & ... \\
  B1604$-$00  & ~~0.805 & ~34$\pm$13 & T & ... \\
  B1642$-$03  & ~~6.050 & ~13$\pm$4~ & S$_t$ & ... \\
  B1718$-$32  & ~~1.175 & ~23$\pm$9~ & T$_{1/2}$ & ... \\
  B1732$-$07  & ~~3.250 & ~20$\pm$8~ & T & ... \\
   B1737+13   & ~~0.555 & $\sim$90 & M & Drift \\
   B1737$-$39 & ~~2.835 & ~10$\pm$4~ & T & ... \\
  B1745$-$12  & ~~3.910 & ~~7$\pm$1~ & T & ... \\ 
  B1822$-$09  & ~22.800 & ~46$\pm$1~ & T$_{1/2}$ & Mode \\
  B1845$-$01  & ~~3.615 & ~20$\pm$8~ & $_c$T & ... \\
%   B1907+10   & ~22.850 & ~14$\pm$4~ & T$_{1/2}$ & ... \\
   B1917+00   & ~~0.735 & ~11$\pm$0.4 & T & ... \\
   B1929+10   & ~19.650 & ~12$\pm$1~ & S$_t$ & ... \\
   B1946+35   & ~~3.775 & ~50$\pm$10 & T & ... \\
% &  &  &  &  \\
\tableline
\end{tabular}}
\end{center}
\end{table*}

\noindent
We have identified eighteen pulsars with periodic amplitude modulation which 
are reported in Table \ref{tabamp}. The Table also describes different physical 
characteristics of each pulsar including $\dot{E}$, modulation periodicity 
($P_A$), profile classification, and presence of subpulse drifting and/or mode 
changing. These periodicities were measured in the works of \citet{wel06,
wel07,bas16} and this work, but they were not recognised as a separate 
phenomenon in previous studies. 

Some of the interesting physical behaviour of periodic amplitude 
modulation are summarized below. The pulsar B0823+26 has periodic amplitude
modulation during its B mode, while the Q mode shows periodic nulling 
\citep{bas19c}. The two modes in PSR B1822$-$09 have periodic amplitude 
modulation with different periodicities \citep{lat12,her17,yan19}. PSR B1737+13
shows drifting in its conal components \citep{for10}. Periodic amplitude 
modulation is seen primarily in profiles with central LOS traverse, where 
usually prominent core emission is present. The number of pulsars belonging to 
different profile types are estimated to be S$_t$ = 3; $_c$T = 1; 
T$_{1/2}$ = 2; T = 10; M = 1; not classified = 1.

\subsection{Unresolved Modulation}\label{sec:unmod}

\begin{table*}
\begin{center}
\caption{Periodic Modulation (Unresolved) 
\label{tabunmod}}
{\begin{tabular}{ccccc}
  PSR & $\dot{E}$ & $P_M$ & Profile &  \\
\tableline
  & (2$\times$10$^{32}$~erg/s) & ($P$) &  &  \\
\tableline
  B0621$-$04  & ~~0.146 & ~73$\pm$2~ & M & Drift \\
  B0756$-$15  & ~~1.005 & ~23$\pm$10 & T$_{1/2}$ & ... \\
  B1114$-$41  & ~~1.87~ & ~36$\pm$14 & S$_t$ & ... \\
  B1510$-$48  & ~~1.94~ & ~37$\pm$4~ & ... & Drift \\
 J1603$-$2531 & ~13.85~ & ~69$\pm$28 & S$_t$ & ... \\
  B1601$-$52  & ~~0.178 & ~21$\pm$7~ & D & ... \\
 J1650$-$1654 & ~~0.118 & ~64$\pm$45 & D & Drift \\
  B1700$-$18  & ~~0.655 & ~43$\pm$15 & ... & Drift \\
  B1702$-$19  & ~30.55~ & ~11$\pm$0.4 & T$_{1/2}$ \\
  B1730$-$37  & ~77.0~~ & ~71$\pm$20 & ... \\
   B1753+52   & ~~0.023 & ~11$\pm$5~  & $_c$Q & ... \\
  B1758$-$03  & ~~0.835 & ~81$\pm$42 & T$_{1/2}$ & ... \\
   B1839+56   & ~~0.066 & ~40$\pm$20 & ... & ... \\
   B1839+09   & ~~3.88~ & ~37$\pm$3~ & S$_t$ & ... \\
   B1859+01   & ~19.45~ & ~14$\pm$1~ & T & ... \\
   B1907+10   & ~22.85~ & ~13$\pm$2~ & S$_t$ & ... \\
  B1911$-$04  & ~~1.425 & ~15$\pm$5~ & S$_t$ & ... \\
   B1952+29   & ~~0.004 & ~26$\pm$12 & M/T? & ... \\
   B1953+50   & ~~1.94~ & ~25$\pm$7~ & S$_t$ & ... \\
   B2011+38   & 143.0~~ & ~30$\pm$15 & ... & ... \\
   B2021+51   & ~~4.08~ & ~22$\pm$15 & D & ... \\
   B2106+44   & ~~0.239 & ~19$\pm$7~ & ... & ... \\
   B2255+58   & ~22.75~ & ~10$\pm$1~ & S$_t$ & ... \\
 J2346$-$0609 & ~~0.163 & 100$\pm$30 & D & ... \\
%  &  &  &  &  \\
\tableline
\end{tabular}}
\end{center}
\end{table*}

\noindent
Twenty four pulsars show clear periodicities in their single pulse sequence, 
but their detection sensitivities are insufficient to determine the exact 
nature of modulation. Table \ref{tabunmod} shows basic physical properties of 
these pulsars including $\dot{E}$, modulation periodicity ($P_M$), profile 
classification, and presence of subpulse drifting. Mode changing is not seen in
any of these sources. Four pulsars in this group, PSR B0621$-$04, B1510$-$48, 
J1650$-$1654 and B1700$-$18 also has subpulse drifting. Their profiles show a 
wide variety with S$_t$ = 7; D = 4; $_c$Q = 1; T$_{1/2}$ = 3; T = 1 or 2; M = 1
or 2; not classified = 6. 

There are another additional twenty to thirty pulsars where periodic behaviour 
was also reported in earlier works \citep{wel06,wel07}. In eight of them, PSR 
B0329+54, B0402+61, B0919+06, B0950+08, B1112+50, B1612+07, B1937$-$26 and 
B2217+47 we did not find any clear modulation features in fluctuation spectra. 
In the remaining cases more sensitive observations are required to validate 
presence of periodic behaviour.

\section{Periodic Modulation : Distinct physical phenomenon}\label{sec:disc}

\begin{table*}
\footnotesize
\caption{Comparing Periodic Behaviour in Pulsar Radio Emission}
\centering
\begin{tabular}{cll}
\hline
  &  &  \\
   & ~~~~~Subpulse Drifting & ~~~~~Periodic Modulations \\
\hline
  &  &  \\
 Profile Component & (i) Only seen in conal components. & (i) Simultaneously seen in central core and cones. \\
  & (ii) Different phase variations for different components. & (ii) longitude stationary, similar across all components. \\
  &  &  \\
 Spindown Energy & (i) Seen in pulsars with $\dot{E} <$ 5$\times$10$^{32}$ erg~s$^{-1}$. & (i) Seen in pulsars with wide $\dot{E}$ distribution. \\
  & (ii) $P_3$ weakly anti-correlated with $\dot{E}$. & (ii) $P_M \sim$ 10-200$P$, no dependence of $P_M$ on $\dot{E}$. \\
  &  &  \\
 Origin & (i) Localised in Inner Acceleration Region. & (i) Seen in both poles of pulsars with interpulse emission. \\
  & (ii) Associated with sparking process of plasma generation. & (ii) Largescale variations affecting pulsar magnetosphere. \\
  &  &  \\
\hline
\end{tabular}
\label{tabperbeh}
\end{table*}

\begin{figure*}
\begin{center}
\includegraphics[scale=1.2,angle=0.]{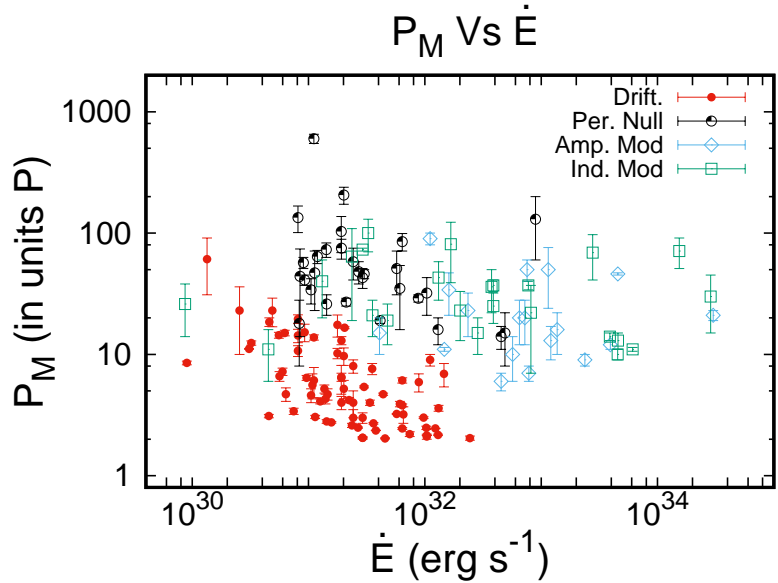}
\end{center}
\caption{The figure shows modulation periodicities ($P_M$) of pulsars 
represented along with their spin-down energy loss ($\dot{E}$). Periodic 
nulling (black open circles), periodic amplitude modulation (blue rhombus) and 
indeterminate periodic modulations (green square) are shown in the figure, 
along with subpulse drifting (red filled circles) for comparison. The three 
periodic modulations overlap along $\dot{E}$ axis, underlying their common 
physical origin. Subpulse drifting on the other hand has different physical 
behaviour. It is seen for a limited region along $\dot{E}$ axis 
($\leq$2$\times$10$^{32}$ erg~s$^{-1}$) and is weakly anti-correlated with 
$\dot{E}$.} 
\label{fig_edotpm}
\end{figure*}

\begin{figure*}
\begin{center}
\begin{tabular}{@{}cr@{}}
{\mbox{\includegraphics[scale=0.6,angle=0.]{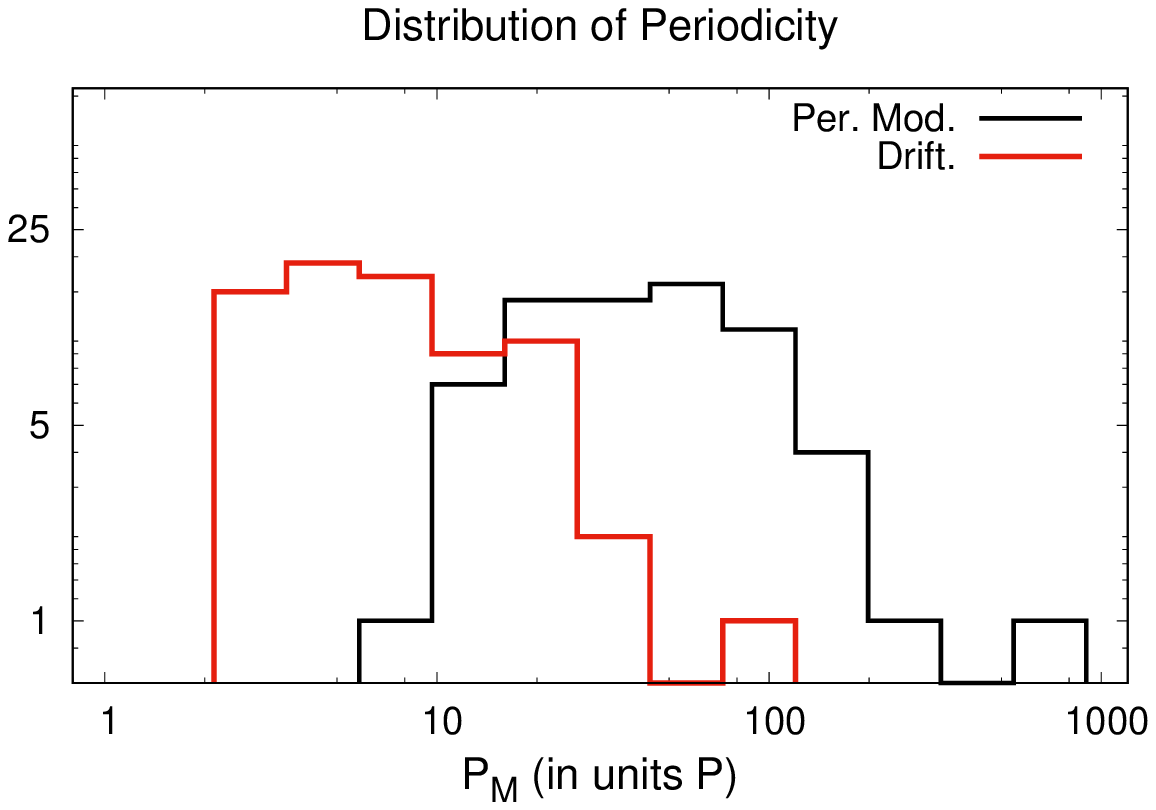}}} &
%\hspace{20px}
{\mbox{\includegraphics[scale=0.6,angle=0.]{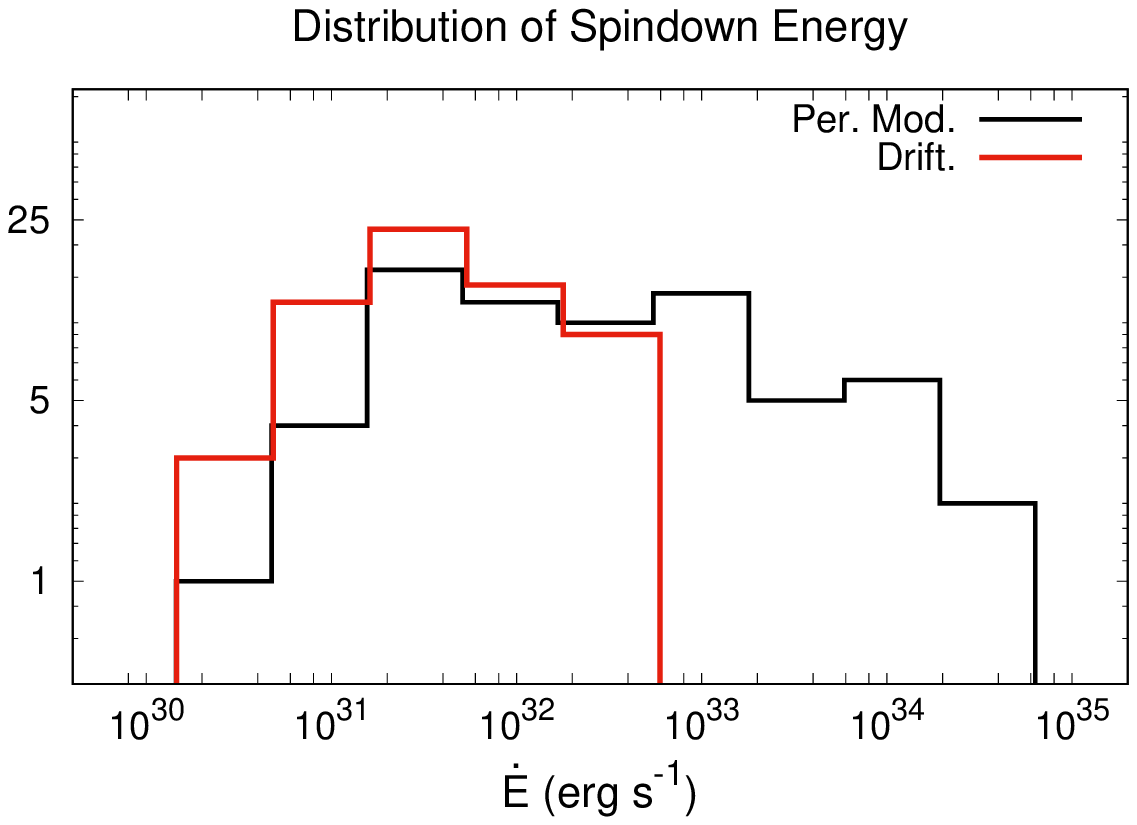}}} \\
%\hspace{20px}
\end{tabular}
\caption{The figure shows distributions of modulation periodicity (left panel) 
and spin-down energy loss (right panel) for subpulse drifting and other 
periodic modulations. A clear distinction is seen between the two populations 
in both plots. Subpulse drifting is seen in pulsars with $\dot{E} <$ 
2$\times$10$^{32}$ erg~s$^{-1}$, while periodic modulations are seen along a 
wider $\dot{E}$ range. Subpulse drifting has usually lower periodicities with 
only a small fraction exceeding 20$P$. On the other hand modulation 
periodicities are usually longer with typical values between 10 and 200$P$.}
\label{fig_hist}
\end{center}
\end{figure*}

\noindent
Major differences between periodic modulations and subpulse drifting are 
summarized in Table \ref{tabperbeh}. We have estimated dependence of 
modulation periodicity ($P_M$) with $\dot{E}$ which is shown in figure 
\ref{fig_edotpm}. The figure represents $P_M$ of three groups, periodic nulling
(black open circles), periodic amplitude modulation (blue rhombus) and 
indeterminate periodic modulation (green square), defined in the previous 
section, along $\dot{E}$ axis. Drifting periodicities (red filled circles) are 
also shown for comparisons. The figure highlights two primary features, a clear
distinction between subpulse drifting and other periodic modulations, and 
overlapping behaviour between the three groups of periodic modulations. 

Physical differences between subpulse drifting and other periodic 
modulations are also seen in their distributions of periodicity and $\dot{E}$.
Periodic modulations (figure \ref{fig_hist}, left panel) have typical 
periodicities between 10-200$P$, the distribution peaking around 50$P$, and 
there are very few exceptions, around five cases, outside this range. On the 
other hand subpulse drifting has periodicity below 20$P$ in most cases, with 
the distribution peaking around 5$P$. Longer periodic drifting is usually seen 
for low $\dot{E}$ pulsars. One of the primary limitations of measuring periodic 
phenomena in pulsars, particularly short periodicities, is the aliasing effect 
around 2$P$. This is reflected in the sharp cutoff at lower part of drifting
periodicity distribution, possibly skewing it towards longer periodicities. 
$\dot{E}$ distributions (figure \ref{fig_hist}, right panel) of the two 
populations also show very different behaviours. Subpulse drifting peaks around 
3$\times$10$^{31}$ erg~s$^{-1}$ with an upper cutoff around 2$\times$10$^{32}$ 
erg~s$^{-1}$. In contrast periodic modulations are seen over a much wider 
$\dot{E}$ range, with a plateau between 10$^{31}$ and 10$^{34}$ erg~s$^{-1}$. 

Both periodic nulling and periodic amplitude modulation share a number of 
identical features. This include lack of any dependence of their periodicities 
on $\dot{E}$, presence across the entire profile including core emission, and 
longitude stationary behaviour in all components. These suggest that their 
underlying physical processes are also likely to be similar. We propose that 
they represent a class of newly emergent emission behaviour in pulsars with 
distinct physical mechanism compared to subpulse drifting.

\begin{figure*}
\begin{center}
\includegraphics[scale=0.75,angle=0.]{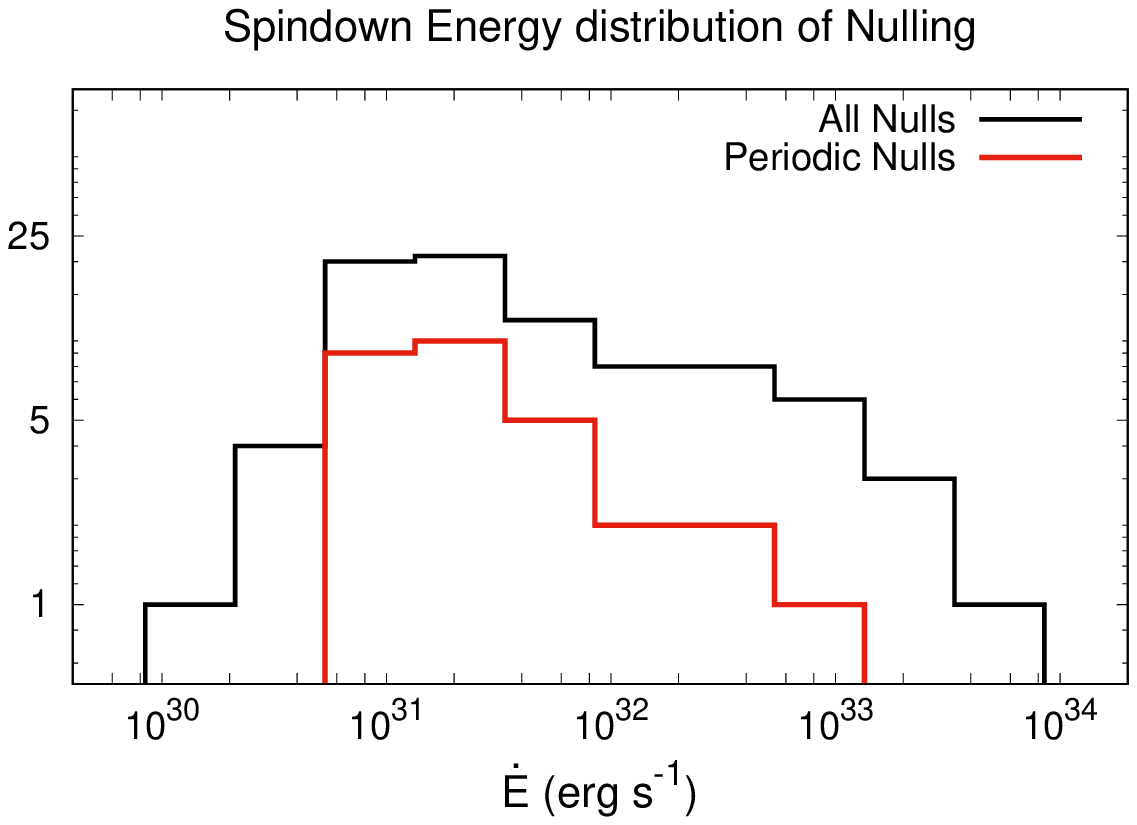}
\caption{The figure shows $\dot{E}$ distribution of pulsars showing the 
presence of nulling in their single pulse sequences. The black histograms shows
the distribution for the entire nulling population, while the red plot 
corresponds to the subgroup showing periodic nulling.}
\label{fig_edotnull}
\end{center}
\end{figure*}

\begin{deluxetable}{ccccccccccc}
\tabletypesize{\footnotesize}
\tablecaption{Nulling in Pulsars \label{taballnull}}
\tablehead{
 \colhead{PSR} & \colhead{$\dot{E}$} & \colhead{NF} & \colhead{Profile} & \colhead{REF} & \colhead{} & \colhead{PSR} & \colhead{$\dot{E}$} & \colhead{NF} & \colhead{Profile} & \colhead{REF} \\
 \colhead{} & \colhead{(2$\times$10$^{32}$~erg/s)} & \colhead{(\%)} & \colhead{} & \colhead{} & \colhead{} & \colhead{} & \colhead{(2$\times$10$^{32}$~erg/s)} & \colhead{(\%)} & \colhead{} & \colhead{}
}
\startdata
  B0031$-$07  & 0.096 & 37.7 & S$_d$ & [1] &  &  B1738$-$08  & 0.053 & 15.8$\pm$1.4 & $_c$Q & [5] \\
   B0045+33   & 0.258 & 21 & ... & [2] &  &  B1742$-$30  & 42.45 & 32.5$\pm$1.2 & M & [5] \\ 
   B0138+59   & 0.042 & 7.6$\pm$1.0 & M & [3] &  &  B1747$-$46  & 0.625 & 2.4$\pm$0.5 & S$_t$ & [5] \\ 
   B0301+19   & 0.096 & 7.4$\pm$0.7 & D & [4] &  &  J1752+2359  & 1.855 & 75 & ... & [15] \\ 
   B0320+39   & 0.005 & 0.7$\pm$0.2 & $_c$T & [3] &  &  B1749$-$28  & 0.009 & 1.0$\pm$0.4 & T & [5] \\ 
   B0525+21   & 0.151 & 28$\pm$2 & D & [4] &  &  B1758$-$03  & 0.835 & 26.9$\pm$1.6 & T & [5] \\ 
  B0628$-$28  & 0.73 & 13.6$\pm$1.9 & S$_d$ & [5] &  & J1808$-$0813 & 0.364 & 10.8$\pm$1.1 & S$_d$ & [5] \\ 
   B0751+32   & 0.071 & 38$\pm$6 & D & [4] &  &  B1809$-$173 & 2.15 & 5.8$\pm$4 & S$_t$ & [10] \\ 
   B0809+74   & 0.015 & 1.2$\pm$0.4 & S$_d$ & [3] &  &  B1813$-$36  & 6.95 & 16.7$\pm$0.7 & T & [5]\\ 
  B0818$-$13  & 0.219 & 0.9$\pm$0.1 & S$_d$ & [3] &  &  J1819+1305  & 0.060 & 41$\pm$6 & $_c$Q & [4] \\ 
   B0820+02   & 0.032 & 0.5$\pm$0.2 & S$_d$ & [3] &  & J1820$-$0509 & 4.795 & 67$\pm$3 & ... & [10] \\ 
   B0823+26   & 2.26 & 3.85$\pm$0.05 & S$_t$ & [6] &  &  B1819$-$22  & 0.041 & 5.5$\pm$0.2 & D & [16] \\ 
   B0834+06   & 0.65 & 4.4$\pm$0.4 & D & [5] &  & J1831$-$1223 & 0.046 & 4$\pm$1 & ... & [10]\\ 
  B0906$-$17  & 2.04 & 12.4$\pm$1.1 & T & [3] &  & J1840$-$0840 & 0.031 & 50$\pm$6 & D & [17] \\ 
 J0930$-$2301 & 0.113 & $>$30 & ... & [7] &  &  B1845$-$19  & 0.058 & 27.2$\pm$1.7 & T & [3] \\ 
  B0932$-$52  & 0.305 & 1.9$\pm$0.1 & S$_d$ & [3] &  &   B1848+12   & 1.30 & 51$\pm$2 & S$_t$ & [4] \\ 
  B0940$-$55  & 15.4 & $<$12.5 & S$_t$ & [8] &  &  J1857$-$1027  & 0.042 & $>$ 30 & T & [3] \\ 
   B0940+16   & 0.014 & 8$\pm$3 & T & [9] &  &  B1857$-$26  & 0.176 & 4.3$\pm$0.5 & M & [3] \\ 
  B0942$-$13  & 0.048 & 14.4$\pm$0.9 & T$_{1/2}$ & [5] &  & J1901$-$0906 & 0.057 & 3.8$\pm$0.7 & D & [5] \\ 
 J1049$-$5833 & 0.082 & 47$\pm$3 & ... & [10] &  &   B1905+39   & 0.057 & 10.0$\pm$0.5 & M & [3] \\ 
   B1112+50   & 0.109 & 34.8$\pm$1.4 & T$_{1/2}$ & [3] &  &   B1907+03   & 0.070 & 4$\pm$0.2 & T/M & [9]\\ 
  B1114$-$41  & 1.87 & 3.3$\pm$0.5 & S$_t$ & [5] &  &  J1920+1040  & 0.118 & 50$\pm$4 & ... & [10] \\ 
   B1133+16   & 0.440 & 13$\pm$2 & D & [5] &  &   B1918+19   & 0.320 & 2.0$\pm$0.3 & $_c$T & [5] \\ 
   B1237+25   & 0.072 & 2.5$\pm$0.1 & M & [5] &  & J1926$-$1314 & 0.063 & 74$\pm$2 & ... & [18] \\ 
  B1322$-$66  & 6.55 & 9.1$\pm$3 & ... & [10] &  &   B1942+17   & 0.018 & 60 & D & [19] \\ 
  B1325$-$49  & 0.037 & 4.2$\pm$0.3 & M & [5] &  &  B1942$-$00  & 0.093 & 21$\pm$1 & D & [9] \\ 
  B1358$-$63  & 5.50 & 1.6$\pm$1 & ... & [10] &  &   B1944+17   & 0.056 & 32.0$\pm$1.6 & $_c$T & [5] \\ 
 J1502$-$5653 & 2.35 & 93$\pm$4 & ... & [10] &  &  B2003$-$08  & 0.046 & 28.6$\pm$3.4 & M & [20] \\ 
   B1508+55   & 2.44 & 5.2$\pm$0.4 & T & [3] &  &   B2034+19   & 0.045 & 44$\pm$4 & T & [4] \\ 
 J1525$-$5417 & 3.085 & 16$\pm$5 & ... & [10] &  &  B2045$-$16  & 0.287 & 8.8$\pm$0.6 & T & [5] \\ 
  B1524$-$39  & 0.267 & 5.1$\pm$1.3 & D & [5] &  &   B2110+27   & 0.298 & 30 & S$_d$ & [2] \\ 
   B1530+27   & 0.108 & 6$\pm$2 & S$_d$ & [9] &  &   B2111+46   & 0.135 & 8.7$\pm$0.5 & T & [3] \\ 
  B1604$-$00  & 0.805 & 0.15$\pm$0.07 & T & [3] &  &   B2122+13   & 0.454 & 22 & D & [2] \\ 
  J1649+2533  & 0.106 & 25$\pm$5 & ... & [4] &  &   B2154+40   & 0.191 & 7.5$\pm$2.5 & $_c$T & [21] \\ 
  B1658$-$37  & 0.149 & 14$\pm$2 & ... & [10] &  &  J2253+1516  & 0.026 & 49 & D & [2] \\ 
 J1702$-$4428 & 0.068 & 26$\pm$3 & ... & [10] &  &   B2303+30   & 0.146 & 5.3$\pm$0.5 & S$_d$ & [5] \\ 
  B1700$-$32  & 0.073 & 0.9$\pm$0.2 & T & [5] &  &   B2310+42   & 0.52 & 4.5$\pm$0.6 & $_c$T & [3] \\
  B1706$-$16  & 4.47 & 31$\pm$2 & S$_t$ & [11] &  &   B2315+21   & 0.069 & 3$\pm$0.5 & $_c$T & [9] \\ 
  B1713$-$40  & 1.04 & 80$\pm$15 & S$_t$ & [12] &  &   B2319+60   & 1.21 & 15.2$\pm$1.1 & $_c$Q & [3] \\ 
 J1725$-$4043 & 0.175 & $<$70 & ... & [13] &  &  B2327$-$20  & 0.206 & 11.2$\pm$1.0 & T & [3] \\ 
 J1727$-$2739 & 0.101 & 68.2$\pm$1.1 & D & [14] &  & J2346$-$0609 & 0.163 & 35.6$\pm$2.2 & D & [5] \\ 
 J1738$-$2330 & 0.218 & $>$ 69 & ... & [13] &  &  &  &  &  &  \\ 
%   &  &  &  &  &  &  &  &  &  &  &  &  &  \\
\hline
\enddata
\tablecomments{1-\cite{hug70}; 2-\cite{red09}; 3-This paper; 4-\cite{her09}; 5-\cite{bas17}; 6-\cite{bas19c}; 7-\cite{kaw18}; 8-\cite{big92}; 9-\cite{wei86}; 10-\cite{wan07}; 11-\cite{nai18}; 12-\cite{ker14}; 13-\cite{gaj12}; 14-\cite{wen16}; 15-\cite{lew04}; 16-\cite{bas18b}; 17-\cite{gaj17}; 18-\cite{ros13}; 19-\cite{lor02}; 20-\cite{bas19b}; 21-\cite{rit76}}
\end{deluxetable}

We have also compared physical properties of pulsars with periodic nulling 
and general nulling population. A detailed literature survey of nulling is 
reported in Table \ref{taballnull}, which lists pulsar names, $\dot{E}$, 
nulling fraction (NF), reference for NF, and profile type. Only traditional 
nulling pulsars are considered for comparisons and more extreme examples like 
intermittent pulsars are not included. We have also excluded sources where only
the upper limits of NF are reported due to lower sensitivity detections. This 
left eighty three pulsars where nulling is unambiguously observed. 

No clear trends emerge for nulling, either in NF or their profile types. 
$\dot{E}$ distribution of all nulling pulsars is shown in figure 
\ref{fig_edotnull} along with the distribution for periodic nulling. There is 
no sharp cutoff seen in the $\dot{E}$ distributions, unlike subpulse drifting. 
However, nulling becomes less prevalent in high $\dot{E}$ pulsars, with no 
nulling seen above 10$^{34}$ erg~s$^{-1}$. The figure also shows that 
distribution of periodic nulling pulsars cannot be distinguished from the 
non-periodic case. 

It is likely that periodic nulling is more prevalent than the twenty nine cases 
reported here, and more sensitive single pulse studies in future will increase 
this number. For example, in three pulsars B1114$-$41, B1758$-$03 and 
J2346$-$0609 presence of periodic behaviour is seen in fluctuation spectra, 
but detection sensitivity of single pulses were not sufficient to ascertain 
periodic nulls. However, there are also several examples where sensitive single
pulse studies could not detect presence of any periodicity associated with 
nulling. We did not detect periodicity in seven out of possible seventeen 
pulsars studied here. \citet[][see Table 2]{bas17} also reported eight pulsars 
without any periodic nulling, despite sensitive single pulse detections. 
Currently, periodic nulling is seen in around 35\% of nulling pulsars.

\section{Summary and Conclusion}\label{sec:sum}
\noindent
In this paper we have carried out a detailed study of periodic amplitude 
modulation and periodic nulling seen in single pulse sequence of many normal 
period ($P >$ 0.1 s) pulsars. A complete list of all possible sources was 
compiled from the literature as well as newer observations using GMRT. We have 
carried out detailed nulling and periodic modulation analysis in sixty two 
pulsars. Our studies found nulling in twenty pulsars and periodic modulations 
in twenty four cases. We detected periodic nulling in ten pulsars, eight of 
which were new detections, expanding this population to twenty nine pulsars. We
have identified eighteen pulsars to exhibit periodic amplitude modulation, not 
associated with nulling, which is the first such categorization of this 
behaviour. Additionally, we have also identified twenty four pulsars with 
periodic modulation, where detection sensitivities were insufficient to 
distinguish between the two phenomena. 

Most periodic modulations have been considered a form of subpulse drifting in a
majority of past studies. We have compared periodic behaviour associated with 
subpulse drifting and other periodic modulations and found them to exhibit 
different physical properties. The different periodic modulations on the other
hand exhibit similar characteristics, like overlapping periodicities along 
$\dot{E}$ and longitude stationary behaviour across all profile components, 
which suggest similar physical mechanisms. We have also found periodic nulling 
to be a significant subset of the nulling phenomenon in general with no clear 
distinctions in physical properties.

\section*{Acknowledgments}
We thank the referee for the comments which helped to improve the paper. DM 
acknowledges funding from the grant ``Indo-French Centre for the Promotion of 
Advanced Research - CEFIPRA". We thank the staff of the GMRT who have made 
these observations possible. The GMRT is run by the National Centre for Radio 
Astrophysics of the Tata Institute of Fundamental Research.

\appendix

\section{Pulse energy and off-pulse distributions for Nulling}\label{sec:app1}
\input{appendix1.tex}

\section{Null length and Burst length hisograms}\label{sec:app2}
\input{appendix2.tex}

\section{Time evolution of fluctuation spectra and Nulling FFT}\label{sec:app3}
\input{appendix3.tex}

\end{document}

%% file: appendix1.tex
\clearpage

%1st set of plots
\begin{figure*}
\begin{center}
\begin{tabular}{@{}cr@{}}
\mbox{\includegraphics[angle=0,scale=0.57]{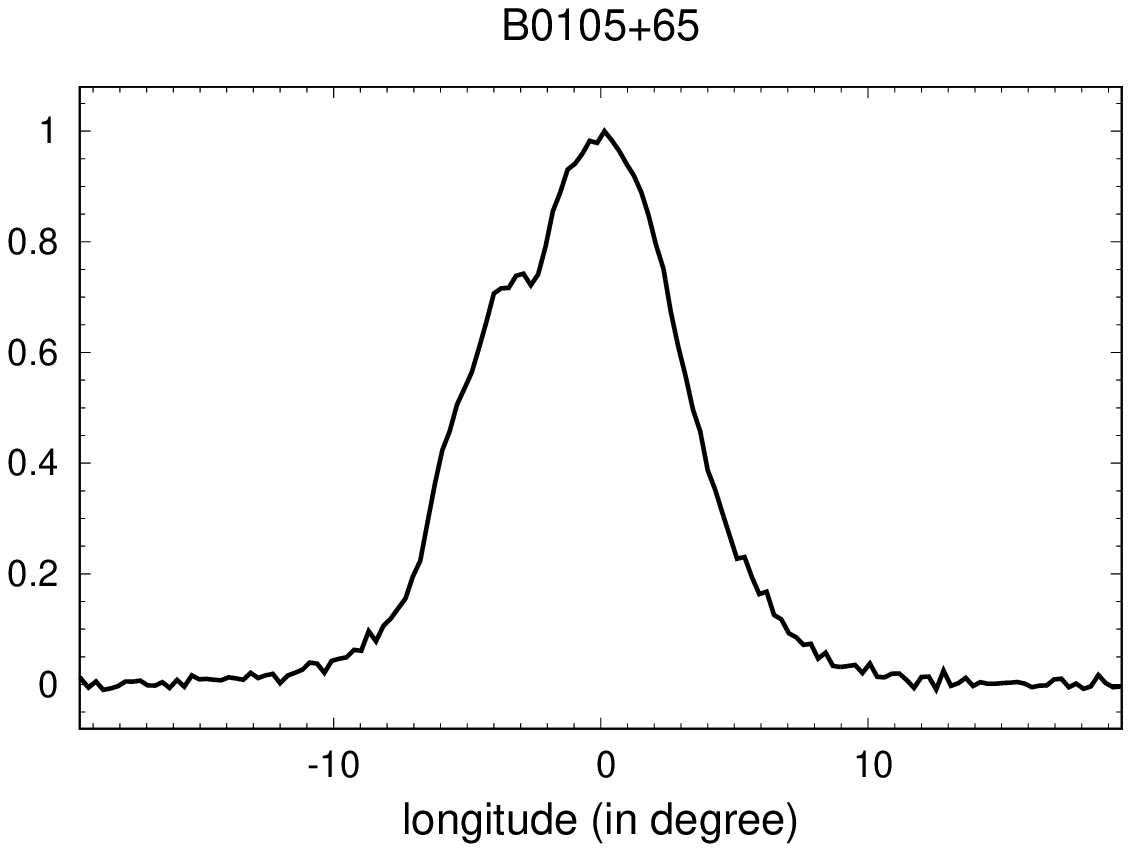}} &
\mbox{\includegraphics[angle=0,scale=0.57]{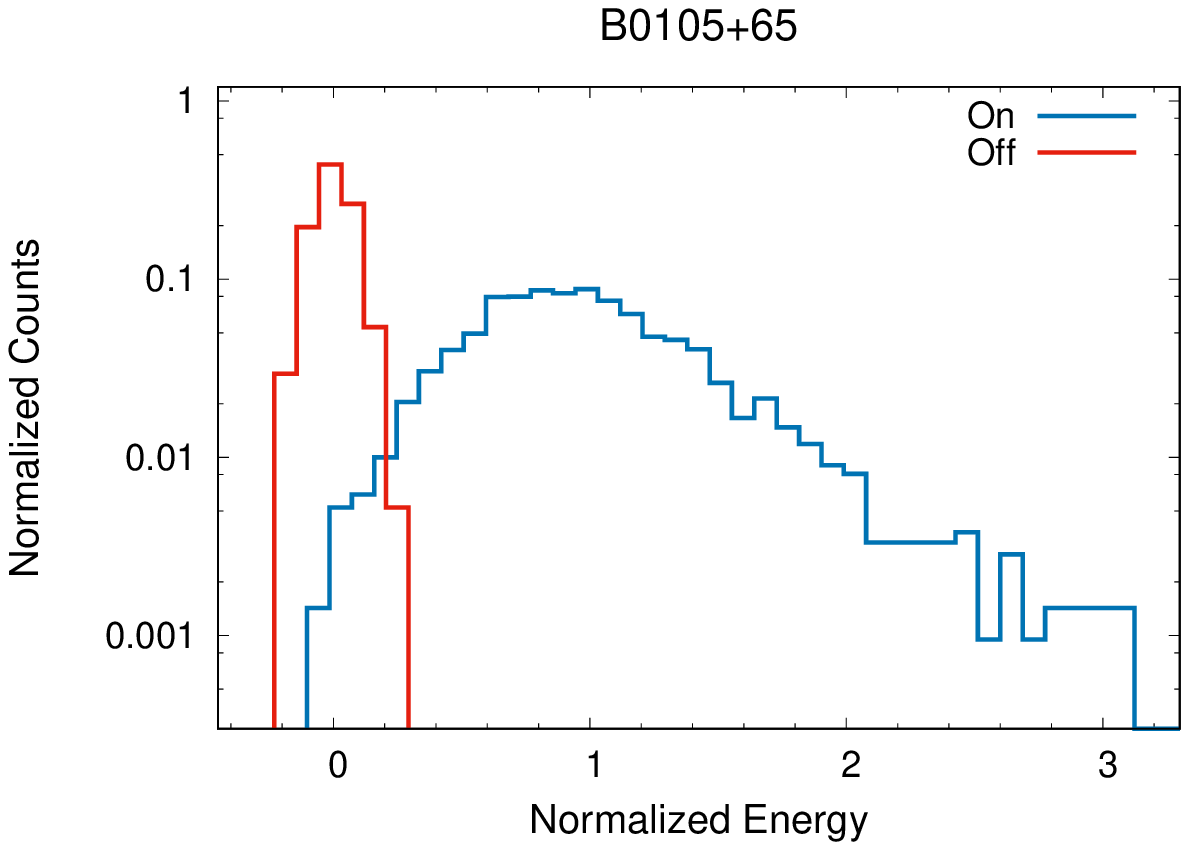}} \\
\mbox{\includegraphics[angle=0,scale=0.57]{J0141+6009_prof.eps}} &
\mbox{\includegraphics[angle=0,scale=0.57]{J0141+6009_energydist.eps}} \\
\mbox{\includegraphics[angle=0,scale=0.57]{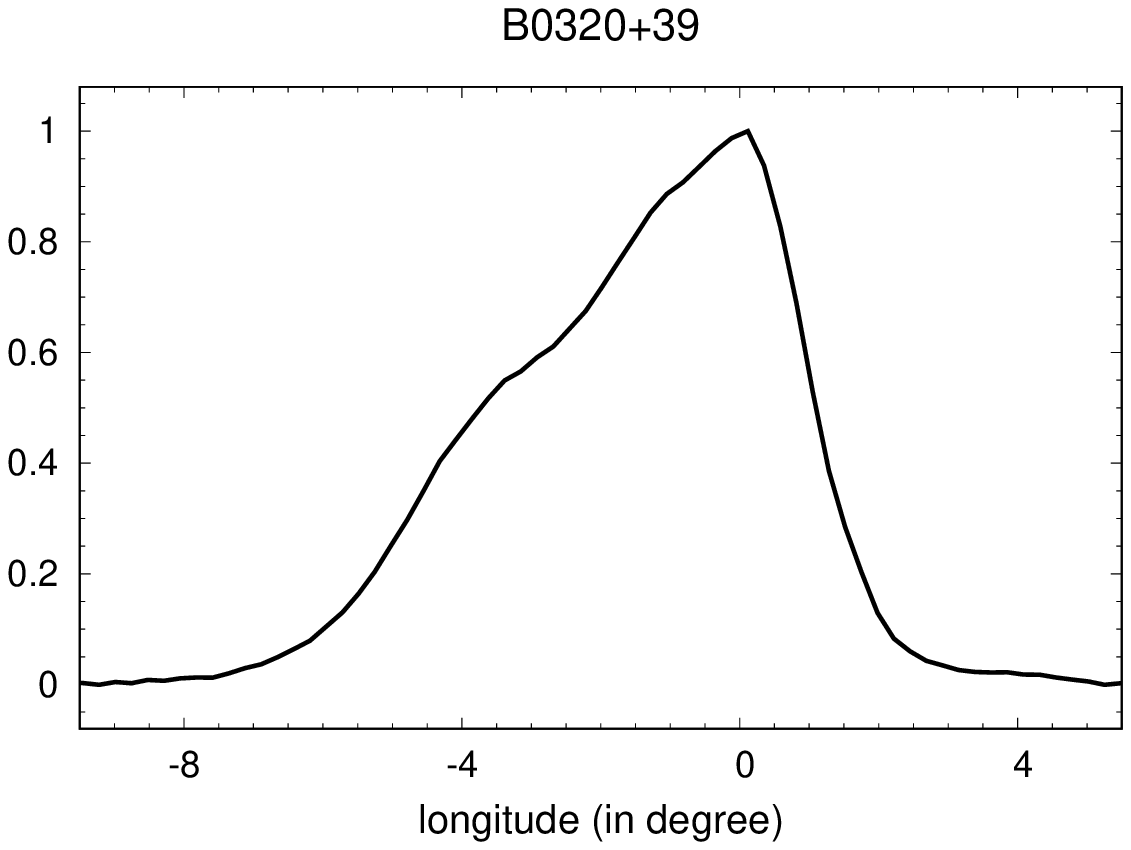}} &
\mbox{\includegraphics[angle=0,scale=0.57]{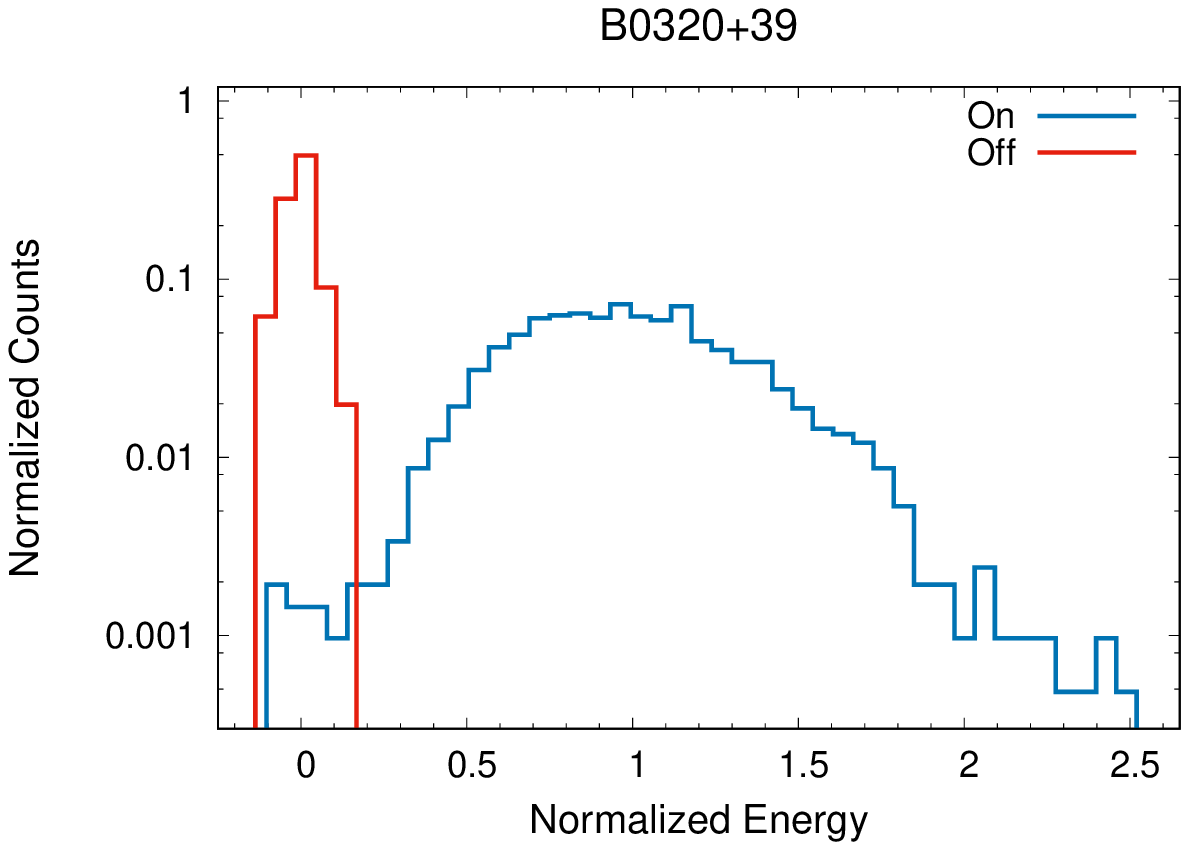}} \\
\mbox{\includegraphics[angle=0,scale=0.57]{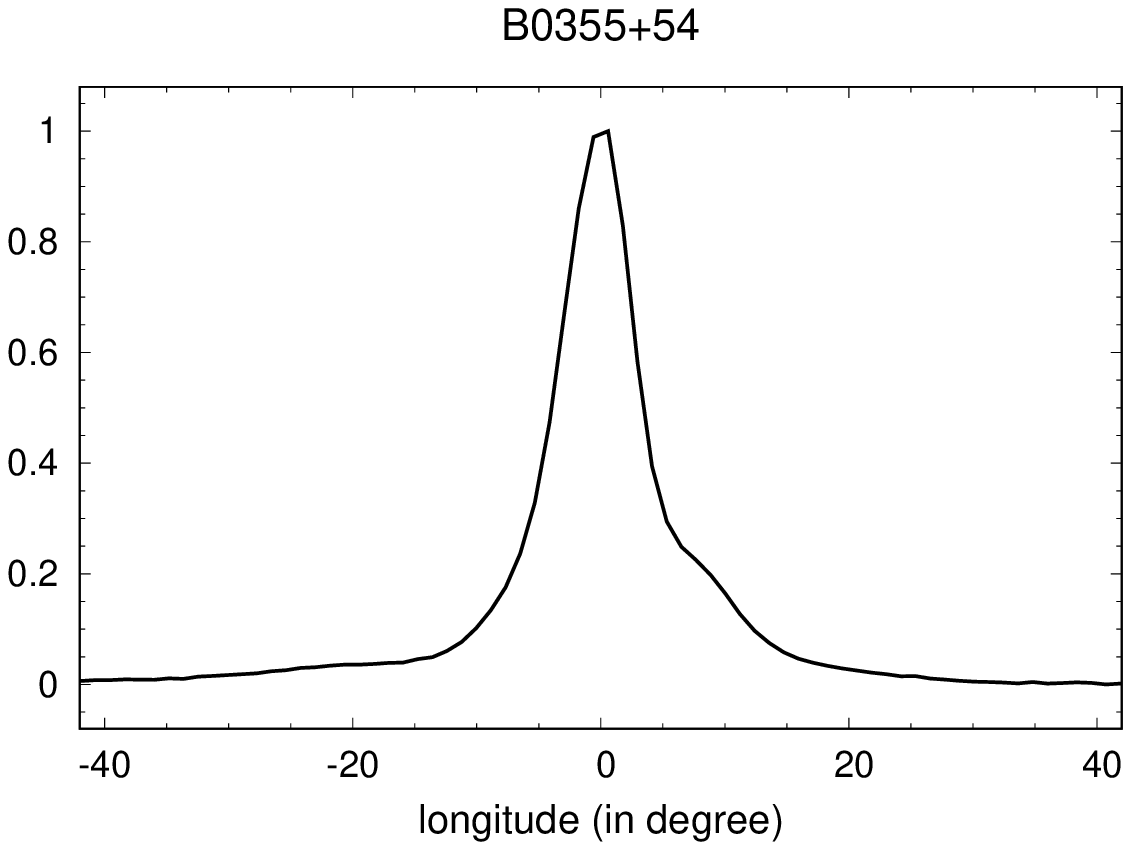}} &
\mbox{\includegraphics[angle=0,scale=0.57]{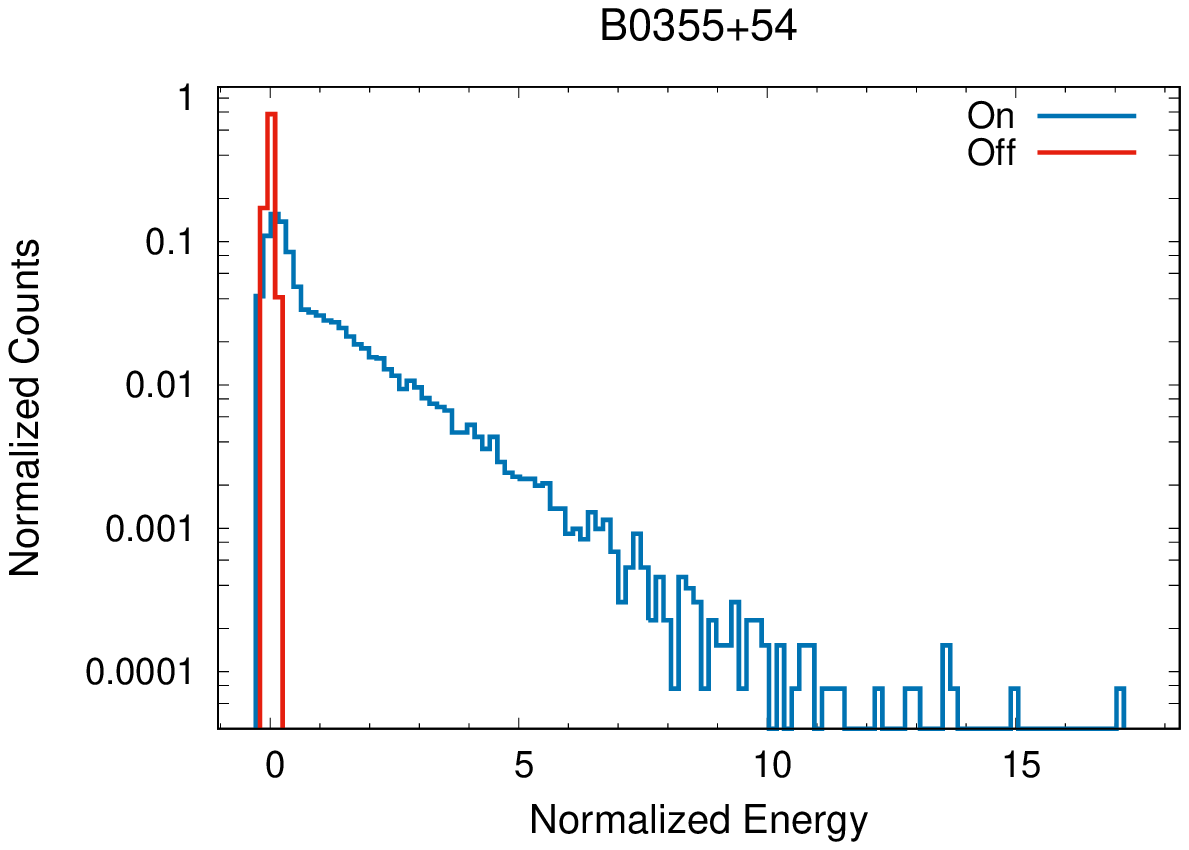}} \\
\end{tabular}
\caption{The pulsar profile and On and Off-pulse energy distributions of the single pulse emission.}
\end{center}
\end{figure*}

\clearpage

%2nd set of plots
\begin{figure*}
\begin{center}
\begin{tabular}{@{}cr@{}}
\mbox{\includegraphics[angle=0,scale=0.57]{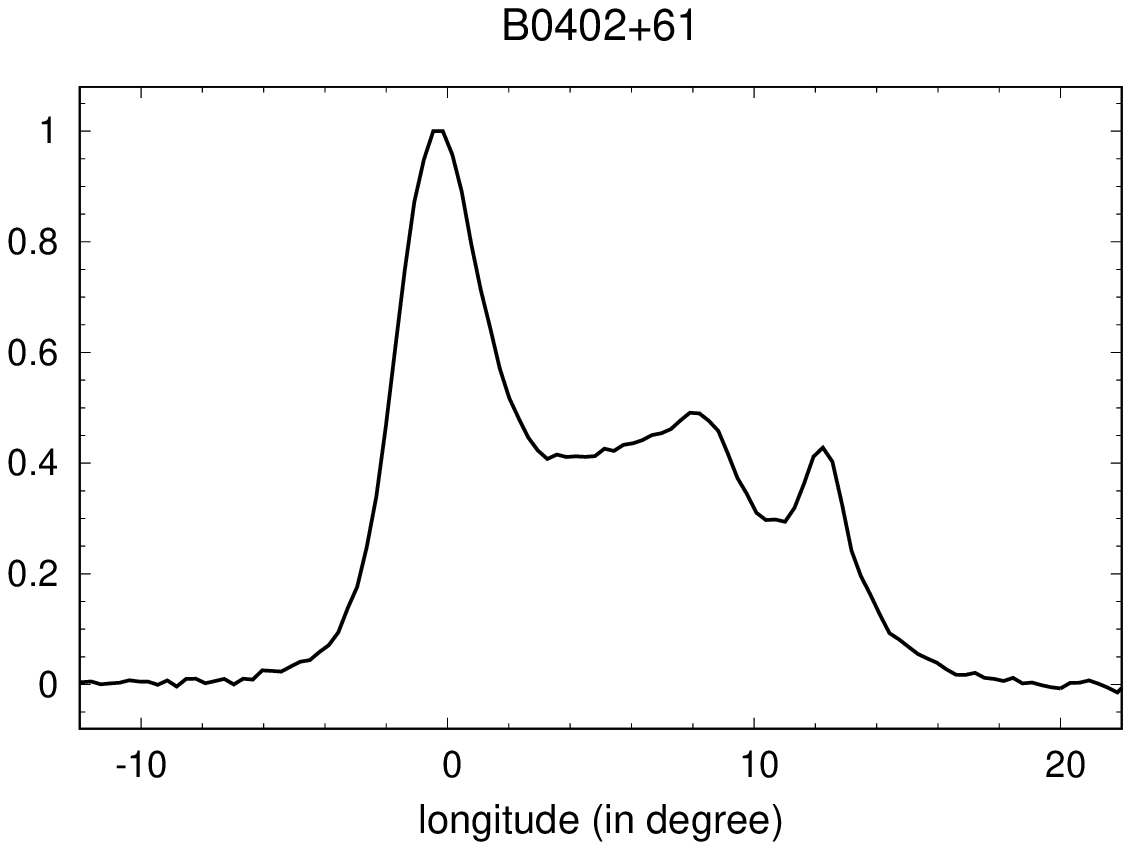}} &
\mbox{\includegraphics[angle=0,scale=0.57]{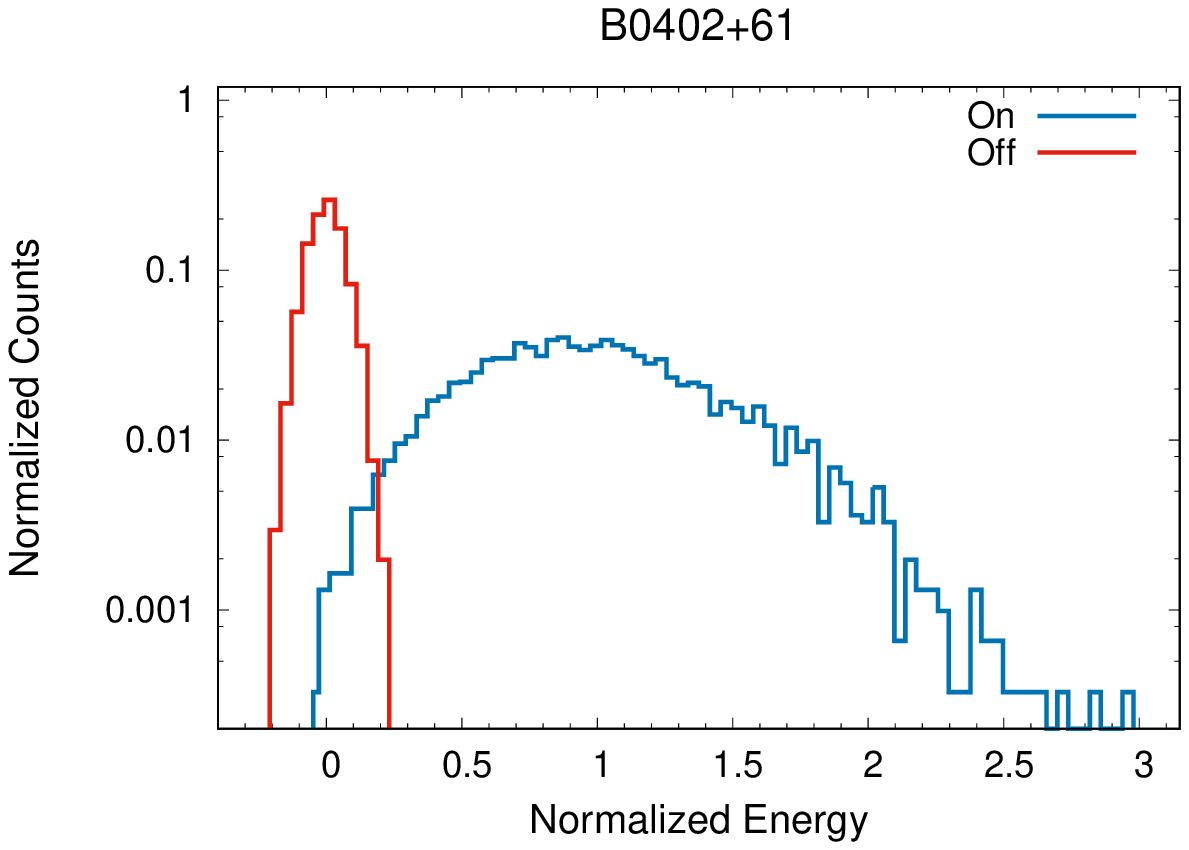}} \\
\mbox{\includegraphics[angle=0,scale=0.57]{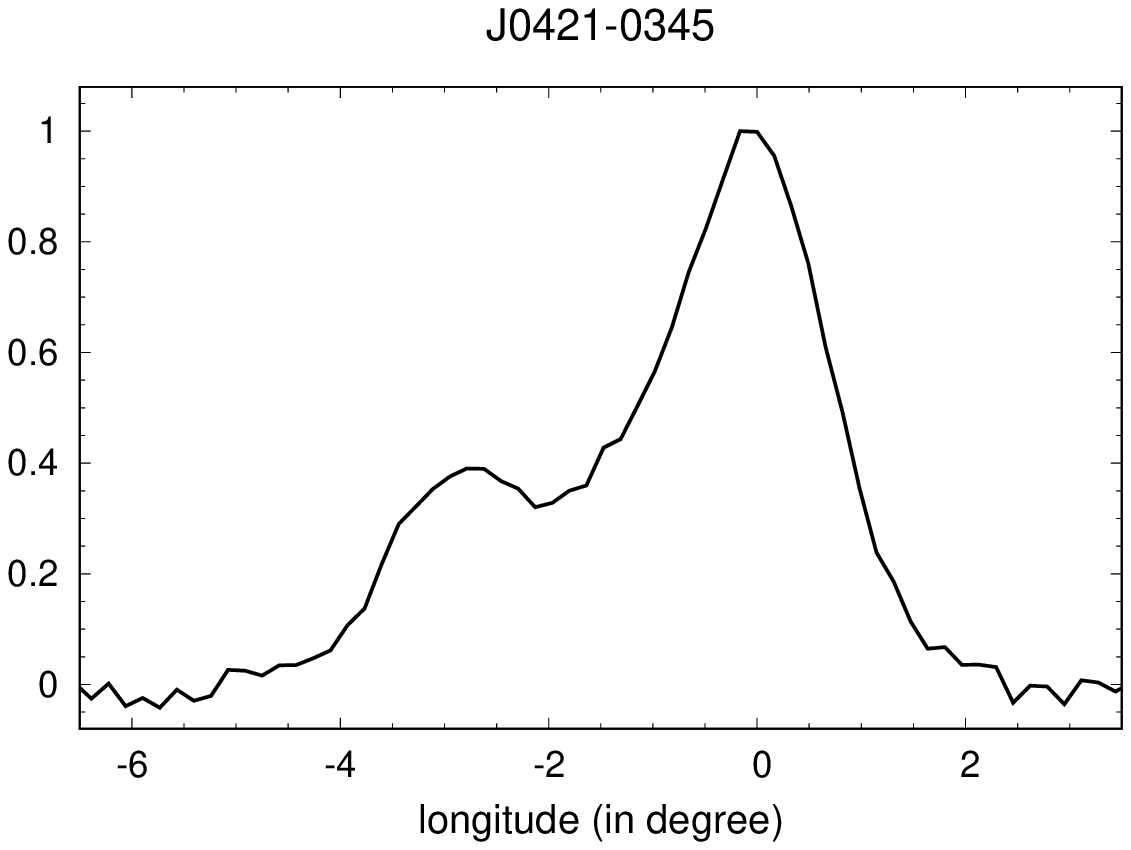}} &
\mbox{\includegraphics[angle=0,scale=0.57]{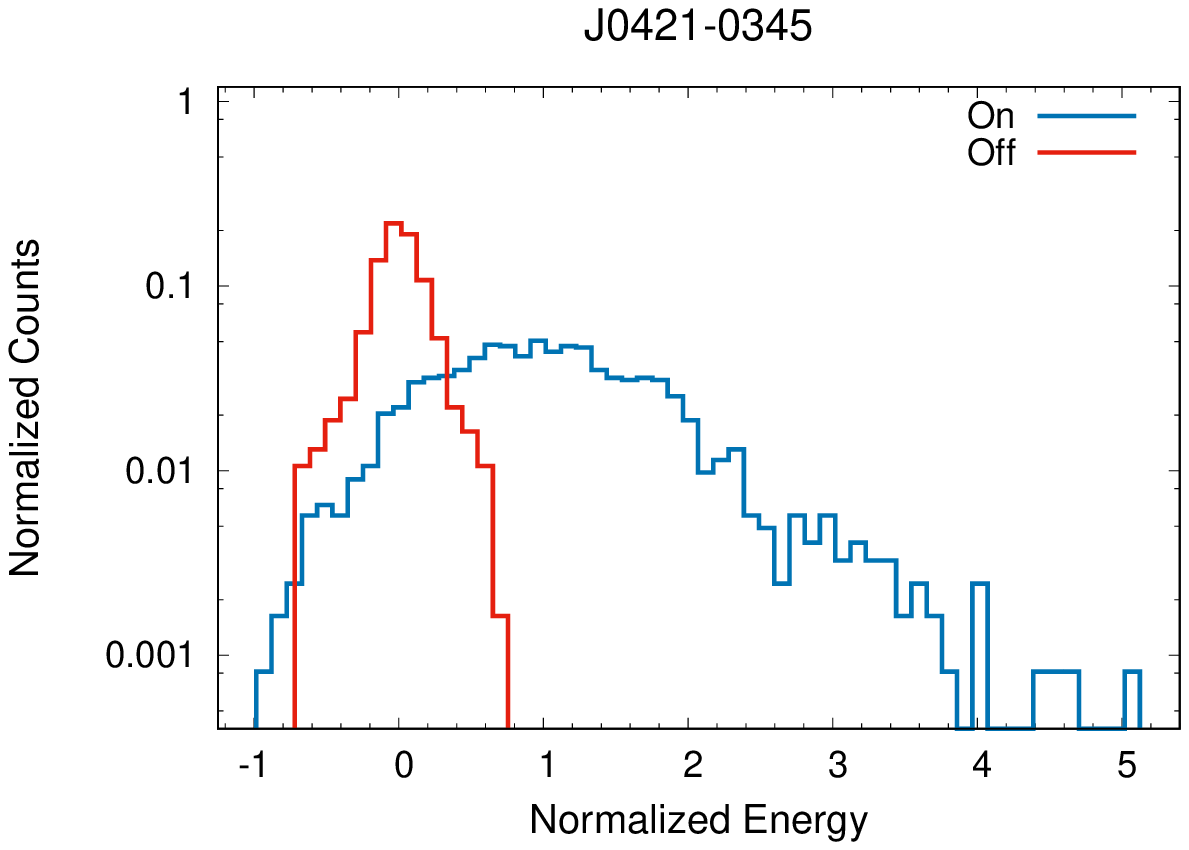}} \\
\mbox{\includegraphics[angle=0,scale=0.57]{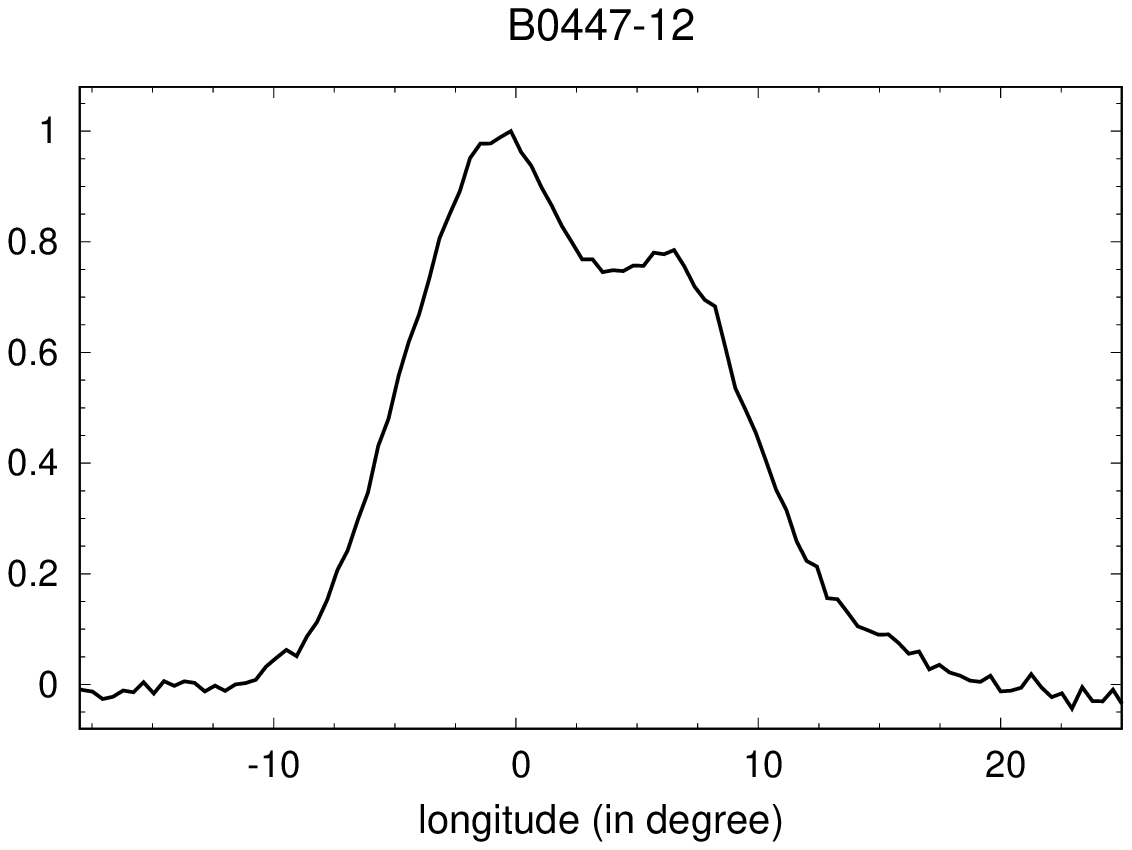}} &
\mbox{\includegraphics[angle=0,scale=0.57]{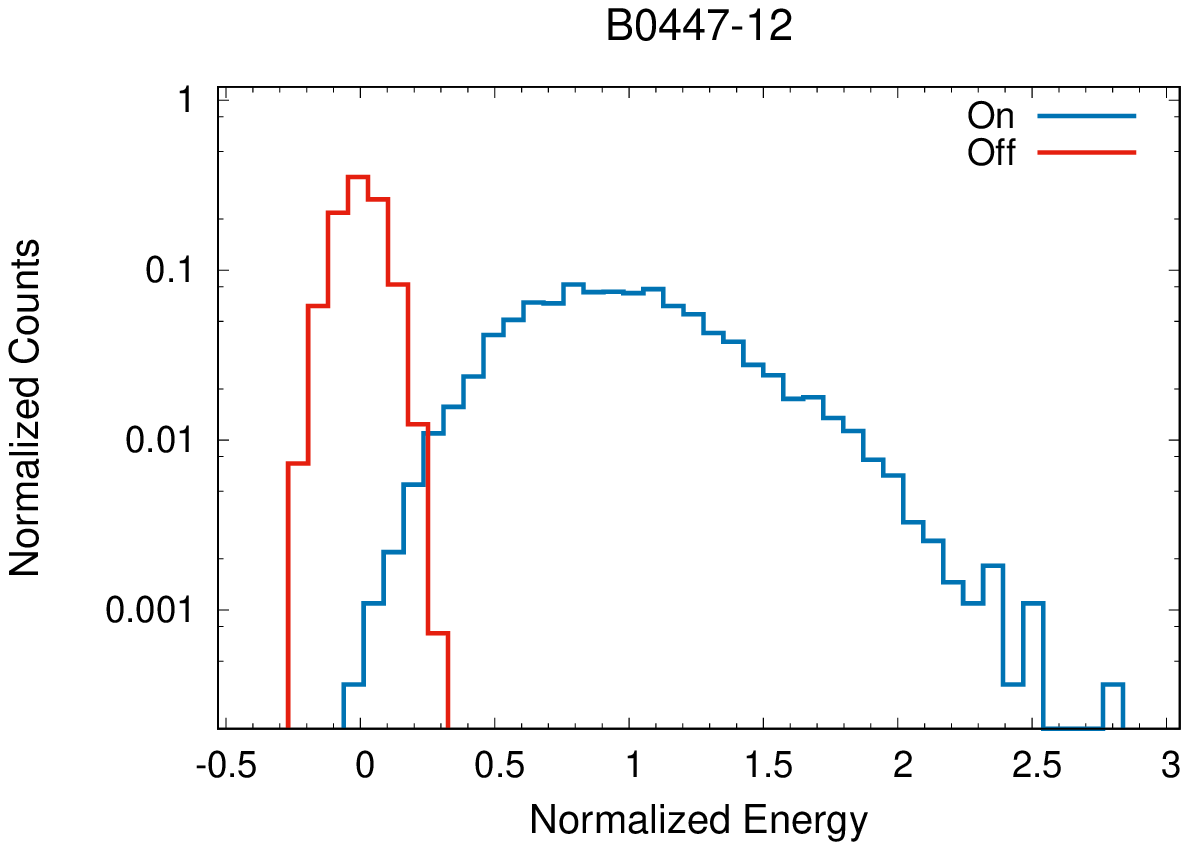}} \\
\mbox{\includegraphics[angle=0,scale=0.57]{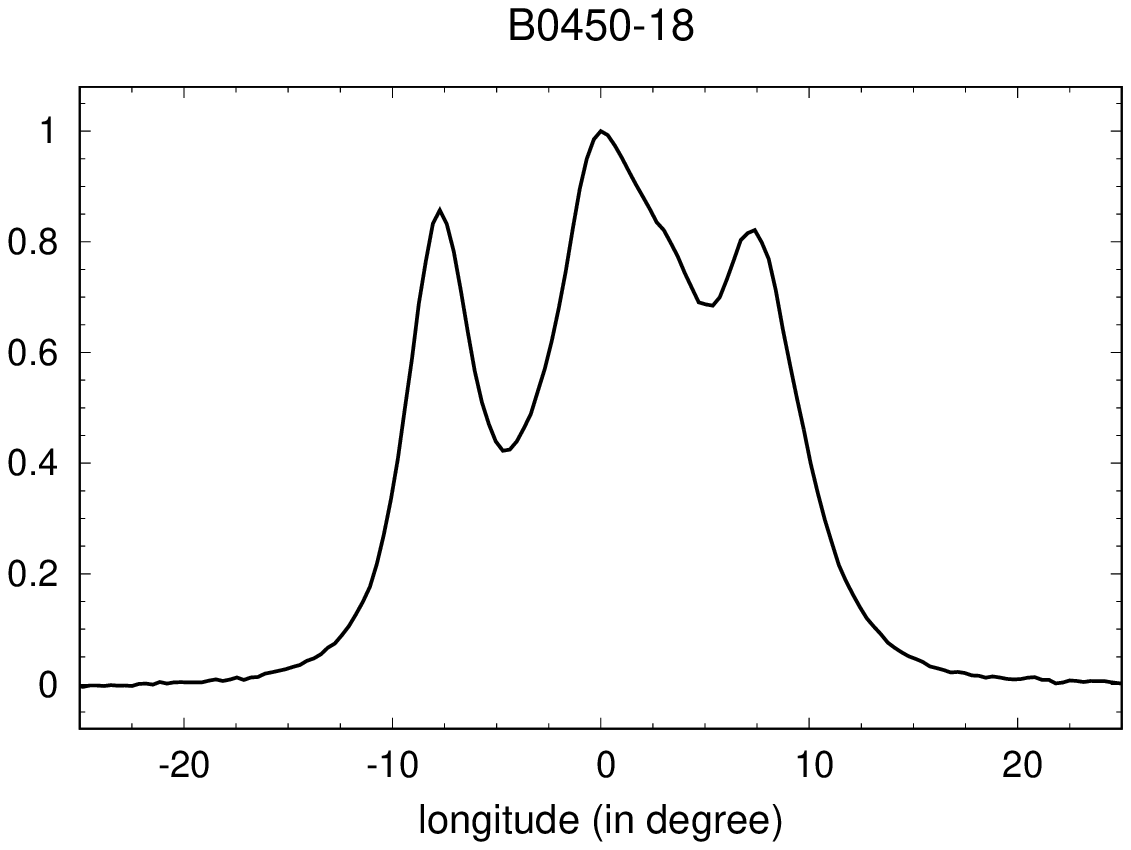}} &
\mbox{\includegraphics[angle=0,scale=0.57]{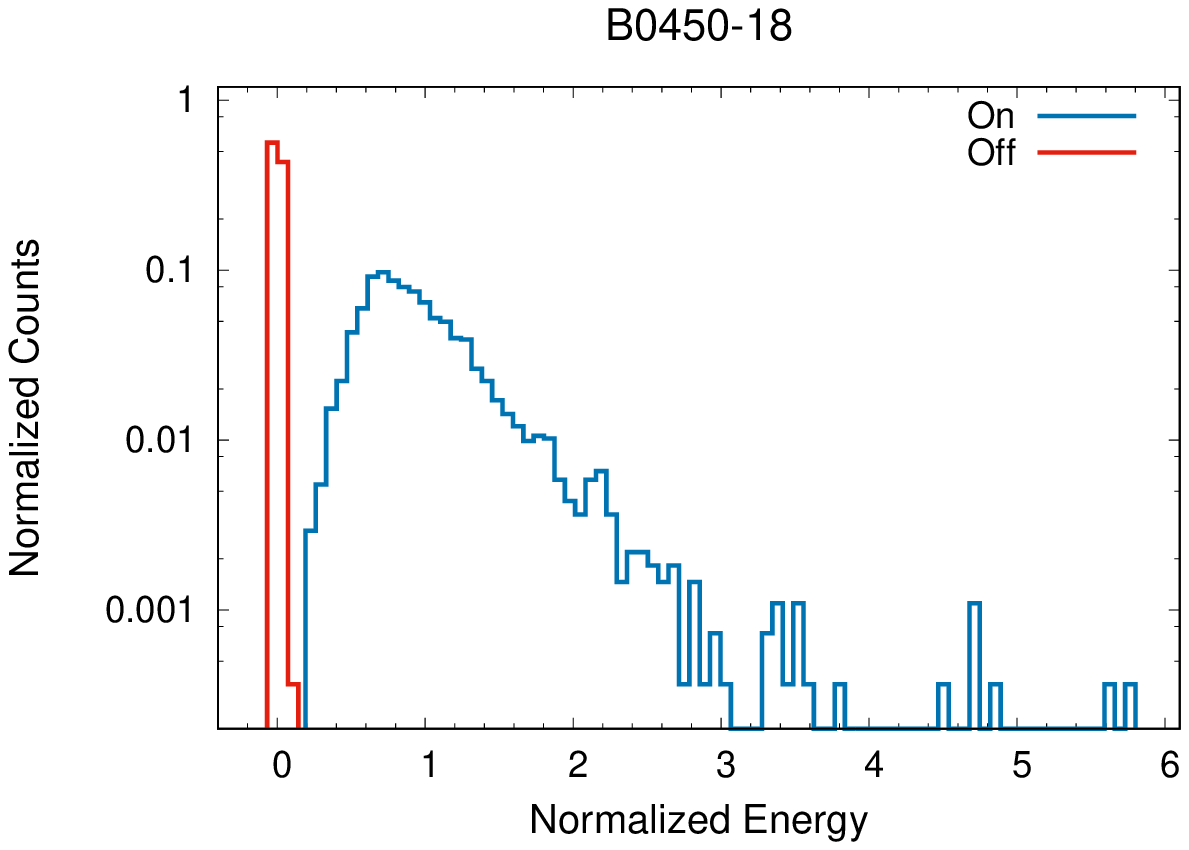}} \\
\end{tabular}
\caption{The pulsar profile and On and Off-pulse energy distributions of the single pulse emission.}
\end{center}
\end{figure*}

\clearpage

%3rd set of plots
\begin{figure*}
\begin{center}
\begin{tabular}{@{}cr@{}}
\mbox{\includegraphics[angle=0,scale=0.57]{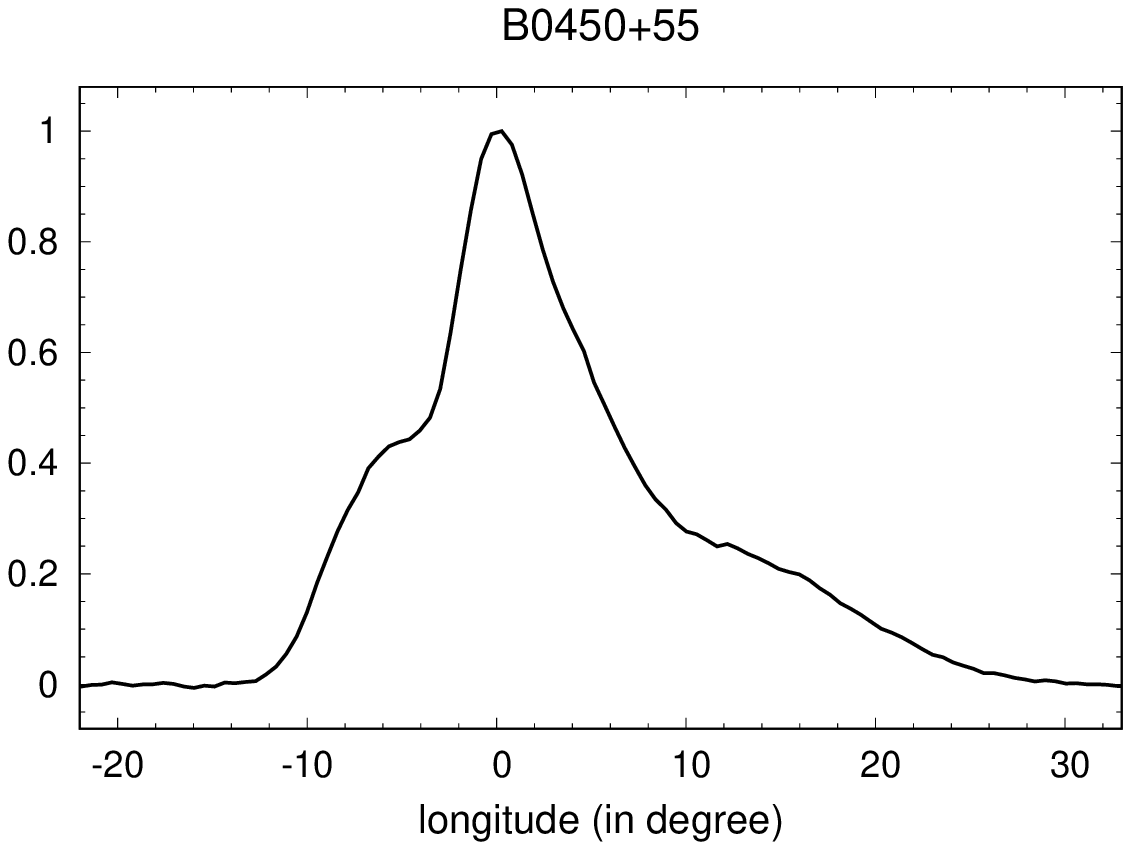}} &
\mbox{\includegraphics[angle=0,scale=0.57]{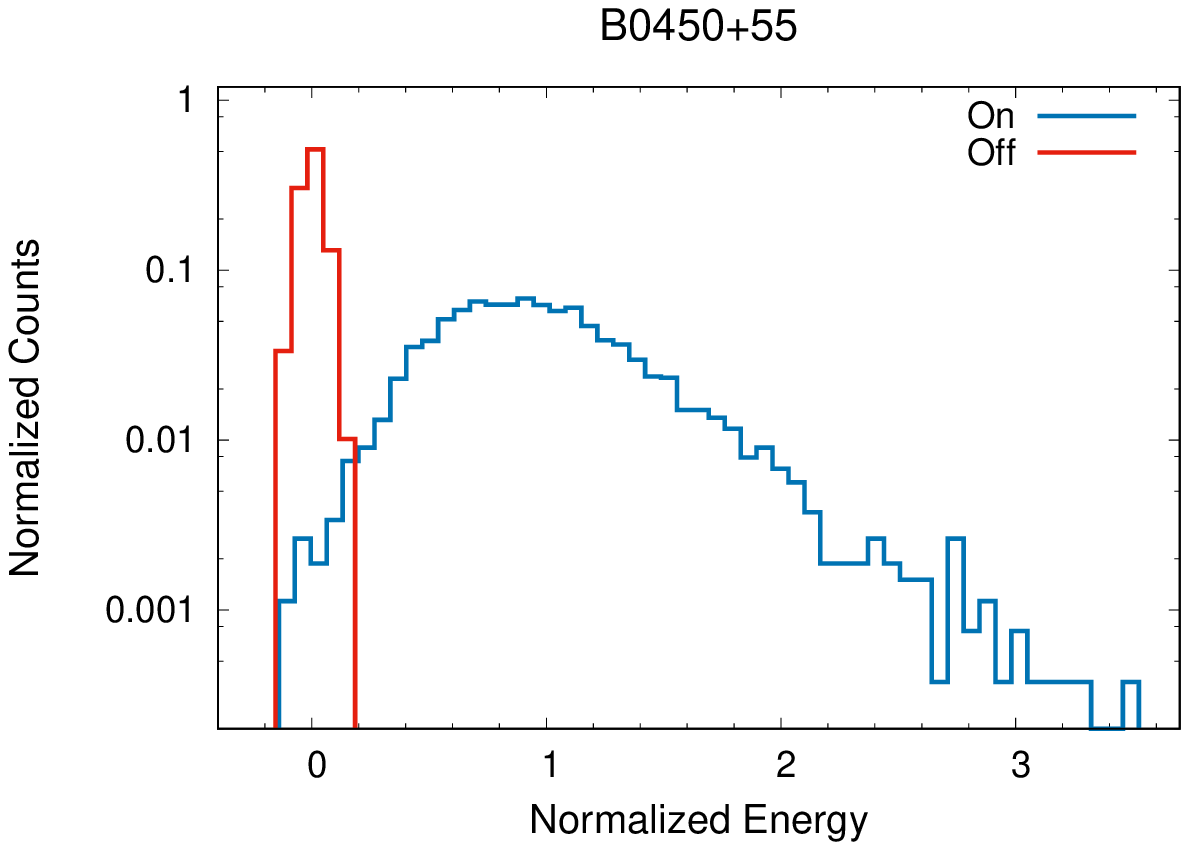}} \\
\mbox{\includegraphics[angle=0,scale=0.57]{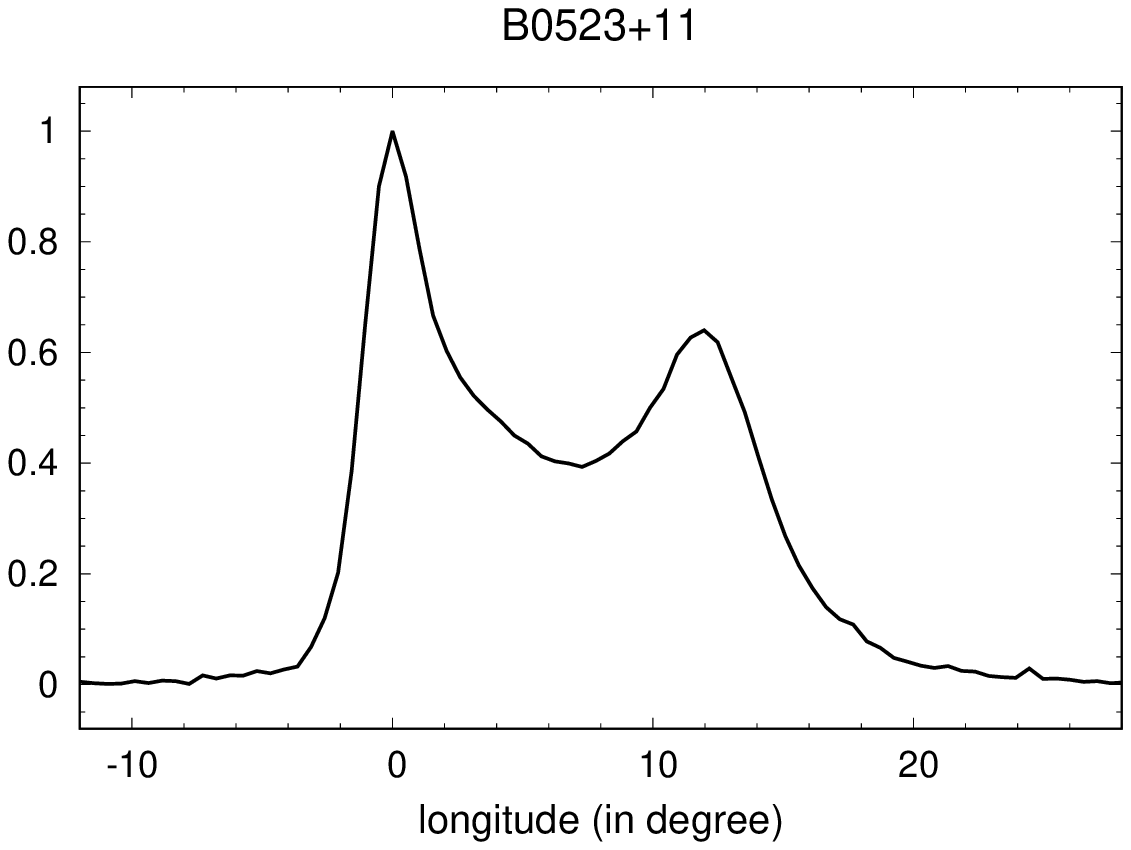}} &
\mbox{\includegraphics[angle=0,scale=0.57]{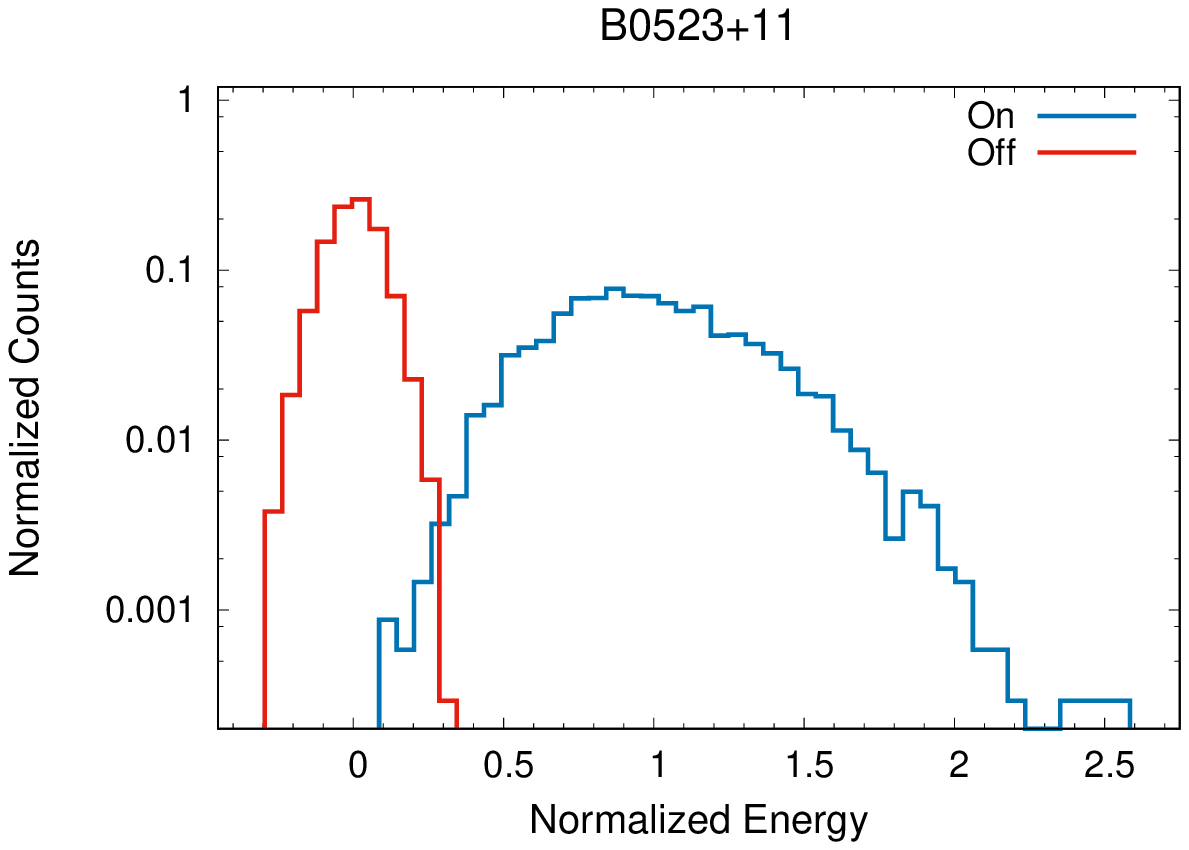}} \\
\mbox{\includegraphics[angle=0,scale=0.57]{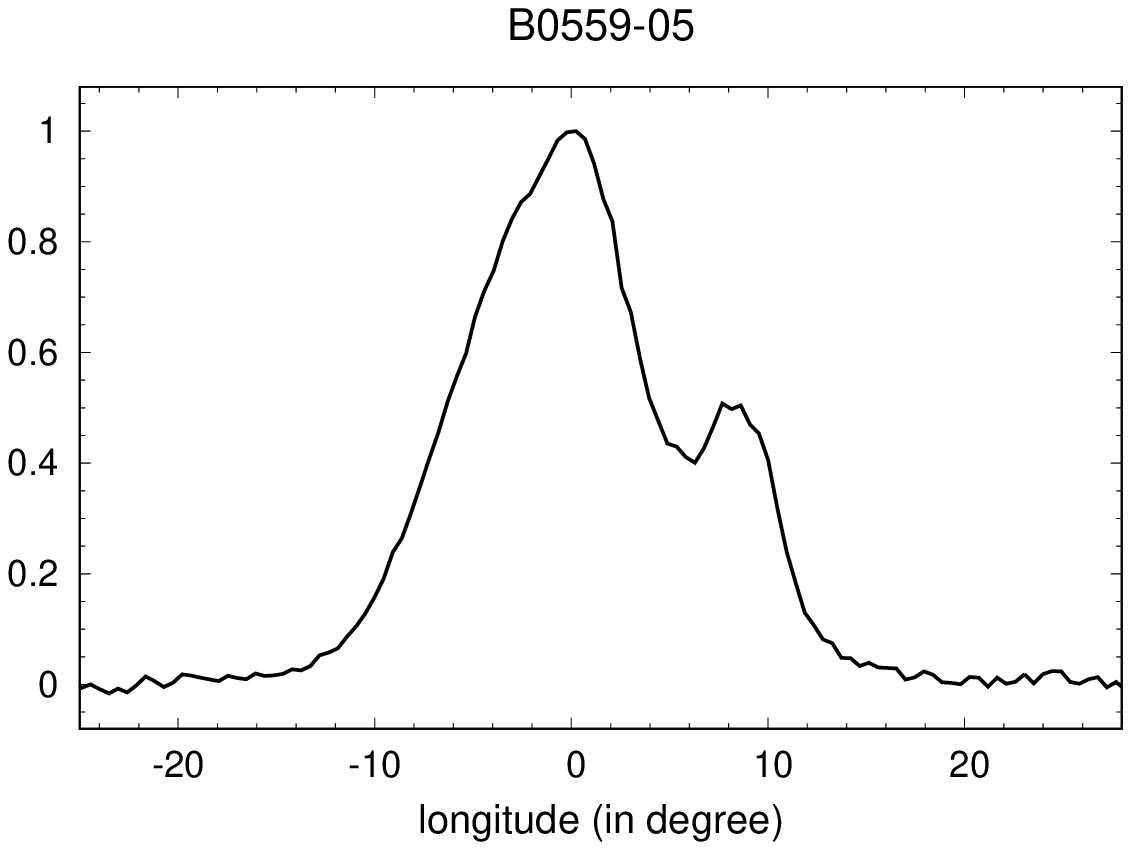}} &
\mbox{\includegraphics[angle=0,scale=0.57]{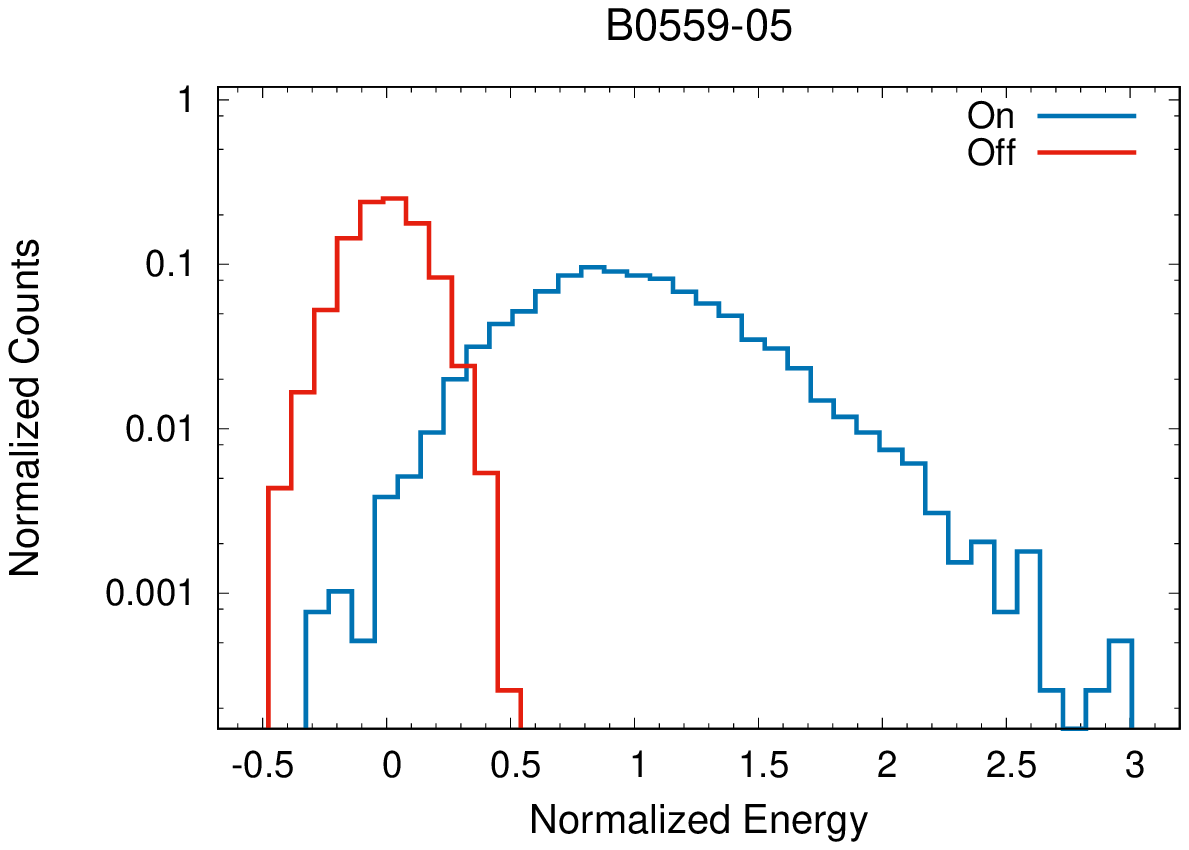}} \\
\mbox{\includegraphics[angle=0,scale=0.57]{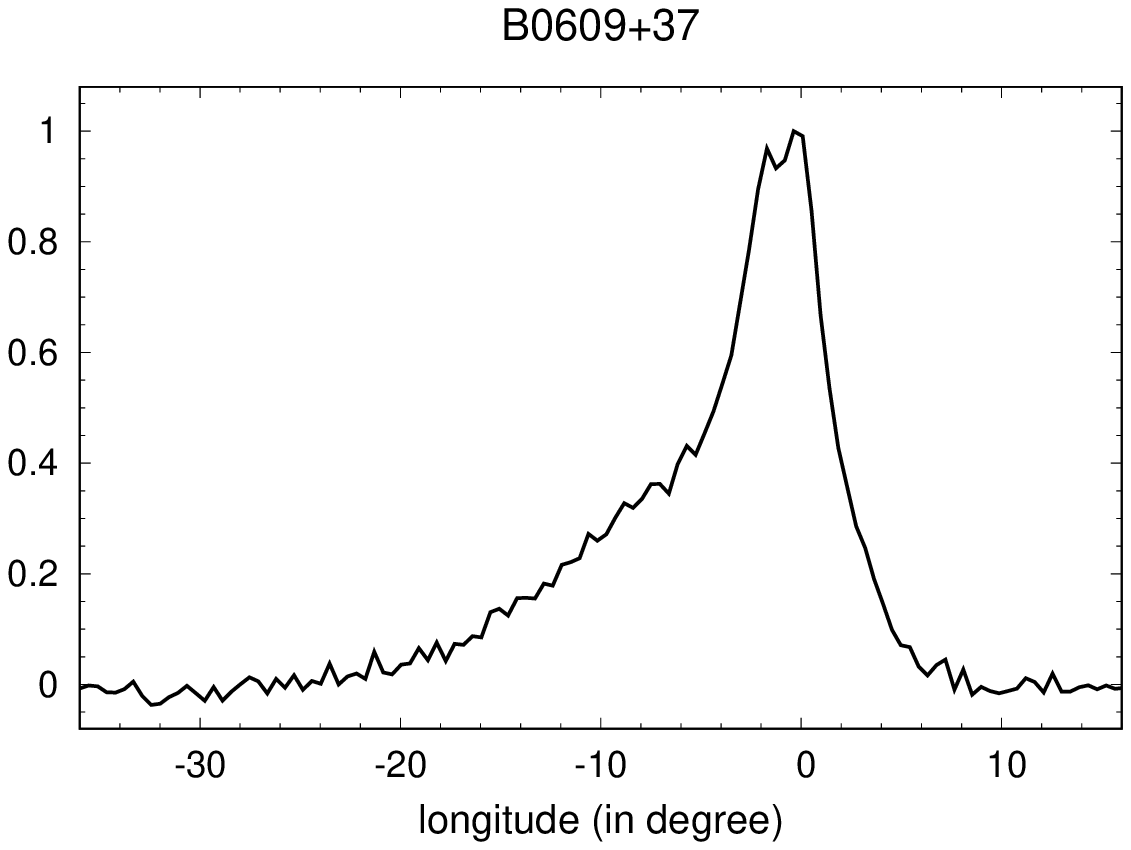}} &
\mbox{\includegraphics[angle=0,scale=0.57]{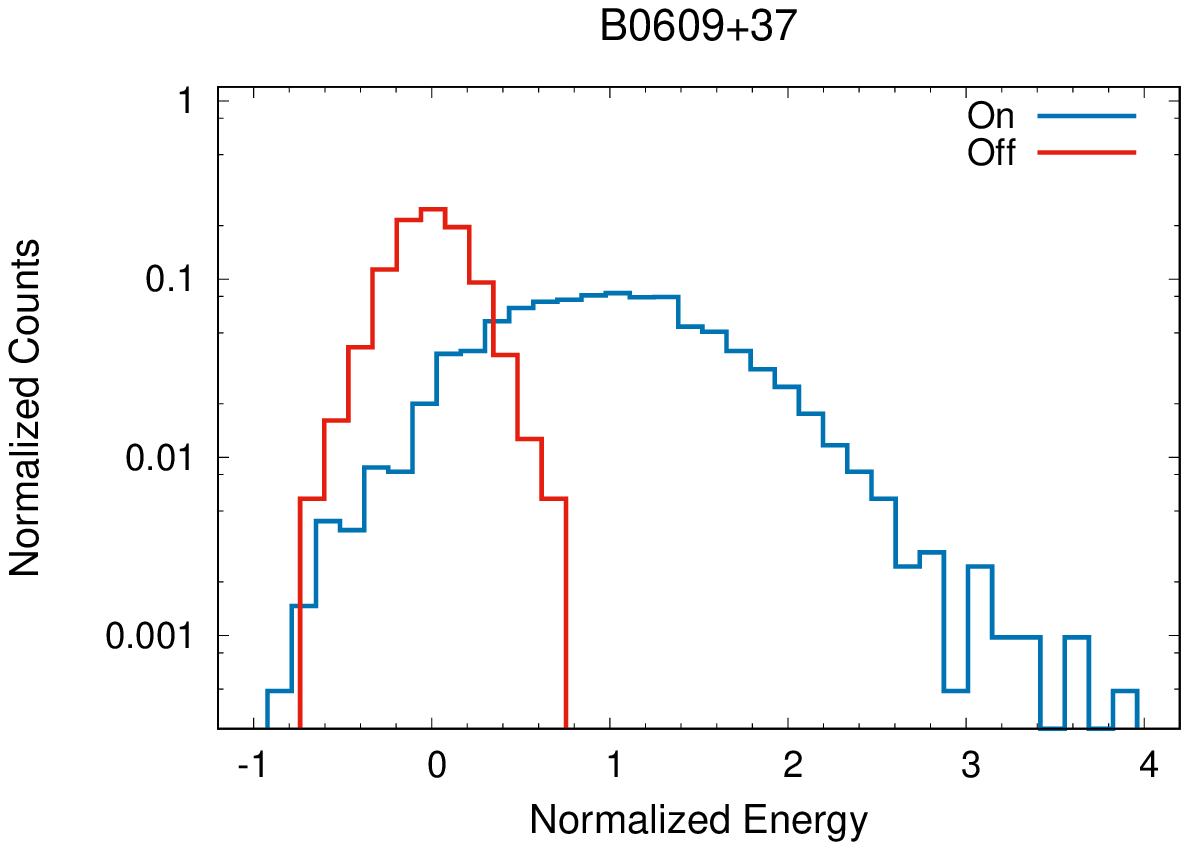}} \\
\end{tabular}
\caption{The pulsar profile and On and Off-pulse energy distributions of the single pulse emission.}
\end{center}
\end{figure*}

\clearpage

%4th set of plots
\begin{figure*}
\begin{center}
\begin{tabular}{@{}cr@{}}
\mbox{\includegraphics[angle=0,scale=0.57]{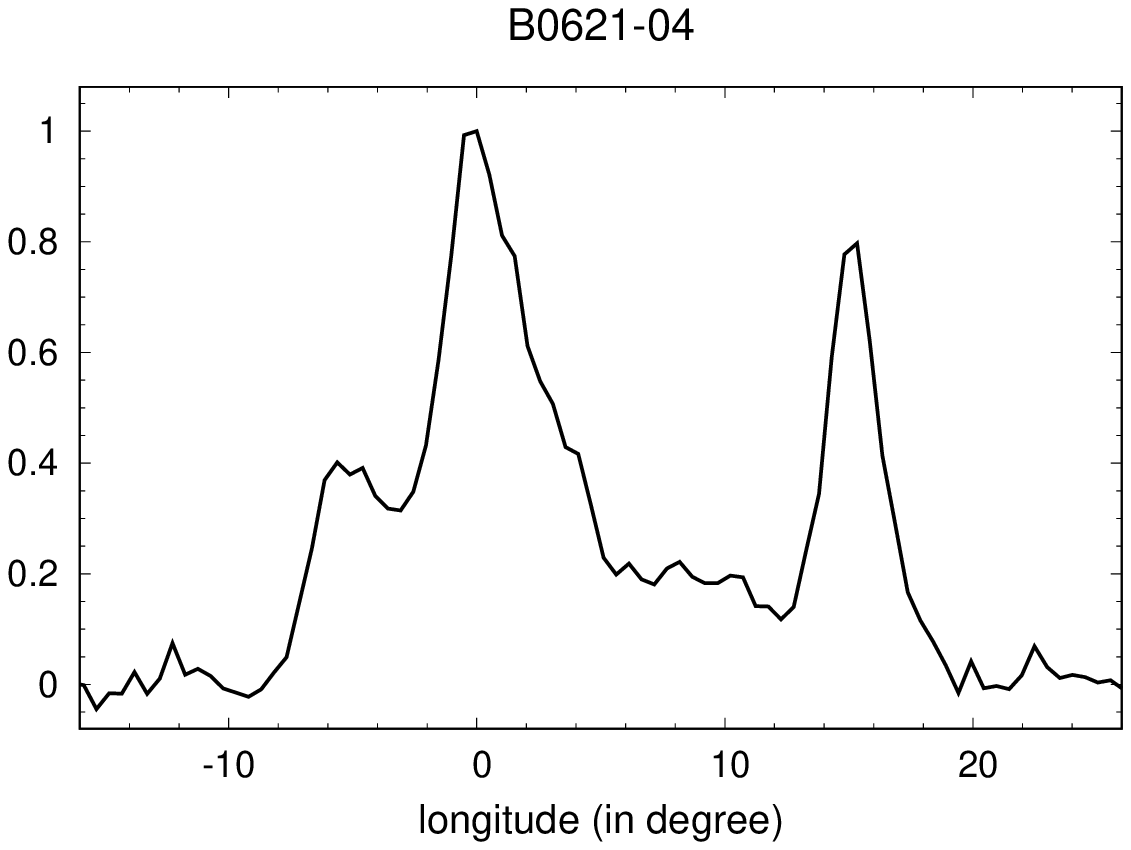}} &
\mbox{\includegraphics[angle=0,scale=0.57]{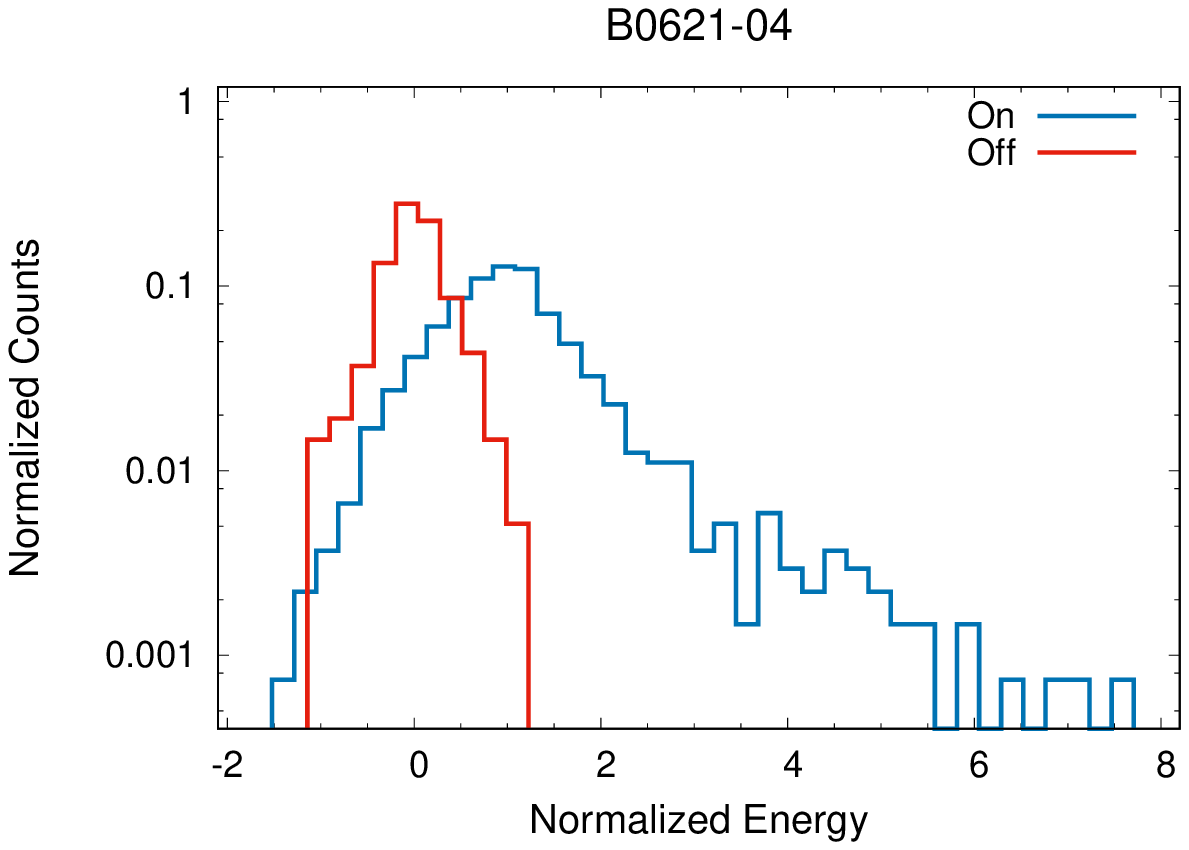}} \\
\mbox{\includegraphics[angle=0,scale=0.57]{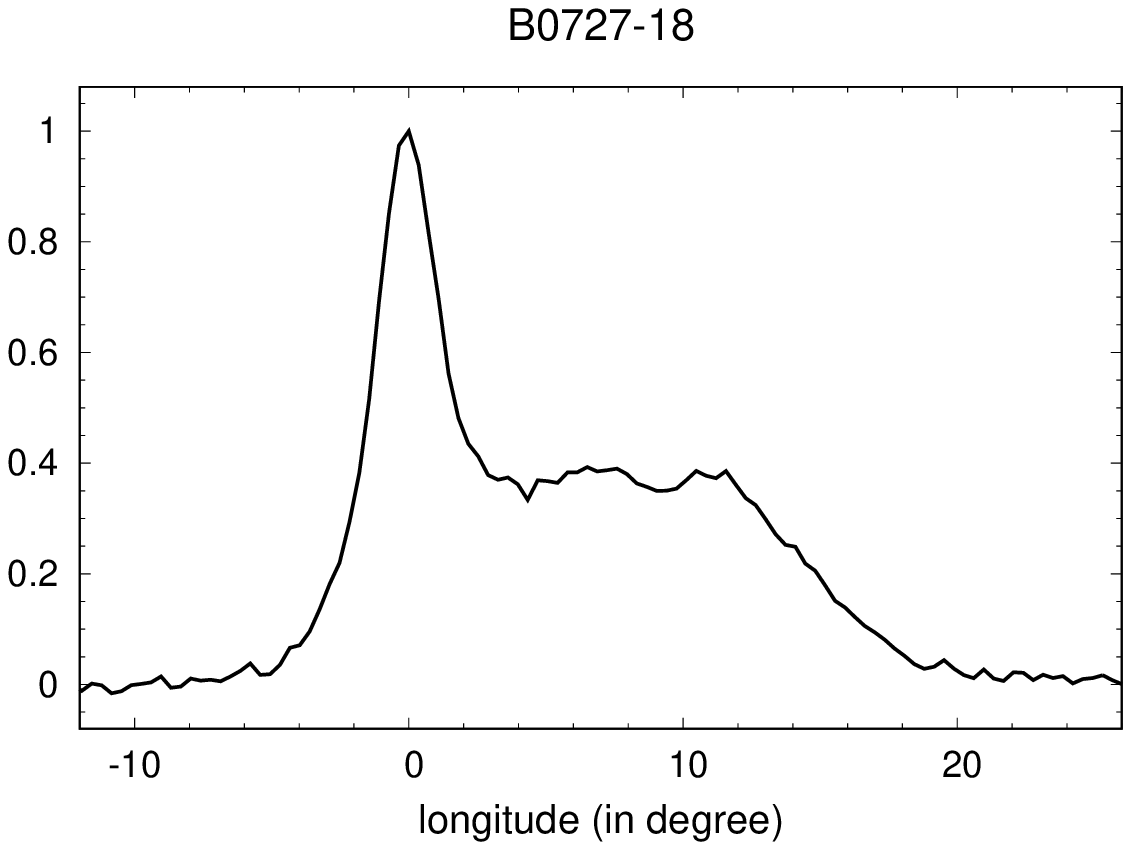}} &
\mbox{\includegraphics[angle=0,scale=0.57]{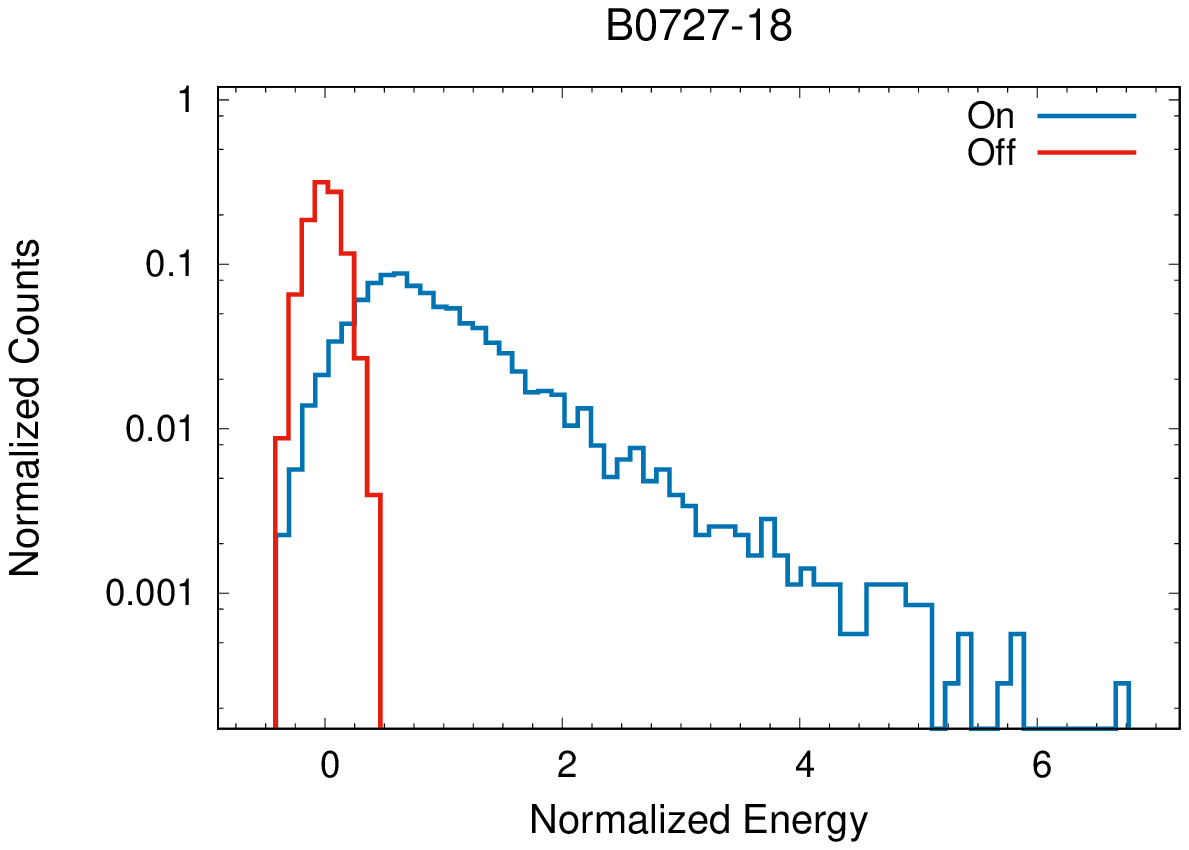}} \\
\mbox{\includegraphics[angle=0,scale=0.57]{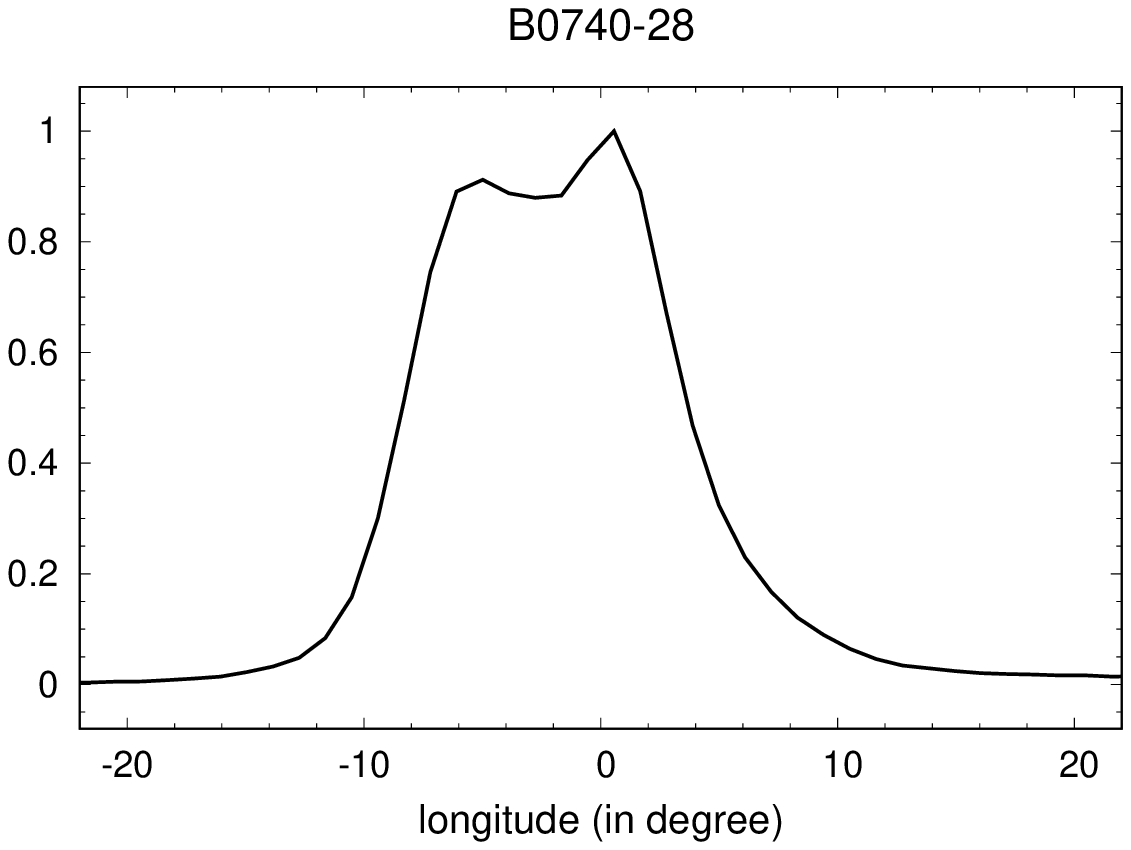}} &
\mbox{\includegraphics[angle=0,scale=0.57]{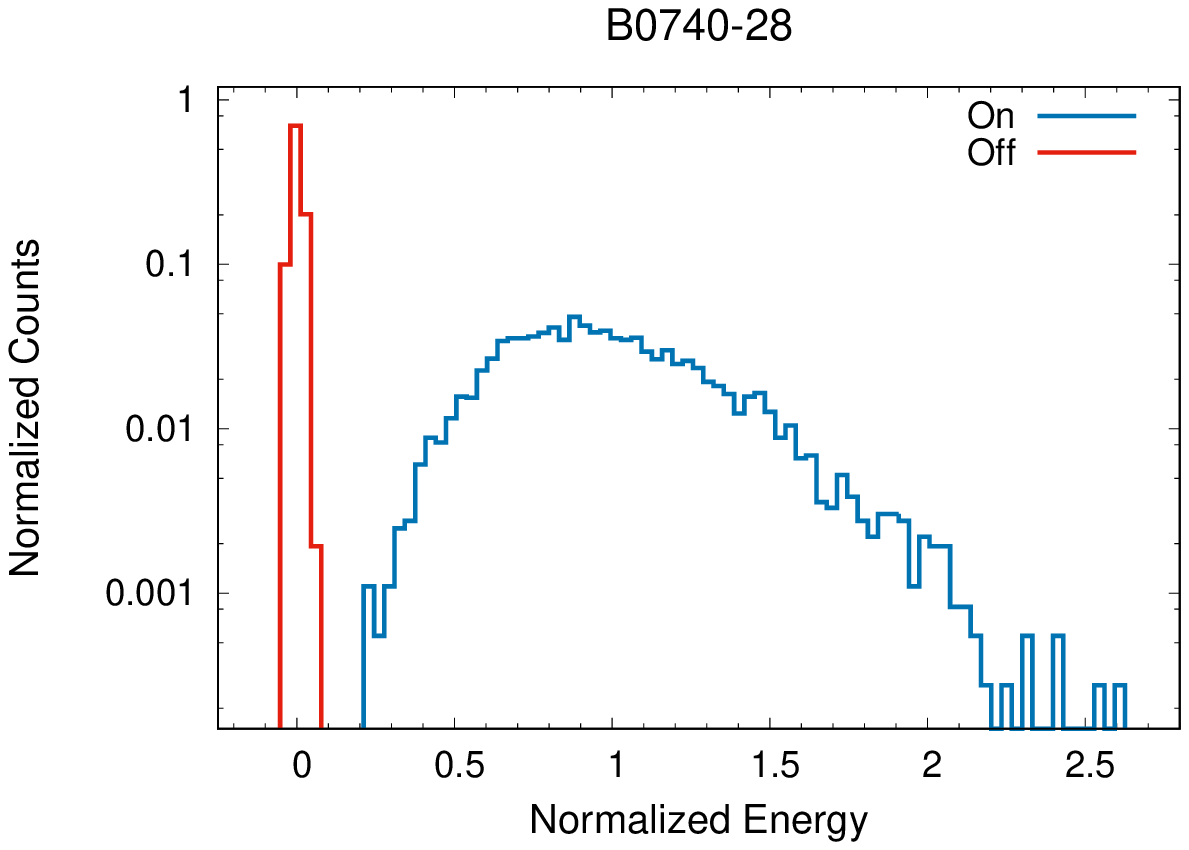}} \\
\mbox{\includegraphics[angle=0,scale=0.57]{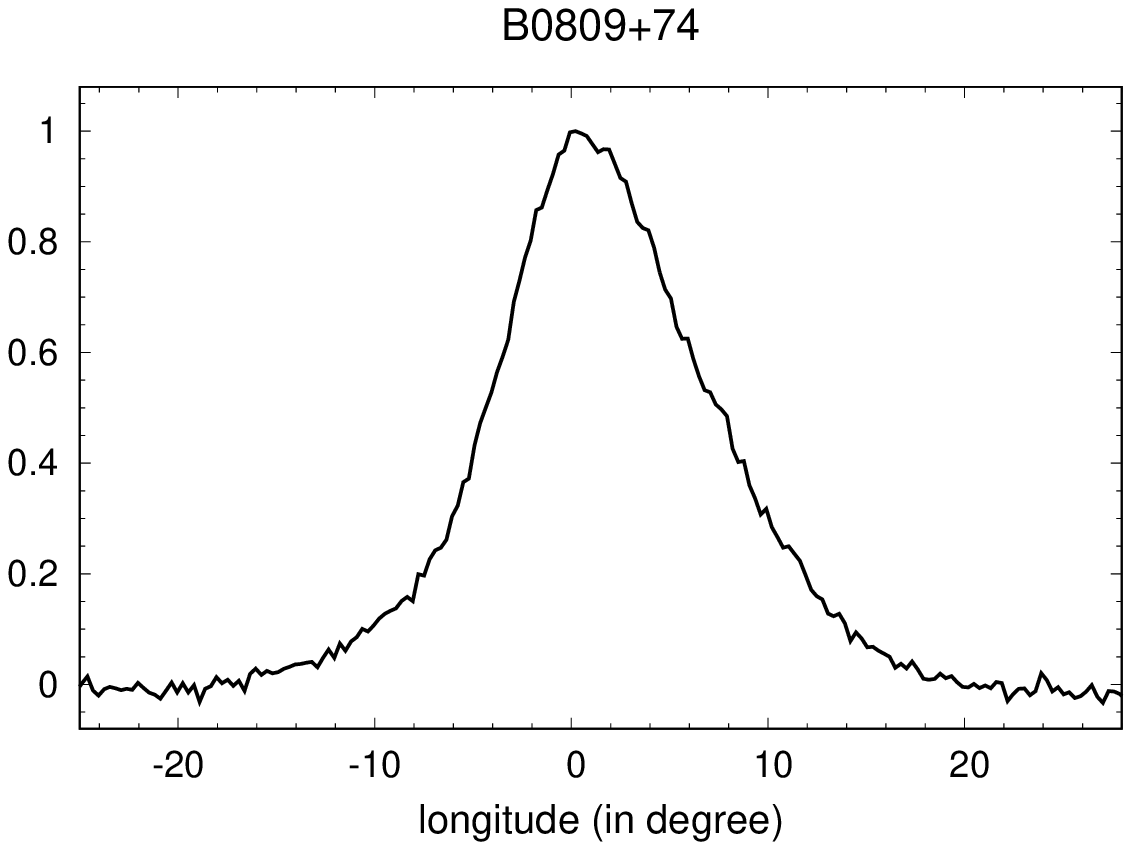}} &
\mbox{\includegraphics[angle=0,scale=0.57]{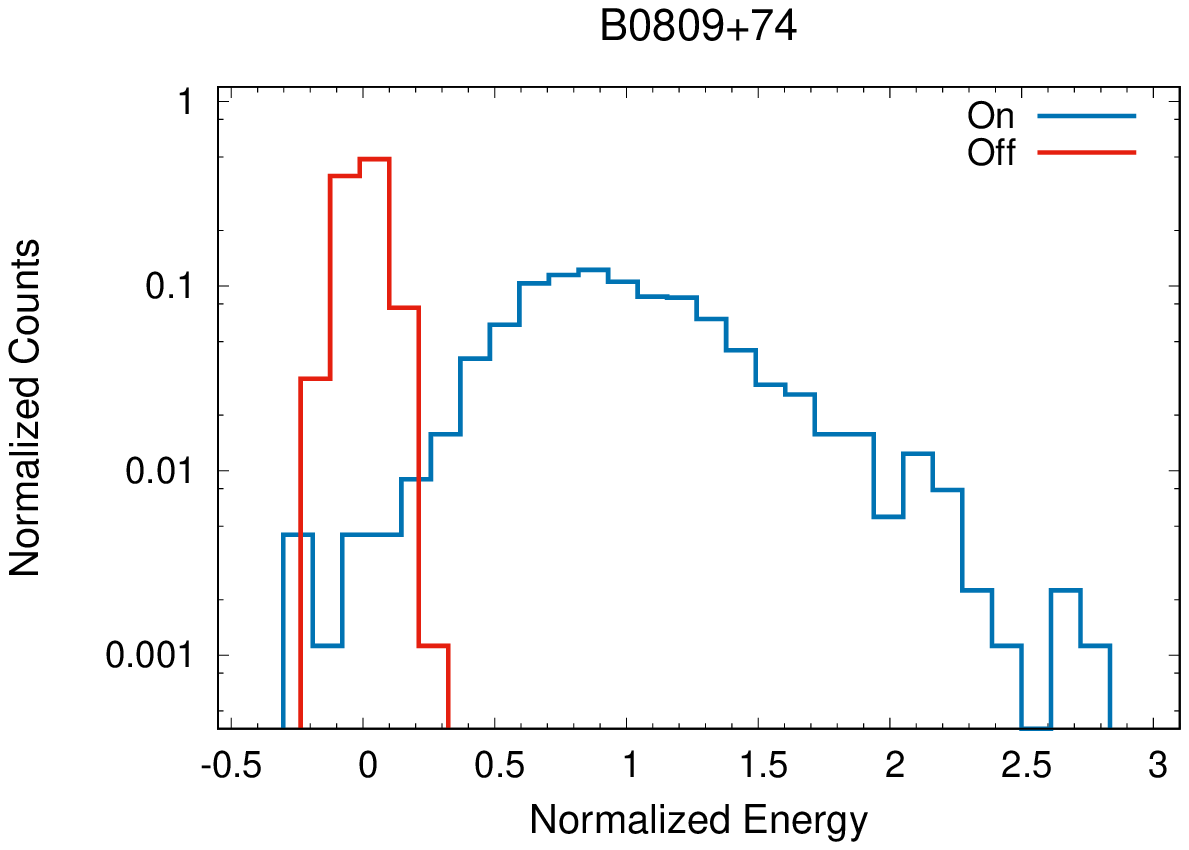}} \\
\end{tabular}
\caption{The pulsar profile and On and Off-pulse energy distributions of the single pulse emission.}
\end{center}
\end{figure*}

\clearpage

%5th set of plots
\begin{figure*}
\begin{center}
\begin{tabular}{@{}cr@{}}
\mbox{\includegraphics[angle=0,scale=0.57]{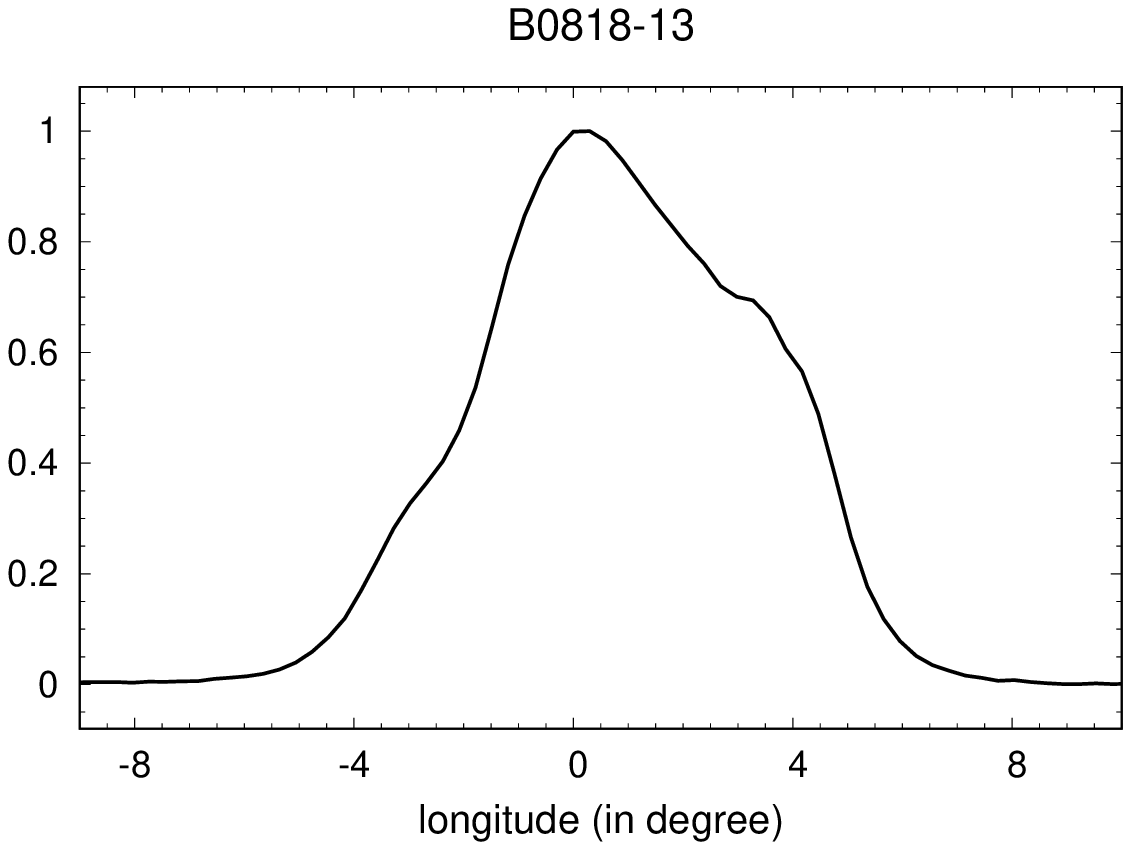}} &
\mbox{\includegraphics[angle=0,scale=0.57]{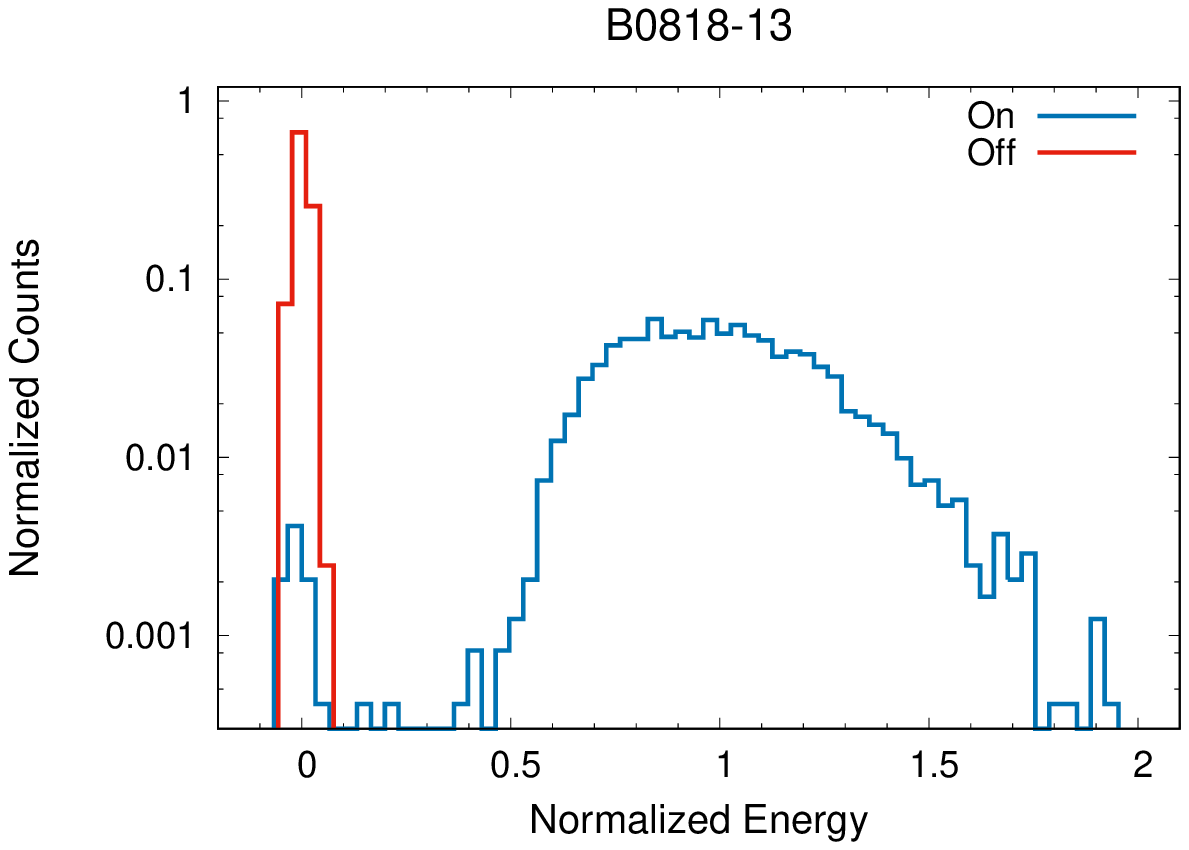}} \\
\mbox{\includegraphics[angle=0,scale=0.57]{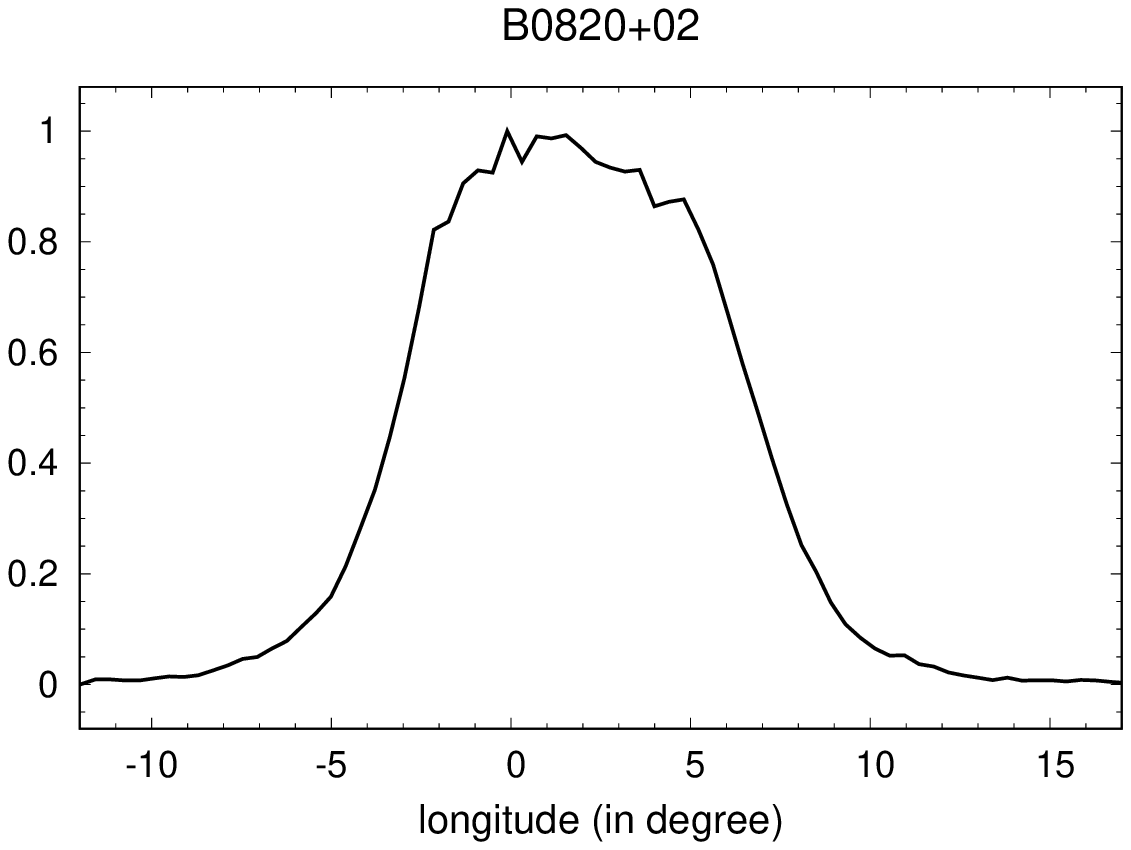}} &
\mbox{\includegraphics[angle=0,scale=0.57]{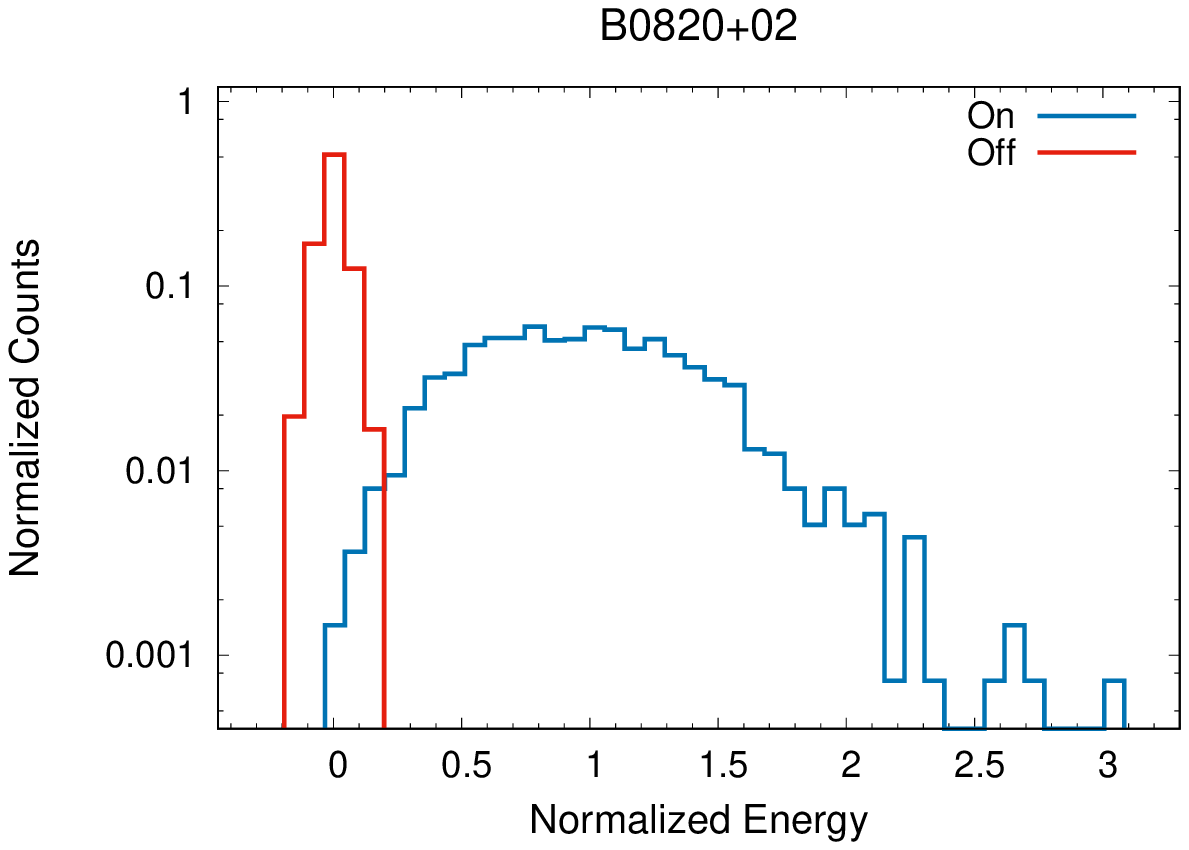}} \\
\mbox{\includegraphics[angle=0,scale=0.57]{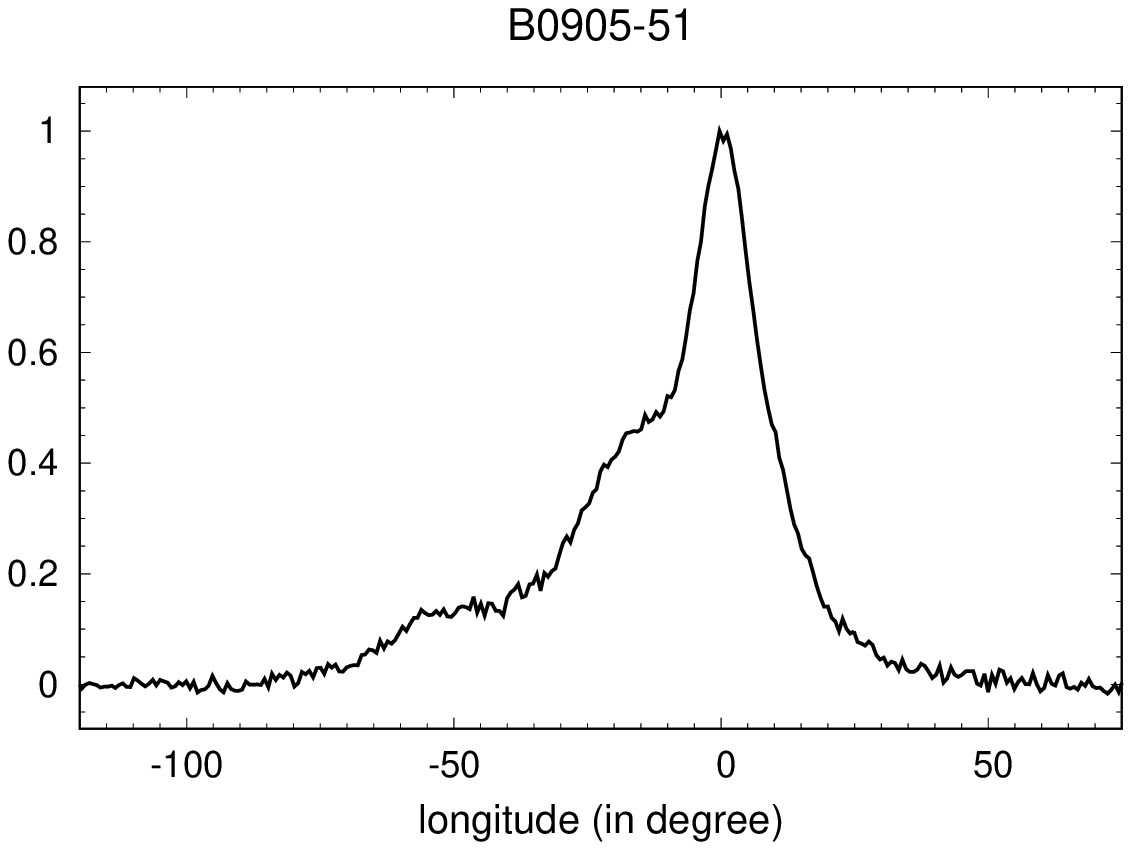}} &
\mbox{\includegraphics[angle=0,scale=0.57]{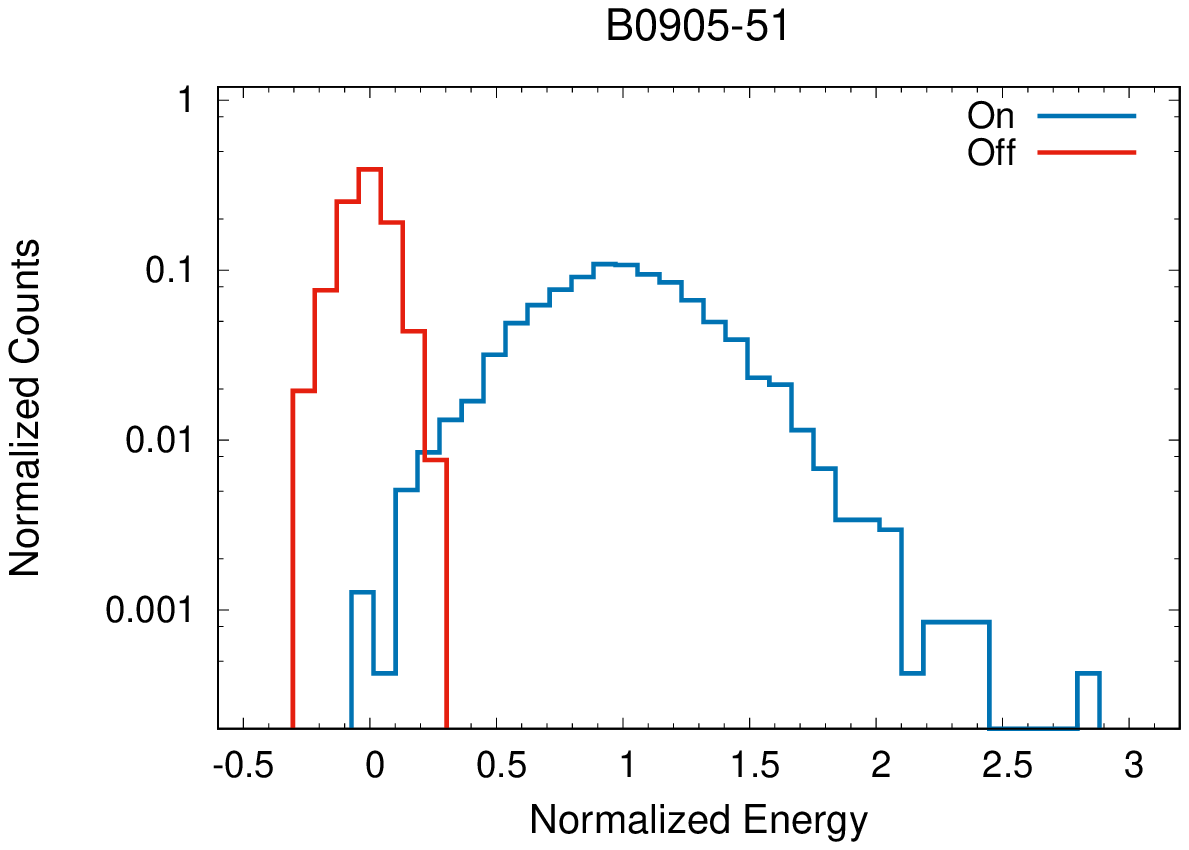}} \\
\mbox{\includegraphics[angle=0,scale=0.57]{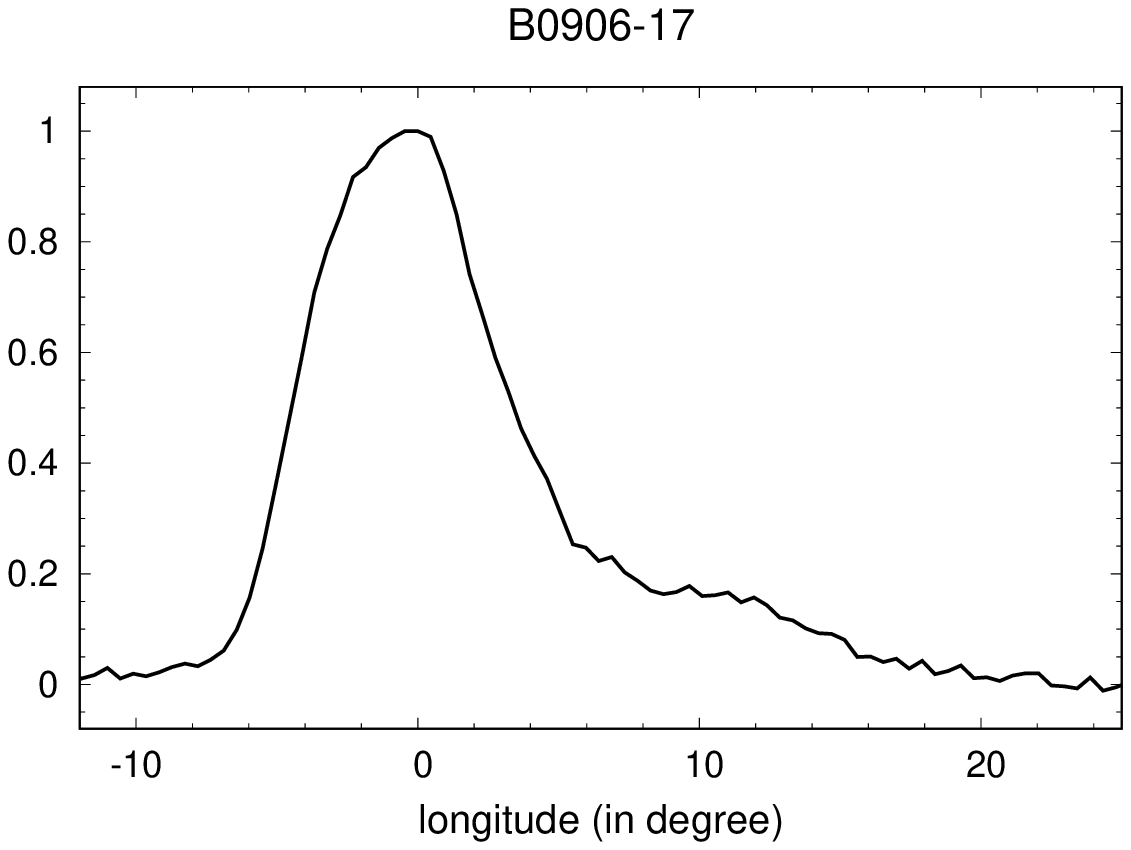}} &
\mbox{\includegraphics[angle=0,scale=0.57]{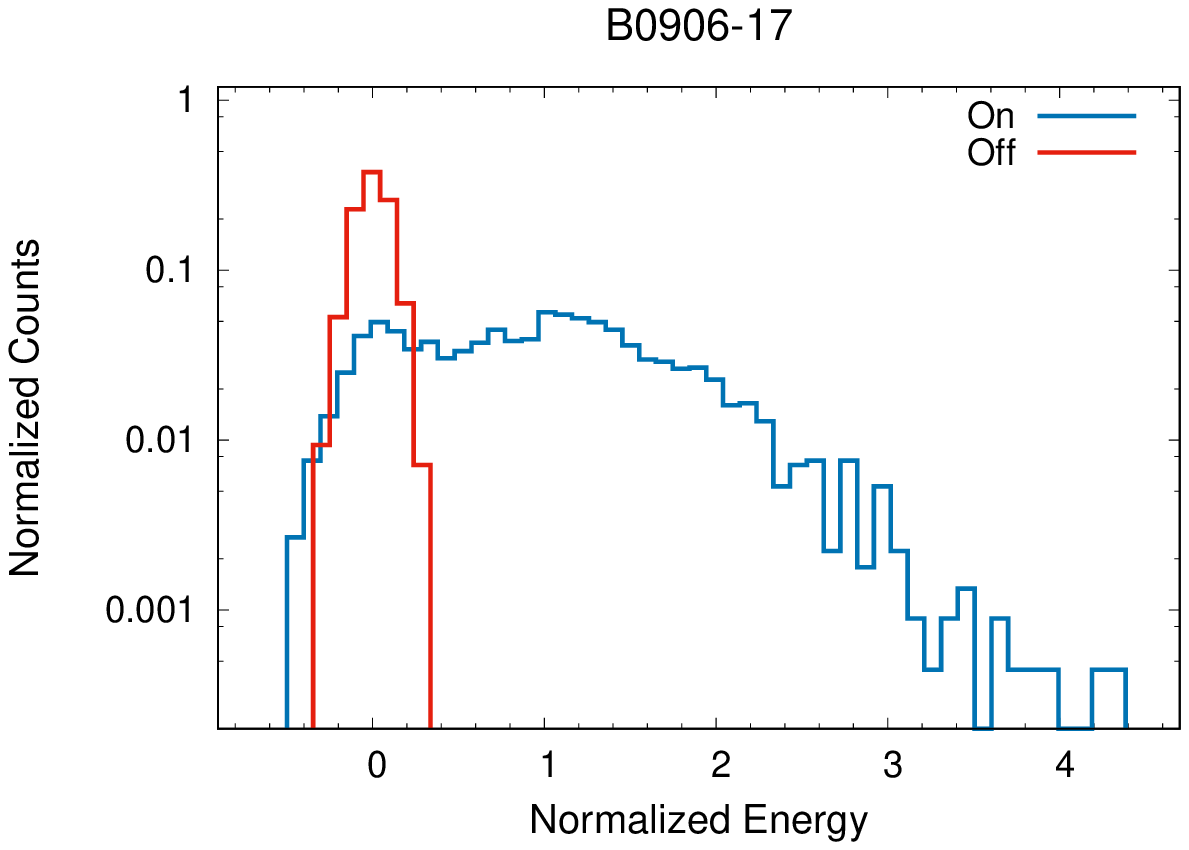}} \\
\end{tabular}
\caption{The pulsar profile and On and Off-pulse energy distributions of the single pulse emission.}
\end{center}
\end{figure*}

\clearpage

%6th set of plots
\begin{figure*}
\begin{center}
\begin{tabular}{@{}cr@{}}
\mbox{\includegraphics[angle=0,scale=0.57]{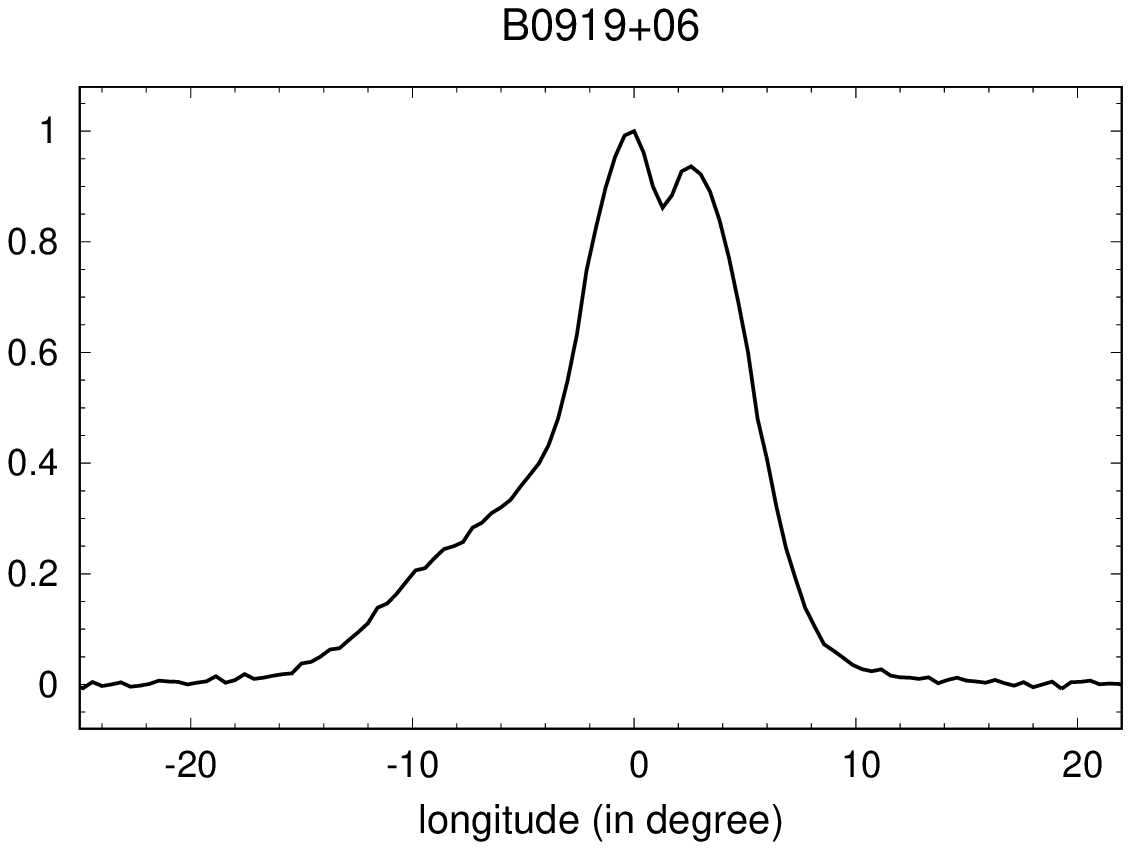}} &
\mbox{\includegraphics[angle=0,scale=0.57]{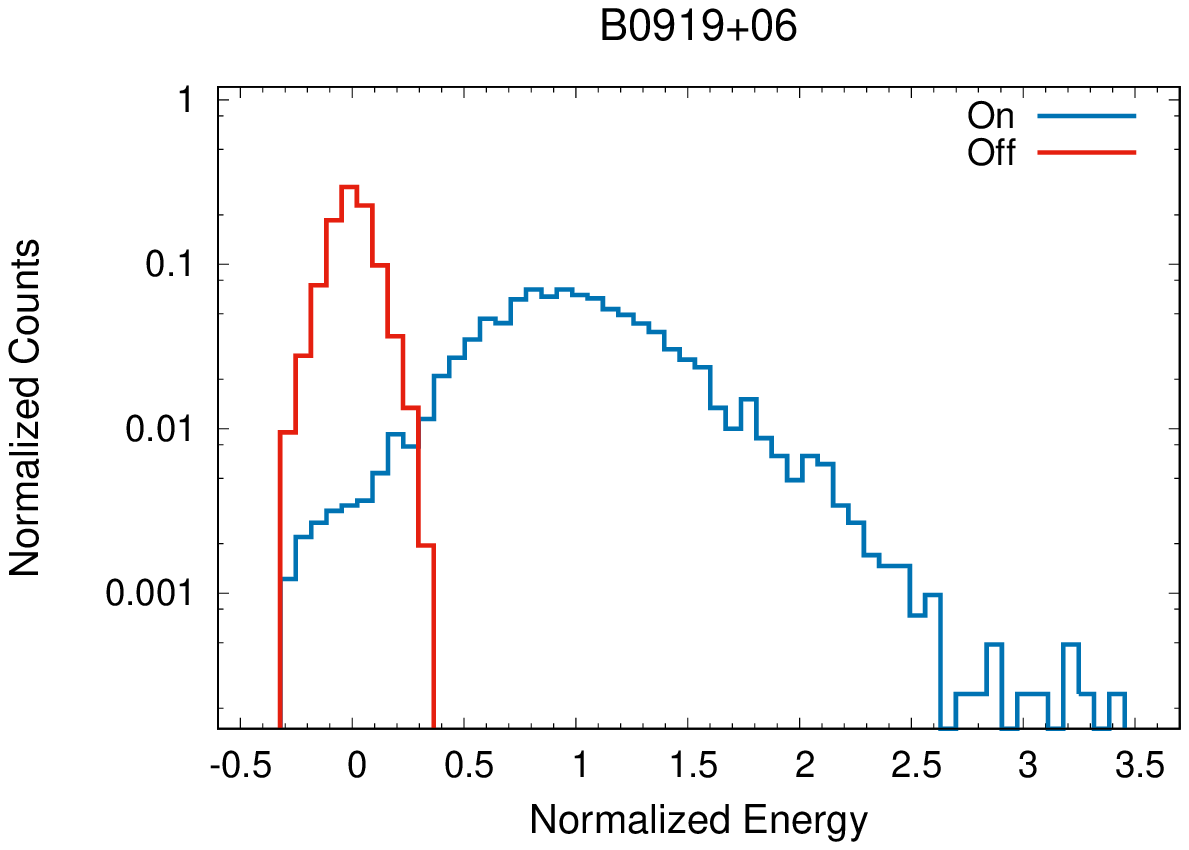}} \\
\mbox{\includegraphics[angle=0,scale=0.57]{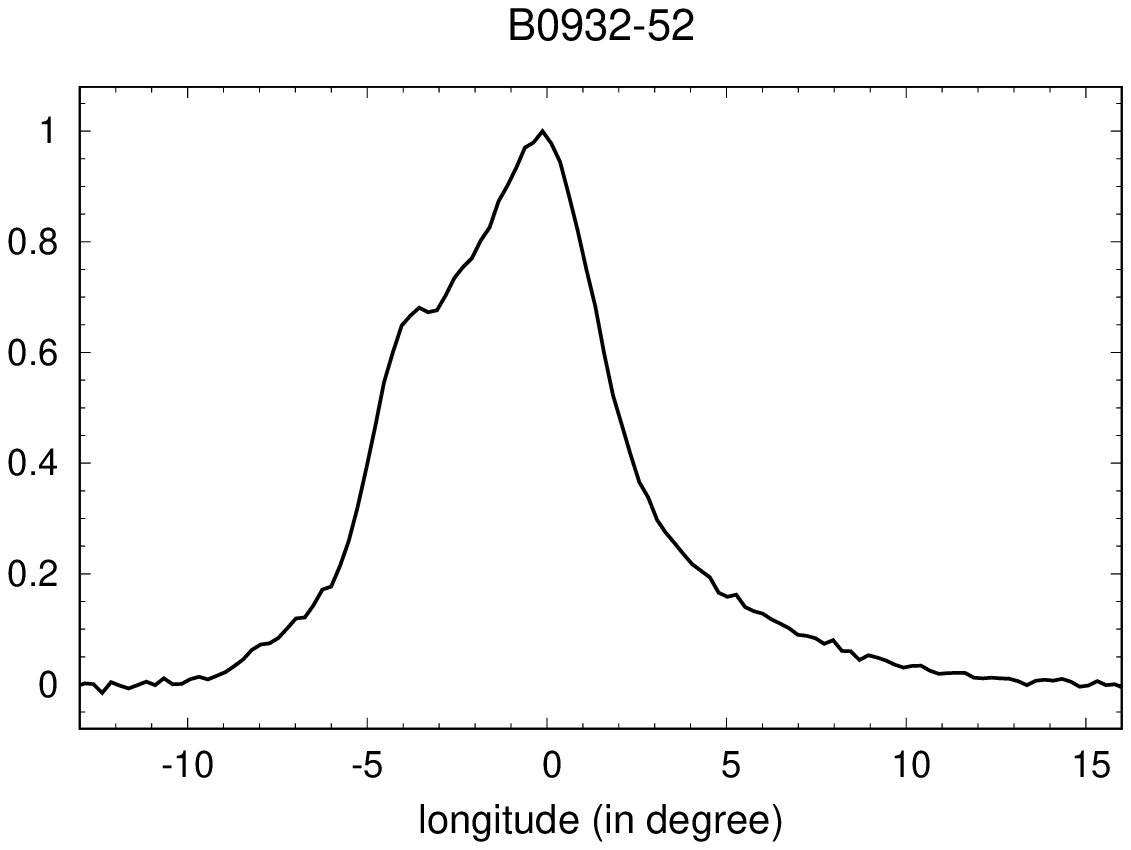}} &
\mbox{\includegraphics[angle=0,scale=0.57]{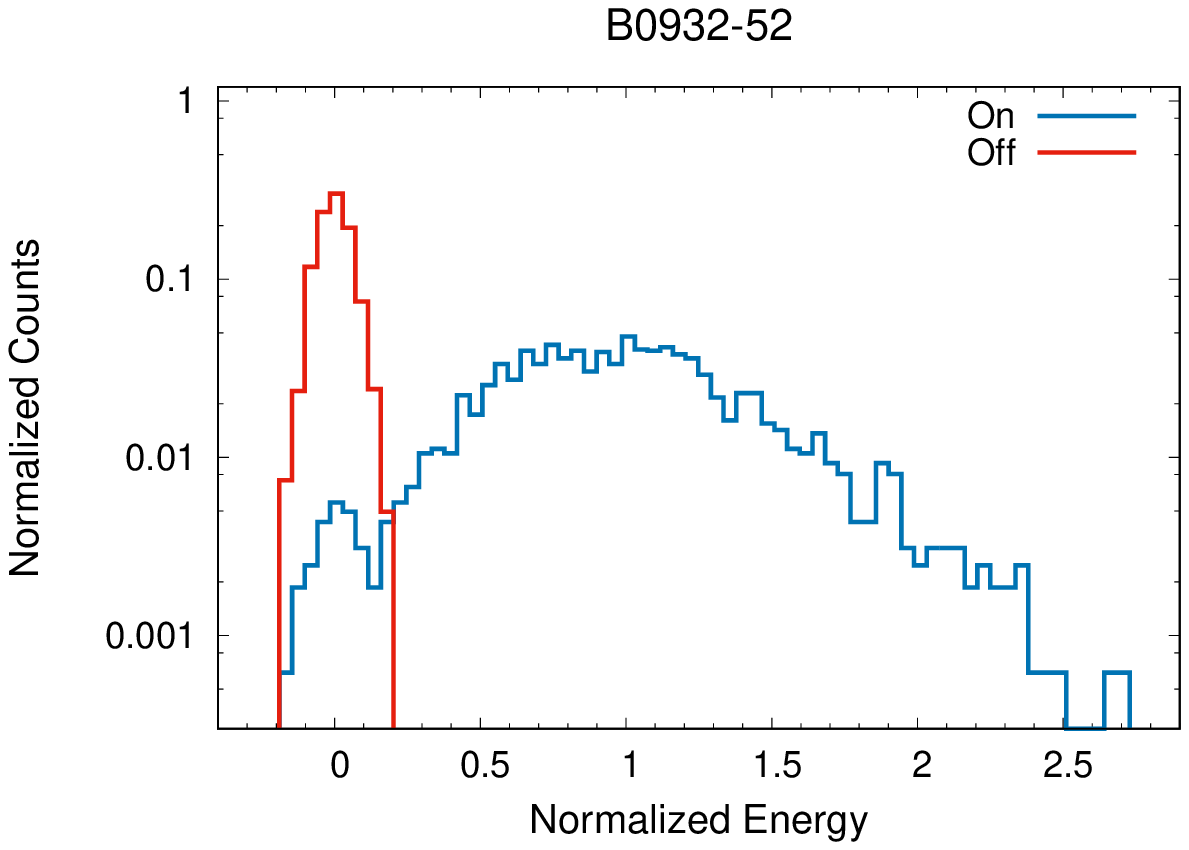}} \\
\mbox{\includegraphics[angle=0,scale=0.57]{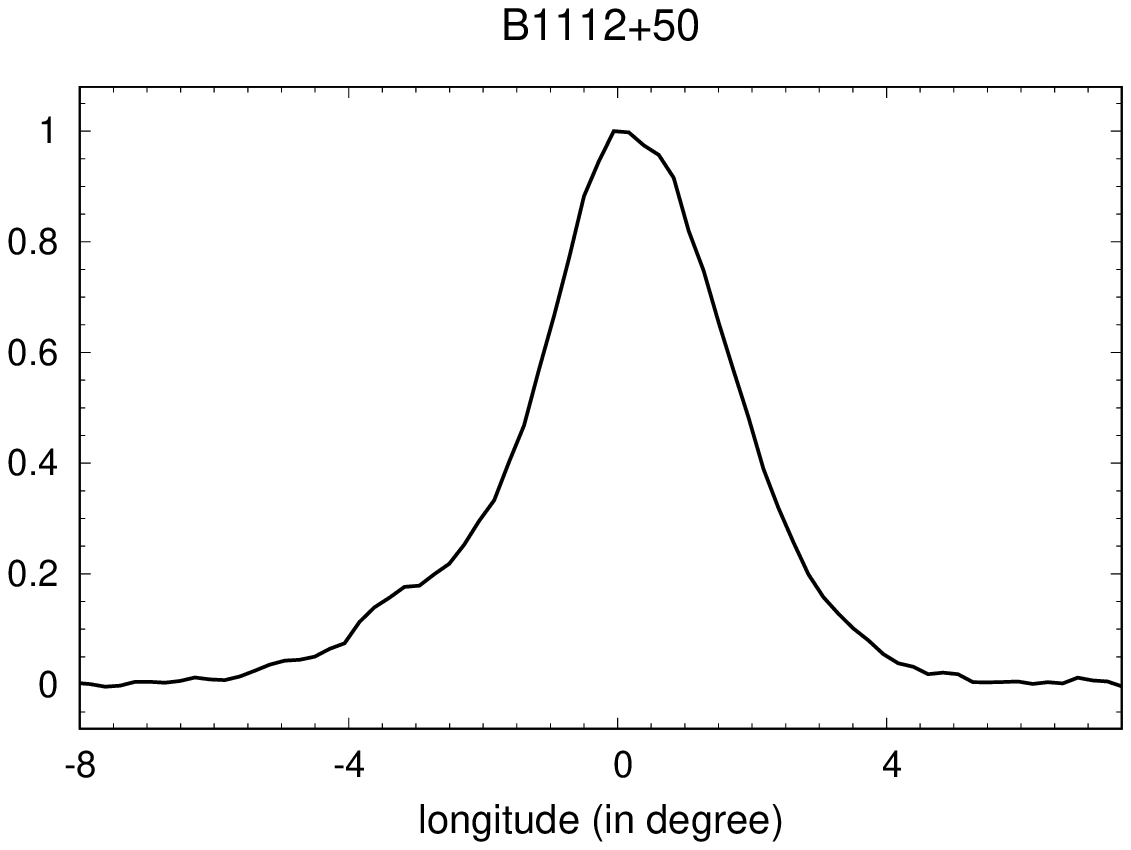}} &
\mbox{\includegraphics[angle=0,scale=0.57]{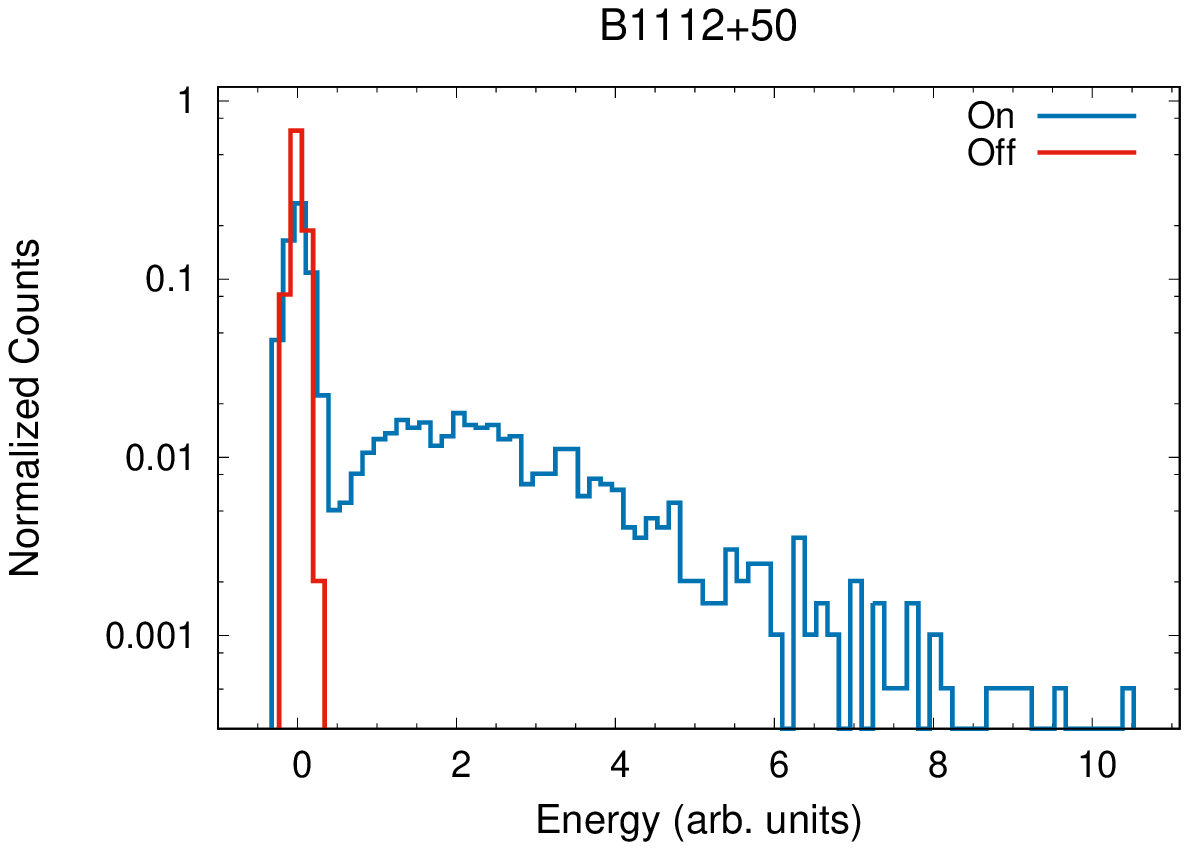}} \\
\mbox{\includegraphics[angle=0,scale=0.57]{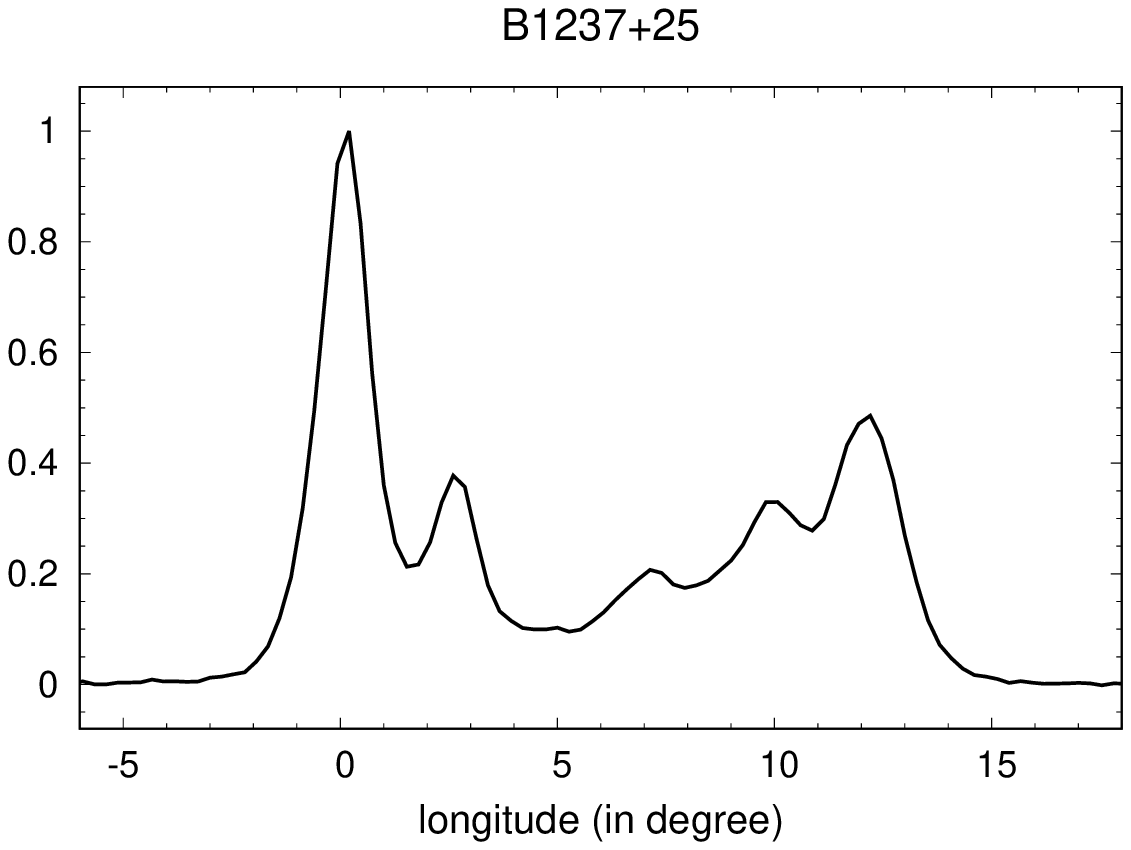}} &
\mbox{\includegraphics[angle=0,scale=0.57]{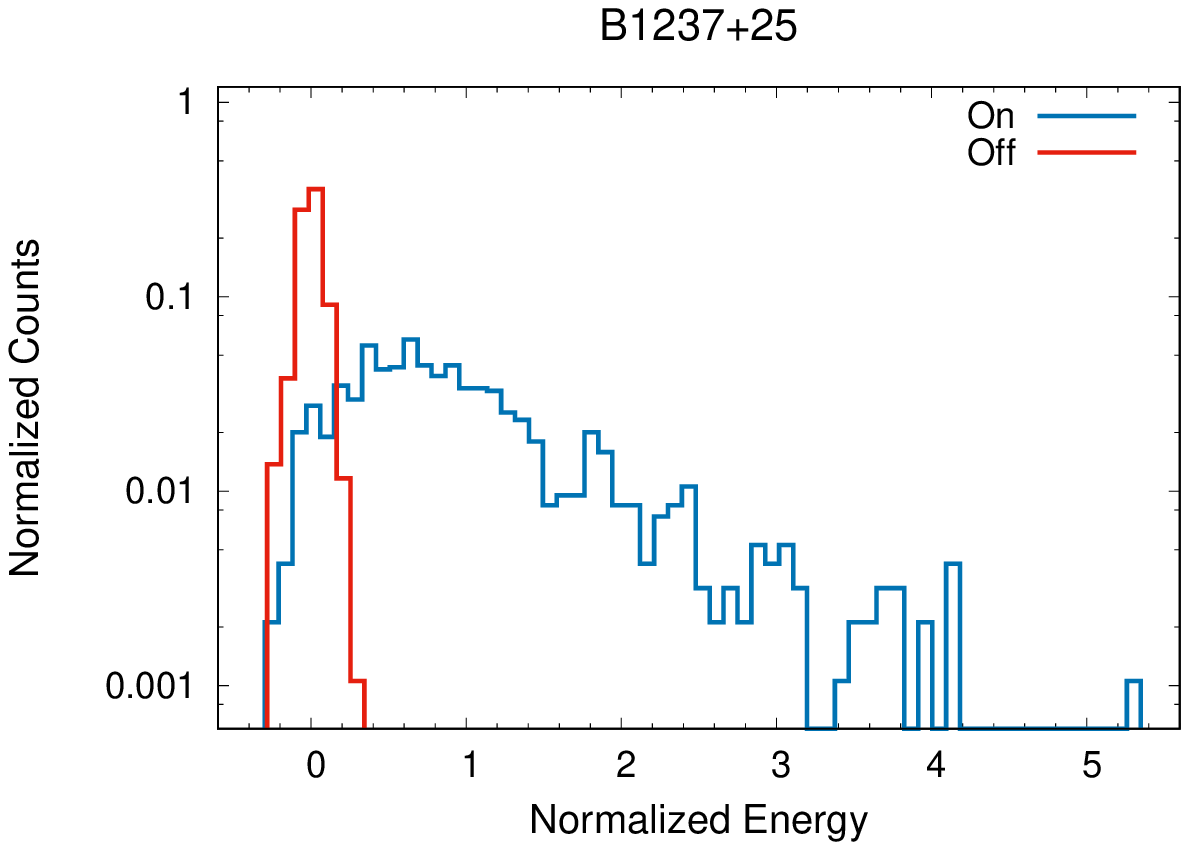}} \\
\end{tabular}
\caption{The pulsar profile and On and Off-pulse energy distributions of the single pulse emission.}
\end{center}
\end{figure*}

\clearpage

%7th set of plots
\begin{figure*}
\begin{center}
\begin{tabular}{@{}cr@{}}
\mbox{\includegraphics[angle=0,scale=0.57]{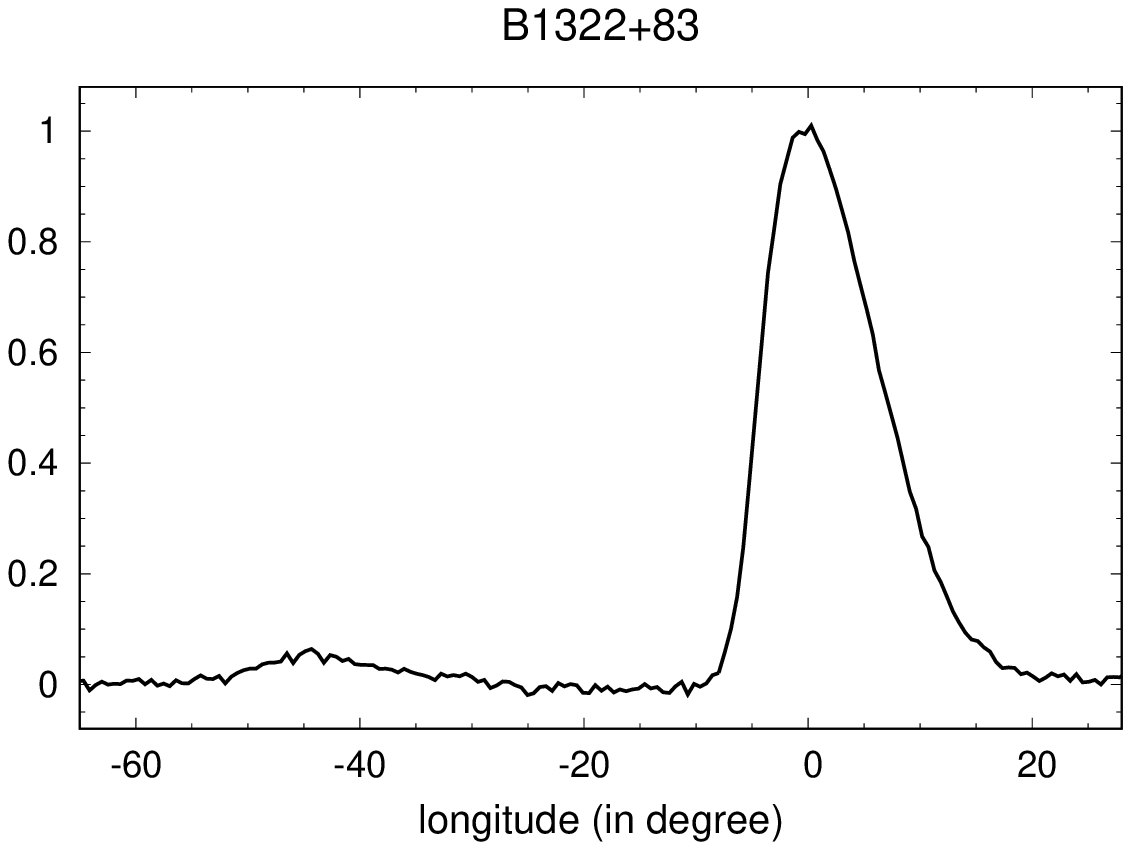}} &
\mbox{\includegraphics[angle=0,scale=0.57]{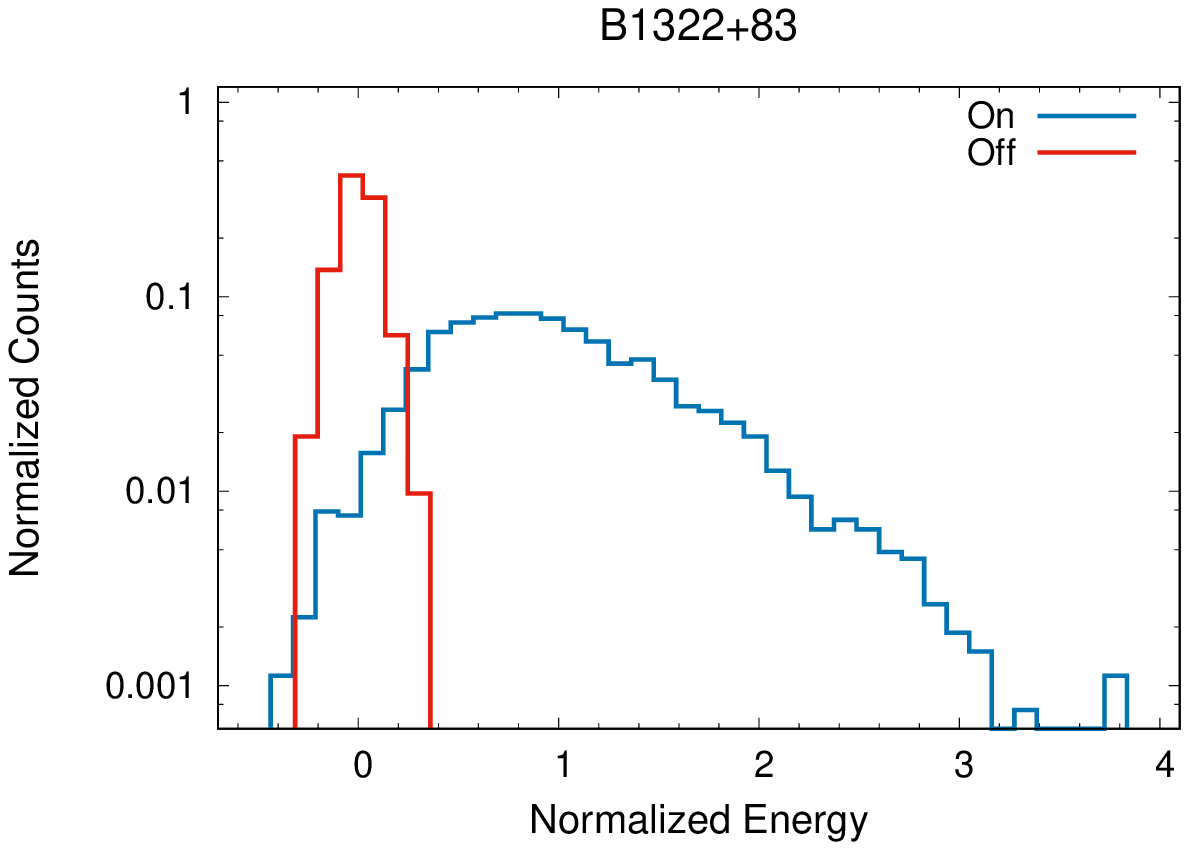}} \\
\mbox{\includegraphics[angle=0,scale=0.57]{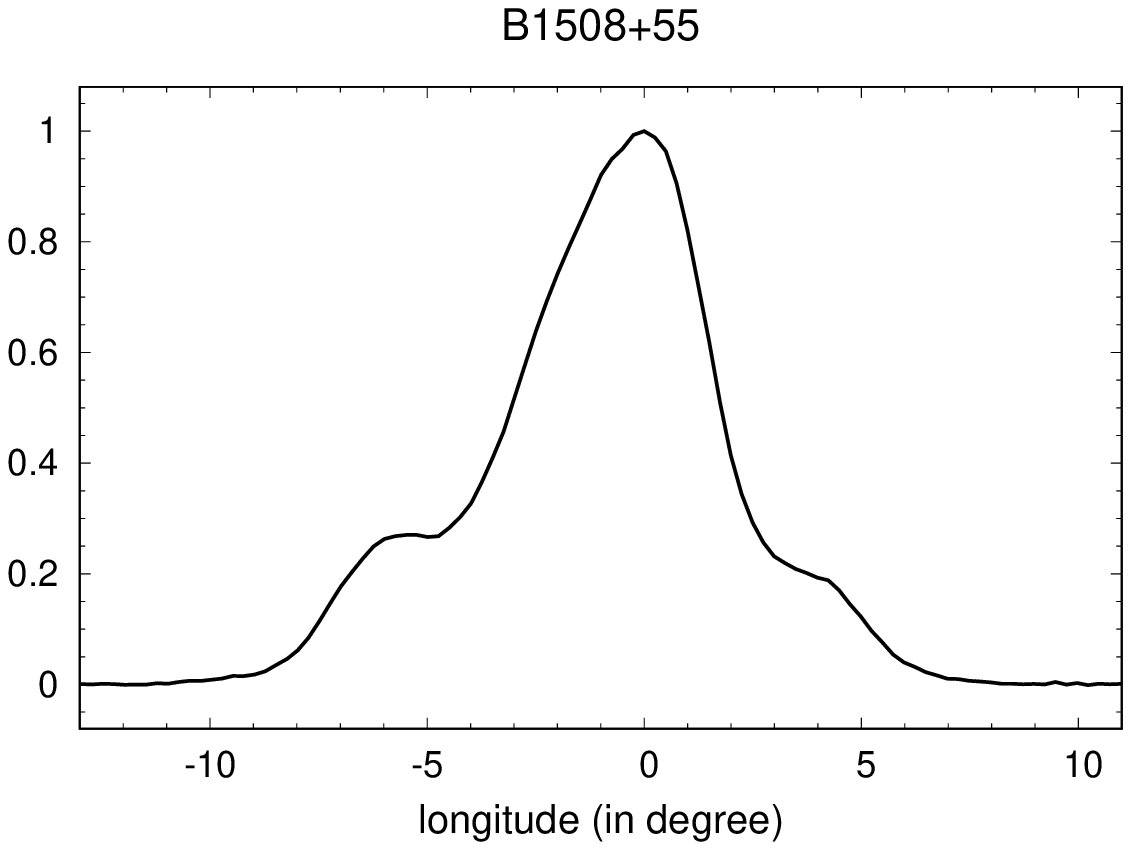}} &
\mbox{\includegraphics[angle=0,scale=0.57]{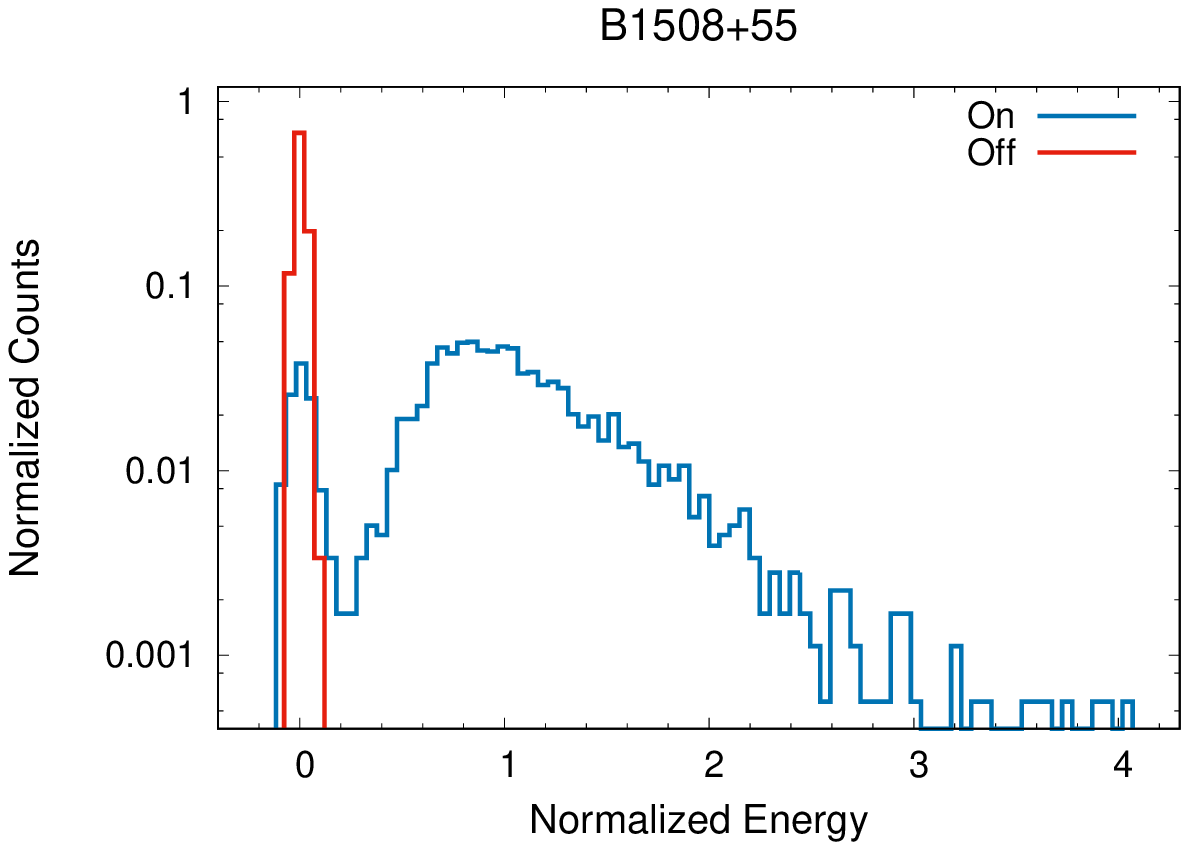}} \\
\mbox{\includegraphics[angle=0,scale=0.57]{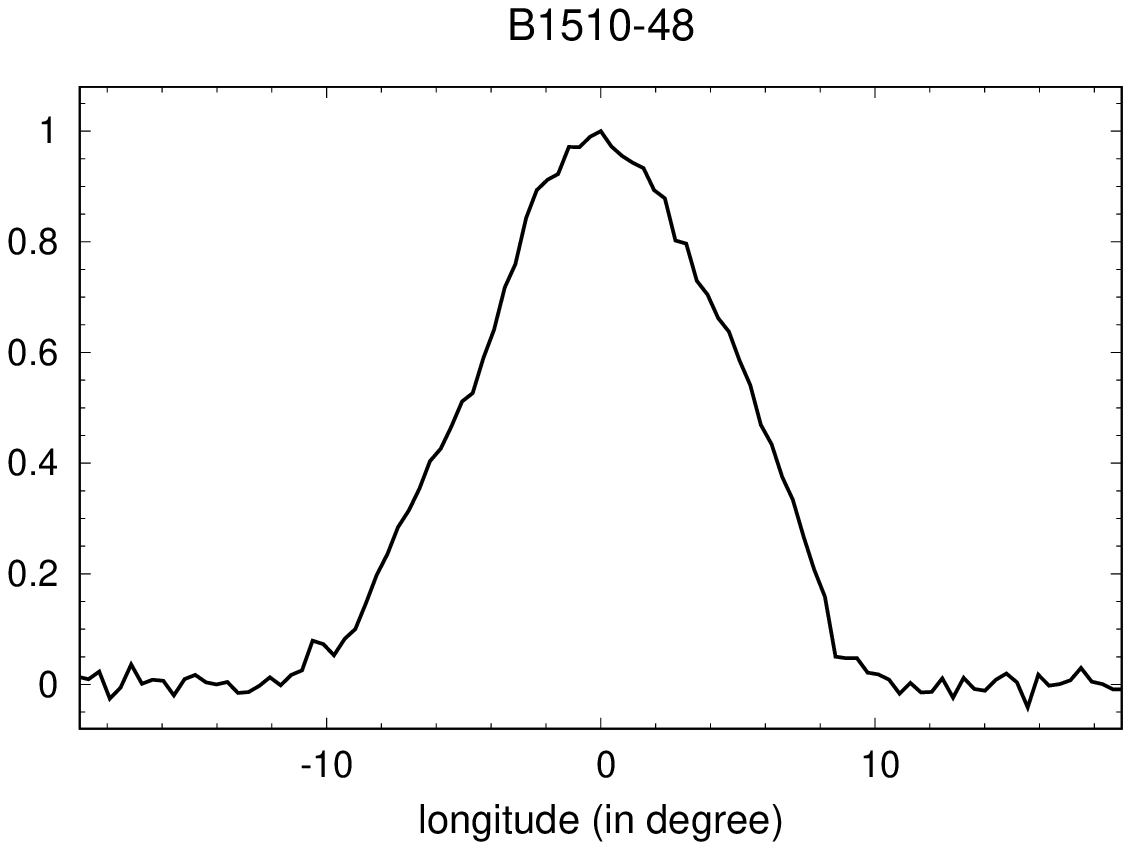}} &
\mbox{\includegraphics[angle=0,scale=0.57]{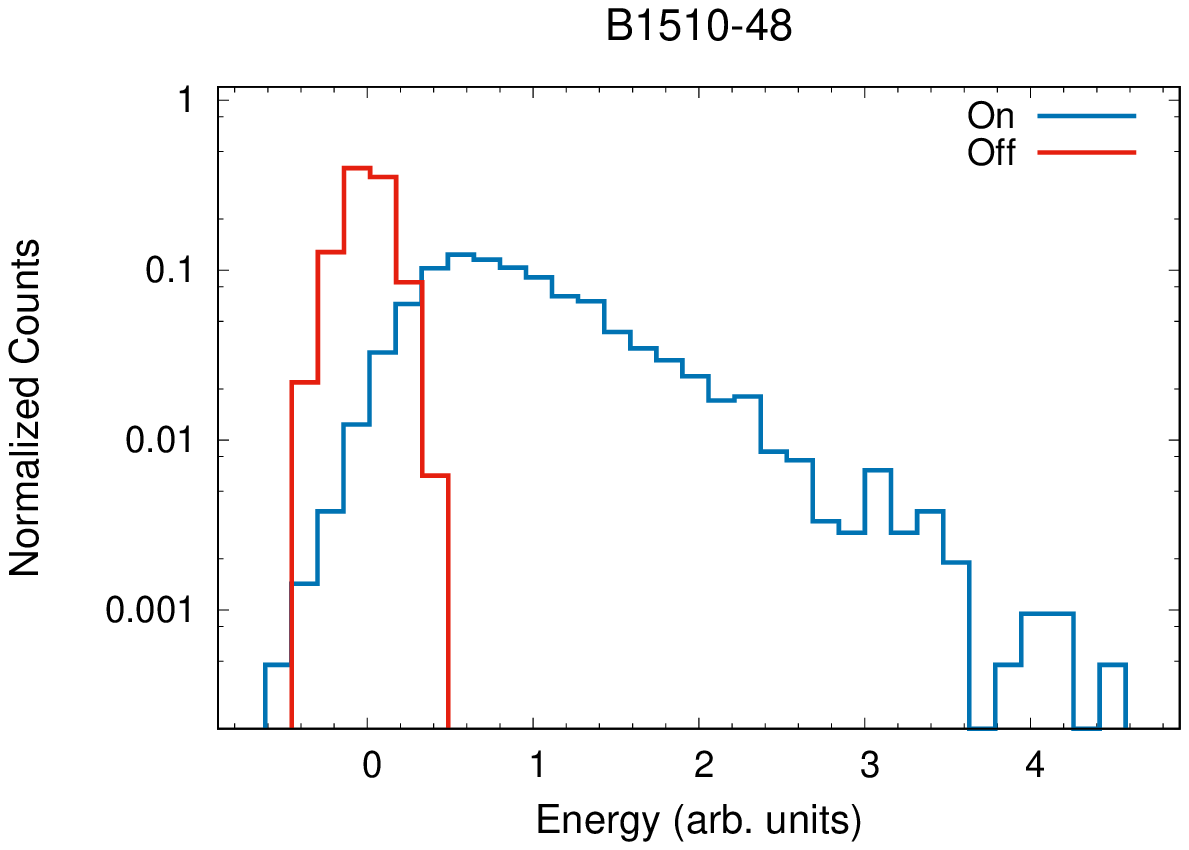}} \\
\mbox{\includegraphics[angle=0,scale=0.57]{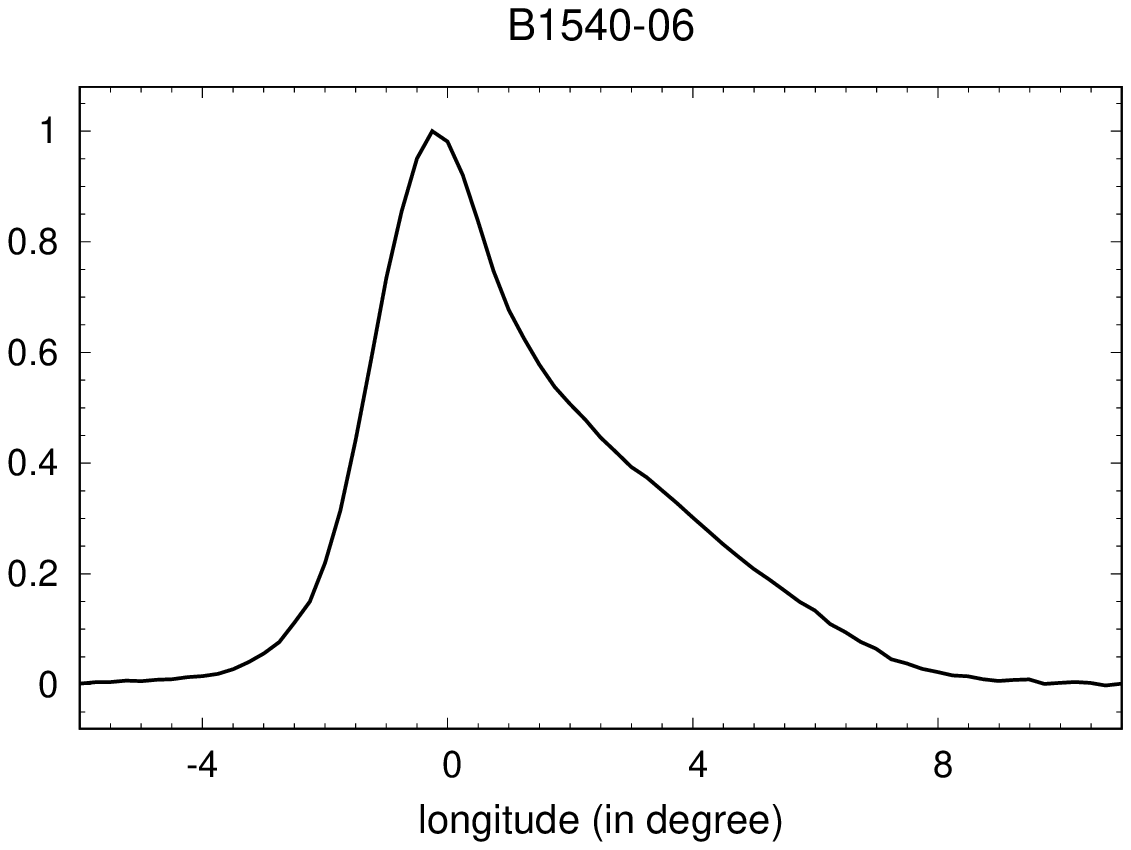}} &
\mbox{\includegraphics[angle=0,scale=0.57]{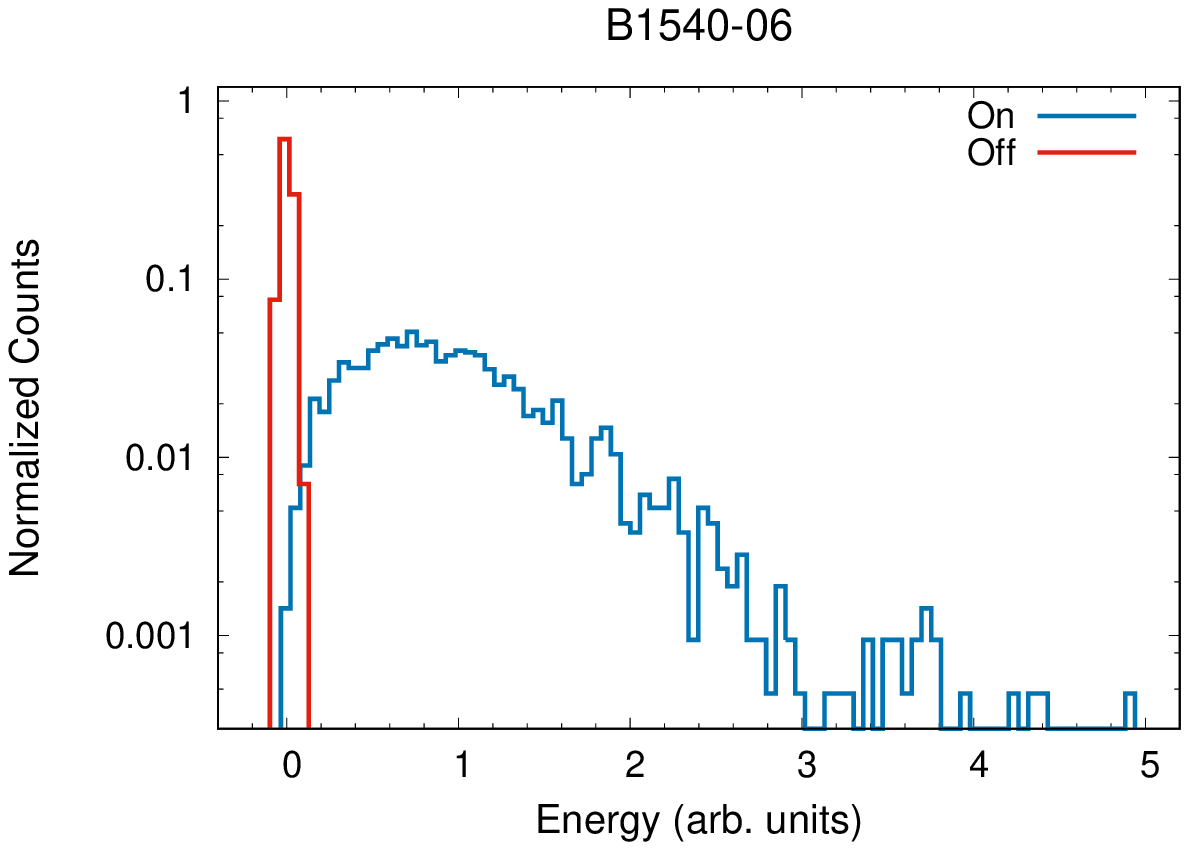}} \\
\end{tabular}
\caption{The pulsar profile and On and Off-pulse energy distributions of the single pulse emission.}
\end{center}
\end{figure*}

\clearpage

%8th set of plots
\begin{figure*}
\begin{center}
\begin{tabular}{@{}cr@{}}
\mbox{\includegraphics[angle=0,scale=0.57]{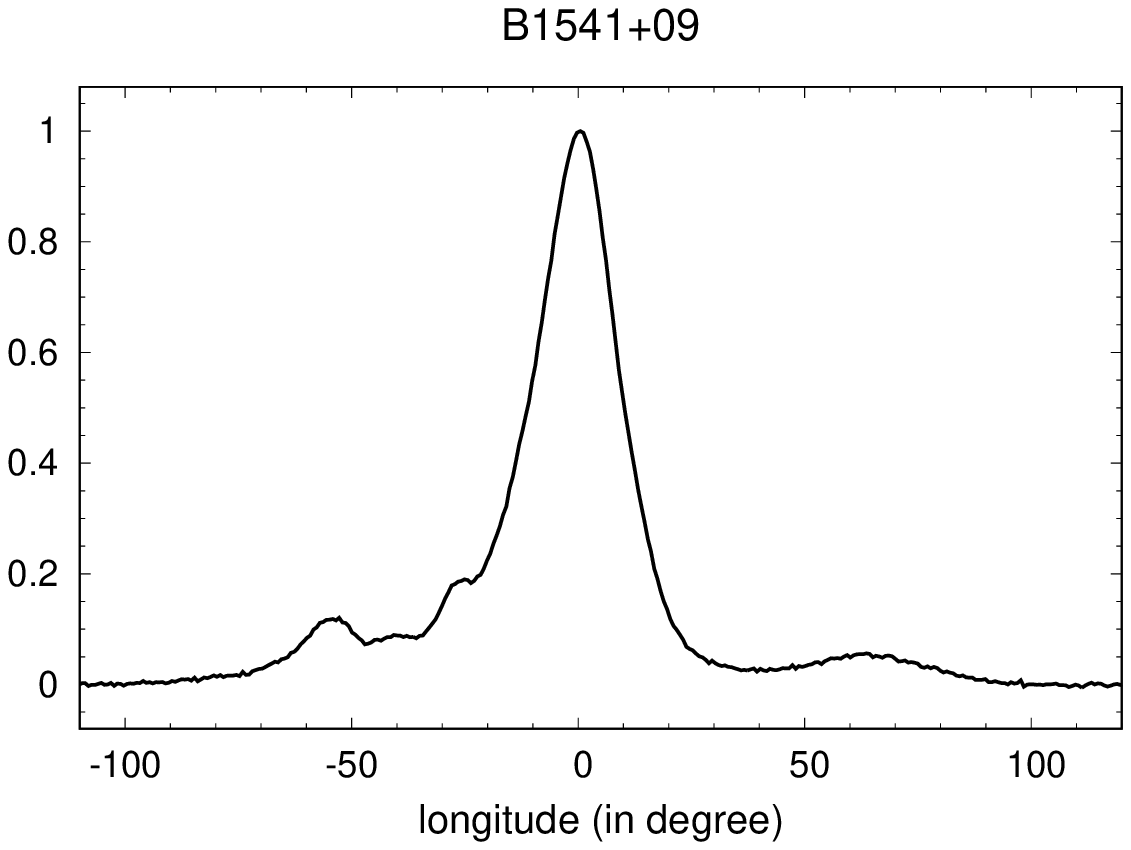}} &
\mbox{\includegraphics[angle=0,scale=0.57]{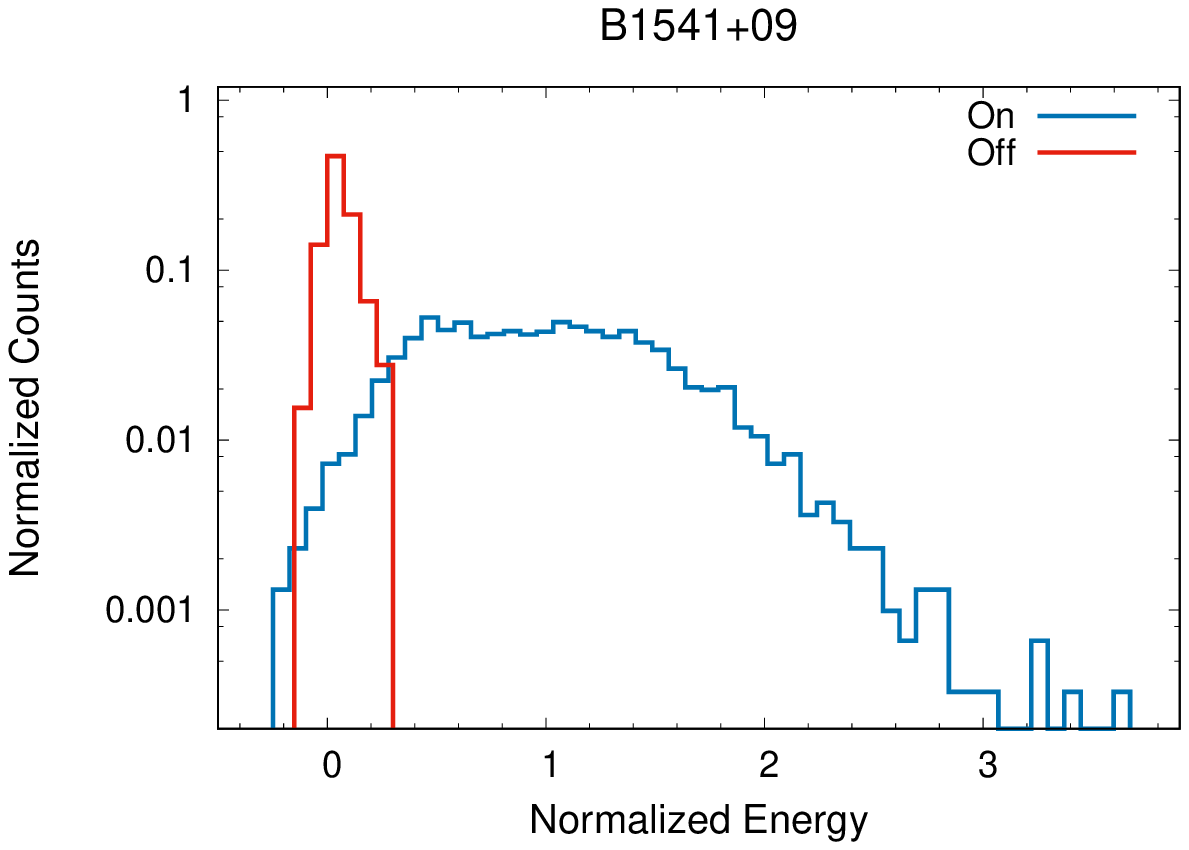}} \\
\mbox{\includegraphics[angle=0,scale=0.57]{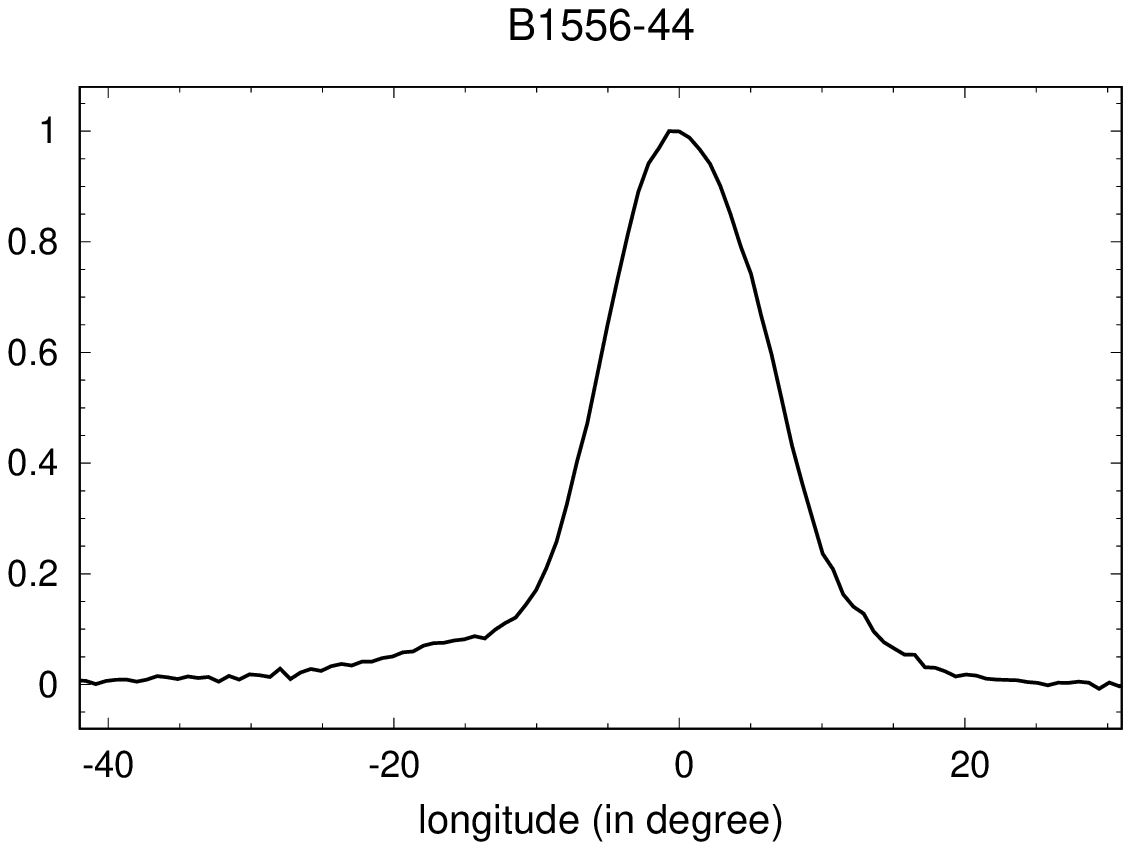}} &
\mbox{\includegraphics[angle=0,scale=0.57]{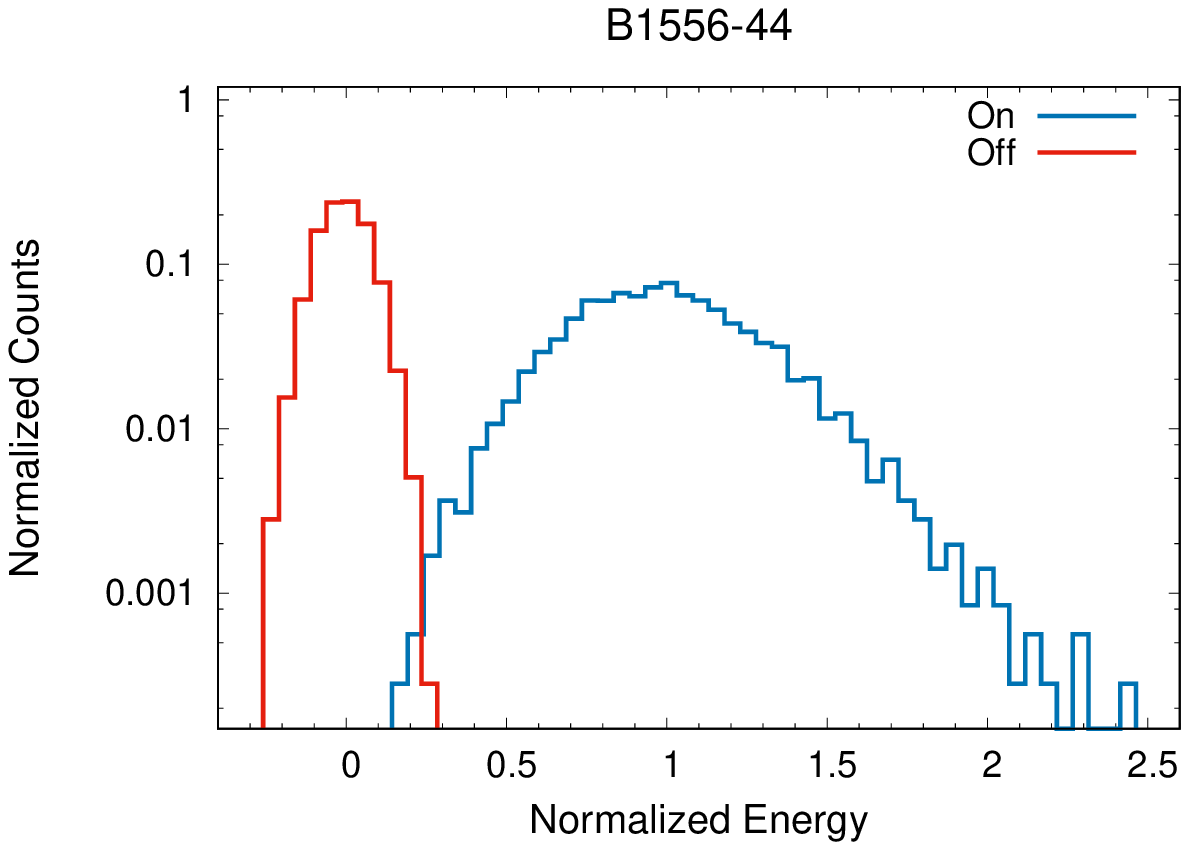}} \\
\mbox{\includegraphics[angle=0,scale=0.57]{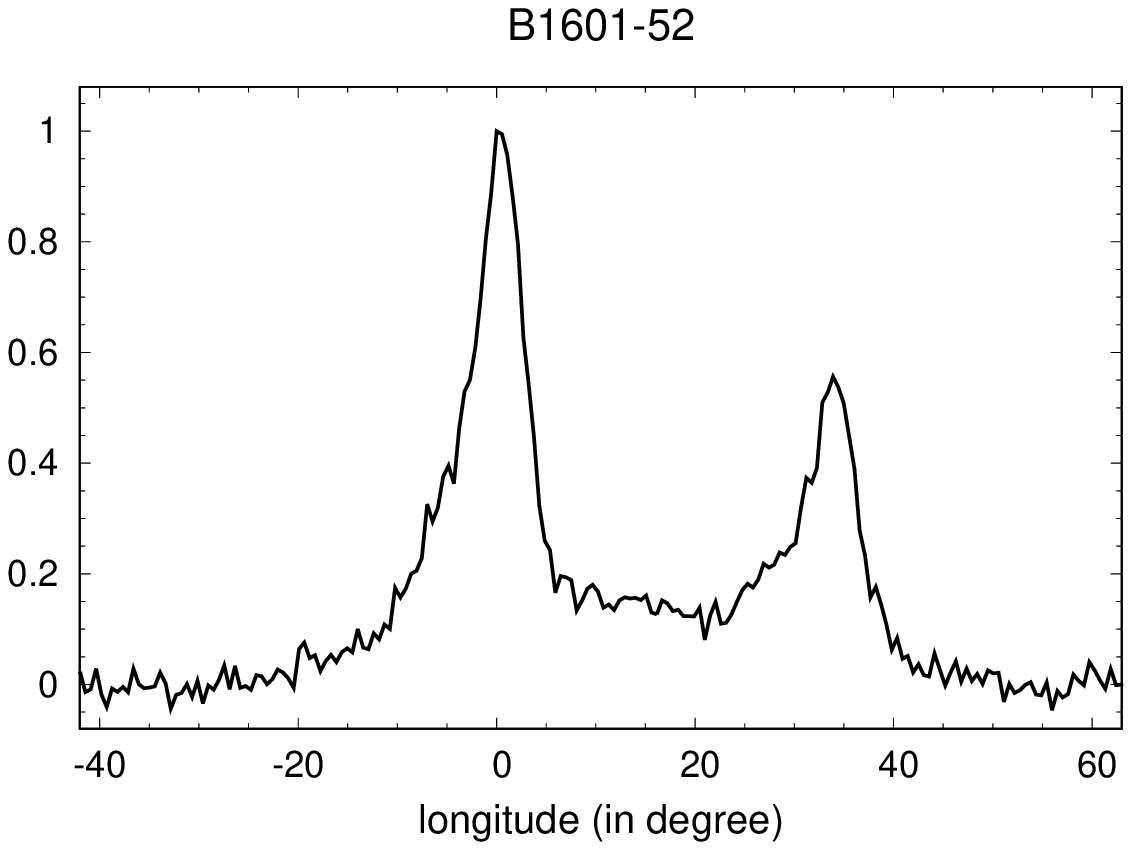}} &
\mbox{\includegraphics[angle=0,scale=0.57]{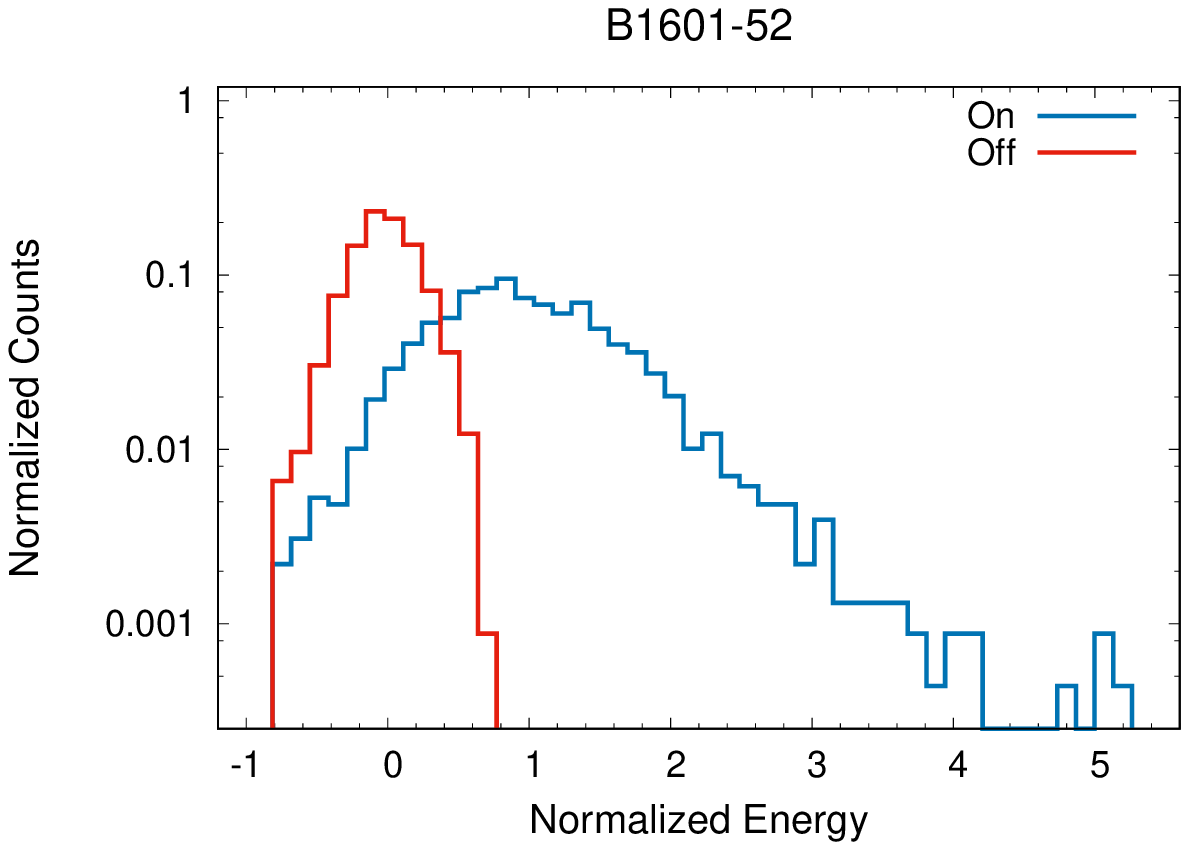}} \\
\mbox{\includegraphics[angle=0,scale=0.57]{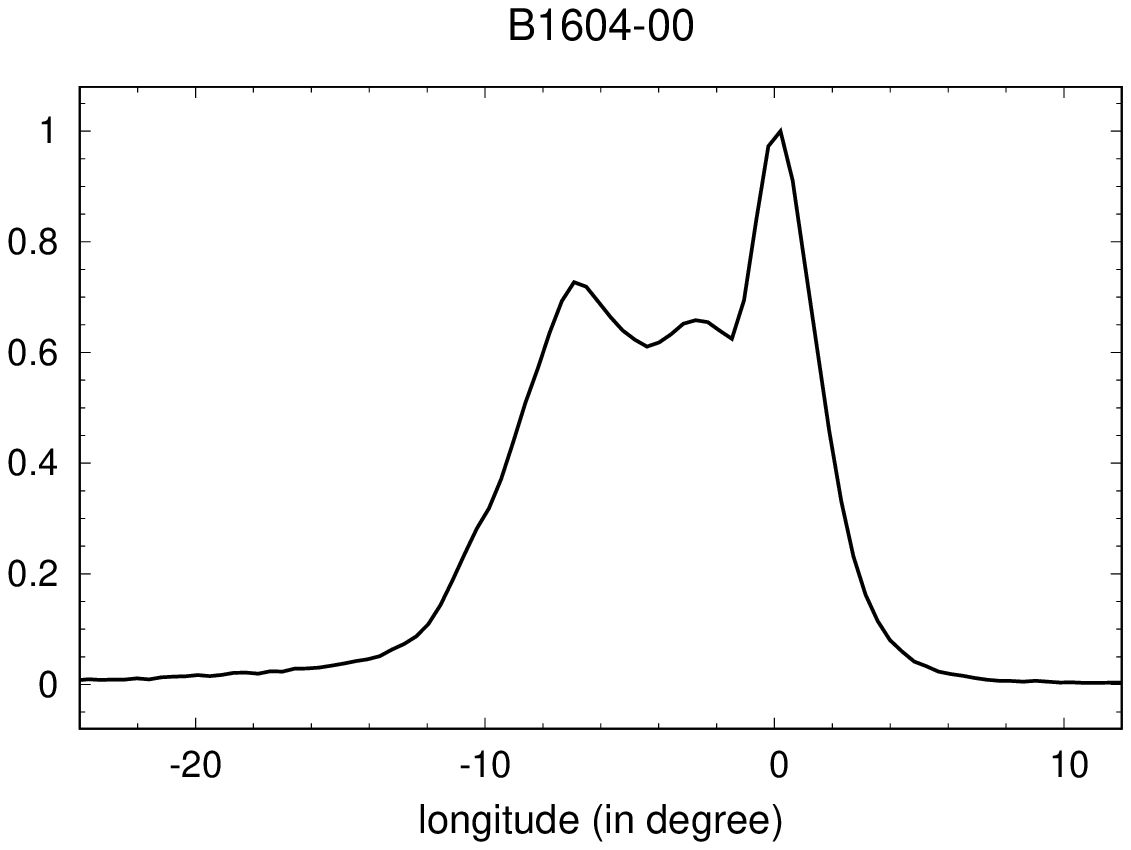}} &
\mbox{\includegraphics[angle=0,scale=0.57]{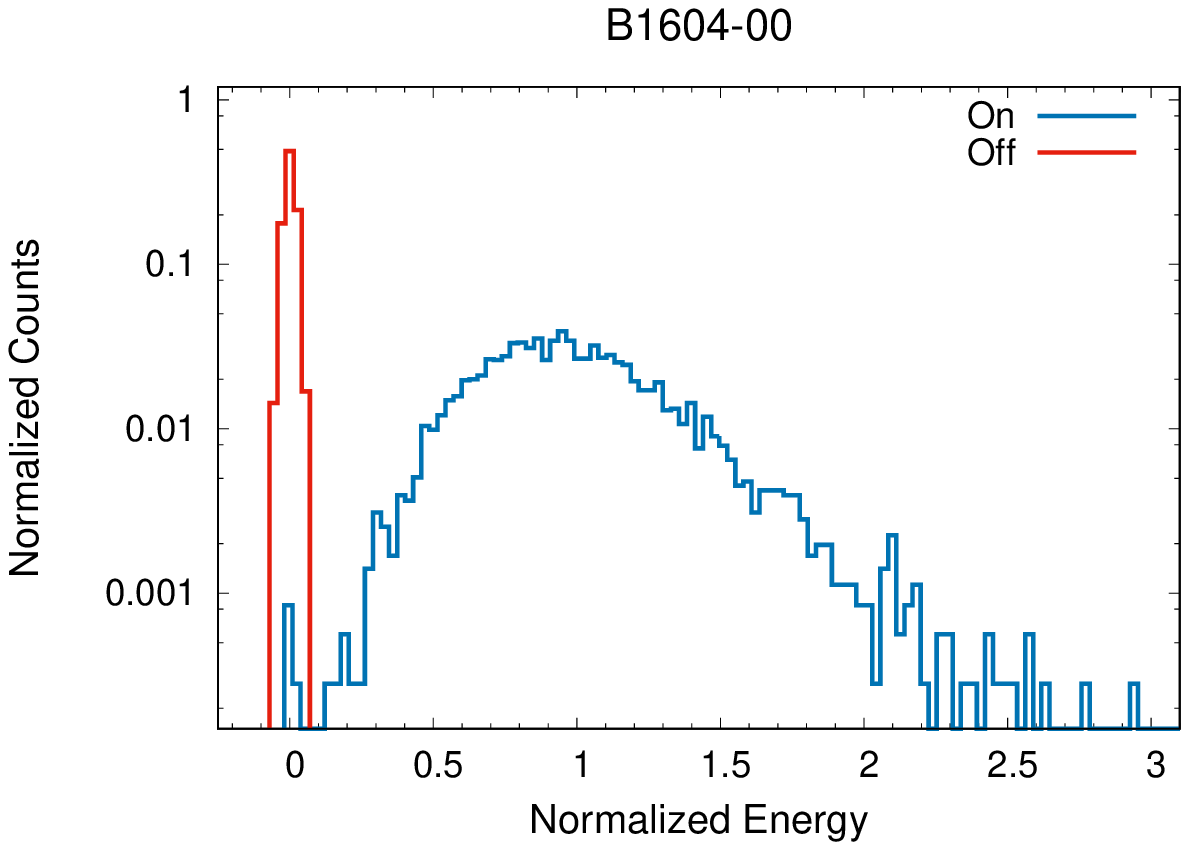}} \\
\end{tabular}
\caption{The pulsar profile and On and Off-pulse energy distributions of the single pulse emission.}
\end{center}
\end{figure*}

\clearpage

%9th set of plots
\begin{figure*}
\begin{center}
\begin{tabular}{@{}cr@{}}
\mbox{\includegraphics[angle=0,scale=0.57]{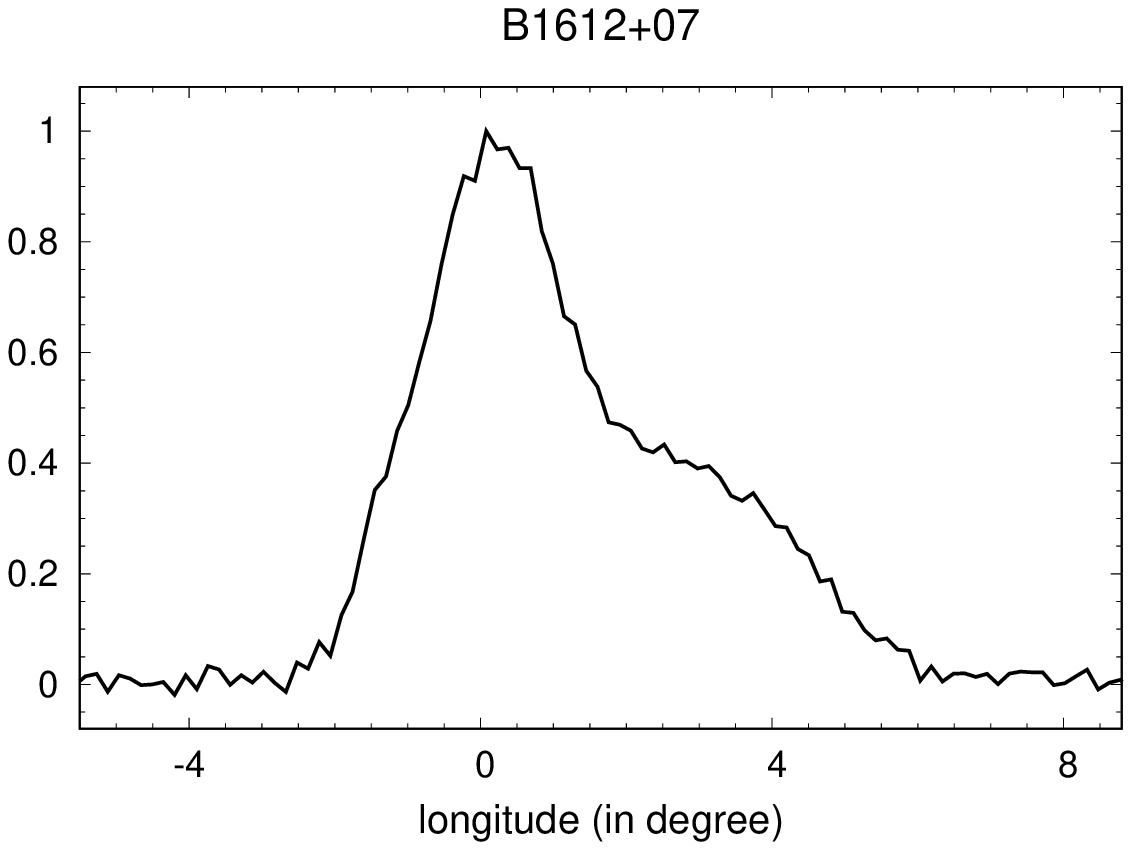}} &
\mbox{\includegraphics[angle=0,scale=0.57]{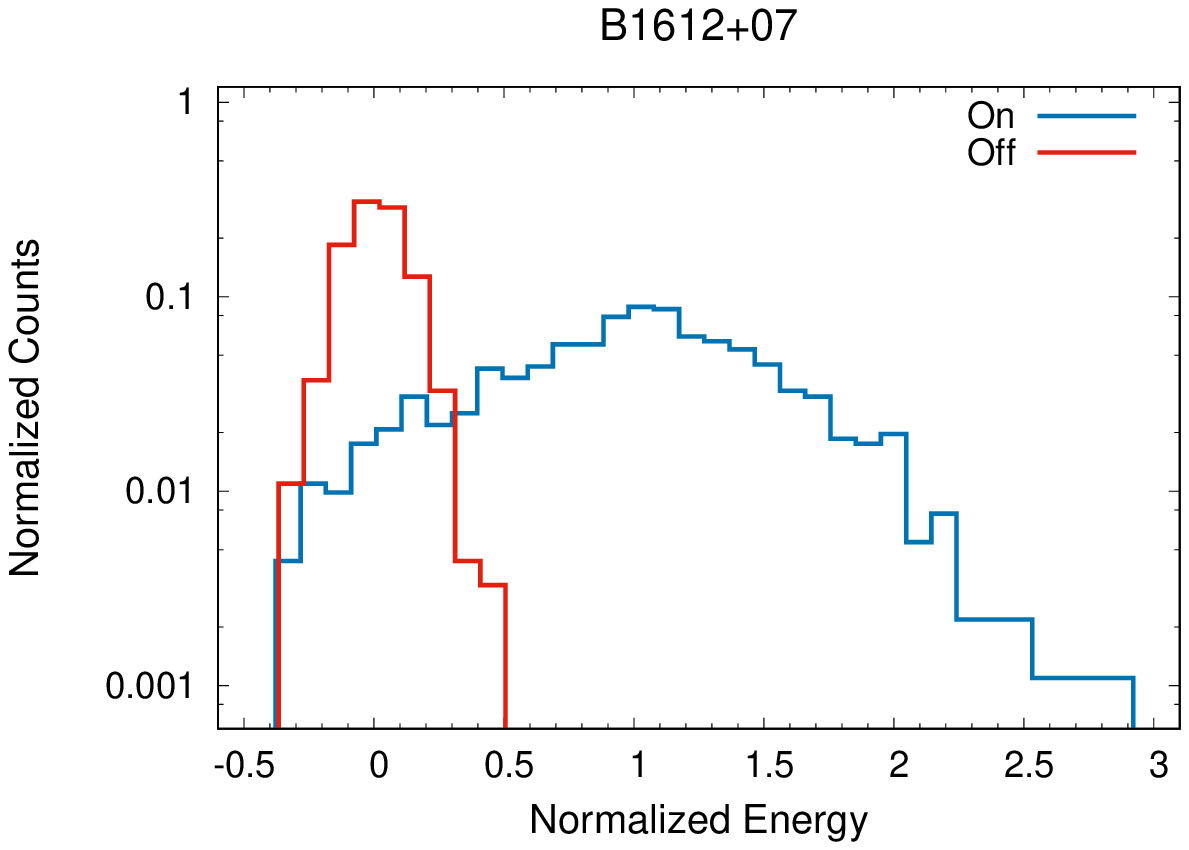}} \\
\mbox{\includegraphics[angle=0,scale=0.57]{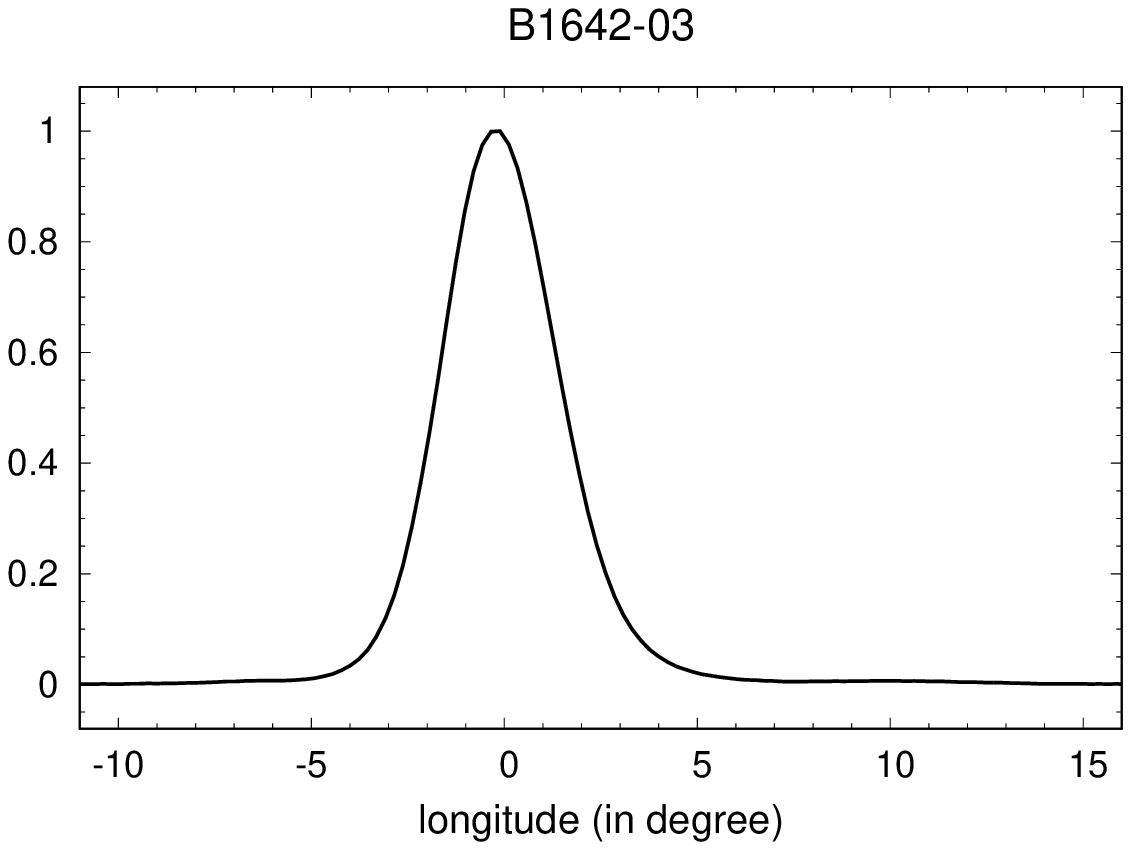}} &
\mbox{\includegraphics[angle=0,scale=0.57]{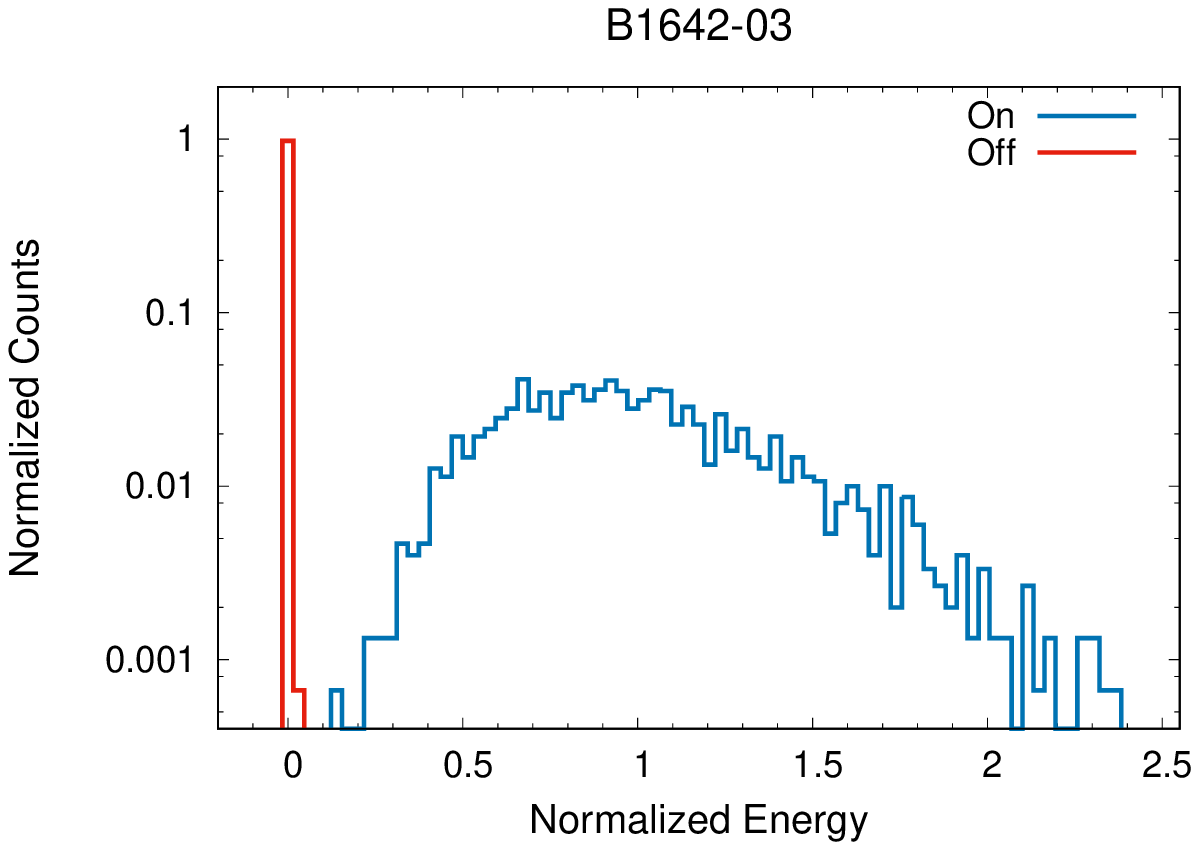}} \\
\mbox{\includegraphics[angle=0,scale=0.57]{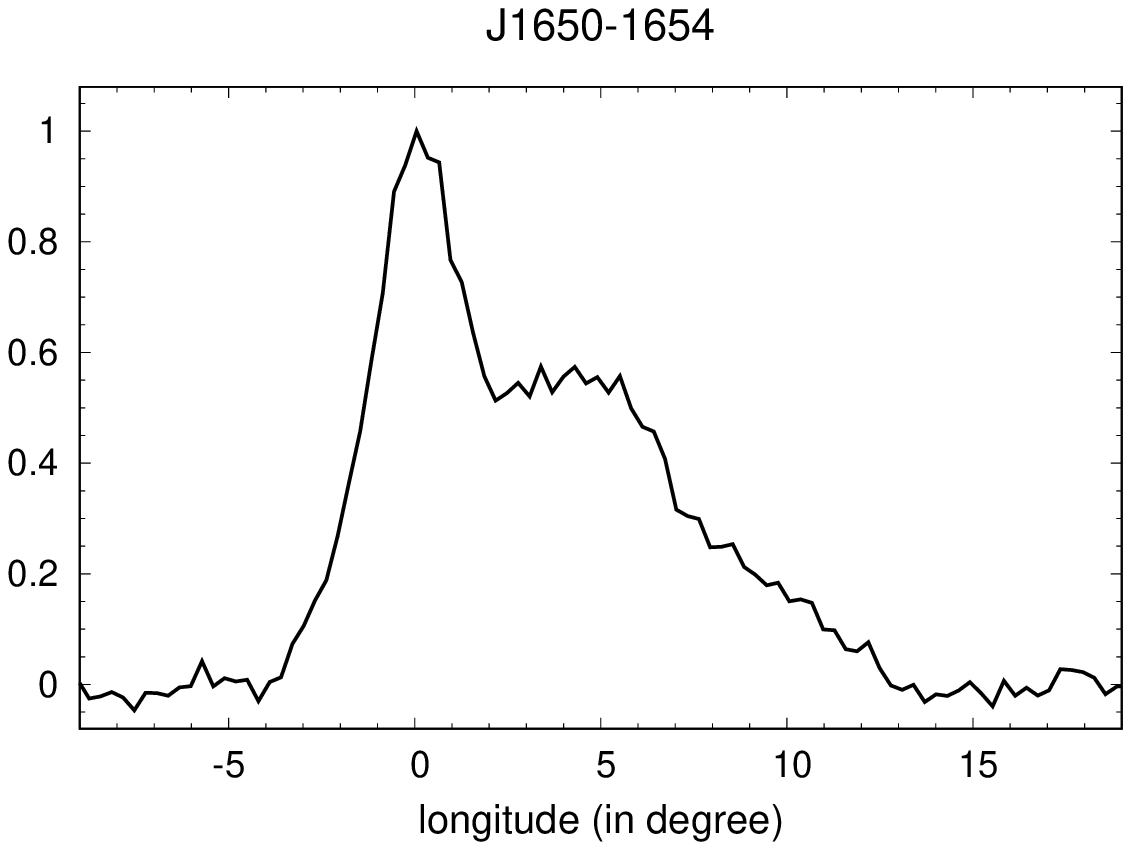}} &
\mbox{\includegraphics[angle=0,scale=0.57]{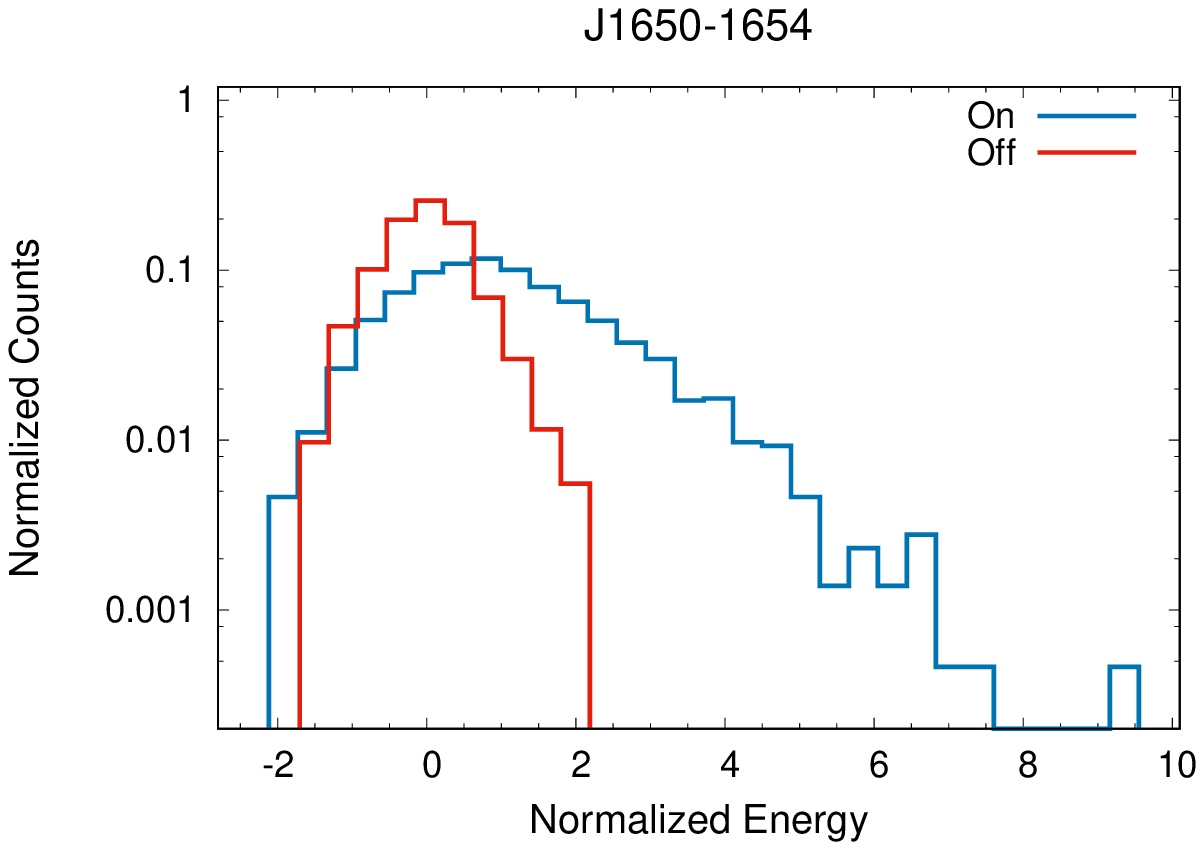}} \\
\mbox{\includegraphics[angle=0,scale=0.57]{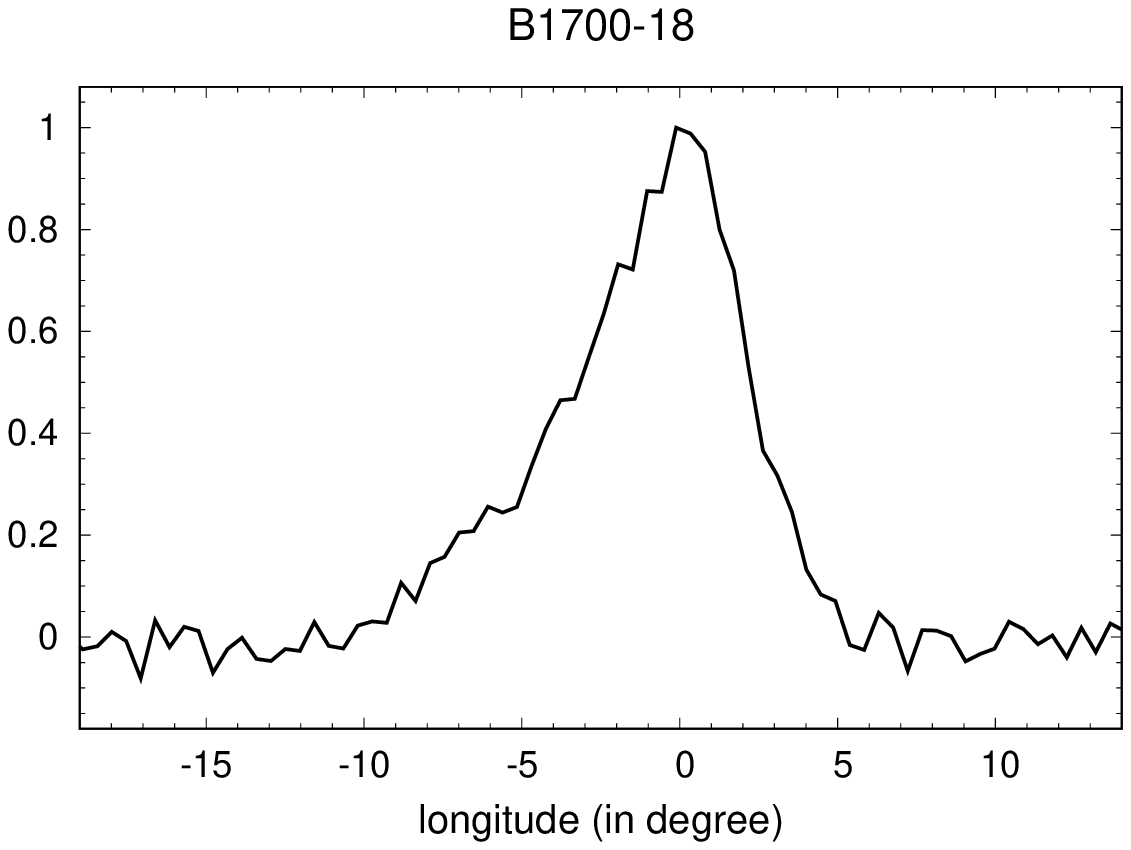}} &
\mbox{\includegraphics[angle=0,scale=0.57]{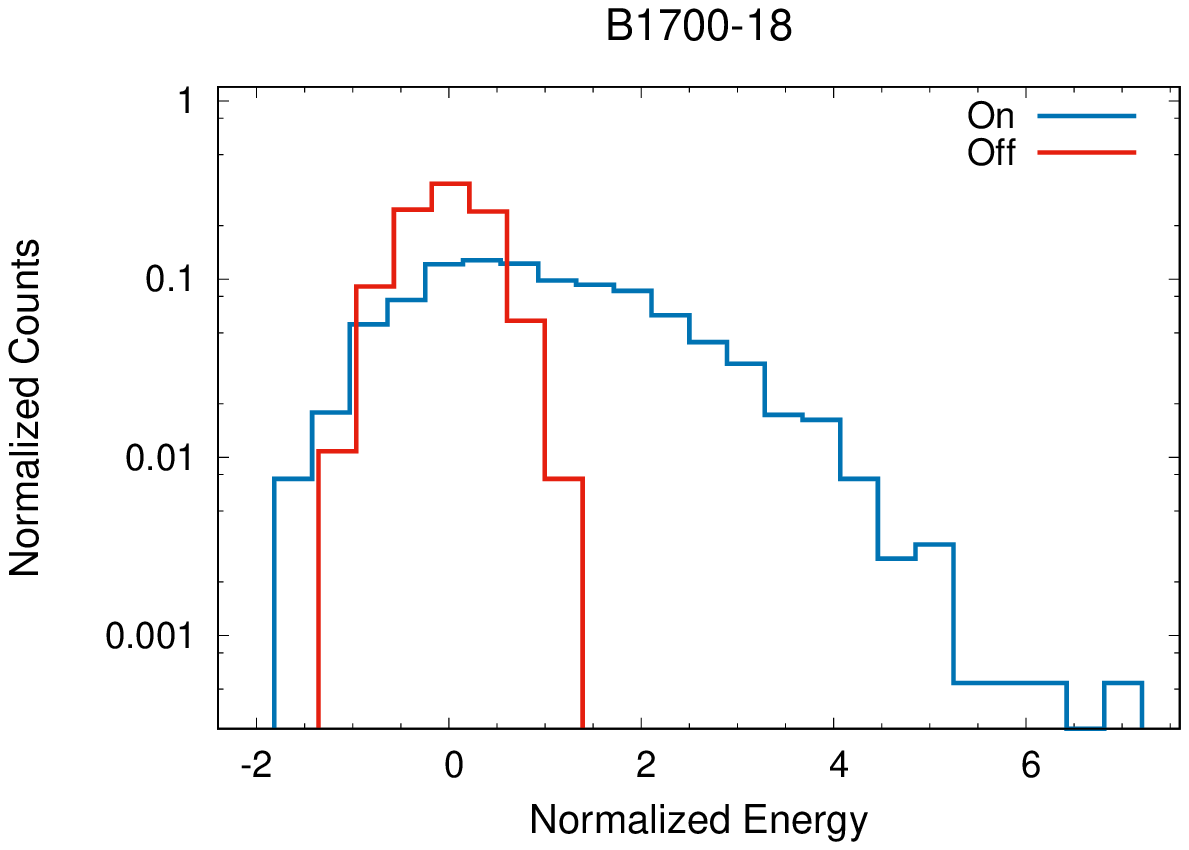}} \\
\end{tabular}
\caption{The pulsar profile On and Off-pulse energy distributions of the single pulse emission.}
\end{center}
\end{figure*}

\clearpage

%10th set of plots
\begin{figure*}
\begin{center}
\begin{tabular}{@{}cr@{}}
\mbox{\includegraphics[angle=0,scale=0.57]{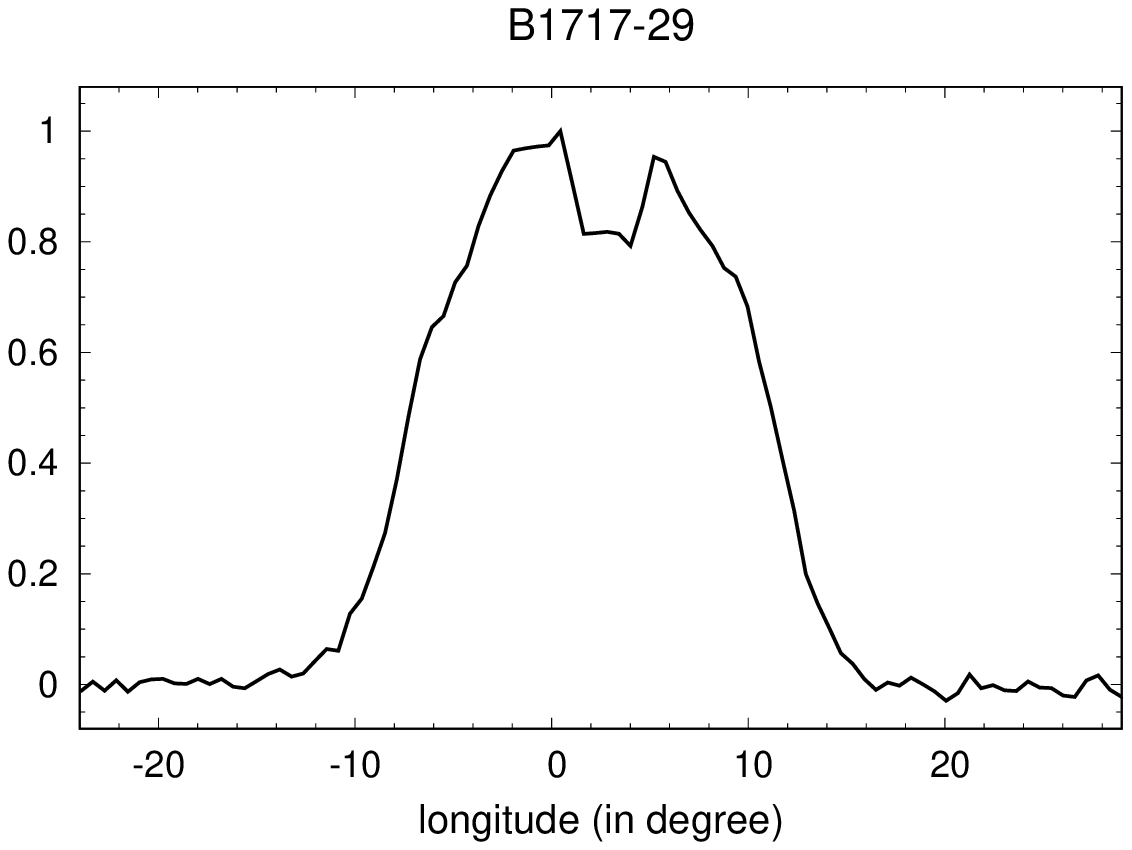}} &
\mbox{\includegraphics[angle=0,scale=0.57]{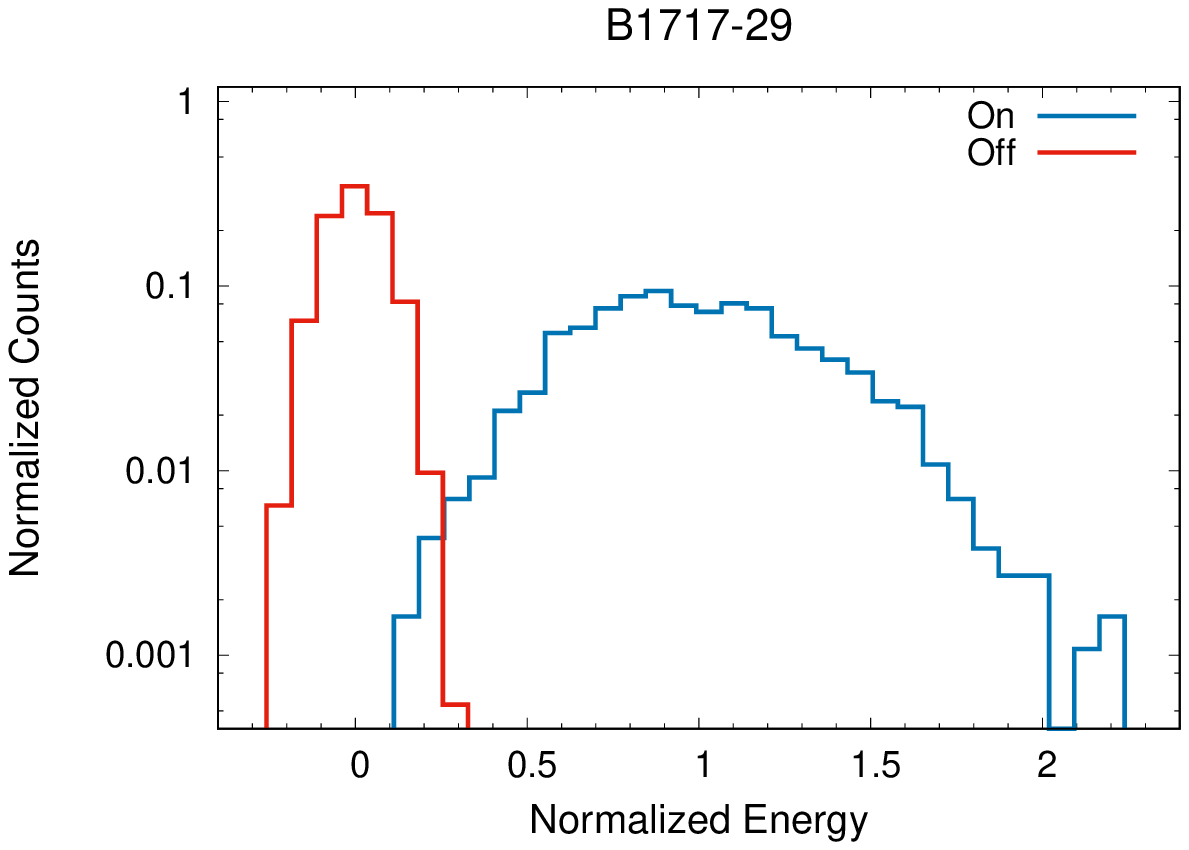}} \\
\mbox{\includegraphics[angle=0,scale=0.57]{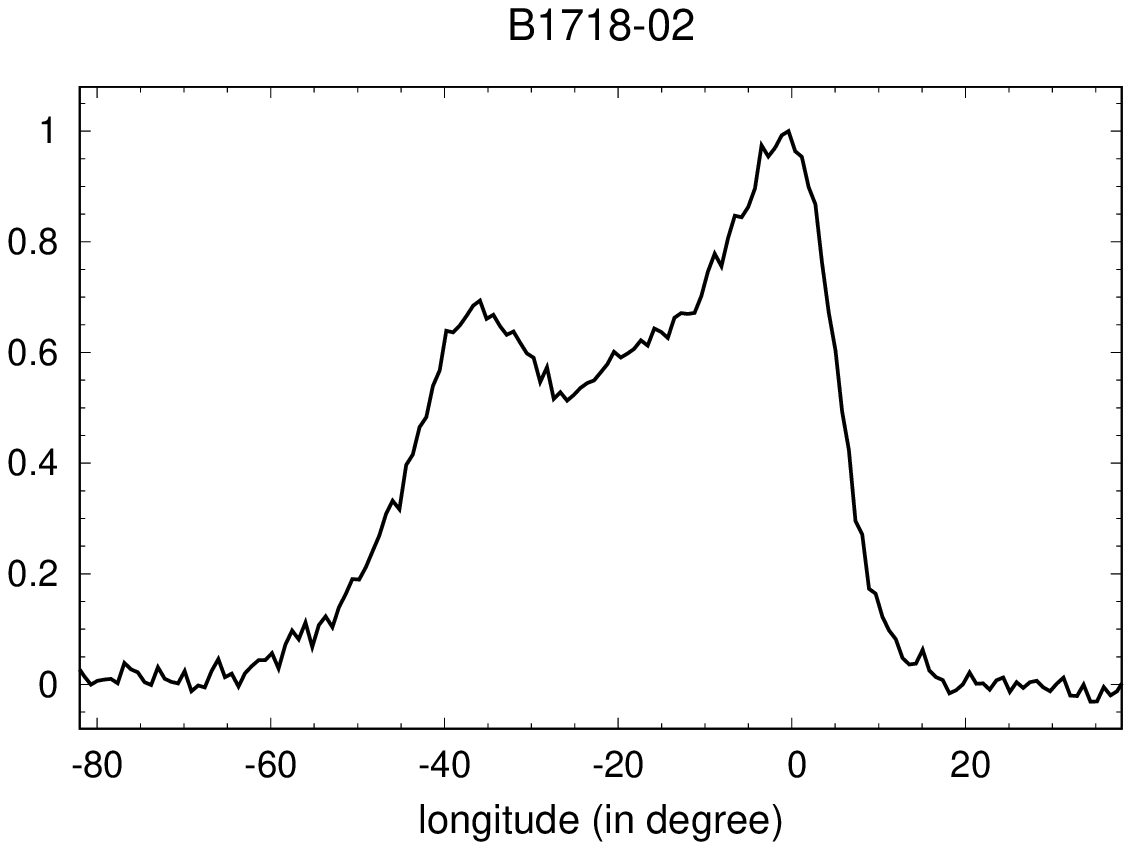}} &
\mbox{\includegraphics[angle=0,scale=0.57]{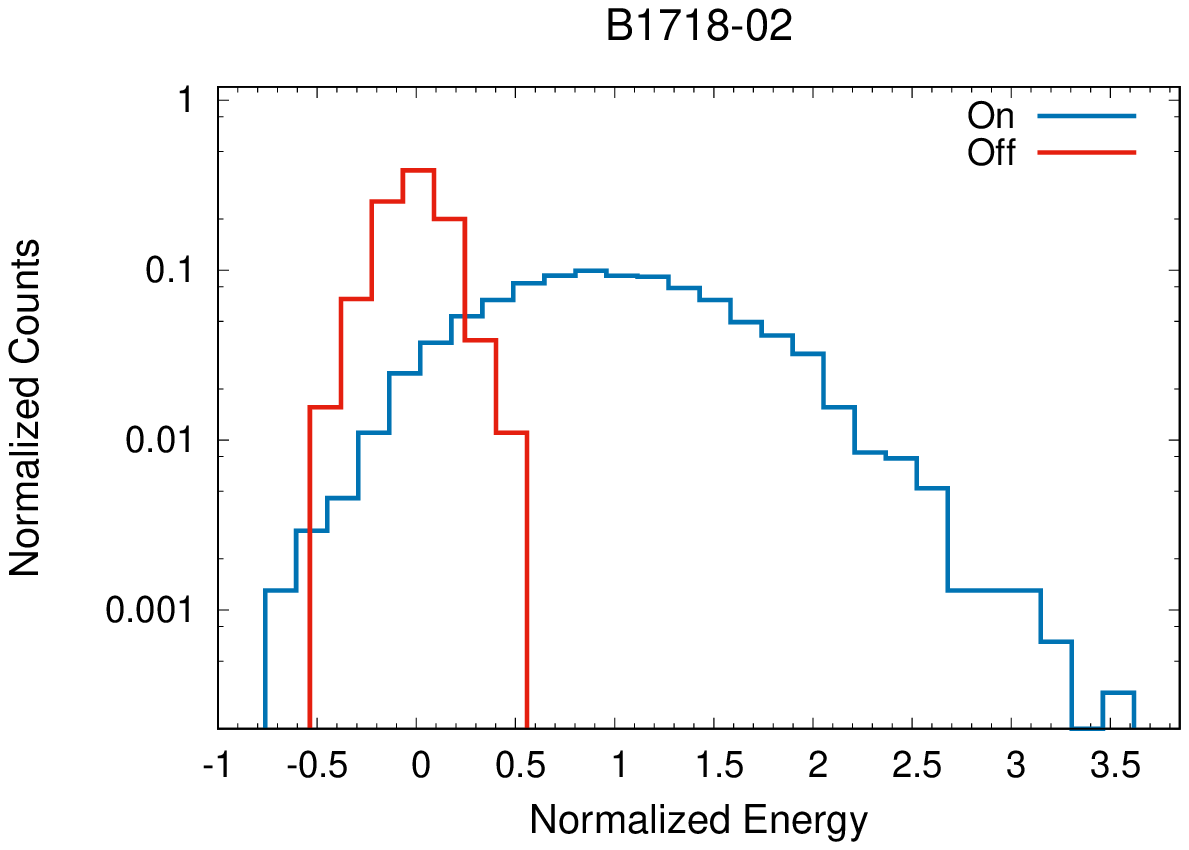}} \\
\mbox{\includegraphics[angle=0,scale=0.57]{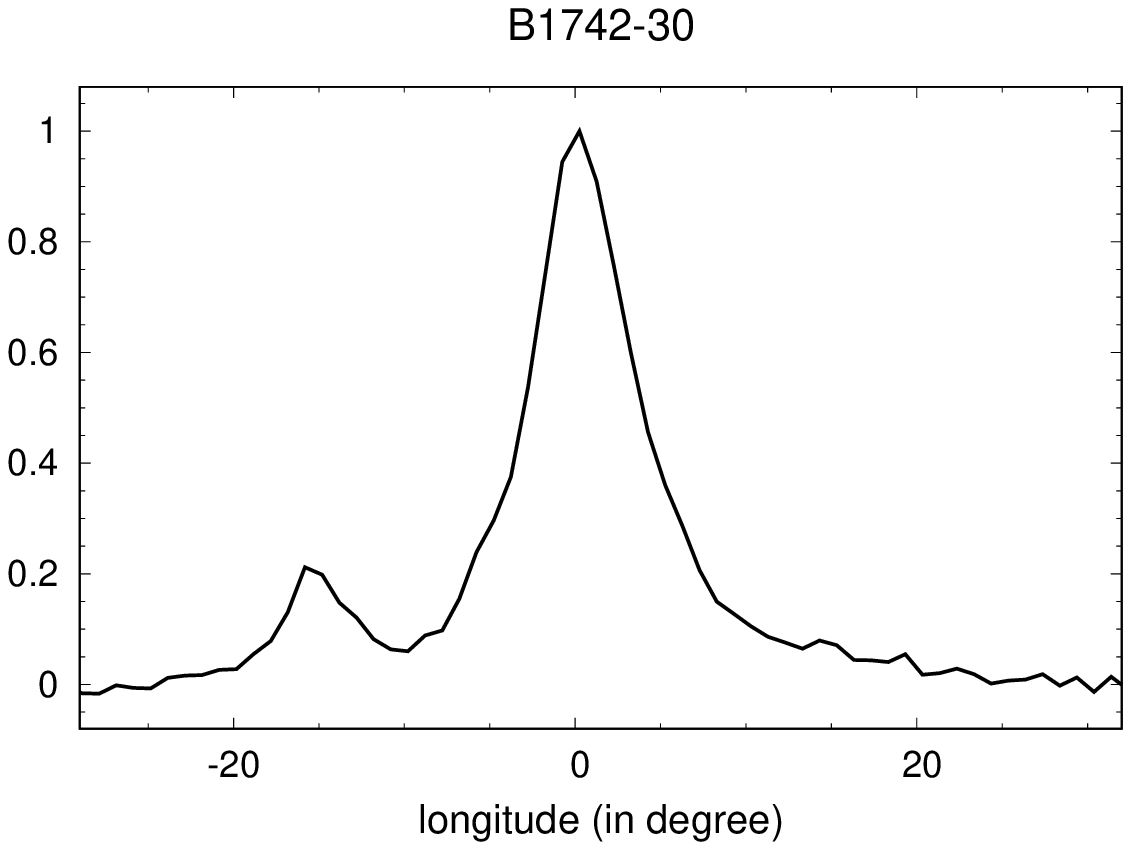}} &
\mbox{\includegraphics[angle=0,scale=0.57]{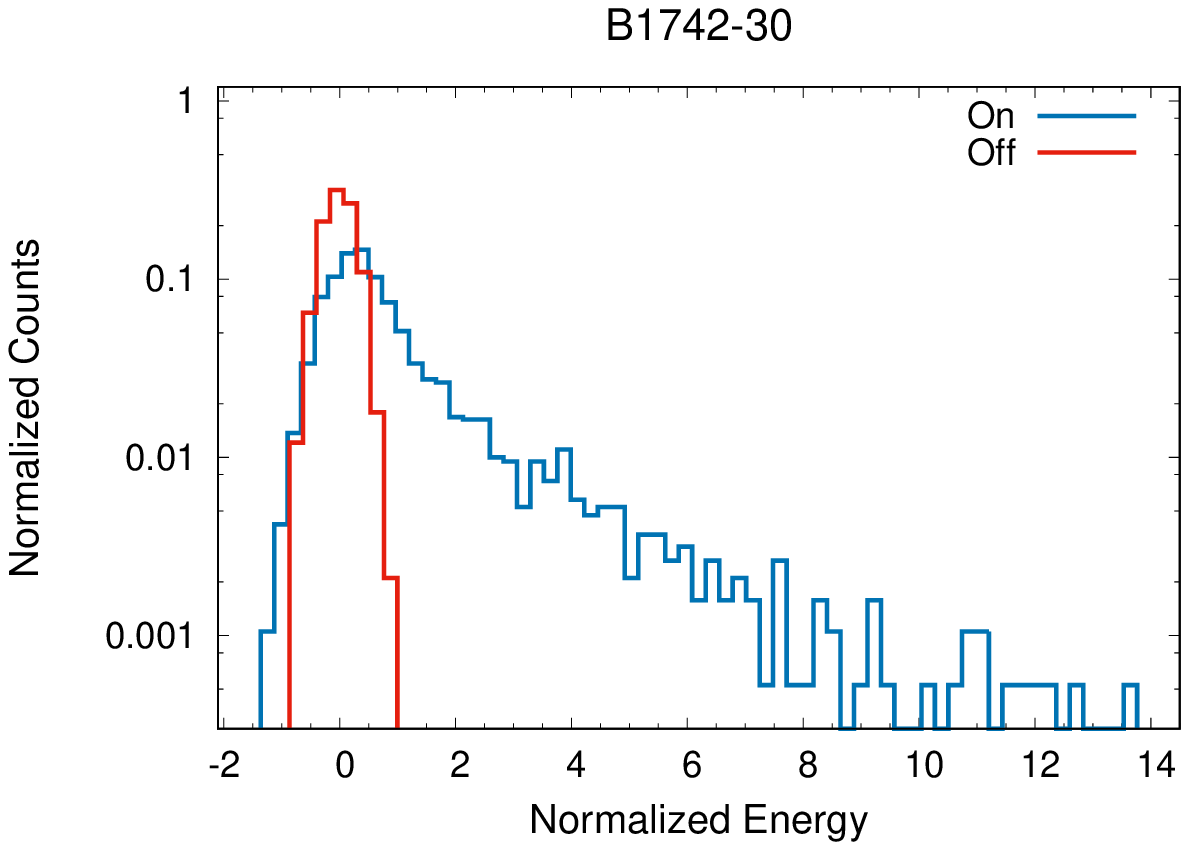}} \\
\mbox{\includegraphics[angle=0,scale=0.57]{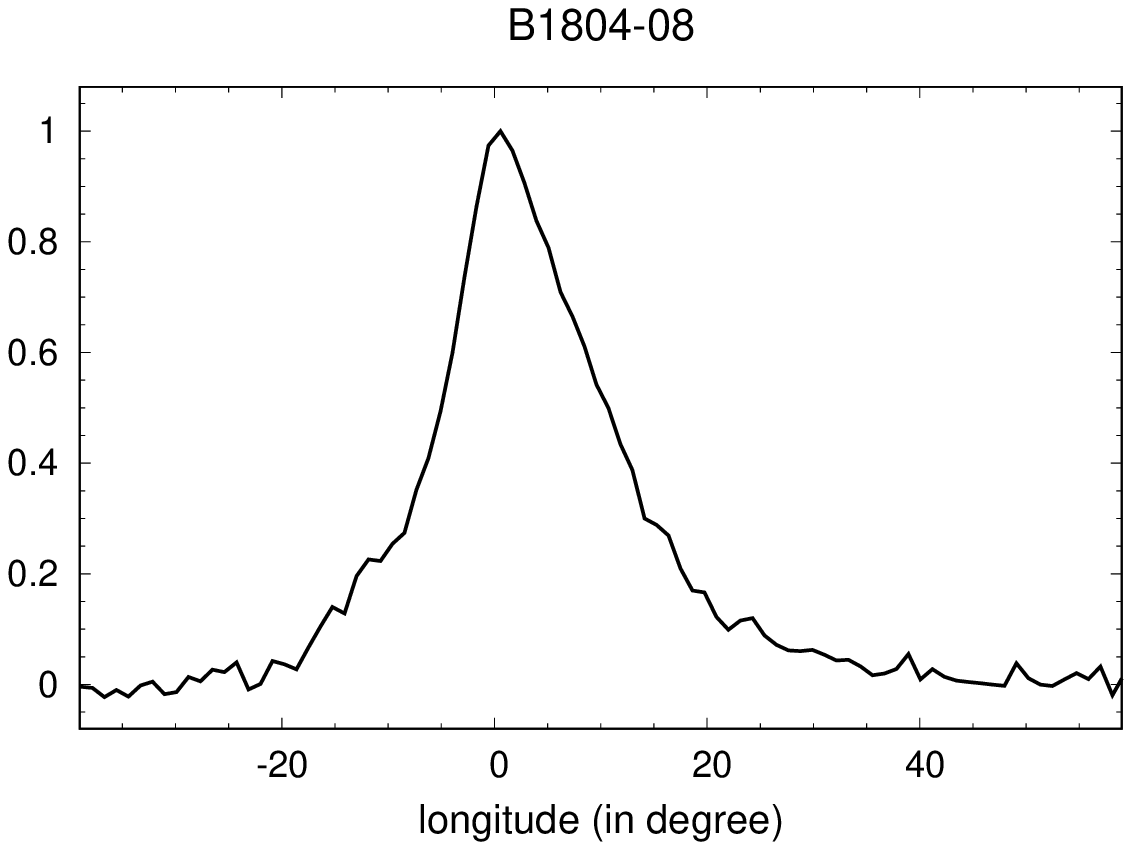}} &
\mbox{\includegraphics[angle=0,scale=0.57]{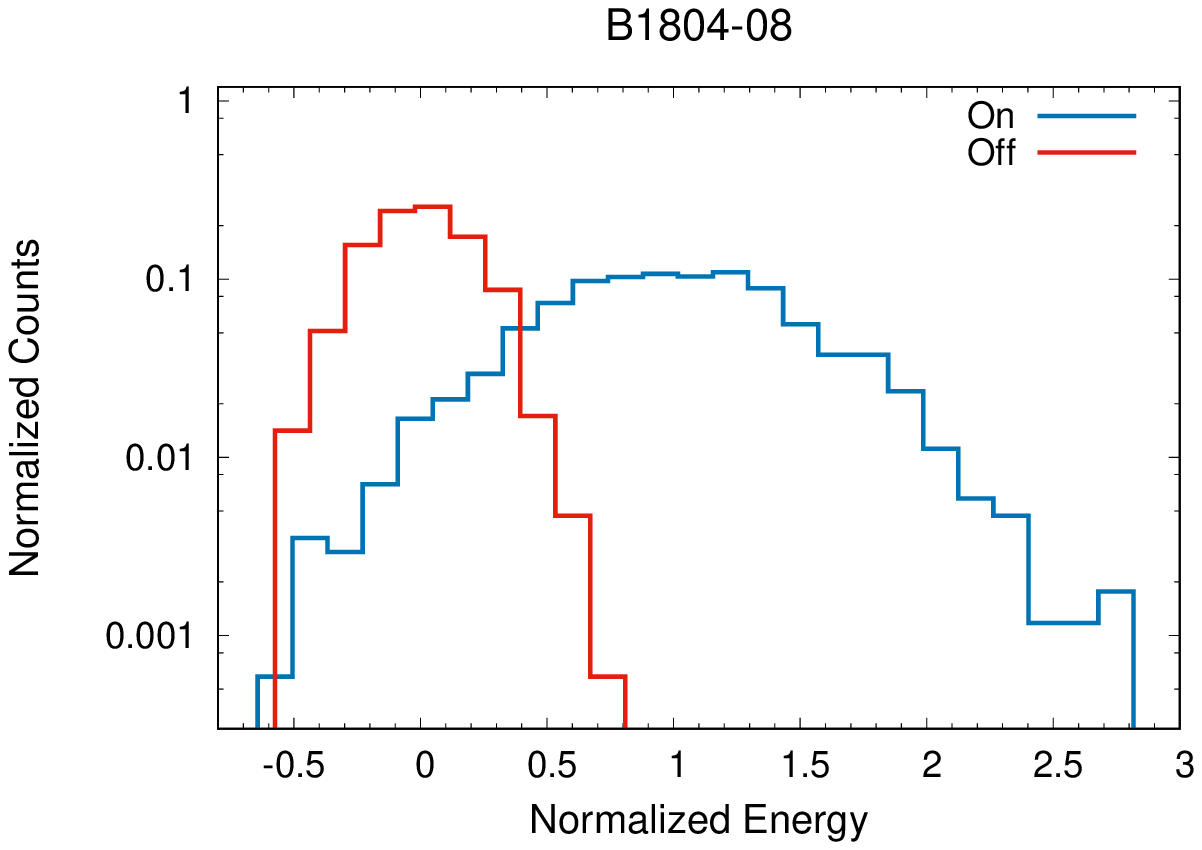}} \\
\end{tabular}
\caption{The pulsar profile On and Off-pulse energy distributions of the single pulse emission.}
\end{center}
\end{figure*}

\clearpage

%11th set of plots
\begin{figure*}
\begin{center}
\begin{tabular}{@{}cr@{}}
\mbox{\includegraphics[angle=0,scale=0.57]{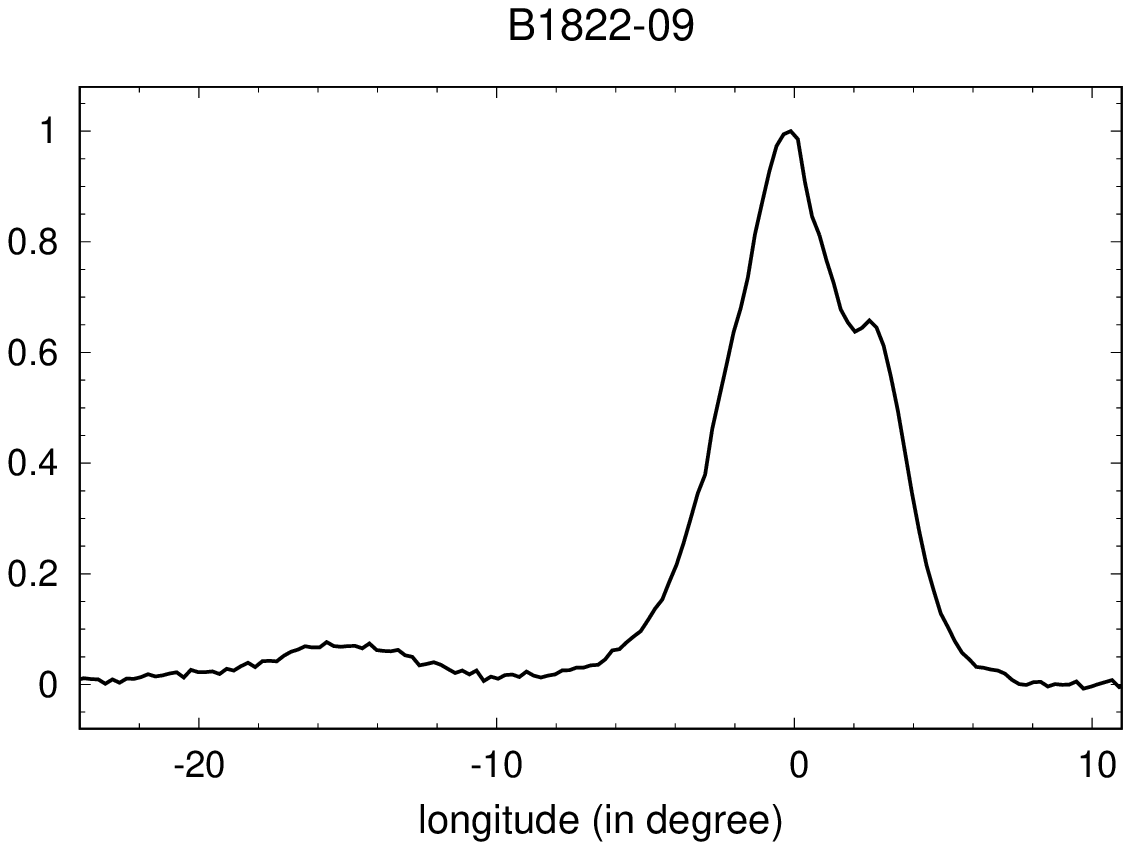}} &
\mbox{\includegraphics[angle=0,scale=0.57]{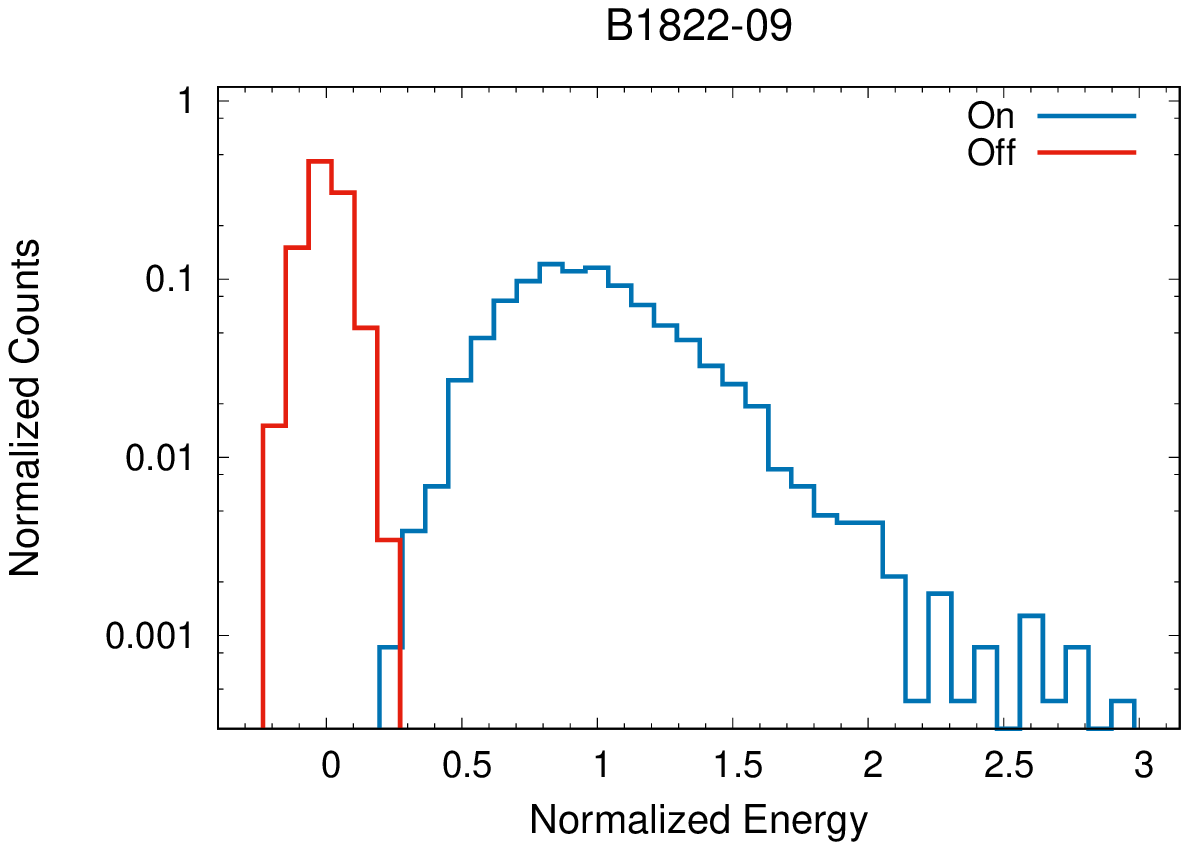}} \\
\mbox{\includegraphics[angle=0,scale=0.57]{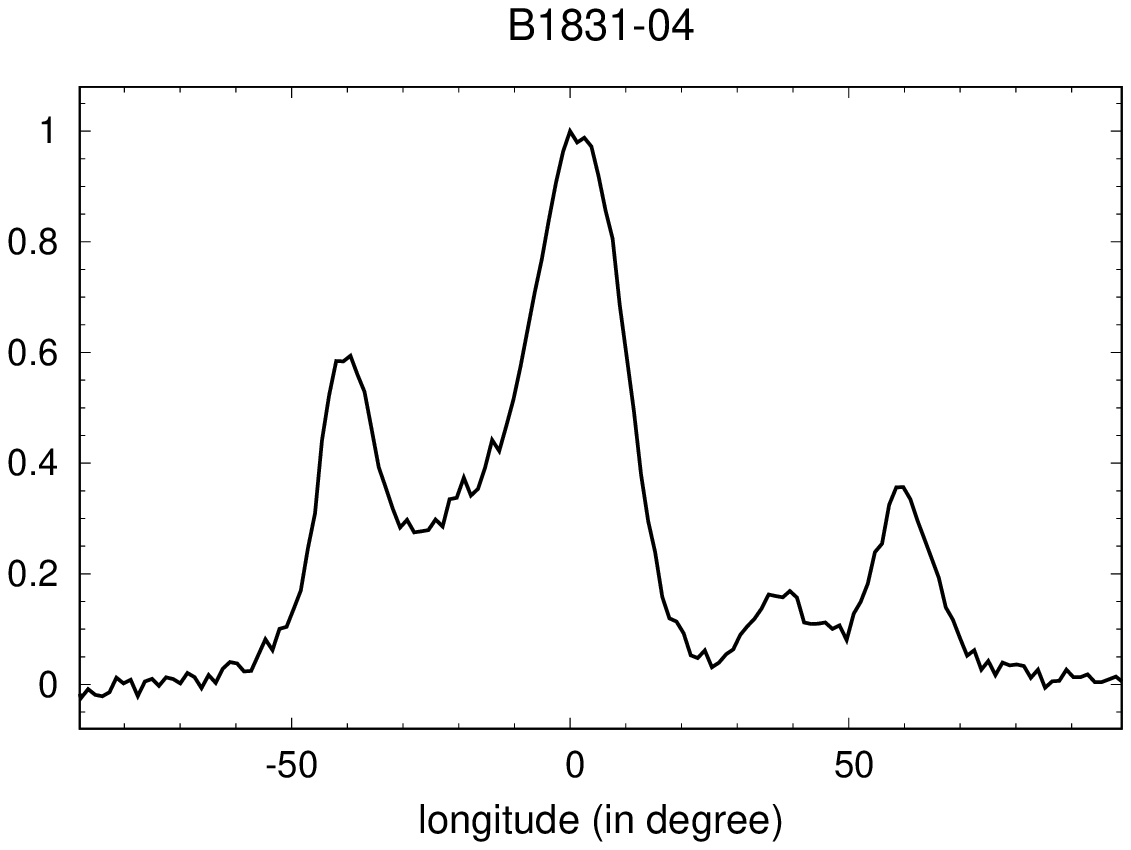}} &
\mbox{\includegraphics[angle=0,scale=0.57]{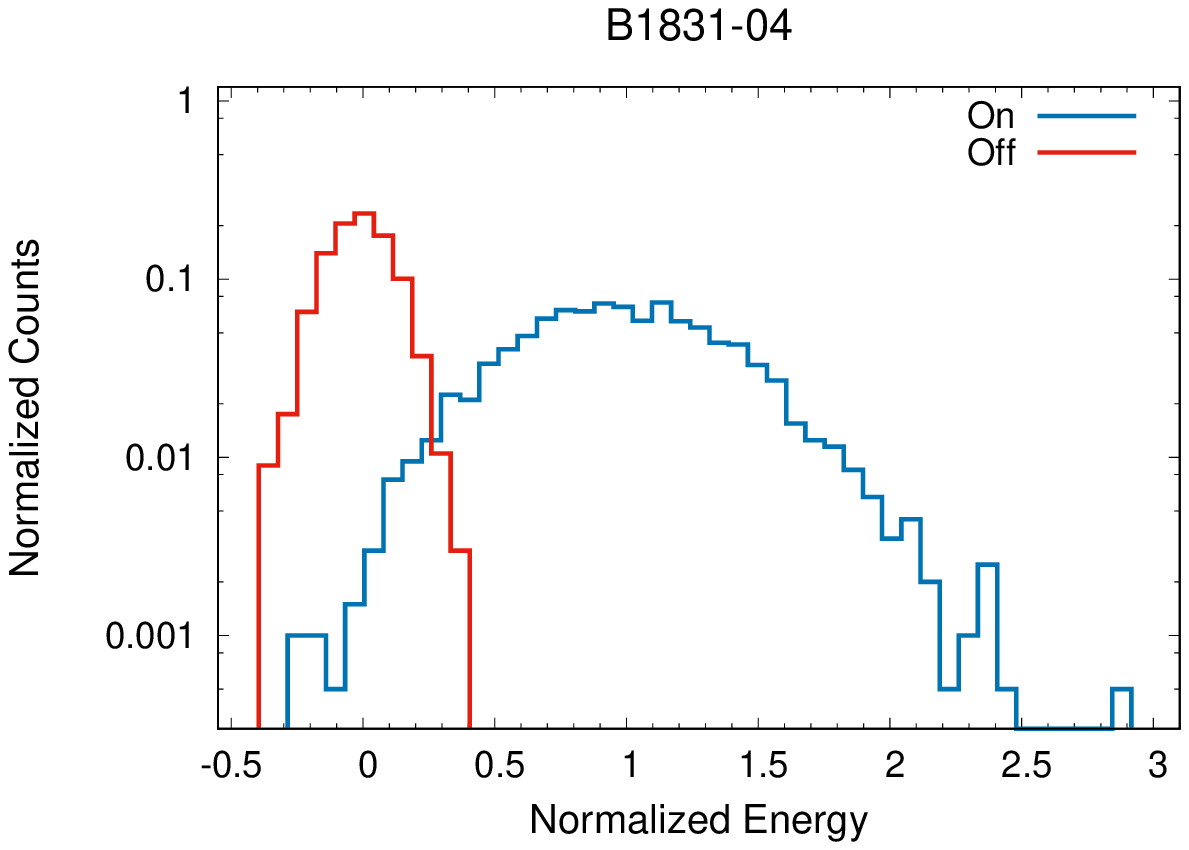}} \\
\mbox{\includegraphics[angle=0,scale=0.57]{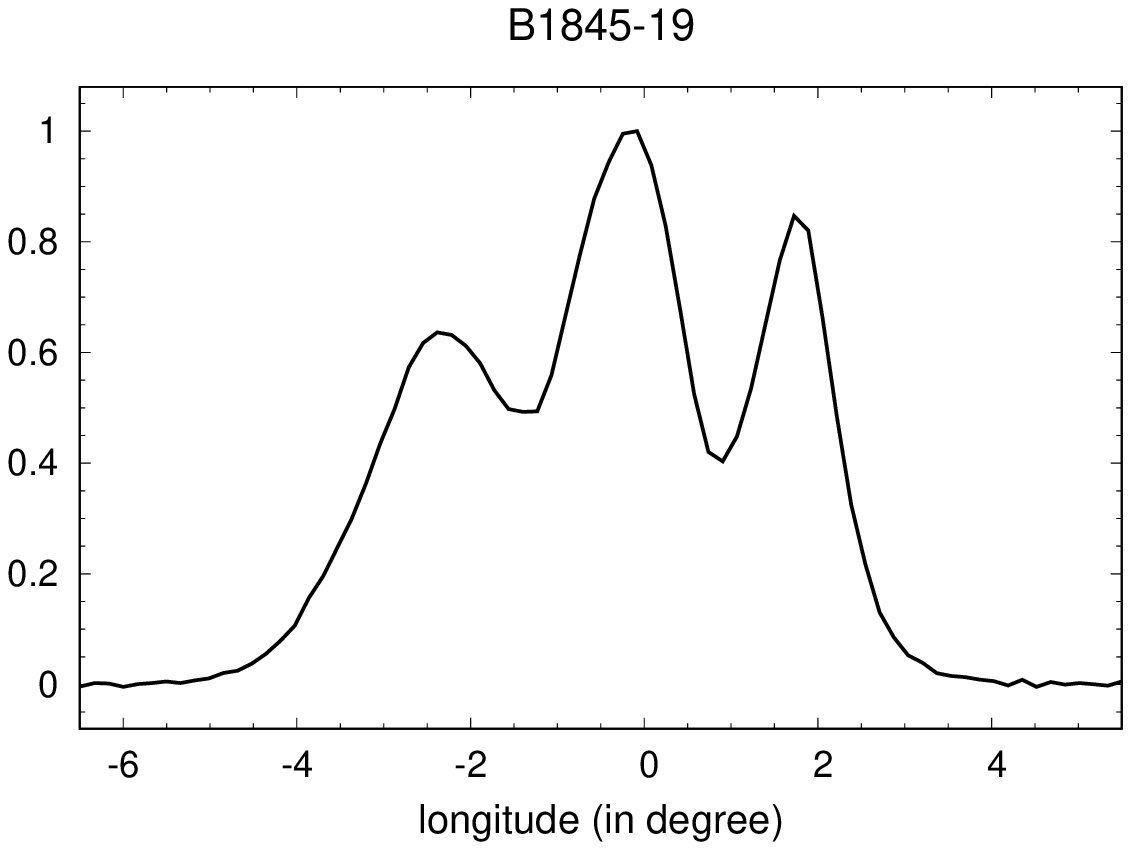}} &
\mbox{\includegraphics[angle=0,scale=0.57]{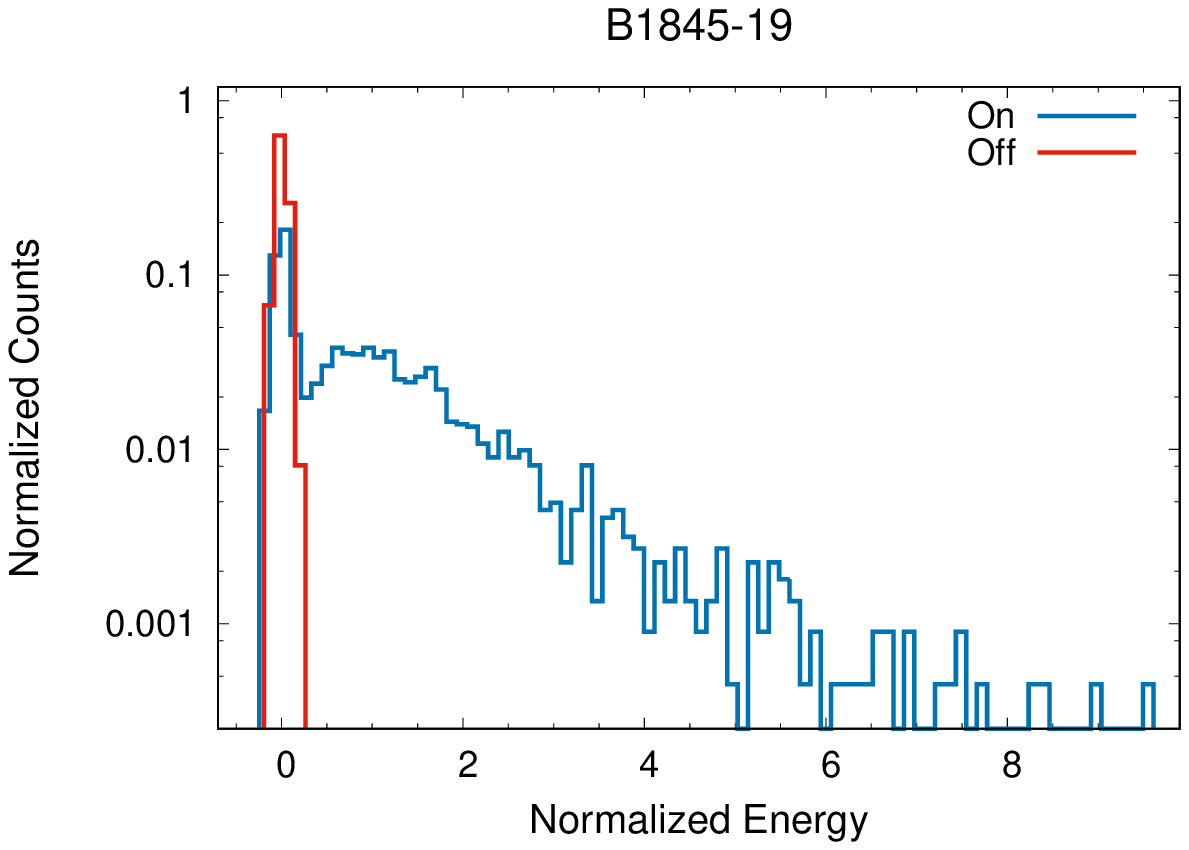}} \\
\mbox{\includegraphics[angle=0,scale=0.57]{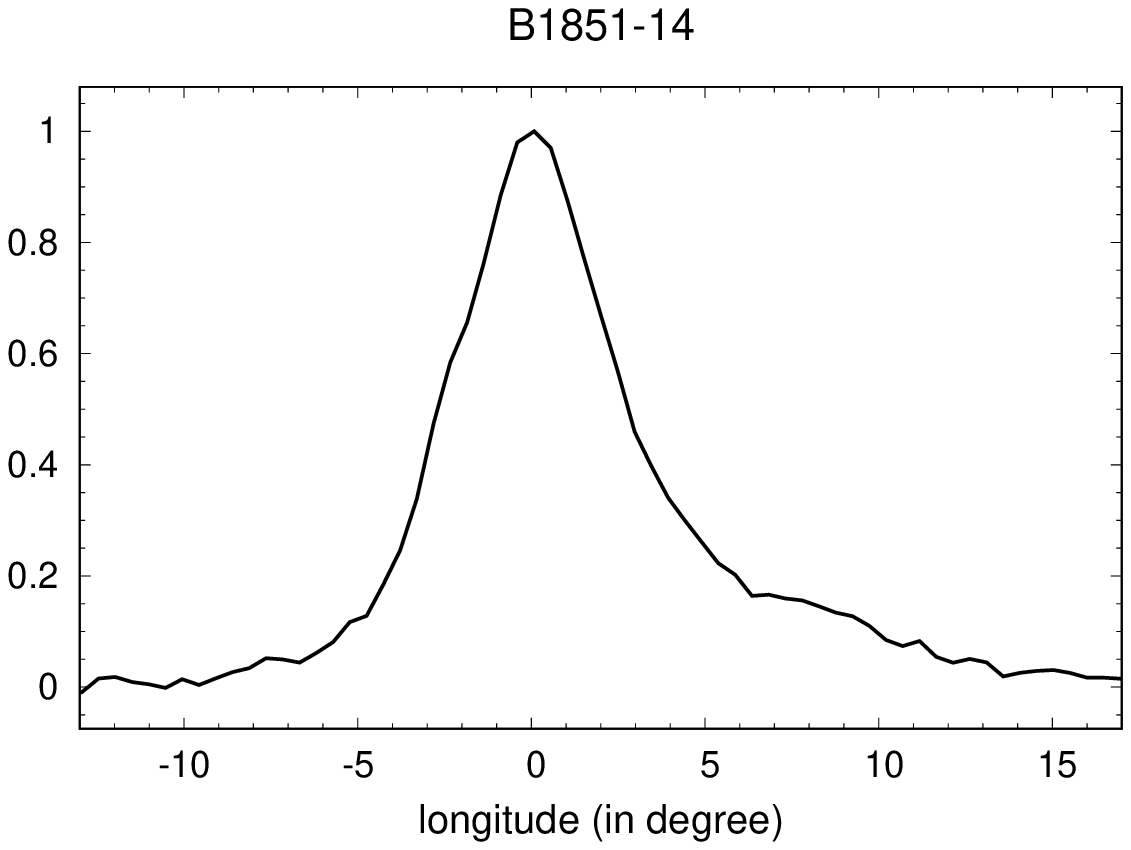}} &
\mbox{\includegraphics[angle=0,scale=0.57]{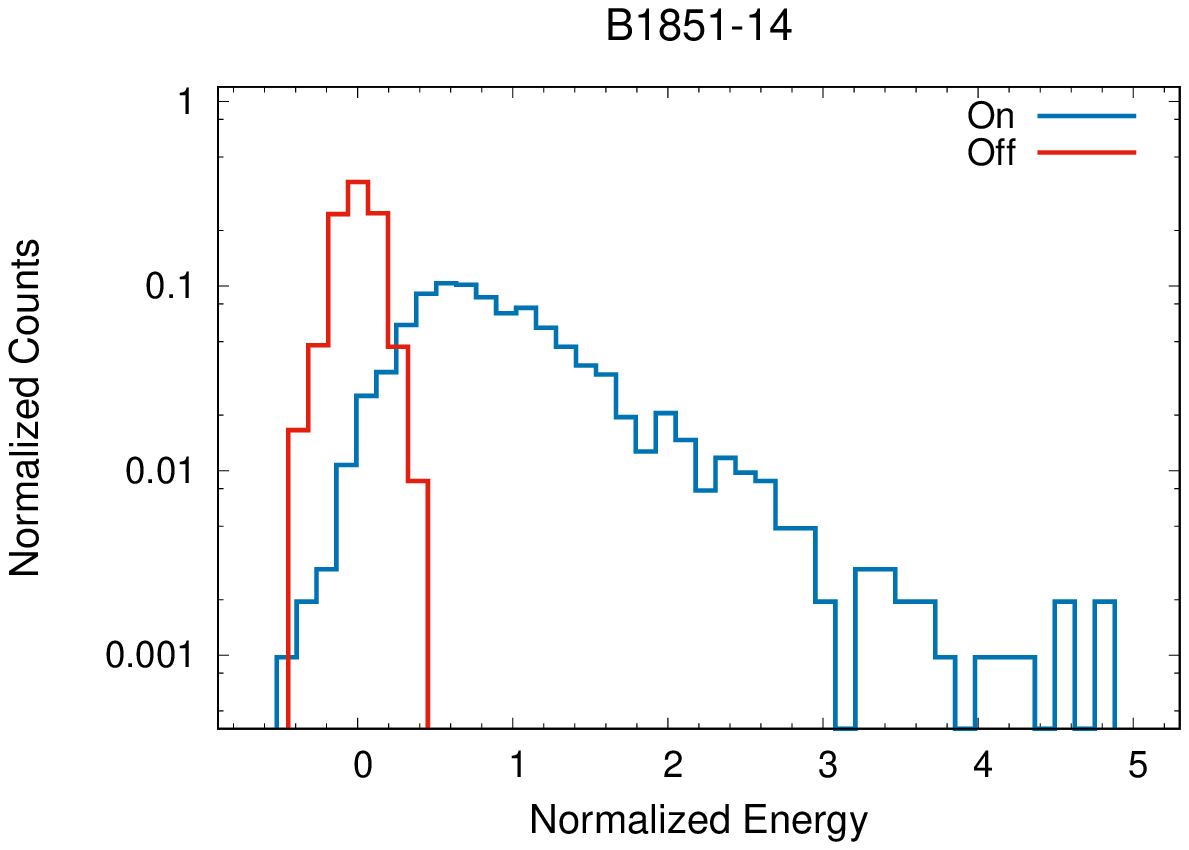}} \\
\end{tabular}
\caption{The pulsar profile On and Off-pulse energy distributions of the single pulse emission.}
\end{center}
\end{figure*}

\clearpage

%12th set of plots
\begin{figure*}
\begin{center}
\begin{tabular}{@{}cr@{}}
\mbox{\includegraphics[angle=0,scale=0.57]{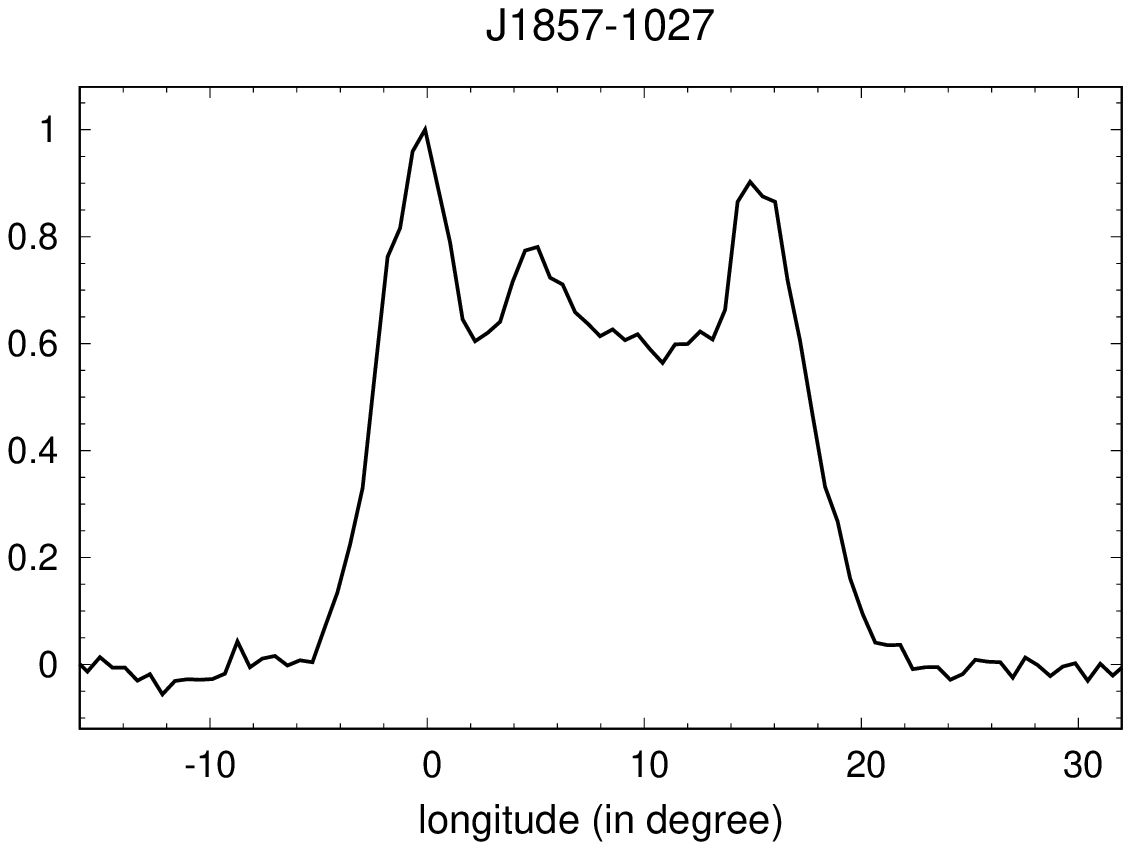}} &
\mbox{\includegraphics[angle=0,scale=0.57]{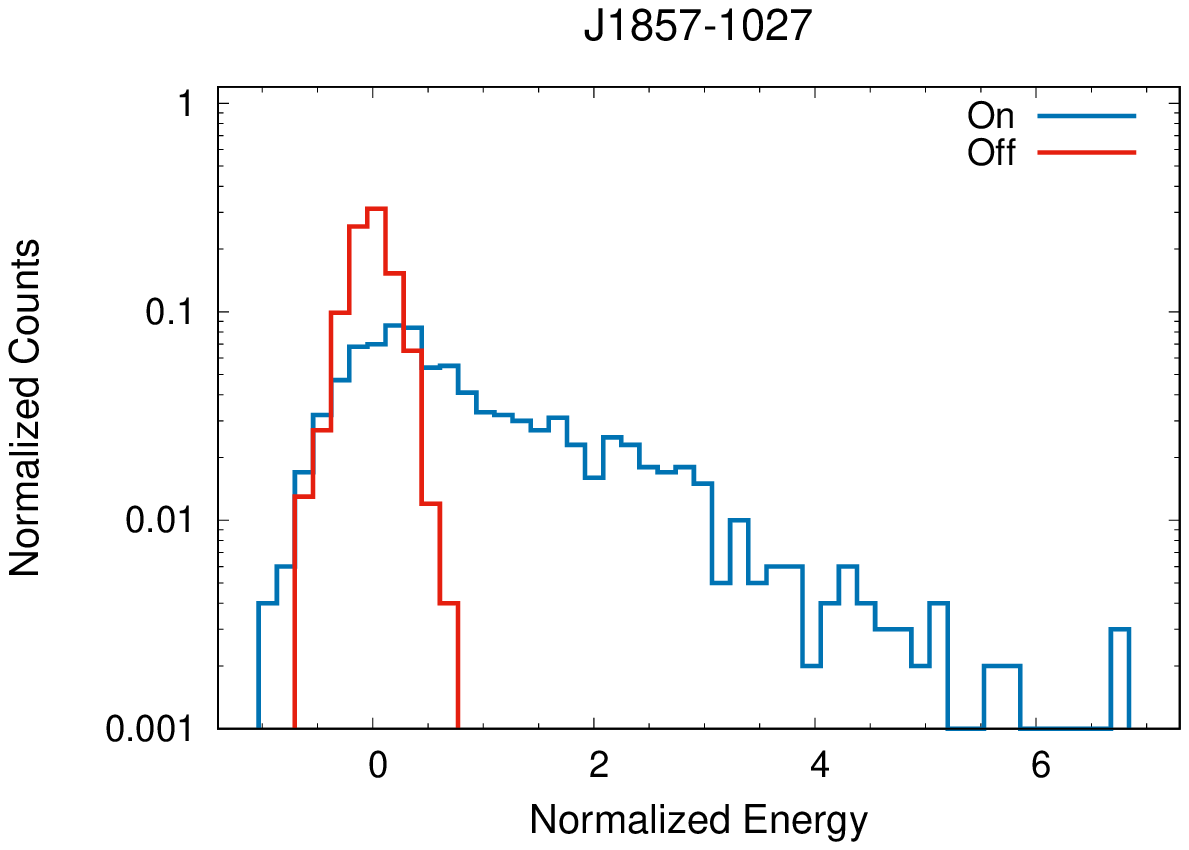}} \\
\mbox{\includegraphics[angle=0,scale=0.57]{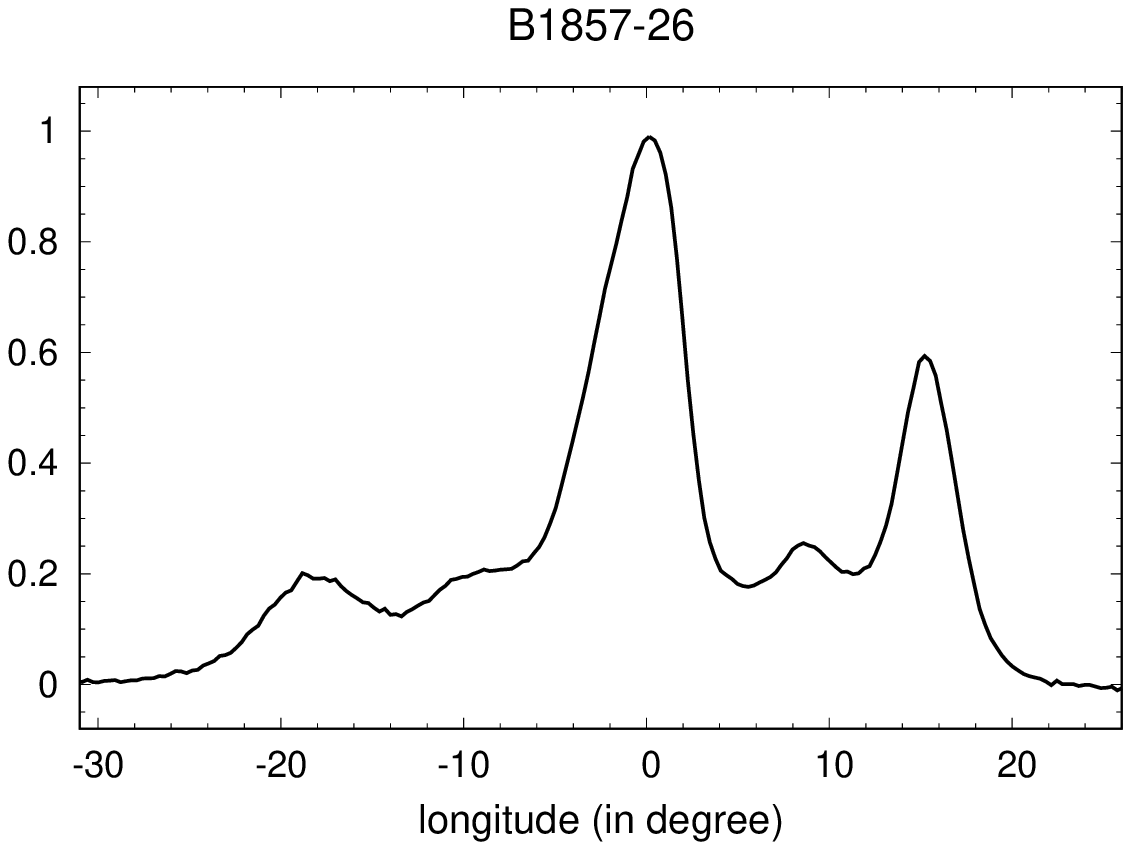}} &
\mbox{\includegraphics[angle=0,scale=0.57]{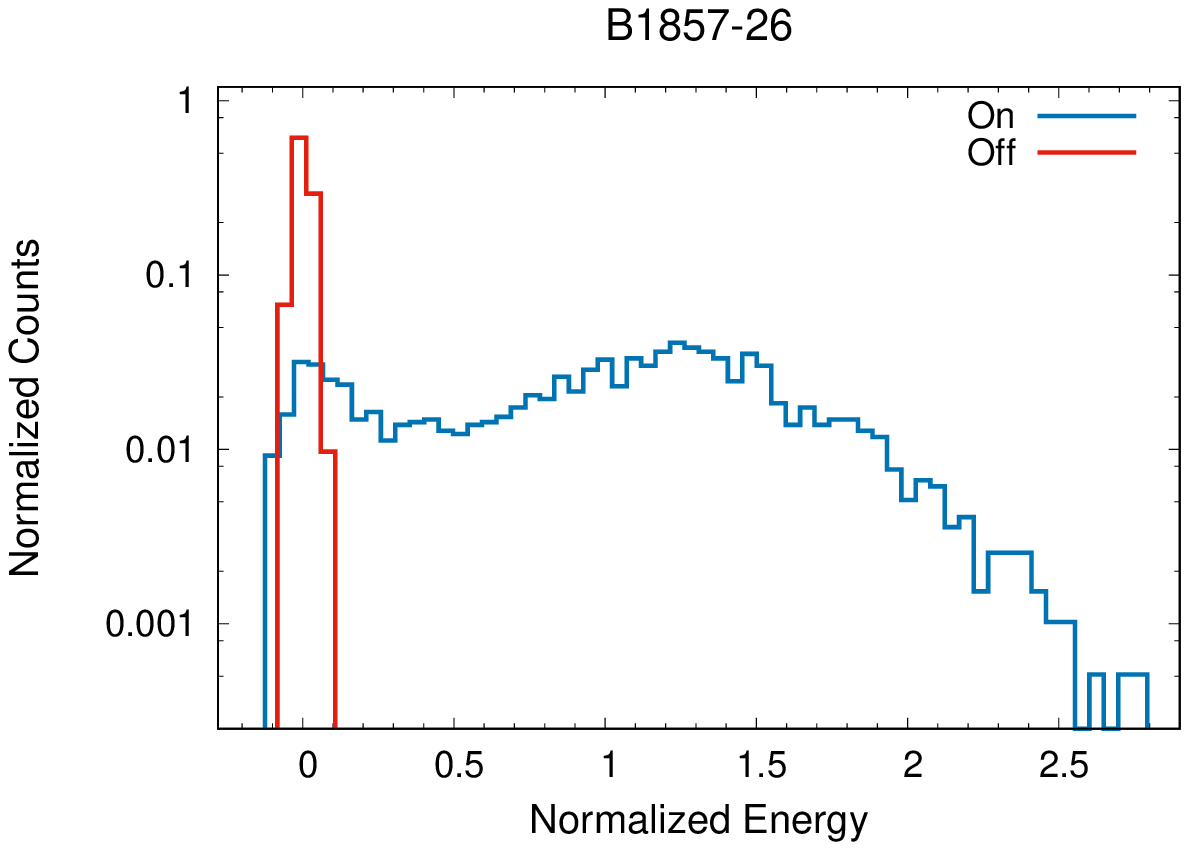}} \\
\mbox{\includegraphics[angle=0,scale=0.57]{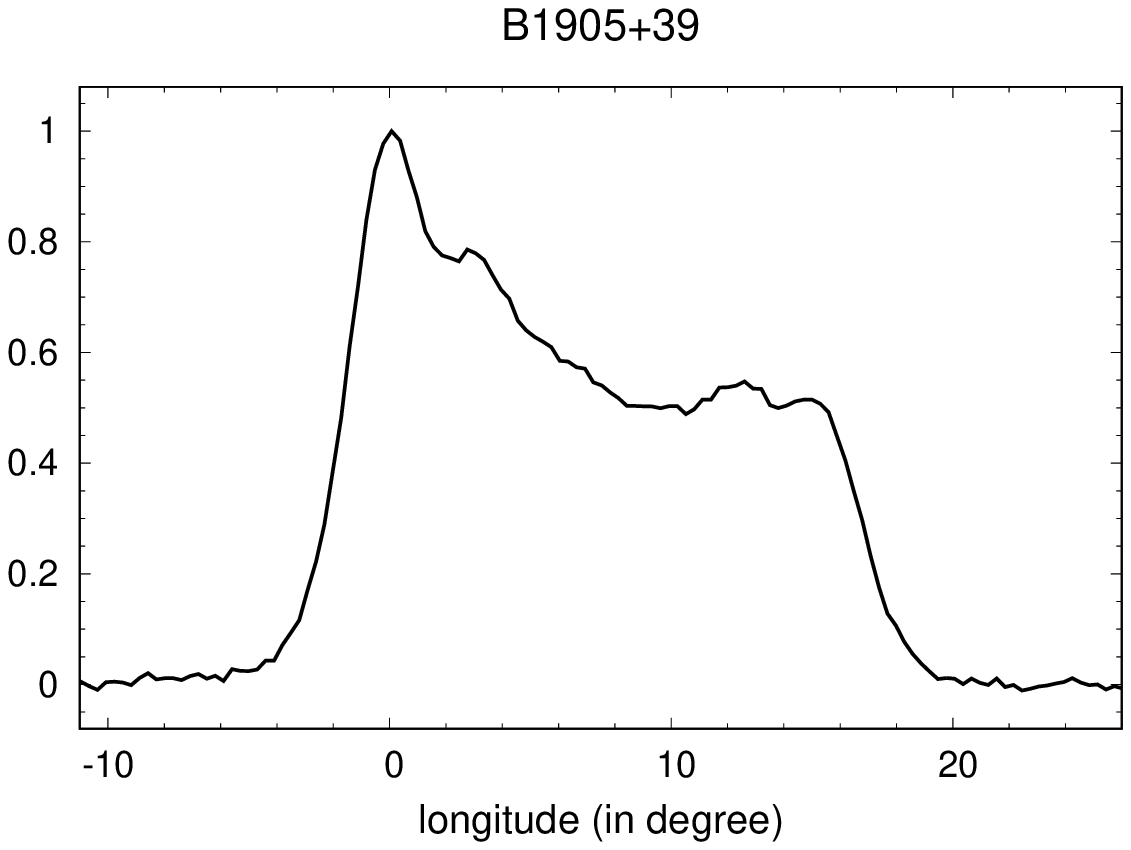}} &
\mbox{\includegraphics[angle=0,scale=0.57]{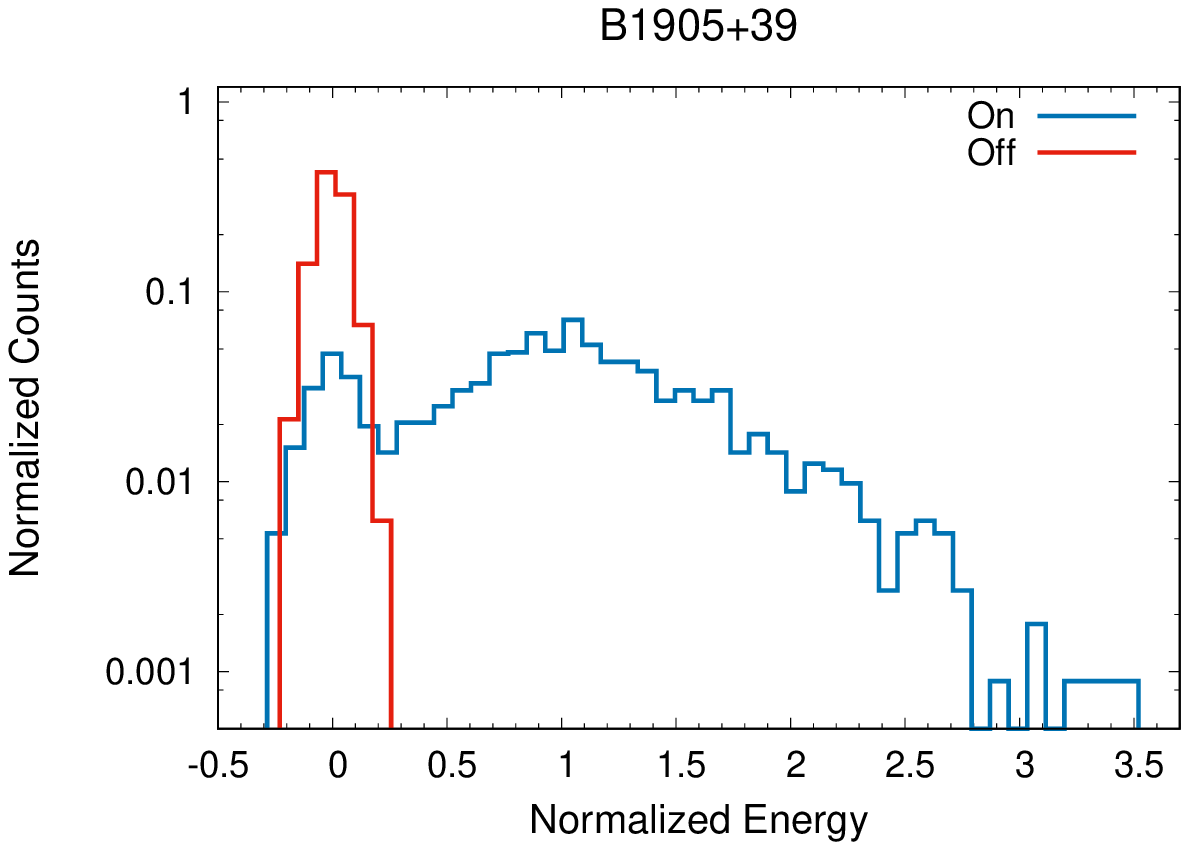}} \\
\mbox{\includegraphics[angle=0,scale=0.57]{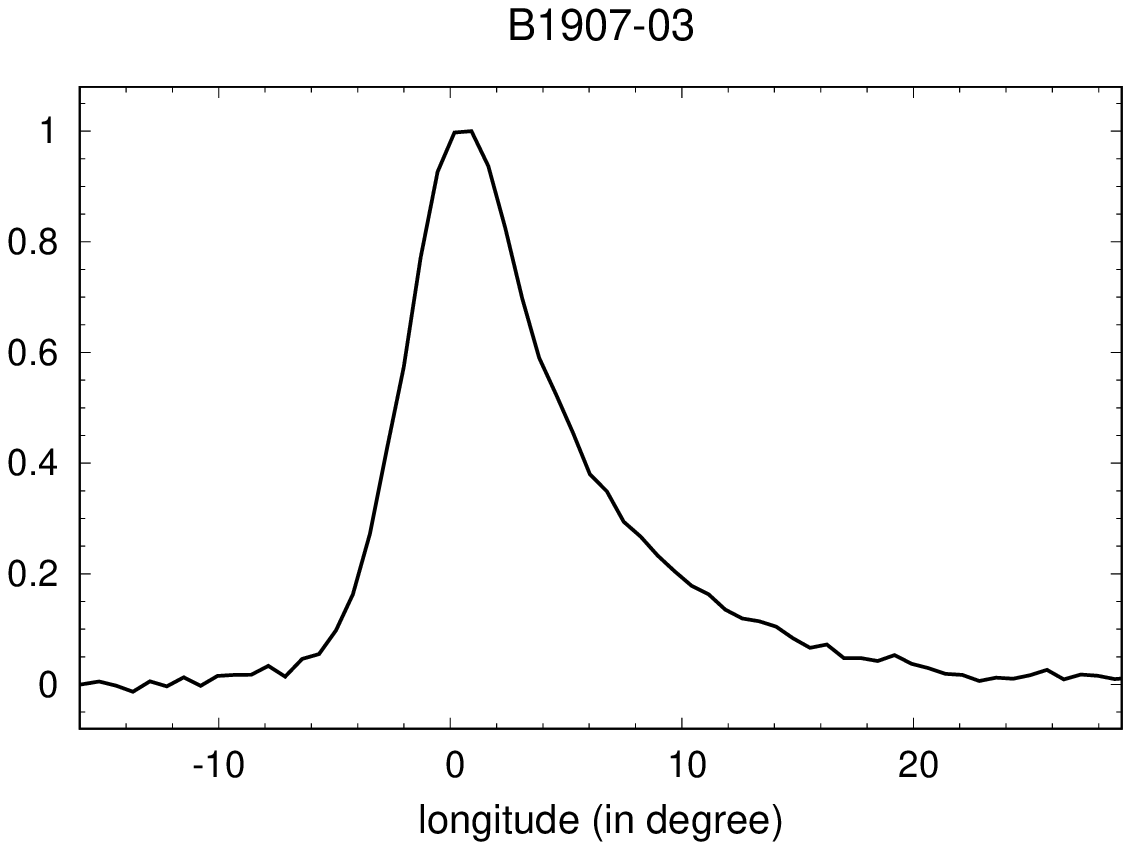}} &
\mbox{\includegraphics[angle=0,scale=0.57]{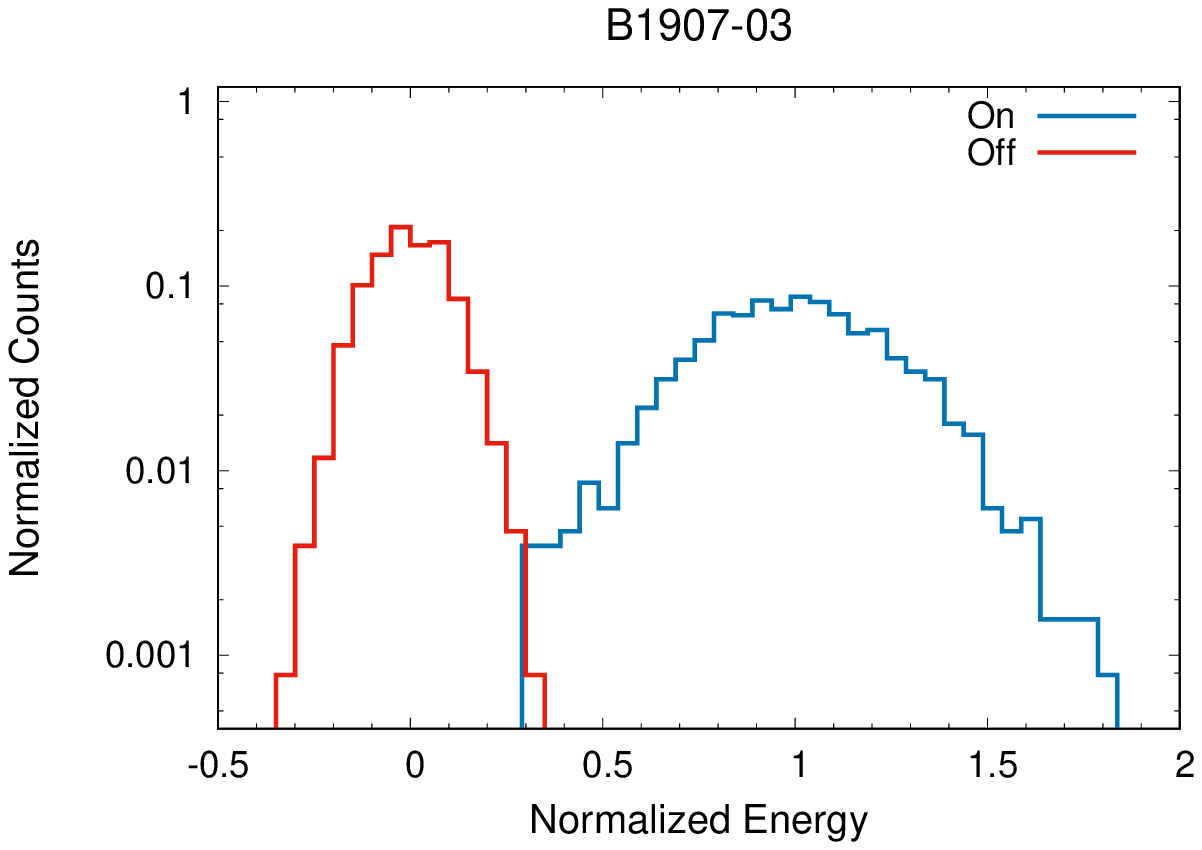}} \\
\end{tabular}
\caption{The pulsar profile On and Off-pulse energy distributions of the single pulse emission.}
\end{center}
\end{figure*}

\clearpage

%13th set of plots
\begin{figure*}
\begin{center}
\begin{tabular}{@{}cr@{}}
\mbox{\includegraphics[angle=0,scale=0.57]{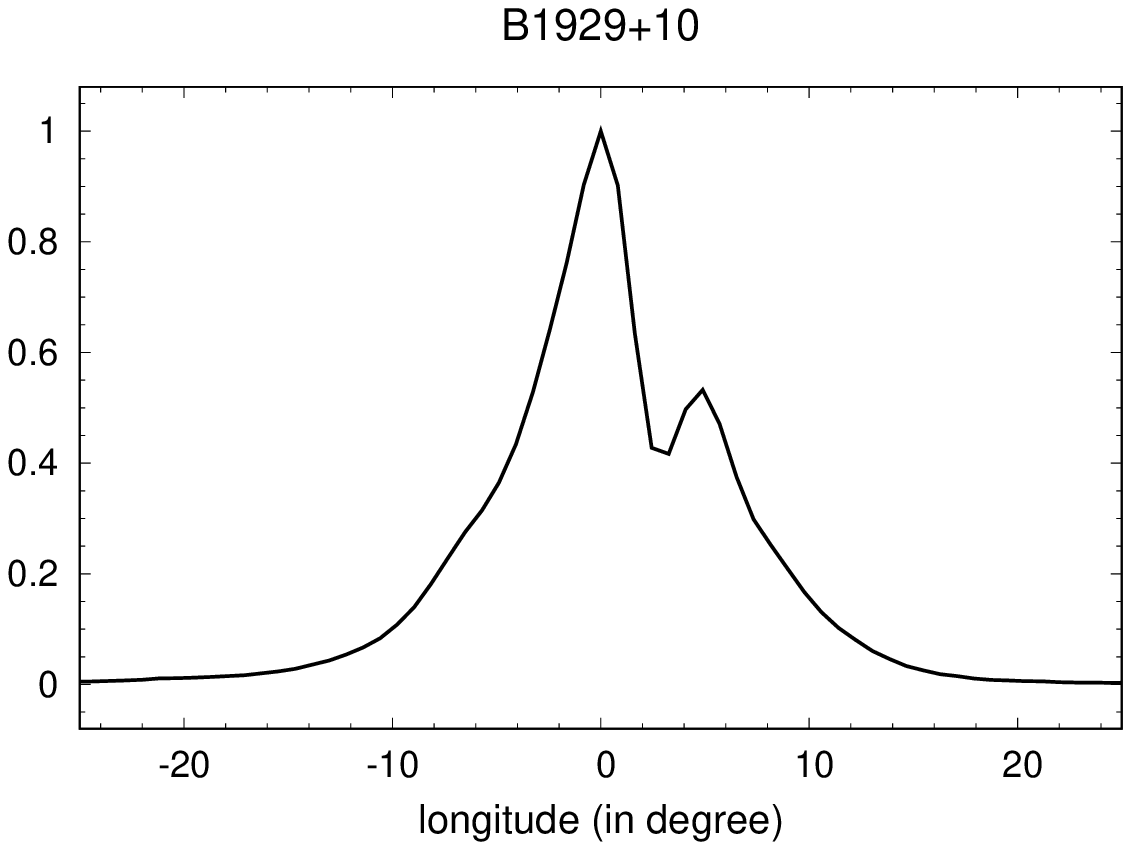}} &
\mbox{\includegraphics[angle=0,scale=0.57]{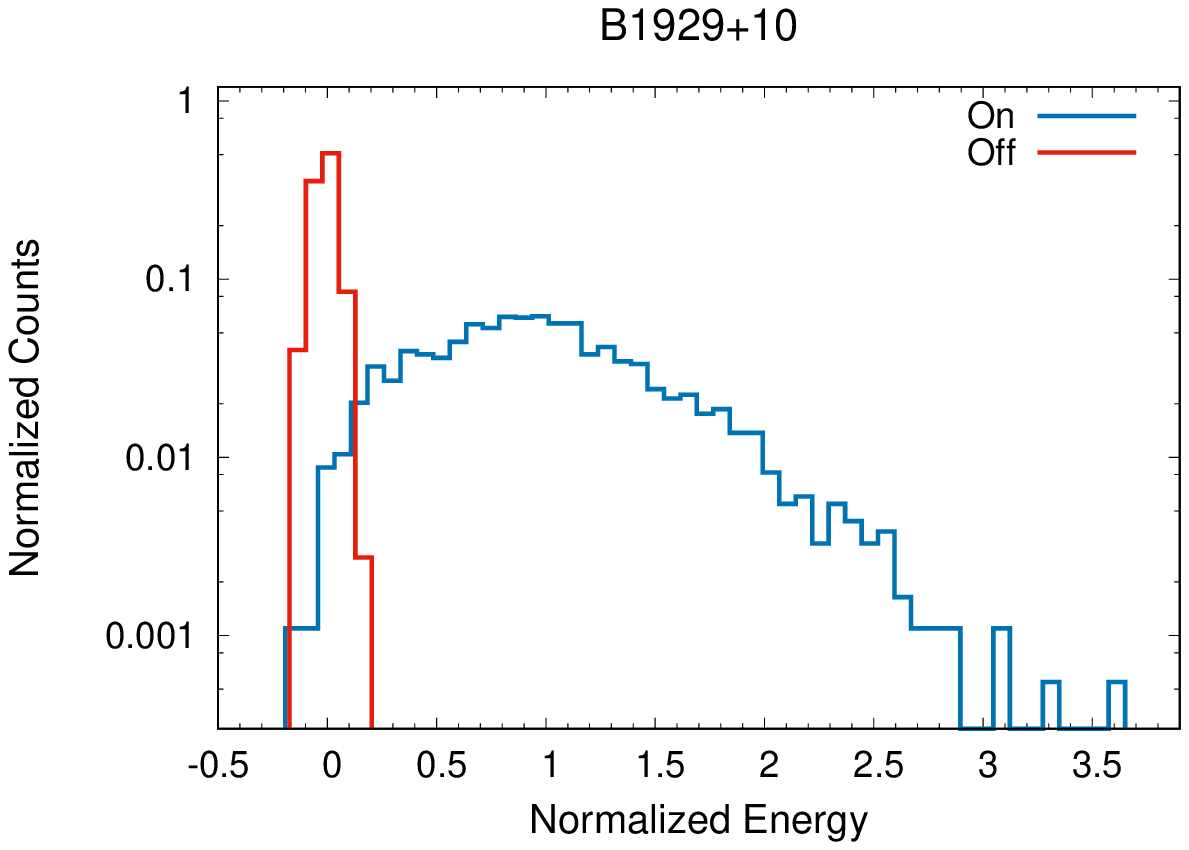}} \\
\mbox{\includegraphics[angle=0,scale=0.57]{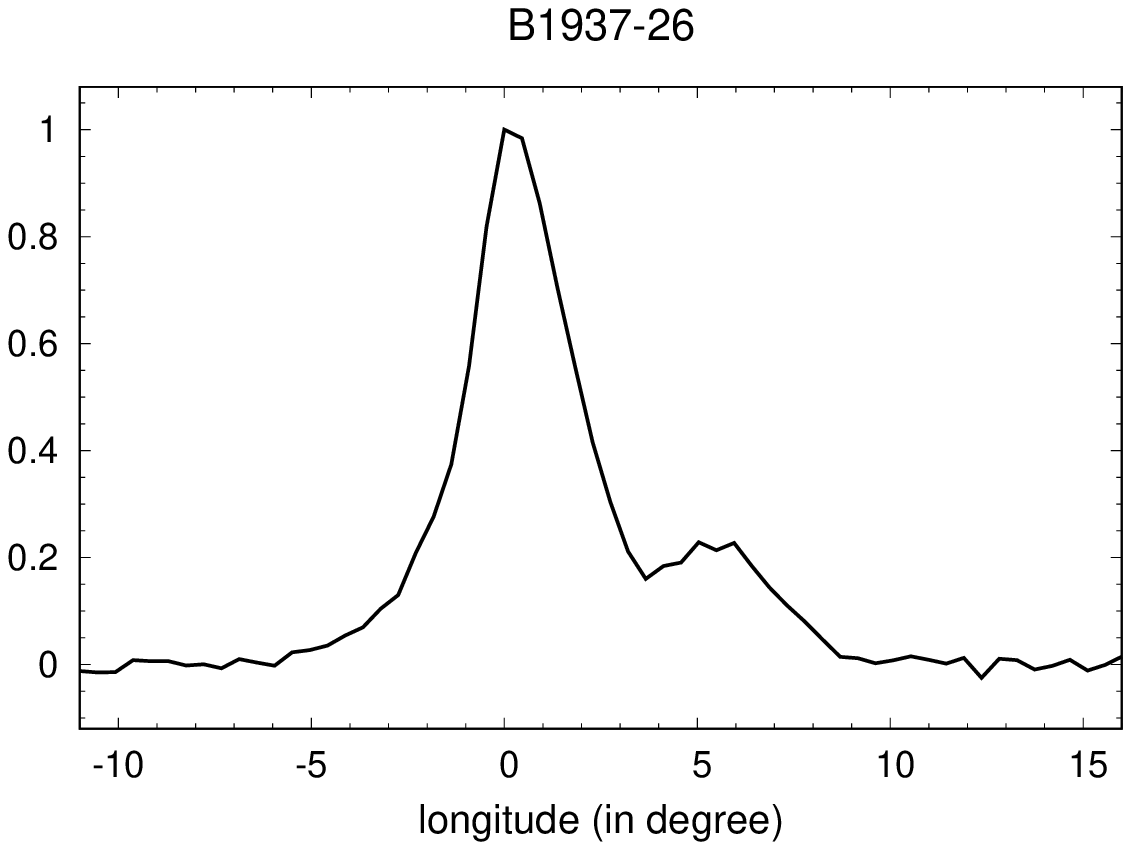}} &
\mbox{\includegraphics[angle=0,scale=0.57]{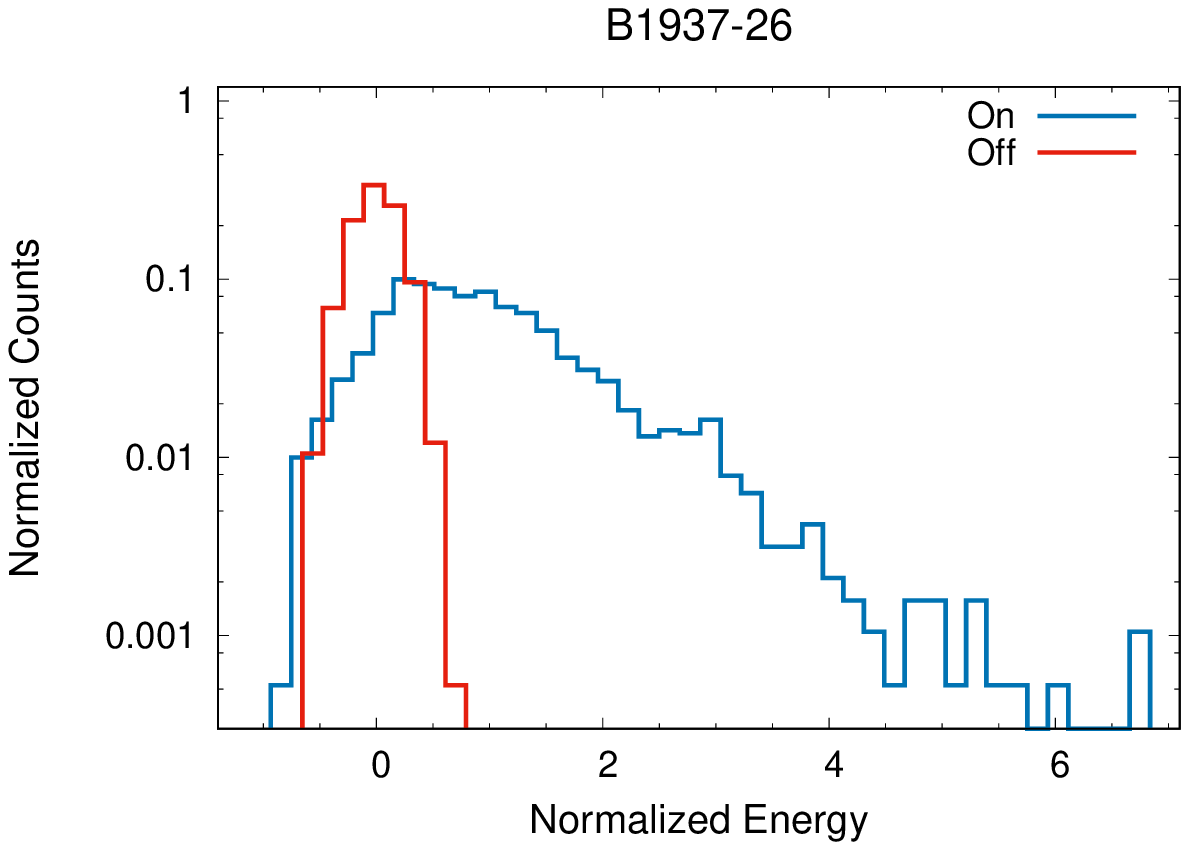}} \\
\mbox{\includegraphics[angle=0,scale=0.57]{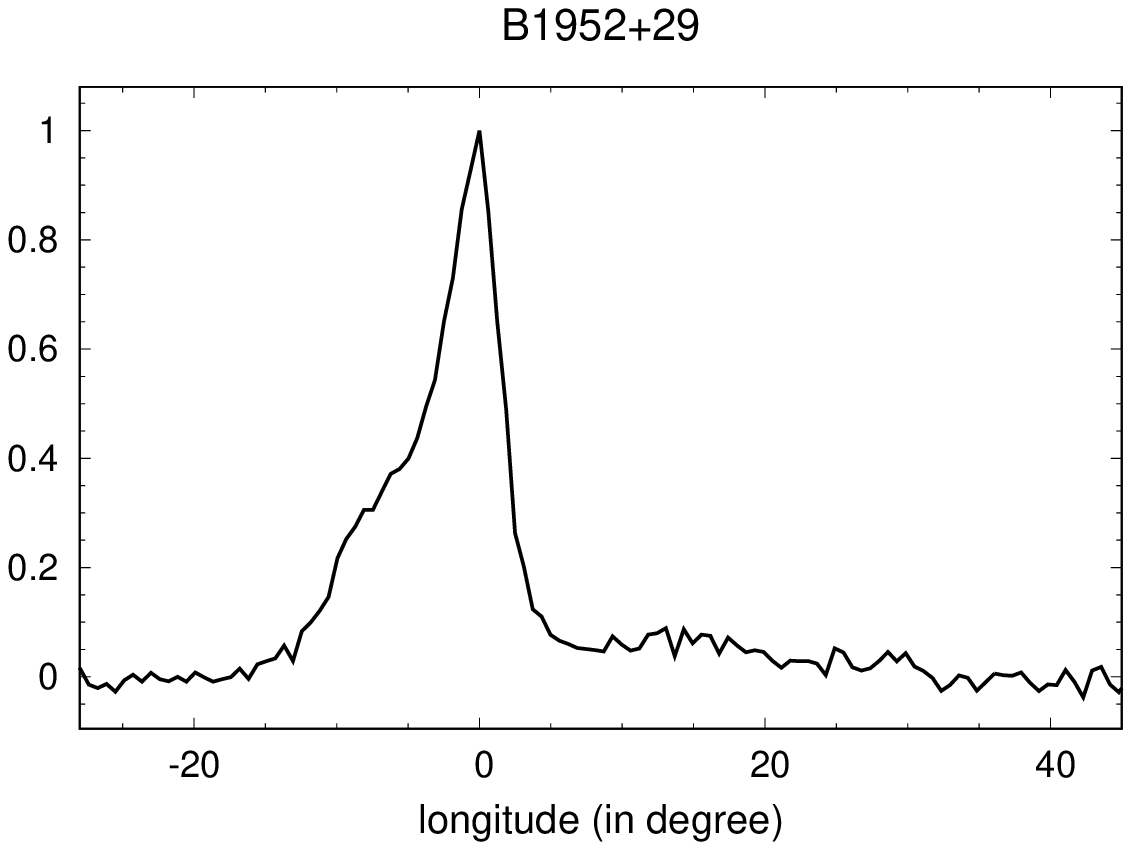}} &
\mbox{\includegraphics[angle=0,scale=0.57]{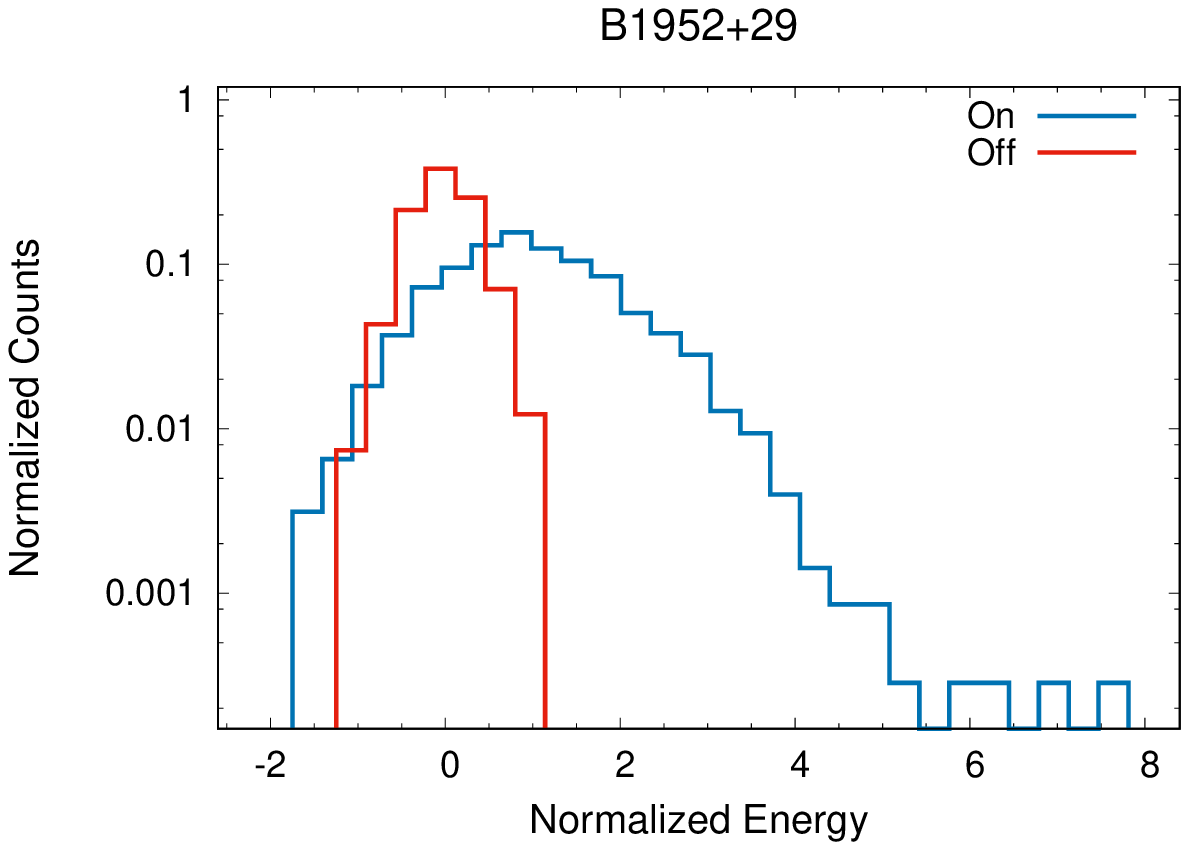}} \\
\mbox{\includegraphics[angle=0,scale=0.57]{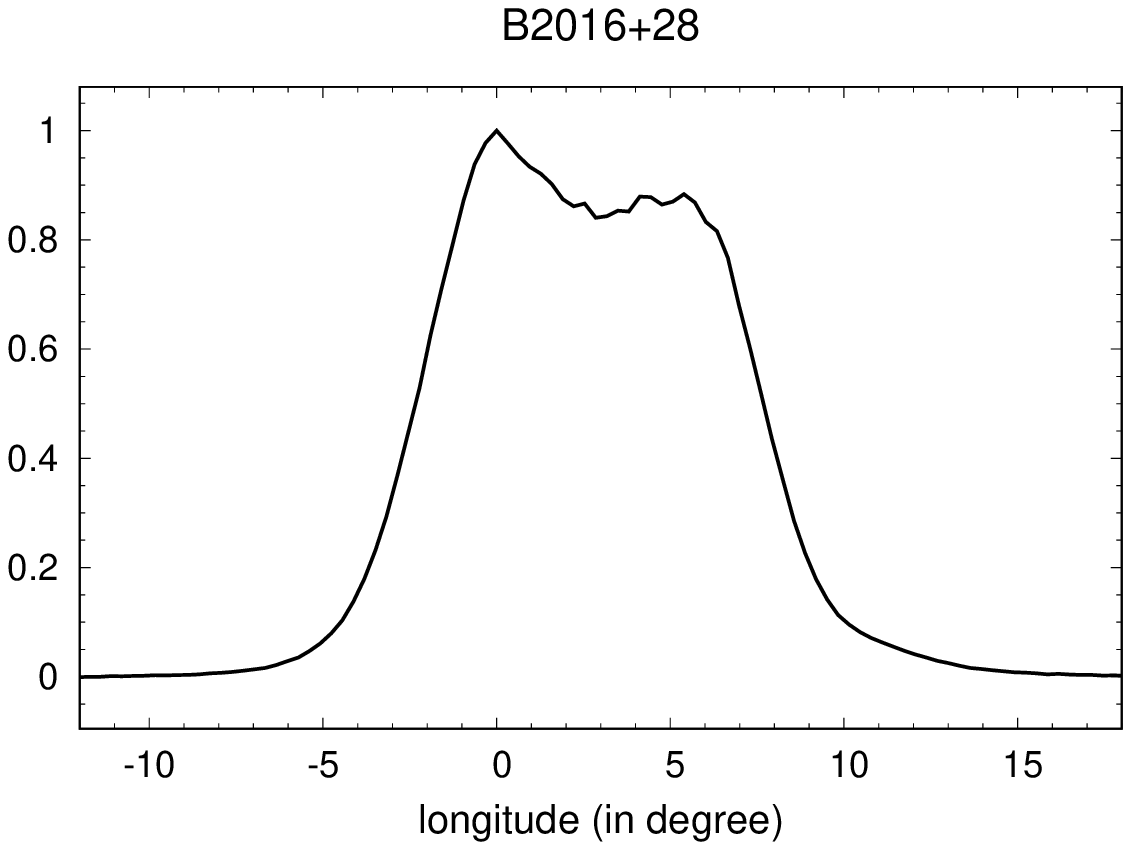}} &
\mbox{\includegraphics[angle=0,scale=0.57]{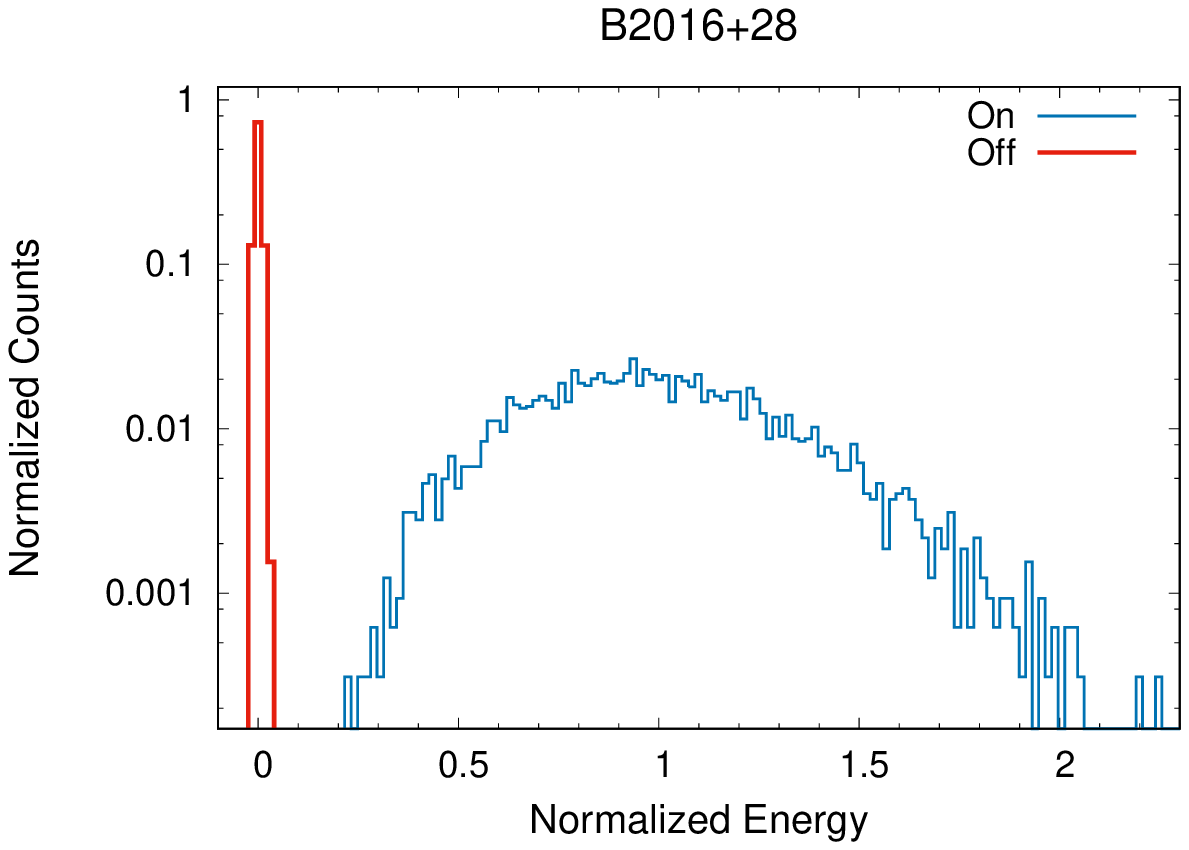}} \\
\end{tabular}
\caption{The pulsar profile On and Off-pulse energy distributions of the single pulse emission.}
\end{center}
\end{figure*}

\clearpage

%14th set of plots
\begin{figure*}
\begin{center}
\begin{tabular}{@{}cr@{}}
\mbox{\includegraphics[angle=0,scale=0.57]{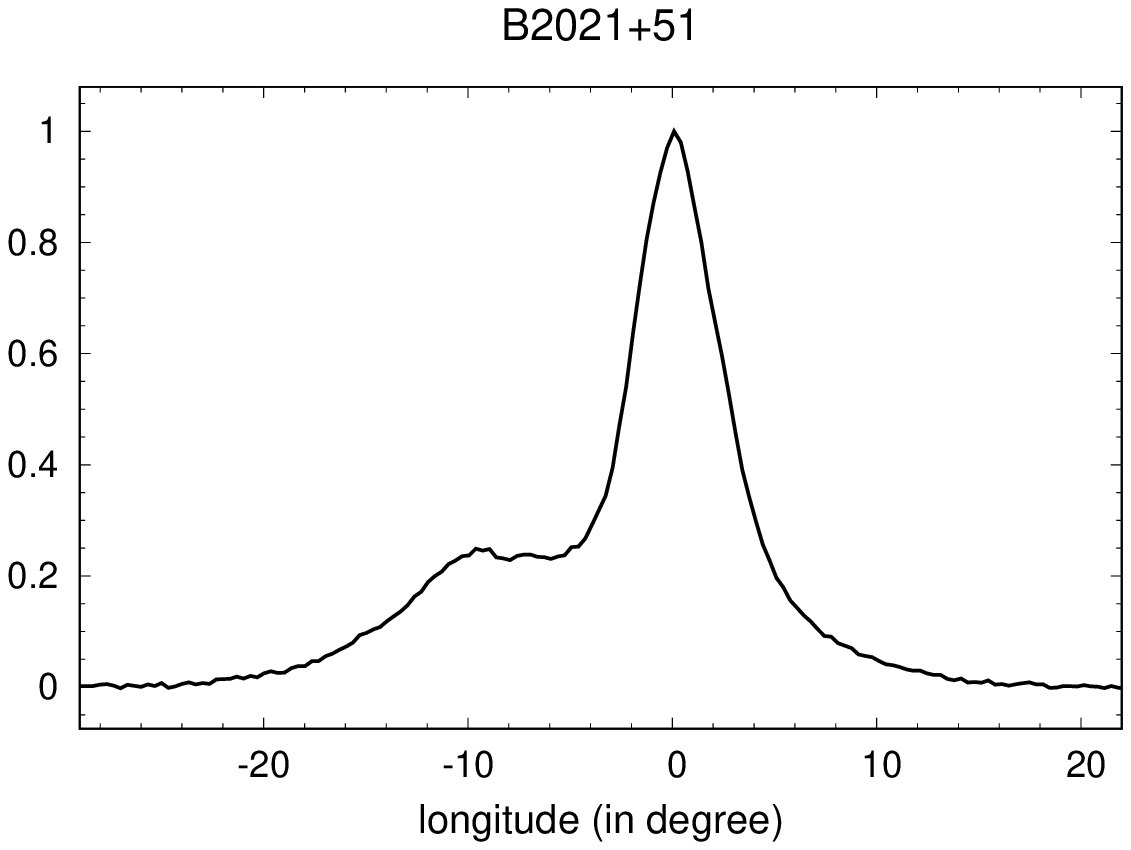}} &
\mbox{\includegraphics[angle=0,scale=0.57]{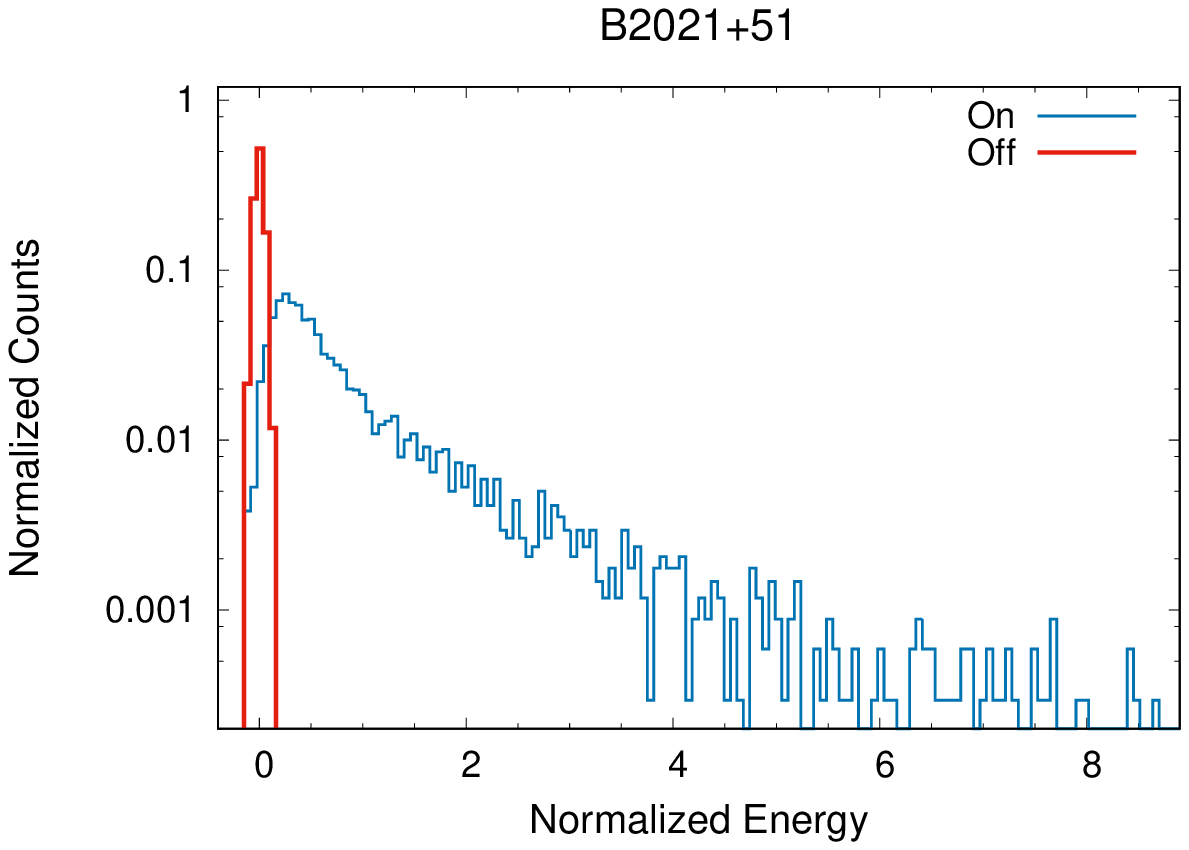}} \\
\mbox{\includegraphics[angle=0,scale=0.57]{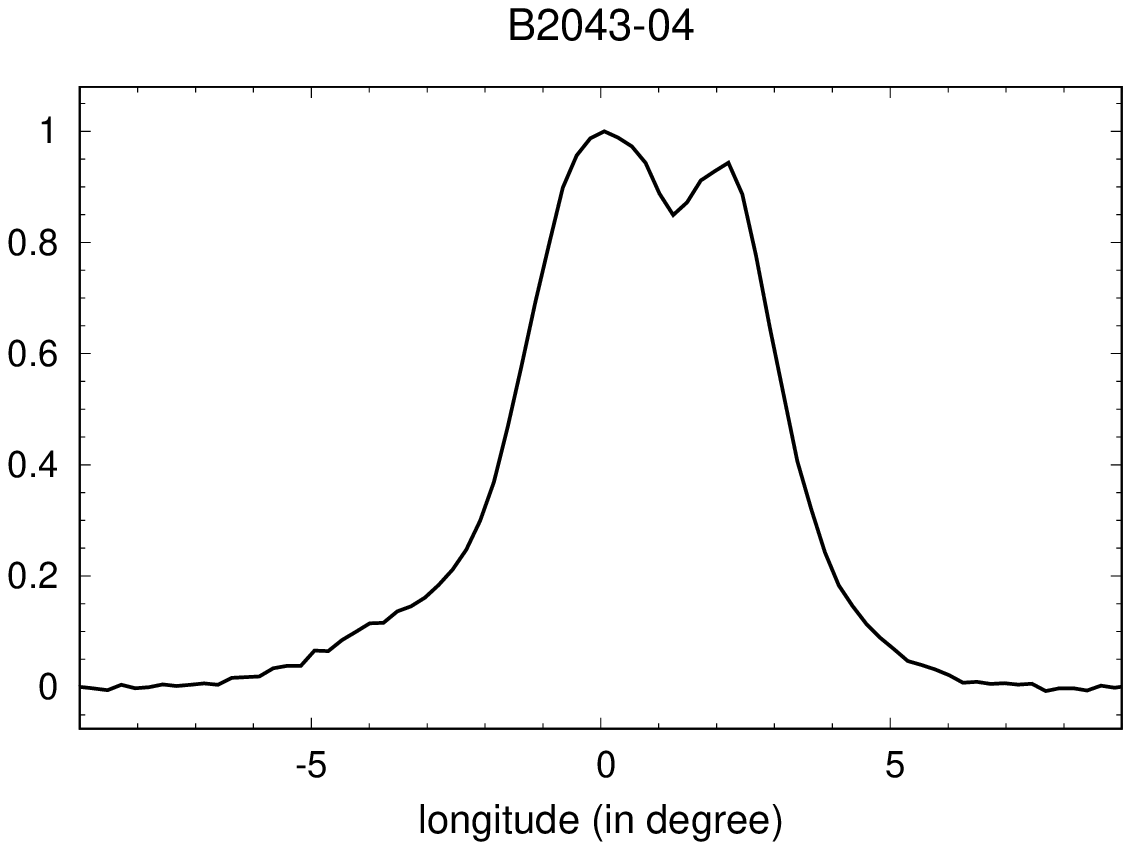}} &
\mbox{\includegraphics[angle=0,scale=0.57]{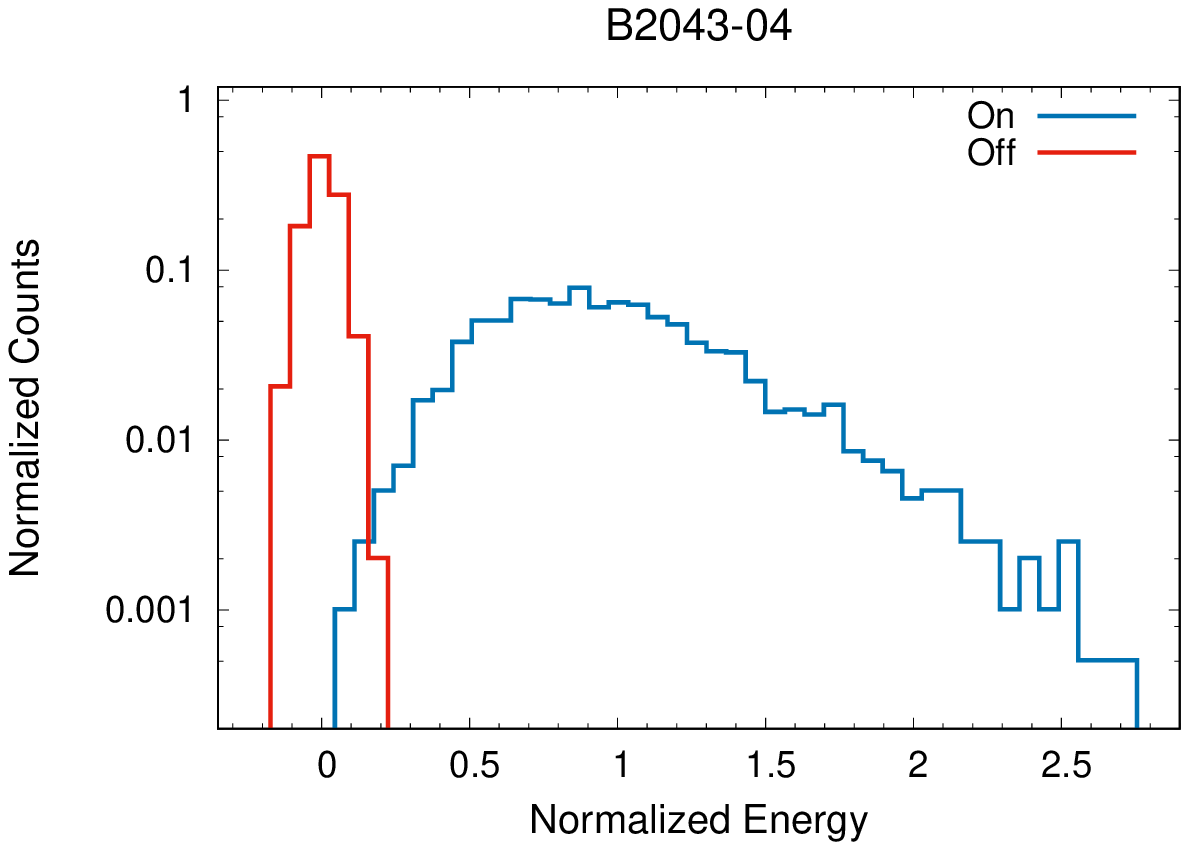}} \\
\mbox{\includegraphics[angle=0,scale=0.57]{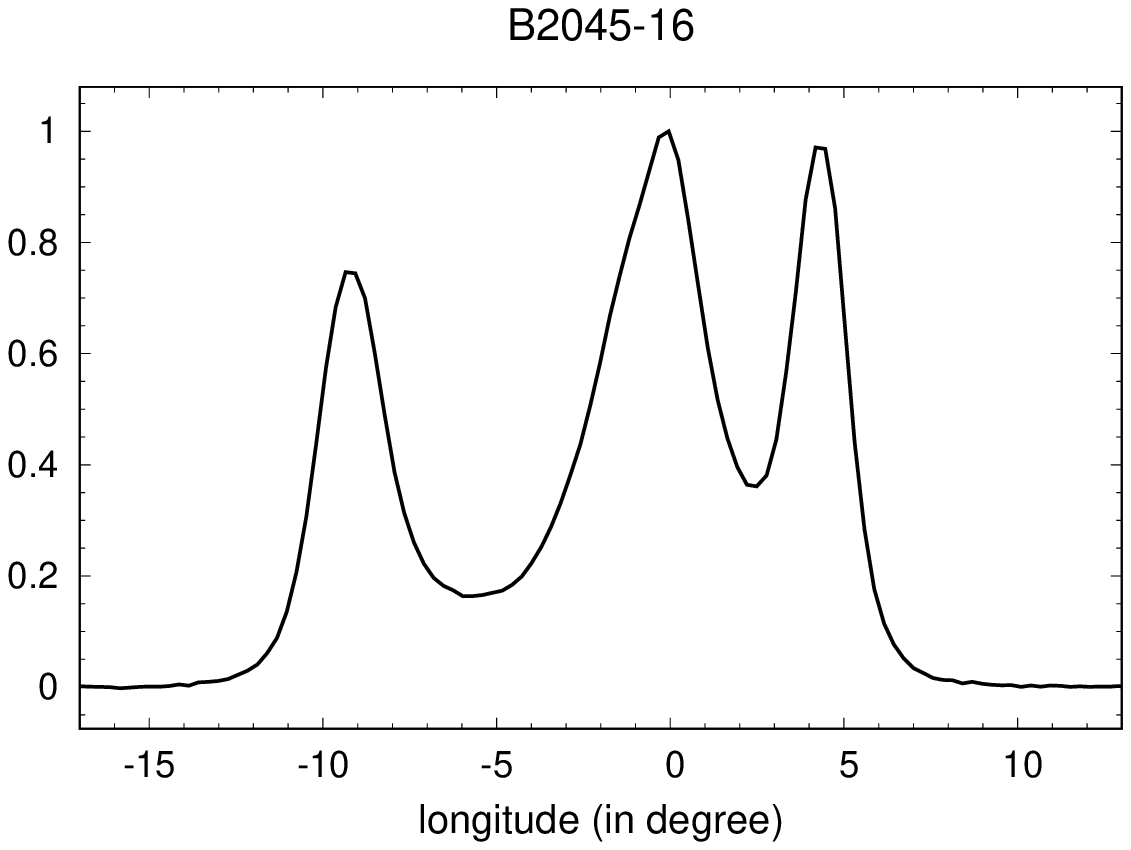}} &
\mbox{\includegraphics[angle=0,scale=0.57]{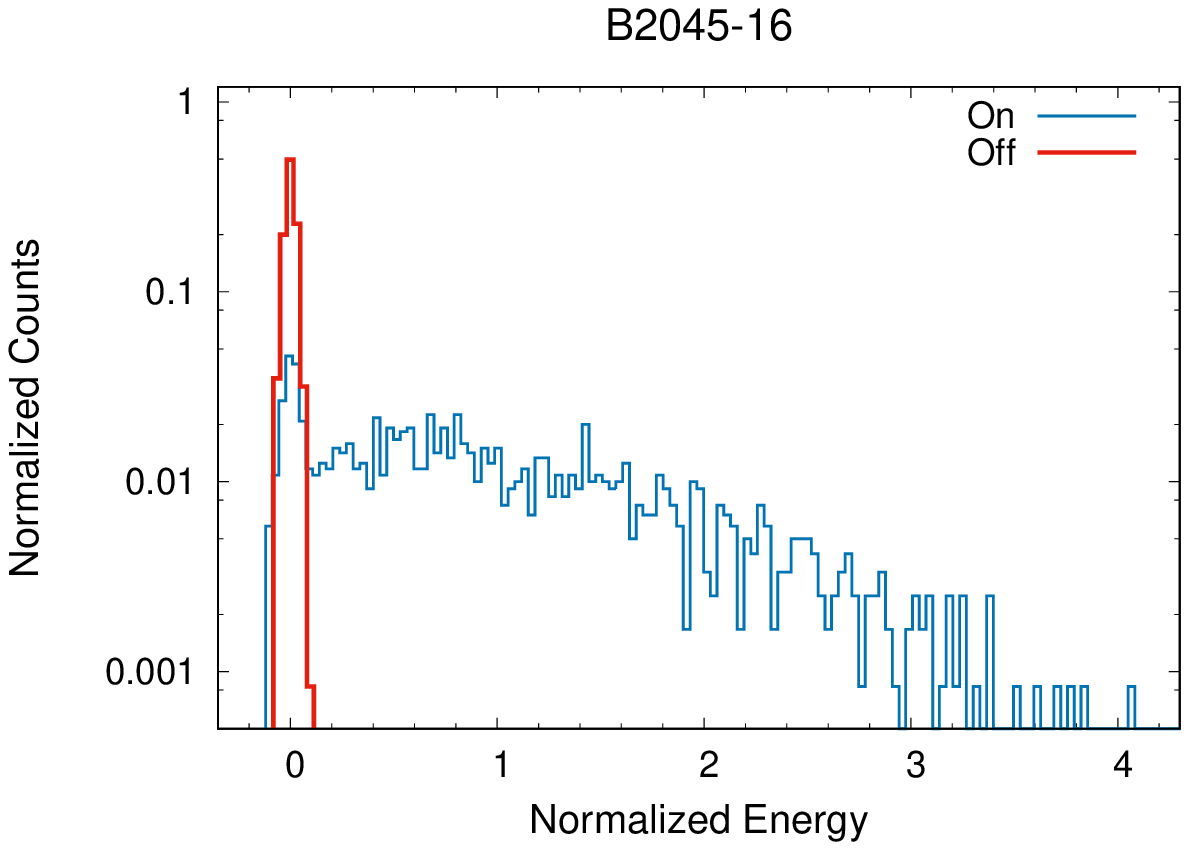}} \\
\mbox{\includegraphics[angle=0,scale=0.57]{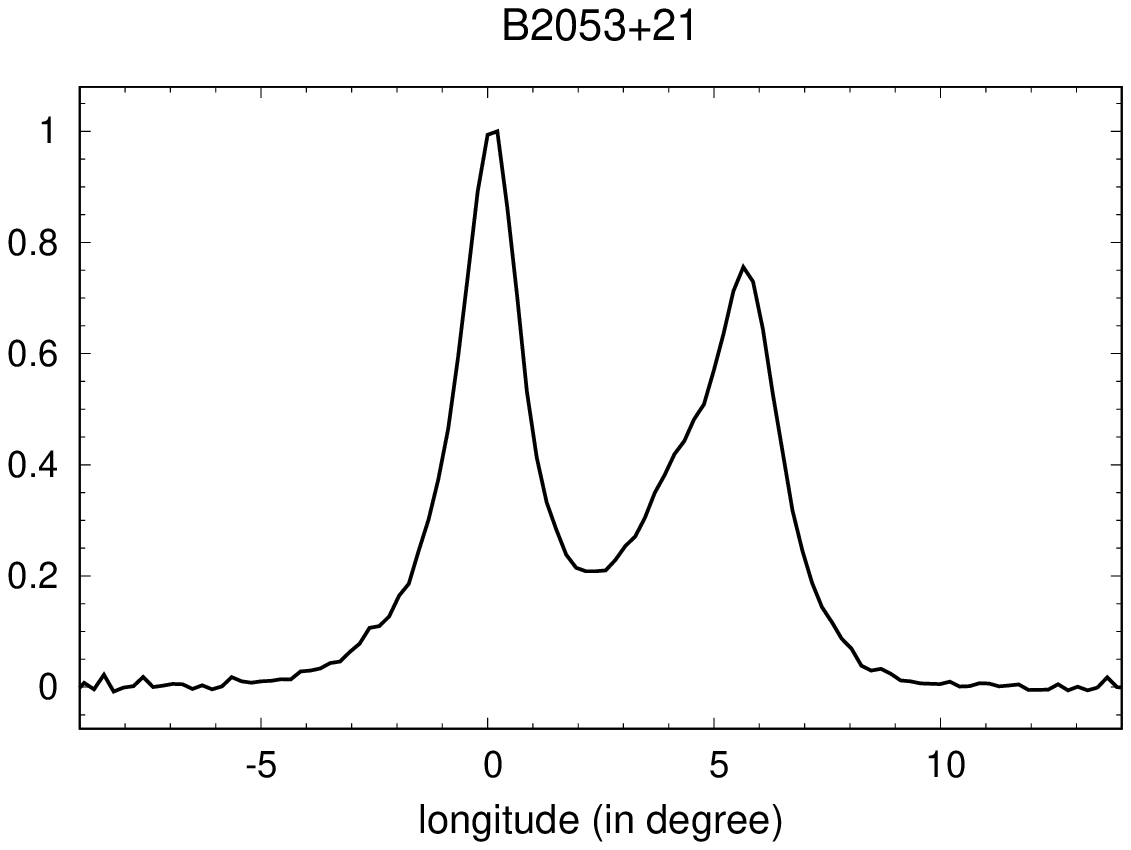}} &
\mbox{\includegraphics[angle=0,scale=0.57]{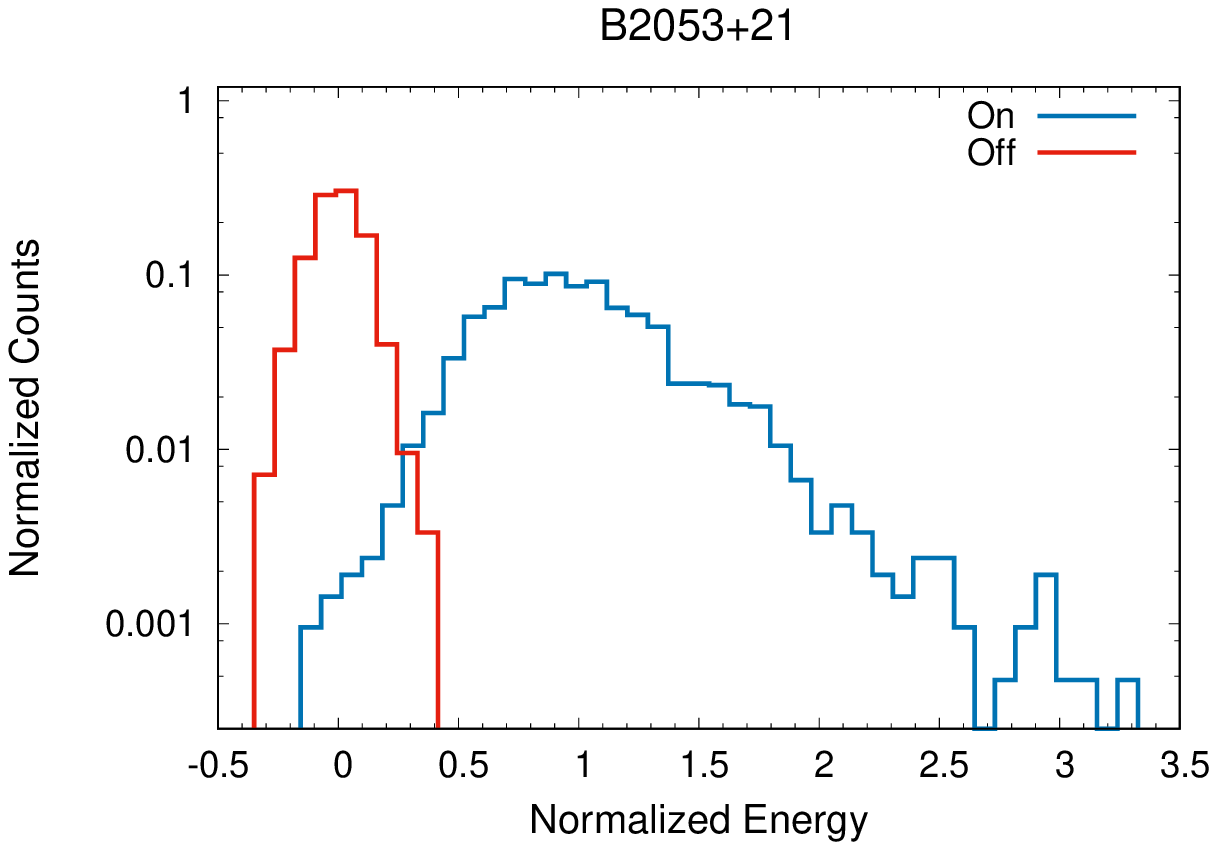}} \\
\end{tabular}
\caption{The pulsar profile On and Off-pulse energy distributions of the single pulse emission.}
\end{center}
\end{figure*}

\clearpage

%15th set of plots
\begin{figure*}
\begin{center}
\begin{tabular}{@{}cr@{}}
\mbox{\includegraphics[angle=0,scale=0.57]{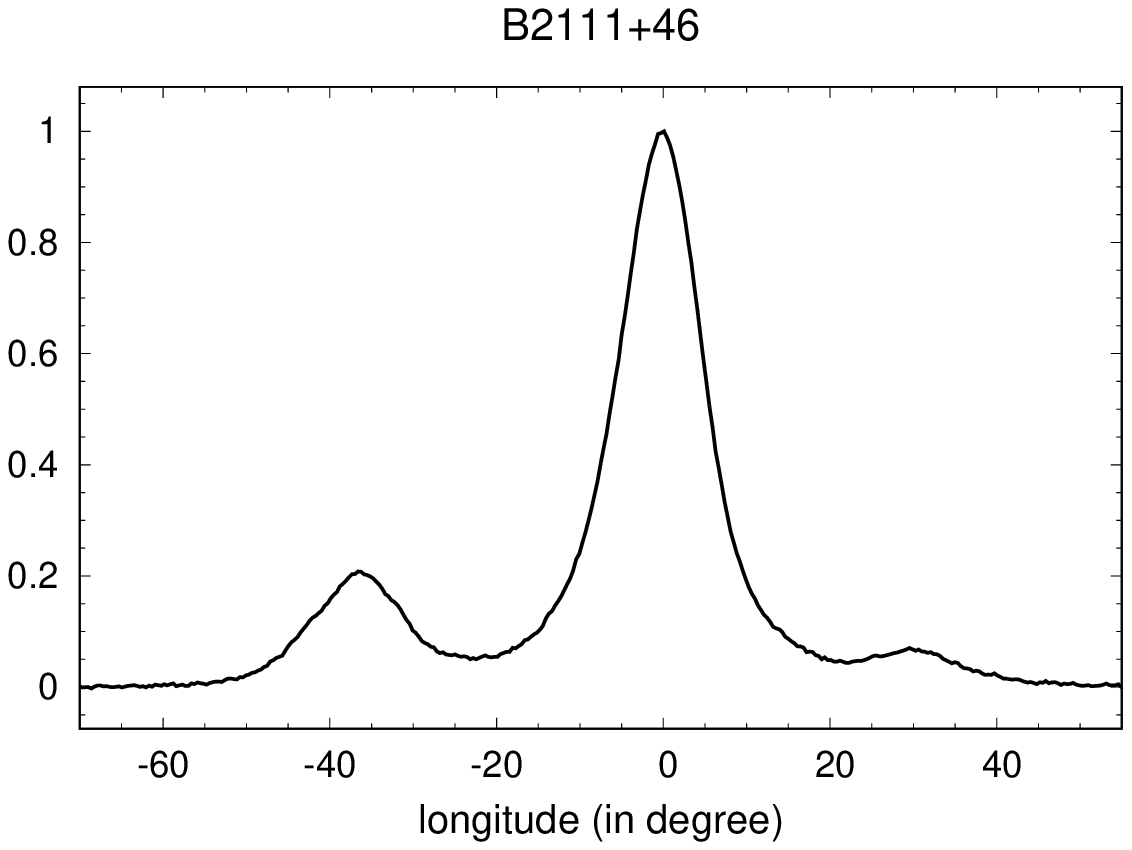}} &
\mbox{\includegraphics[angle=0,scale=0.57]{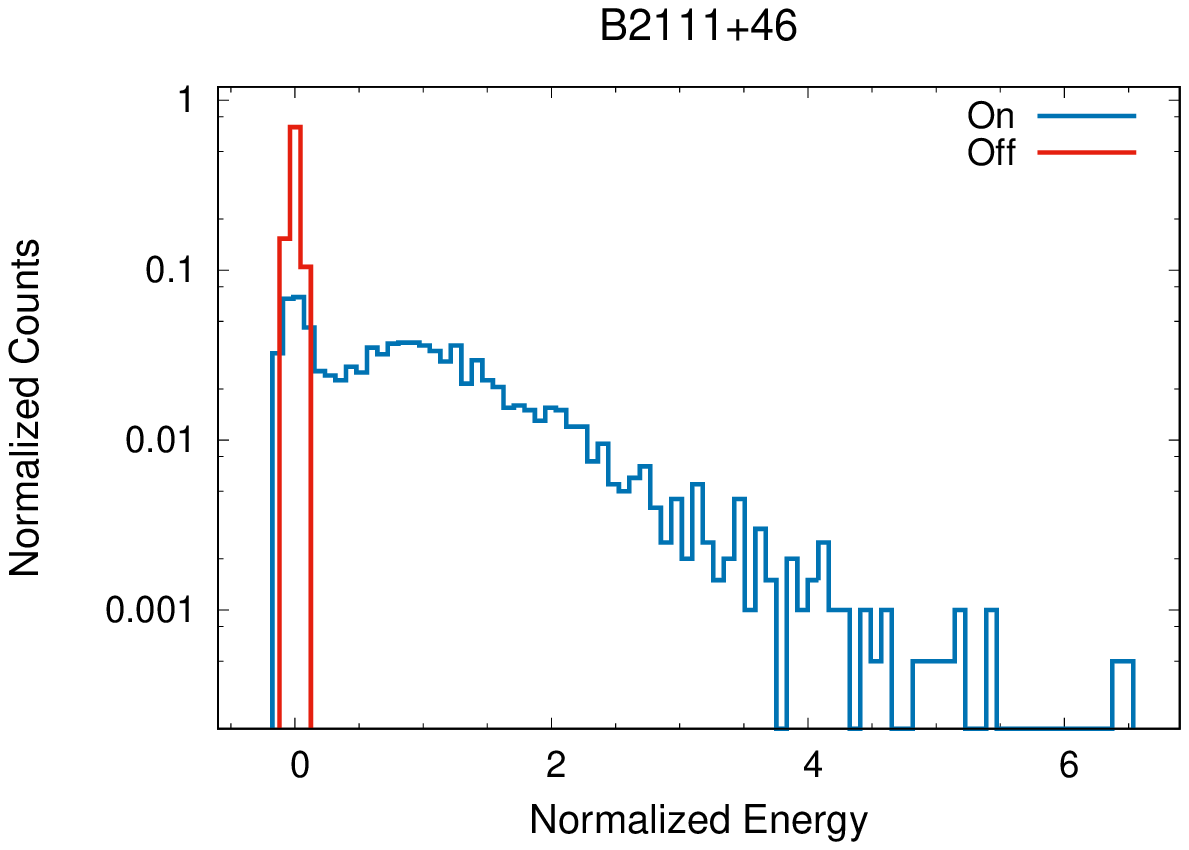}} \\
\mbox{\includegraphics[angle=0,scale=0.57]{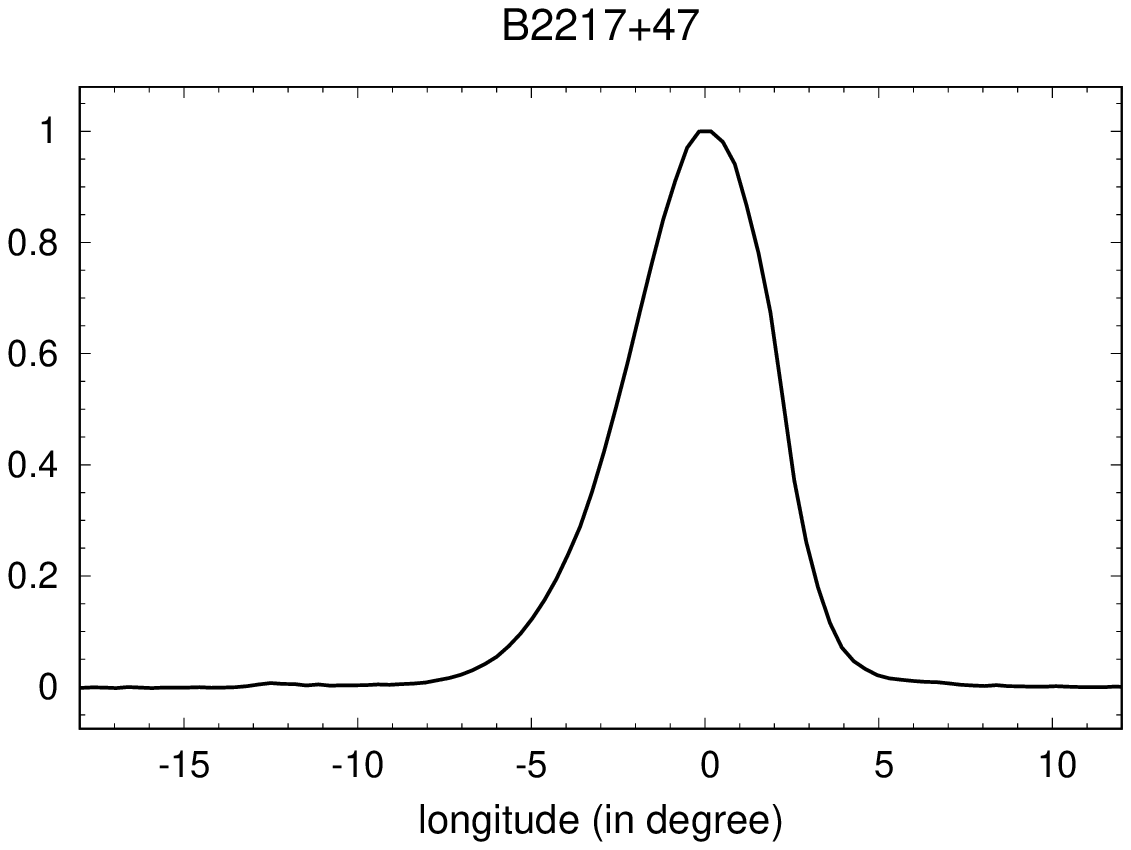}} &
\mbox{\includegraphics[angle=0,scale=0.57]{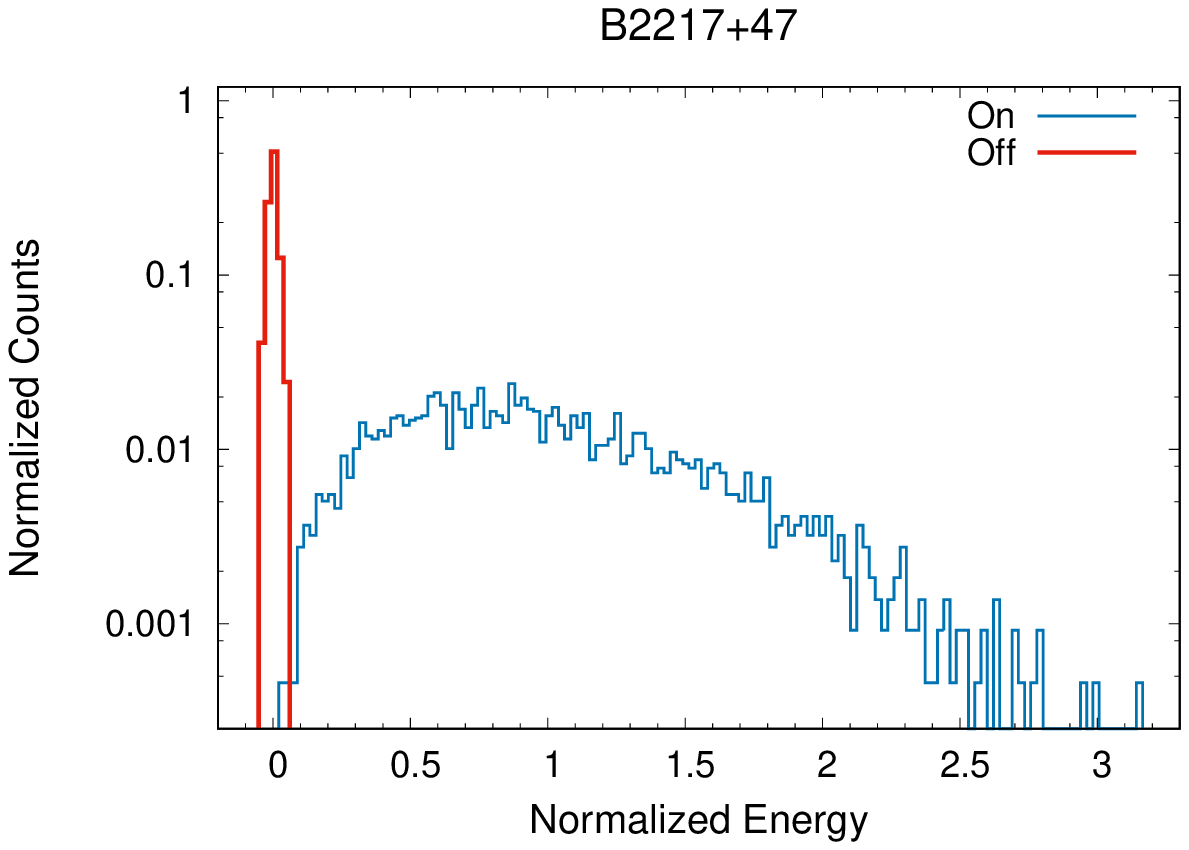}} \\
\mbox{\includegraphics[angle=0,scale=0.57]{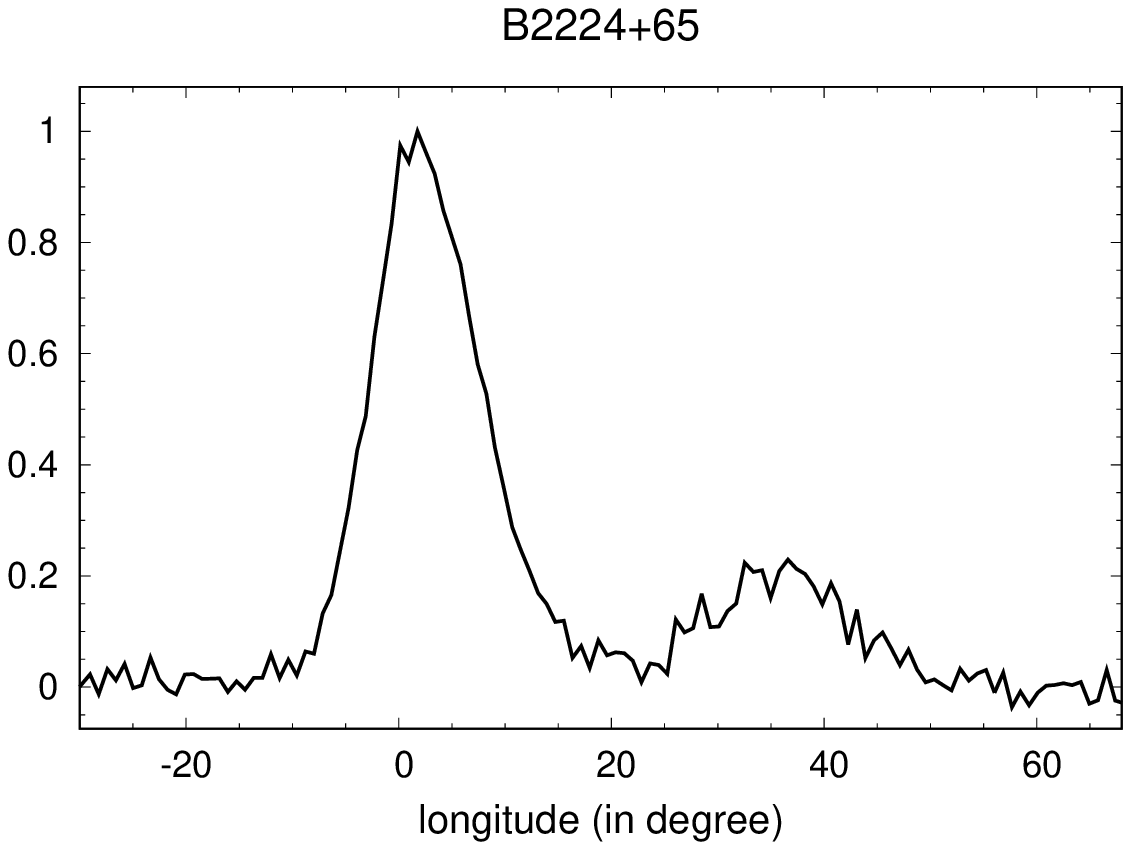}} &
\mbox{\includegraphics[angle=0,scale=0.57]{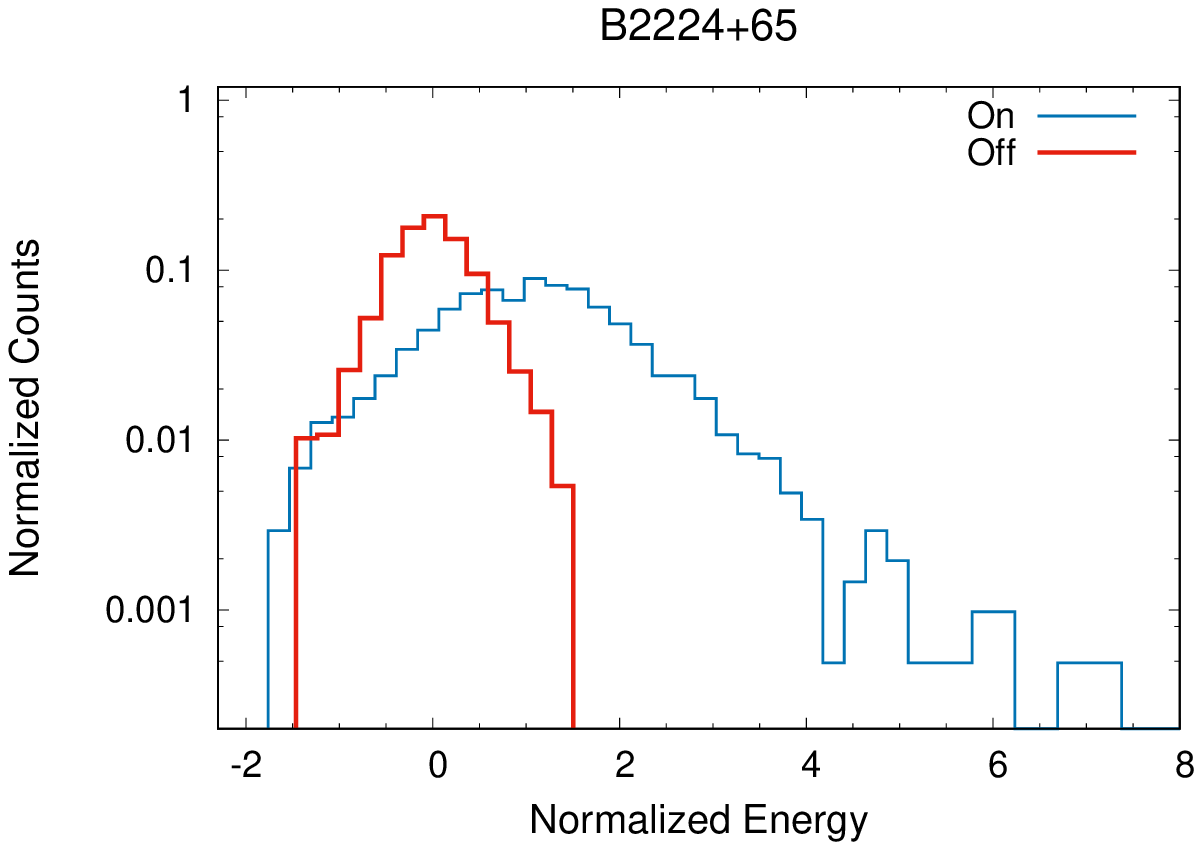}} \\
\mbox{\includegraphics[angle=0,scale=0.57]{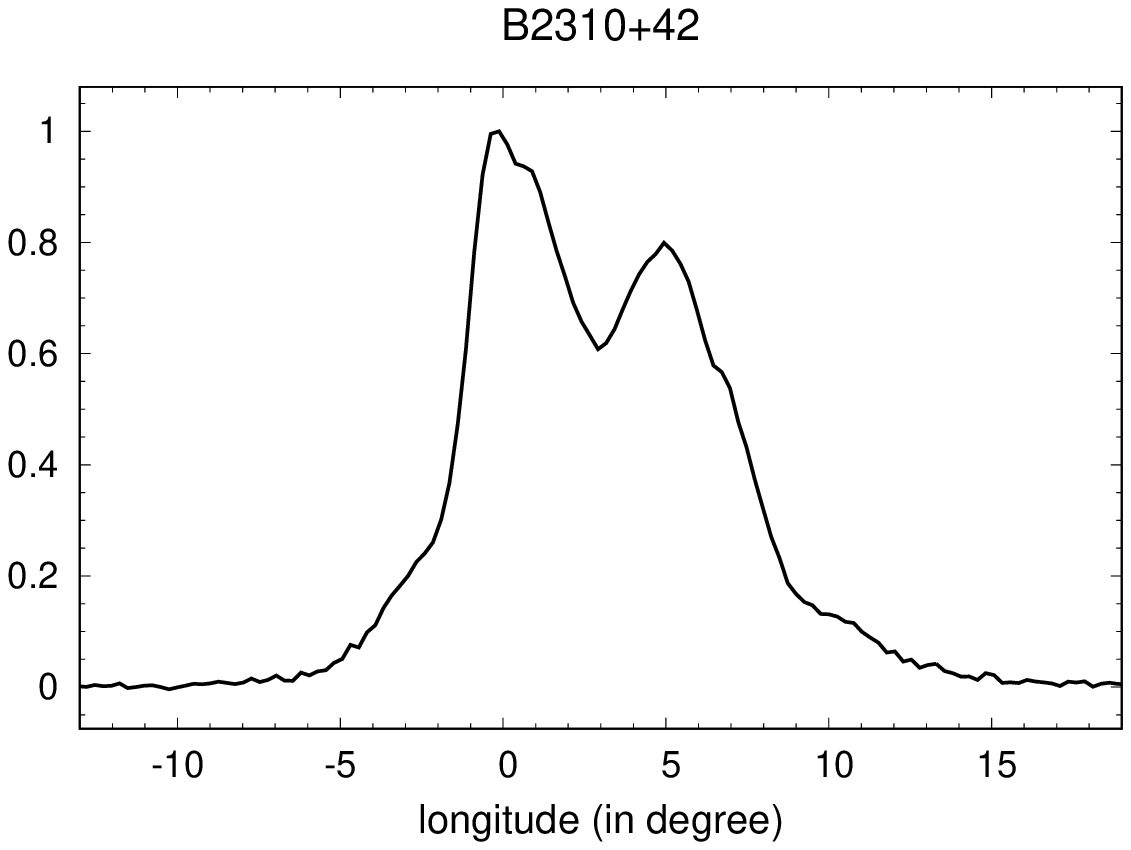}} &
\mbox{\includegraphics[angle=0,scale=0.57]{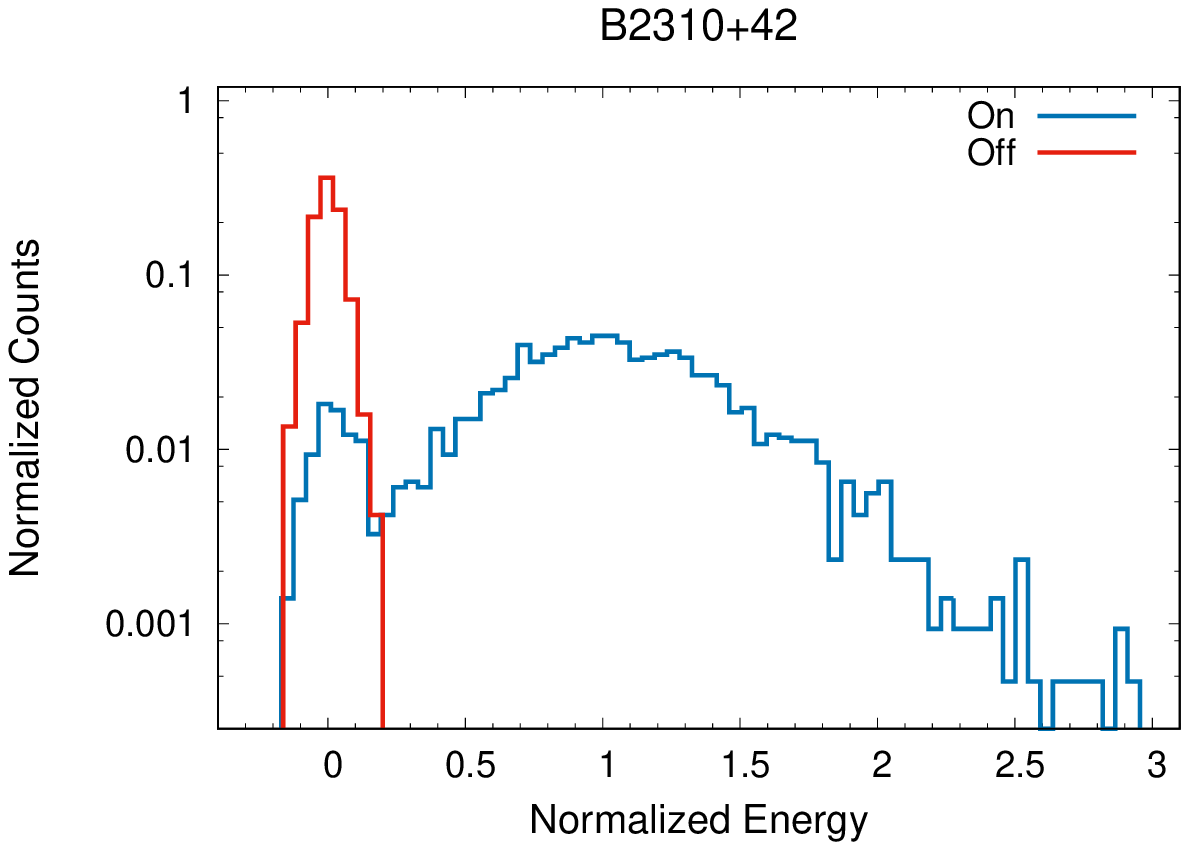}} \\
\end{tabular}
\caption{The pulsar profile On and Off-pulse energy distributions of the single pulse emission.}
\end{center}
\end{figure*}

\clearpage

%16th set of plots
\begin{figure*}
\begin{center}
\begin{tabular}{@{}cr@{}}
\mbox{\includegraphics[angle=0,scale=0.57]{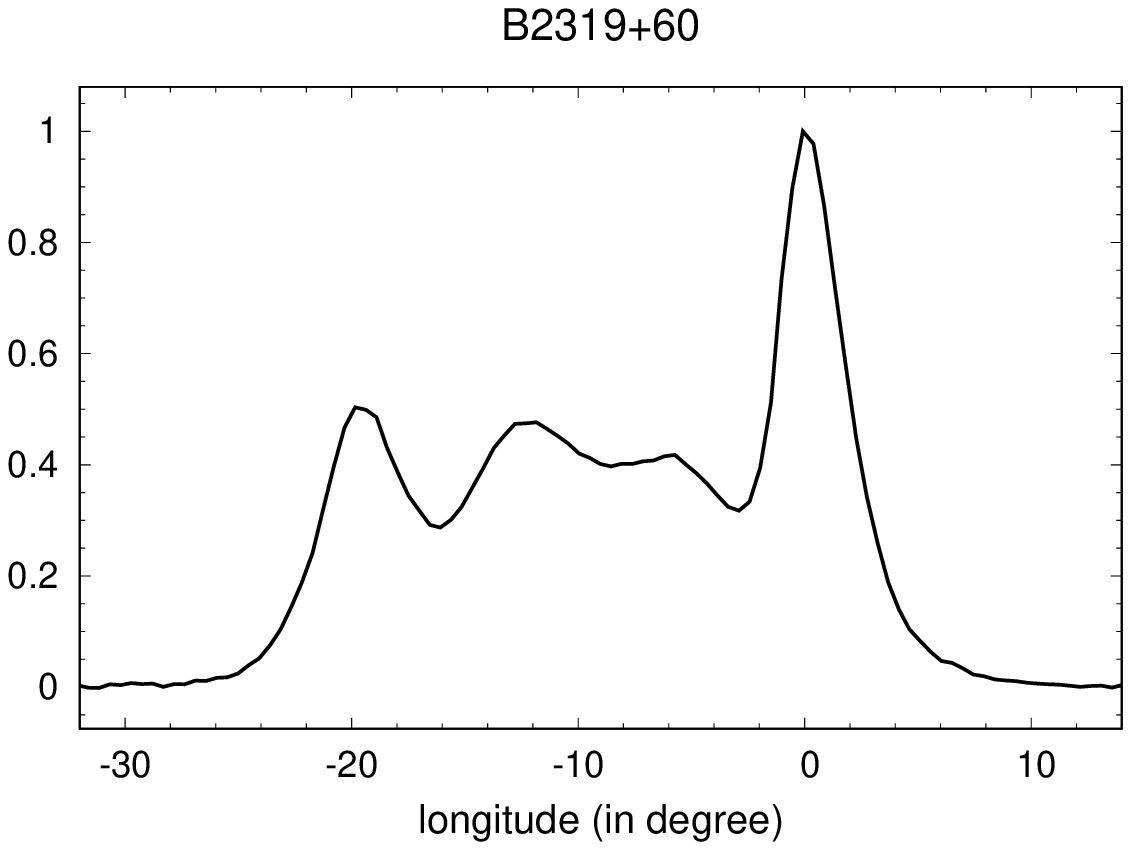}} &
\mbox{\includegraphics[angle=0,scale=0.57]{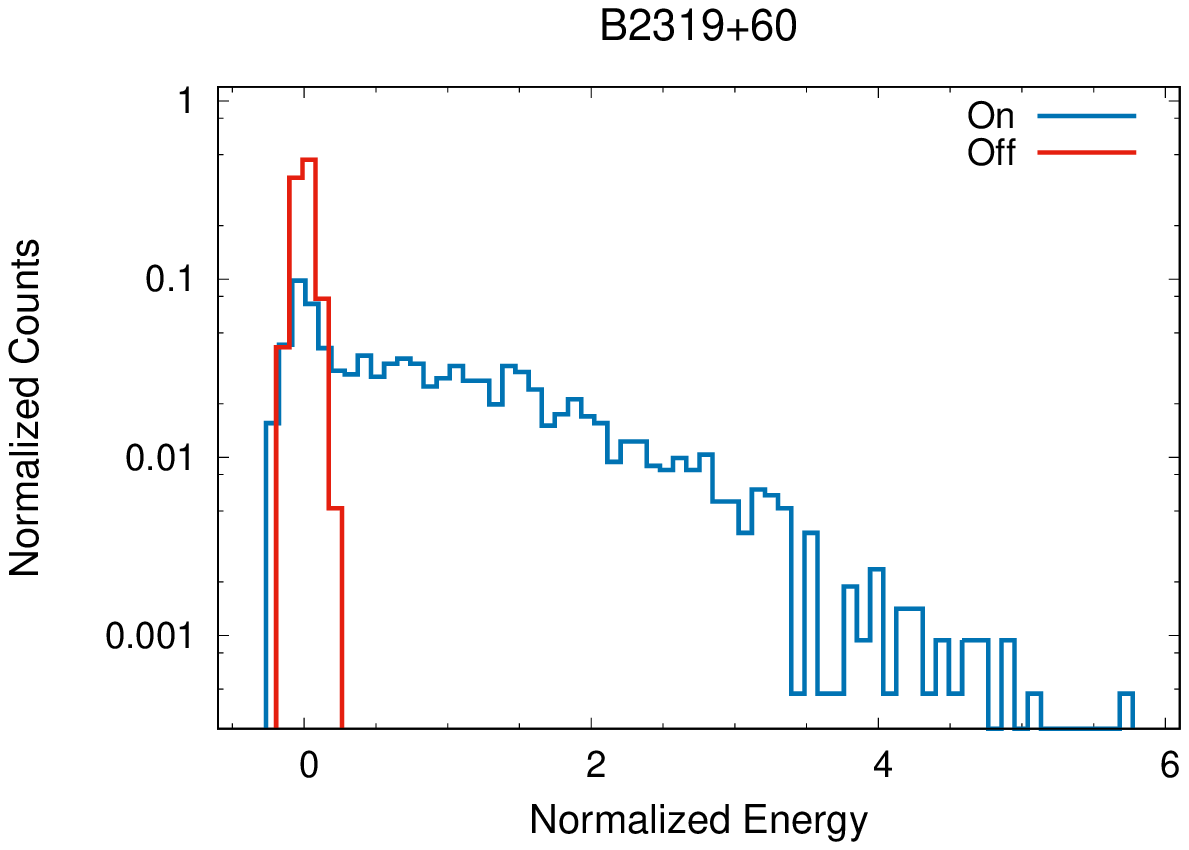}} \\
\mbox{\includegraphics[angle=0,scale=0.57]{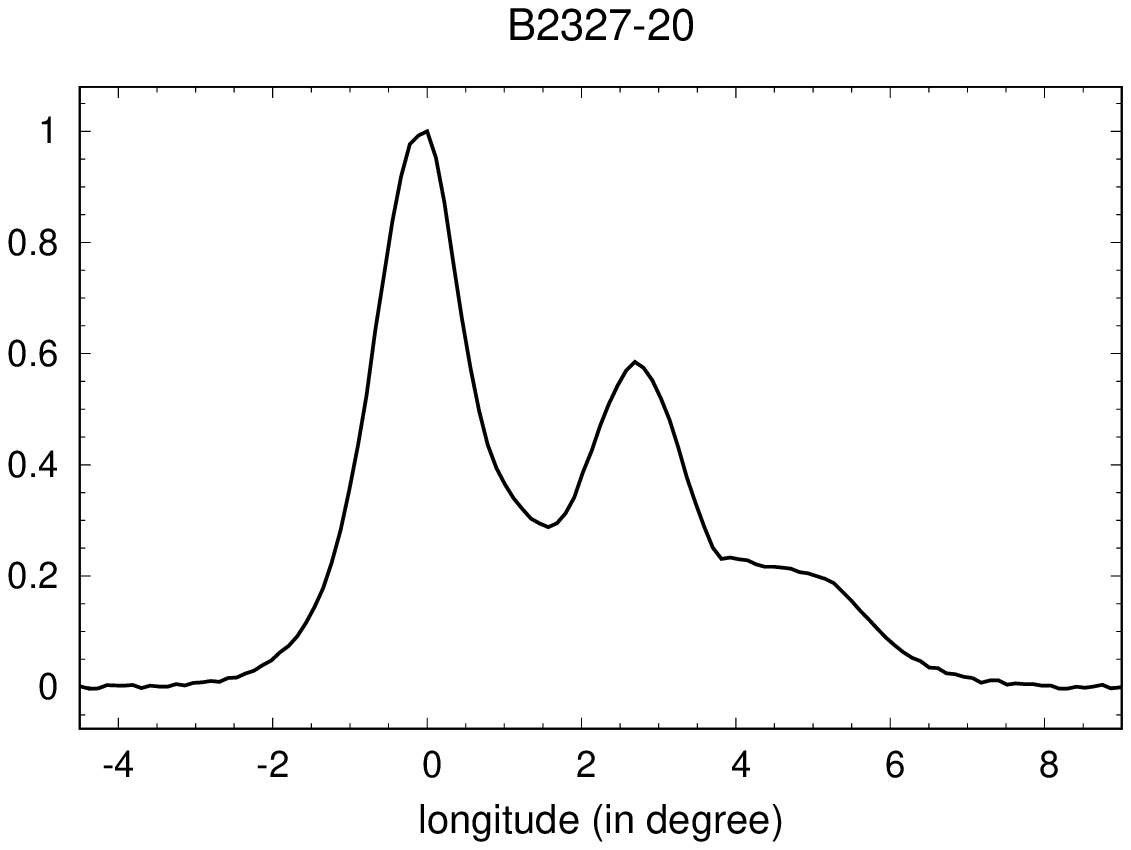}} &
\mbox{\includegraphics[angle=0,scale=0.57]{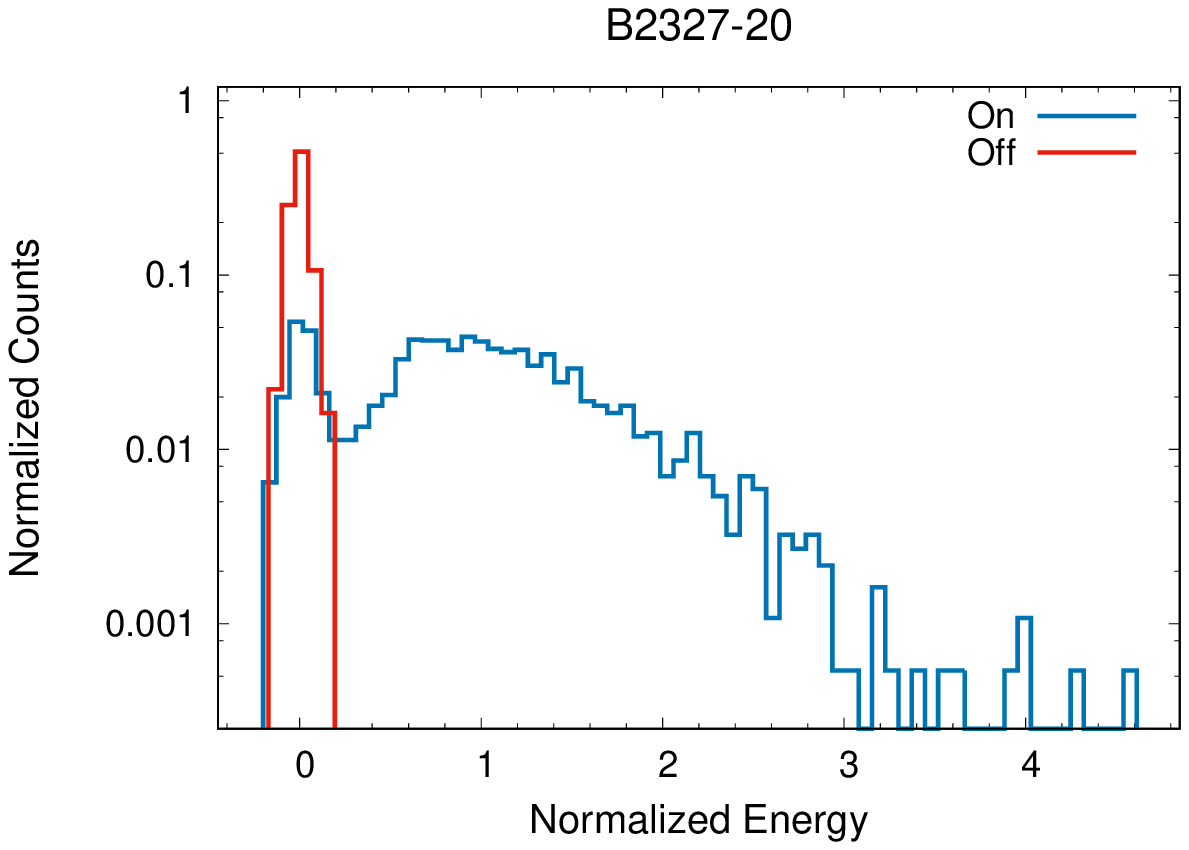}} \\
\end{tabular}
\caption{The pulsar profile On and Off-pulse energy distributions of the single pulse emission.}
\end{center}
\end{figure*}

\clearpage

%% file: appendix2.tex
\clearpage

%1st set of plots
\begin{figure*}
\begin{center}
\begin{tabular}{@{}cr@{}}
\mbox{\includegraphics[angle=0,scale=0.57]{J0141+6009_burlen.eps}} &
\mbox{\includegraphics[angle=0,scale=0.57]{J0141+6009_nullen.eps}} \\
\mbox{\includegraphics[angle=0,scale=0.57]{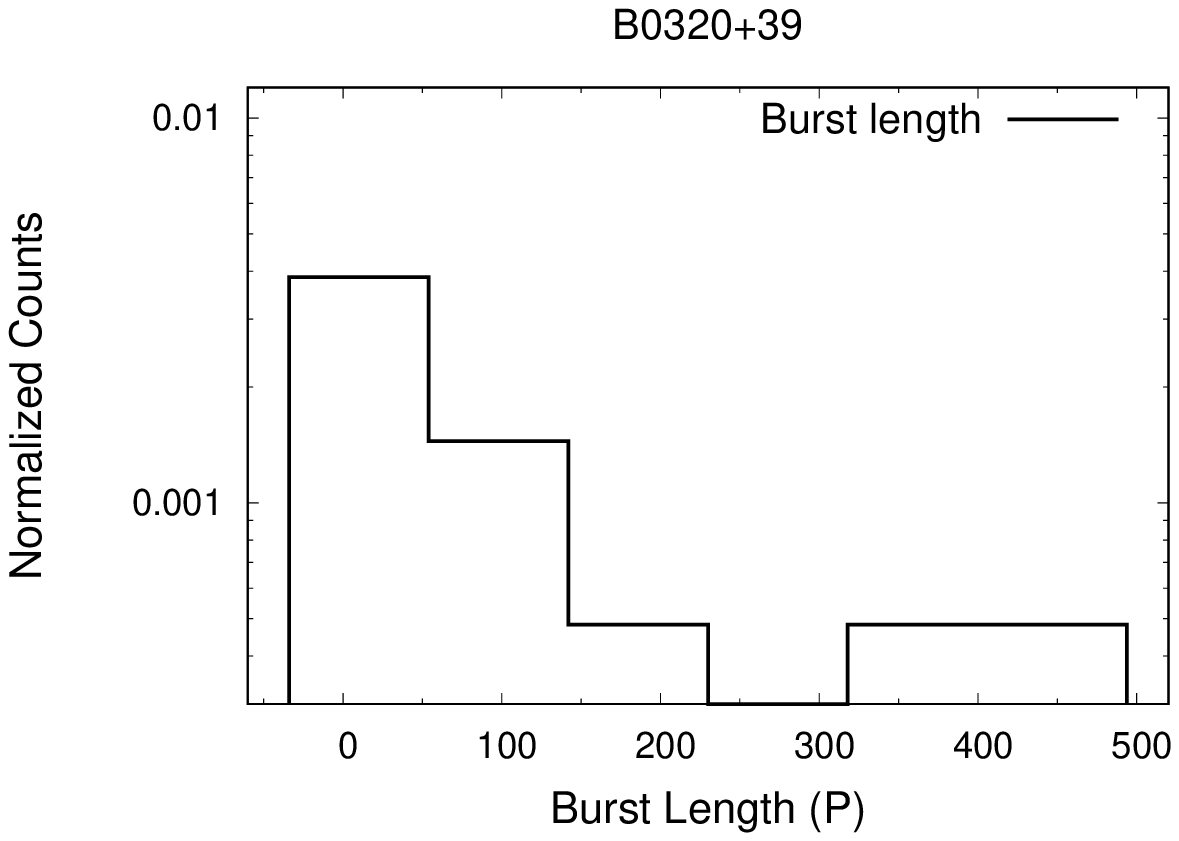}} &
\mbox{\includegraphics[angle=0,scale=0.57]{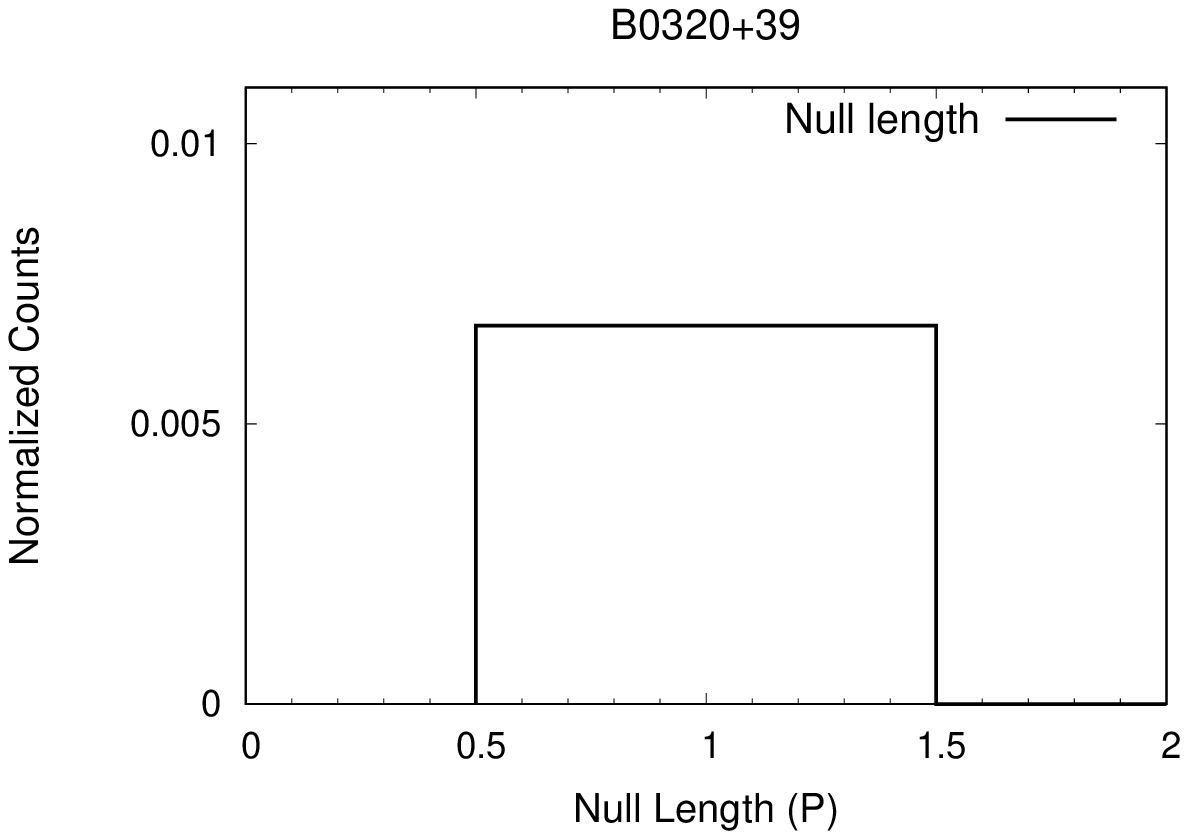}} \\
\mbox{\includegraphics[angle=0,scale=0.57]{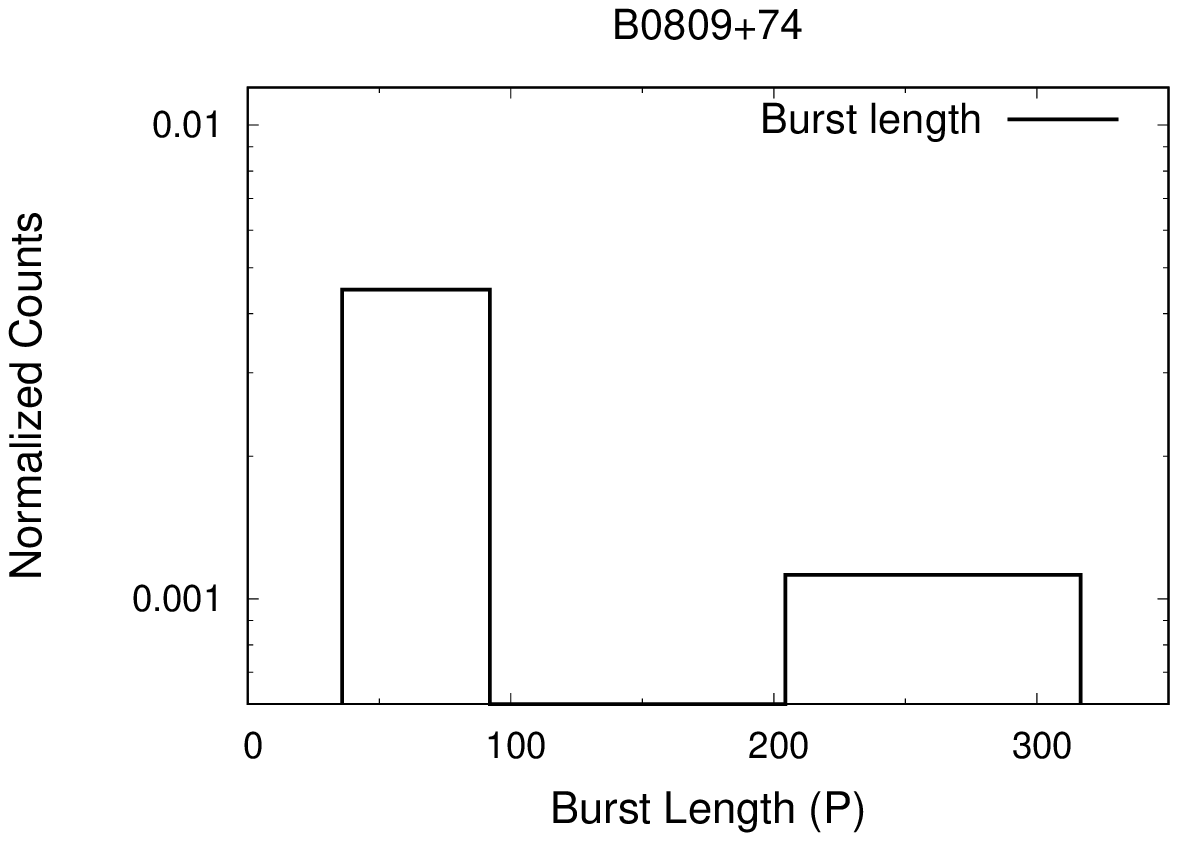}} &
\mbox{\includegraphics[angle=0,scale=0.57]{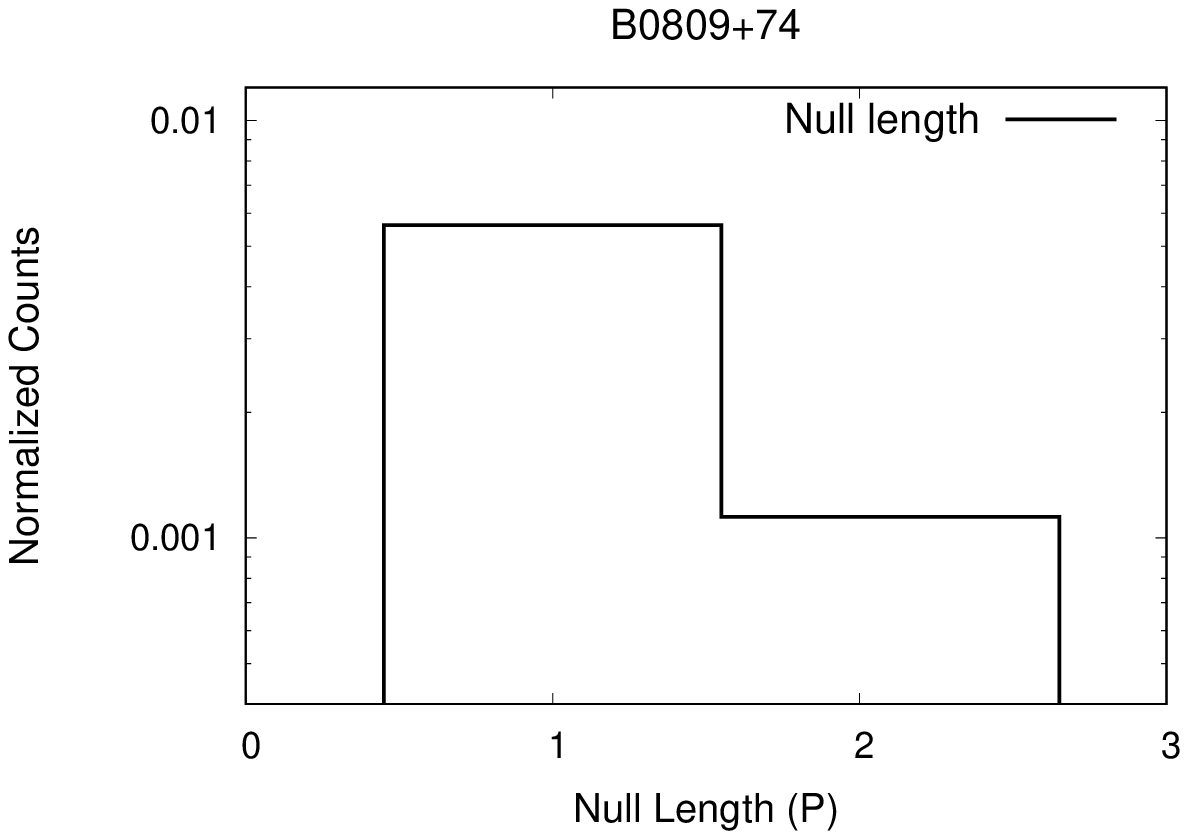}} \\
\mbox{\includegraphics[angle=0,scale=0.57]{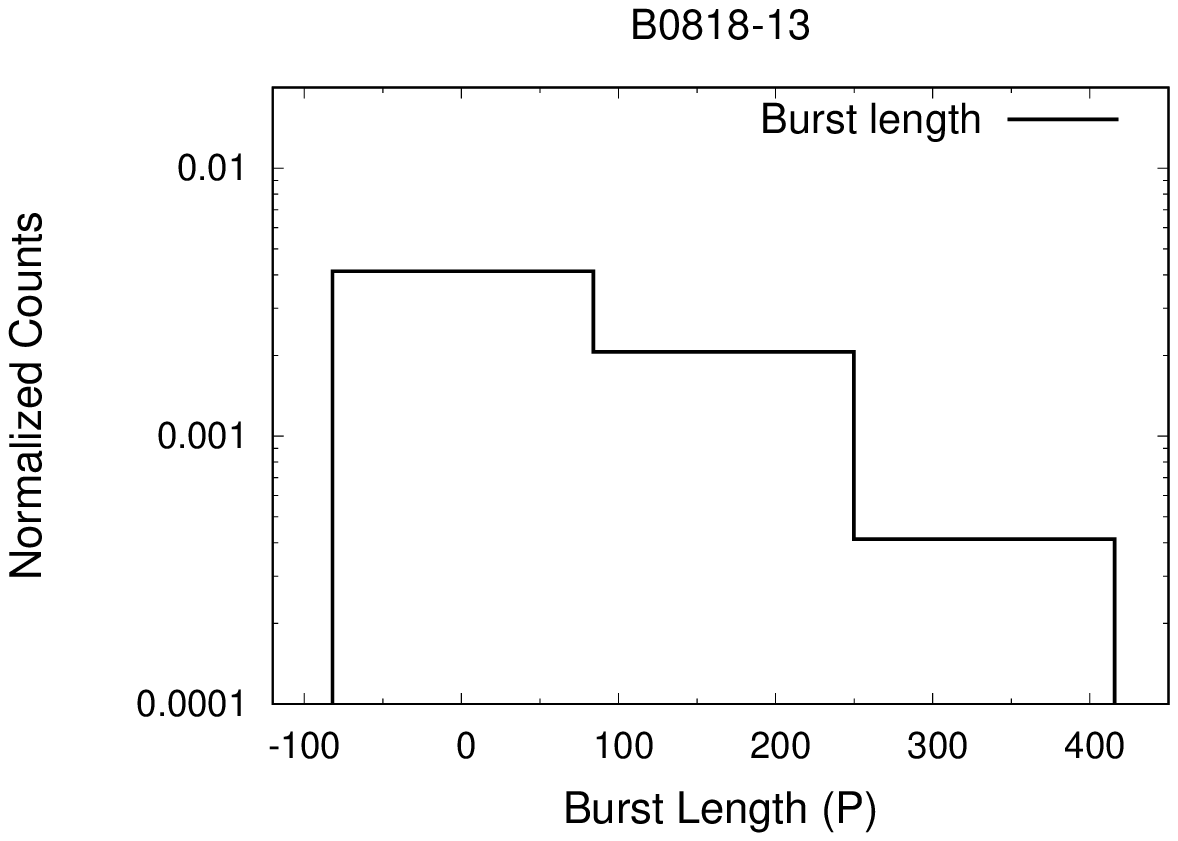}} &
\mbox{\includegraphics[angle=0,scale=0.57]{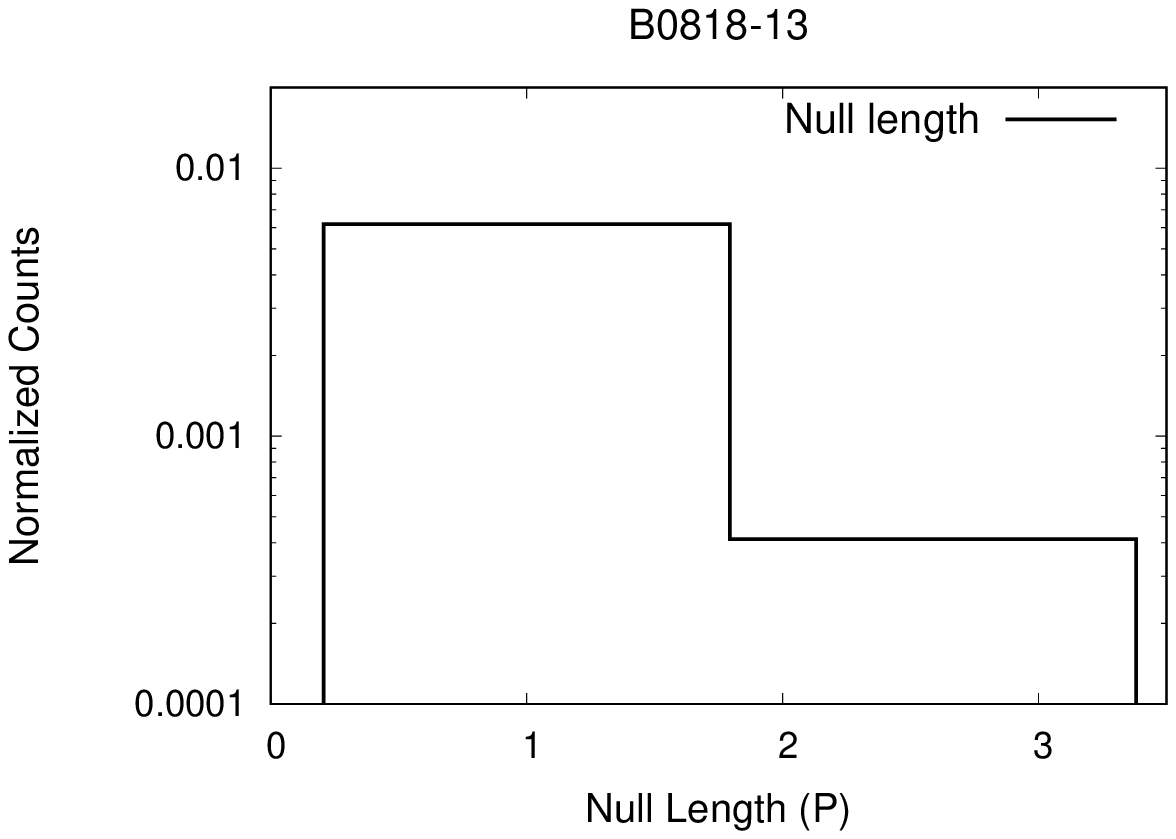}} \\
\end{tabular}
\caption{The Burst length (left panel) and Null length (right panel) distributions.}
\end{center}
\end{figure*}

\clearpage

%2nd set of plots
\begin{figure*}
\begin{center}
\begin{tabular}{@{}cr@{}}
\mbox{\includegraphics[angle=0,scale=0.57]{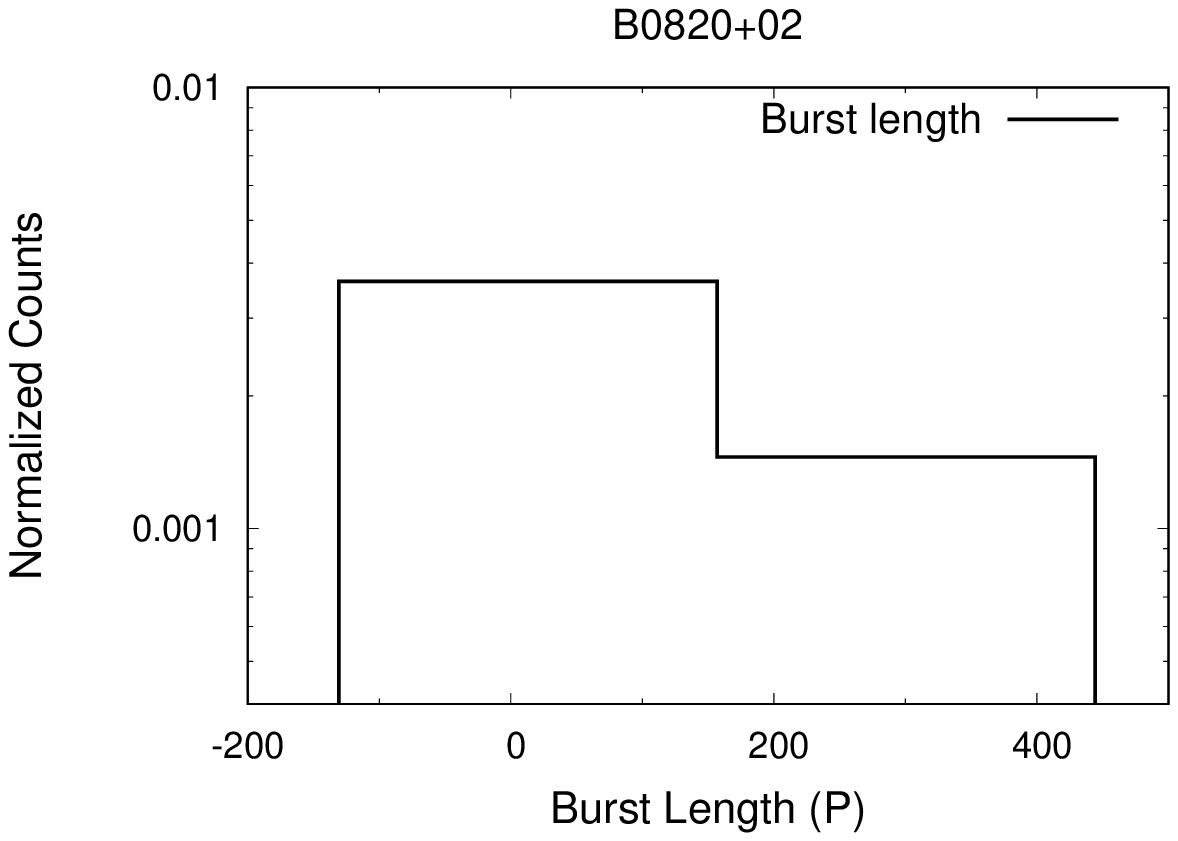}} &
\mbox{\includegraphics[angle=0,scale=0.57]{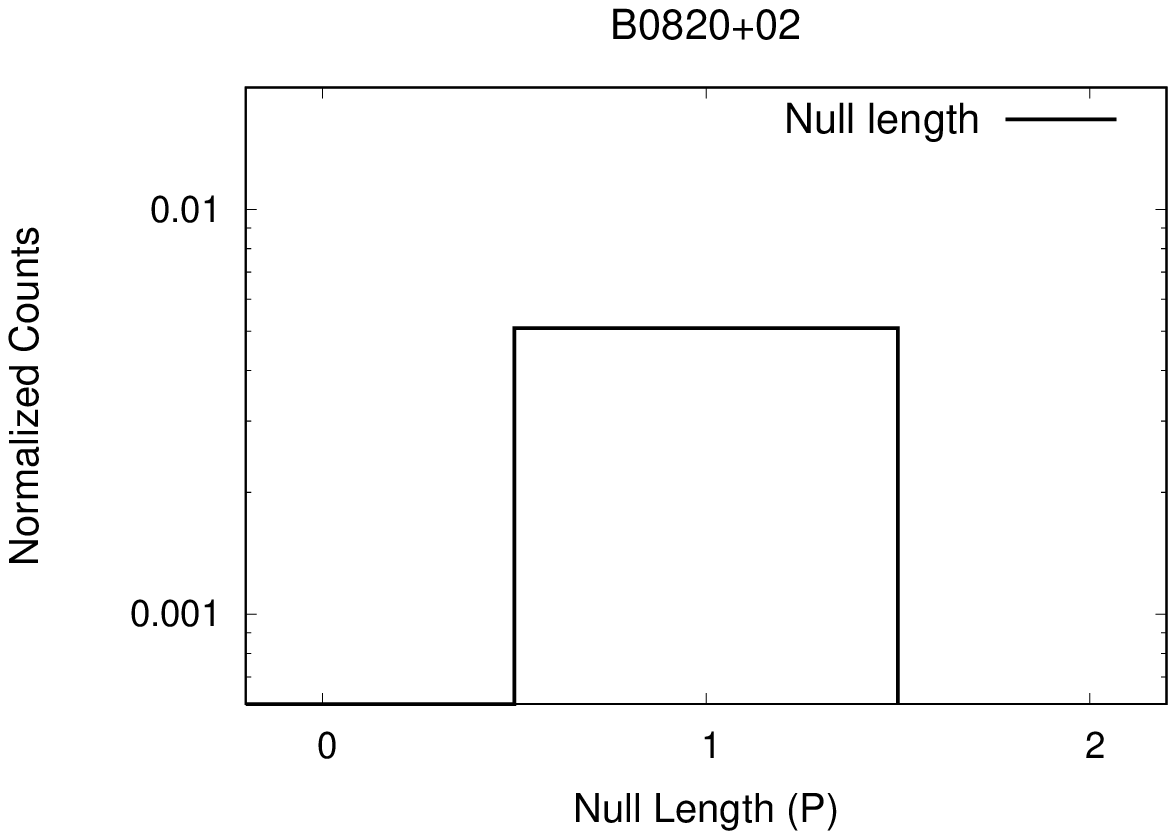}} \\
\mbox{\includegraphics[angle=0,scale=0.57]{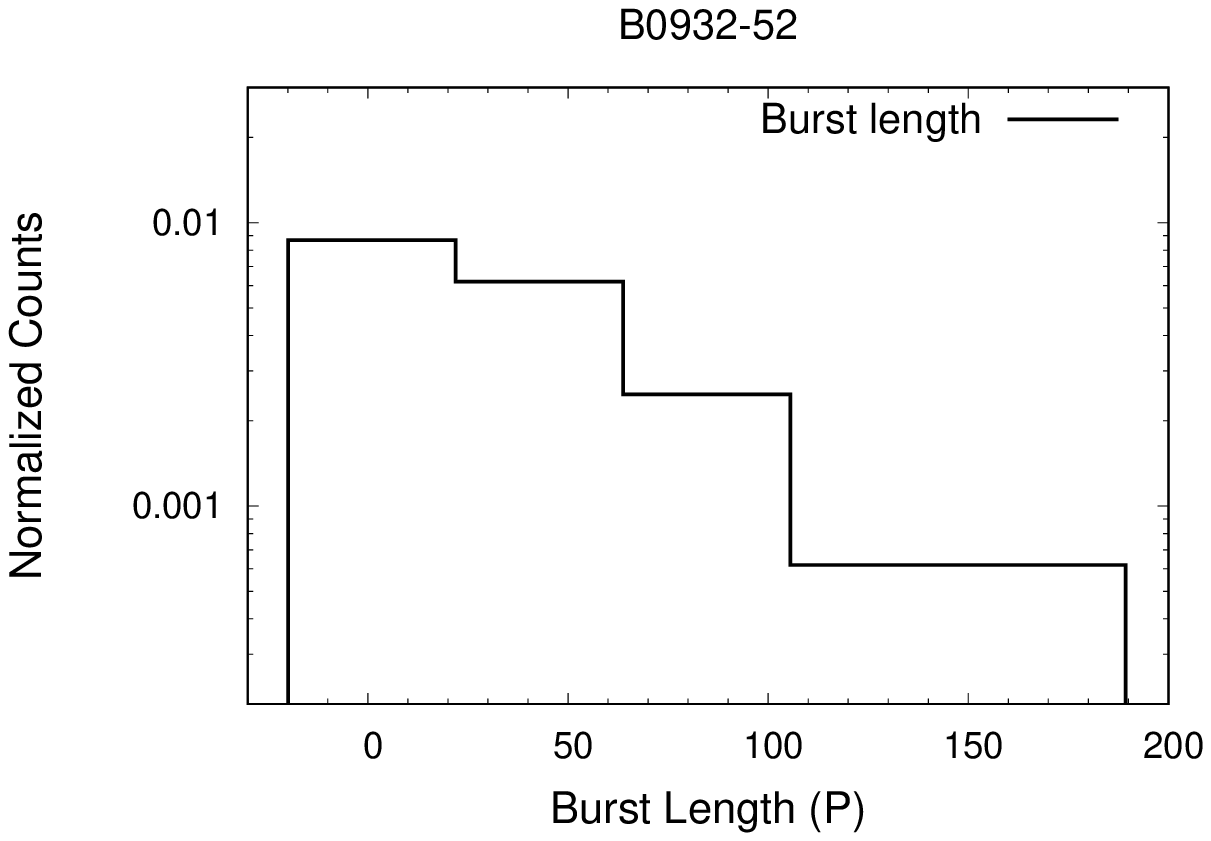}} &
\mbox{\includegraphics[angle=0,scale=0.57]{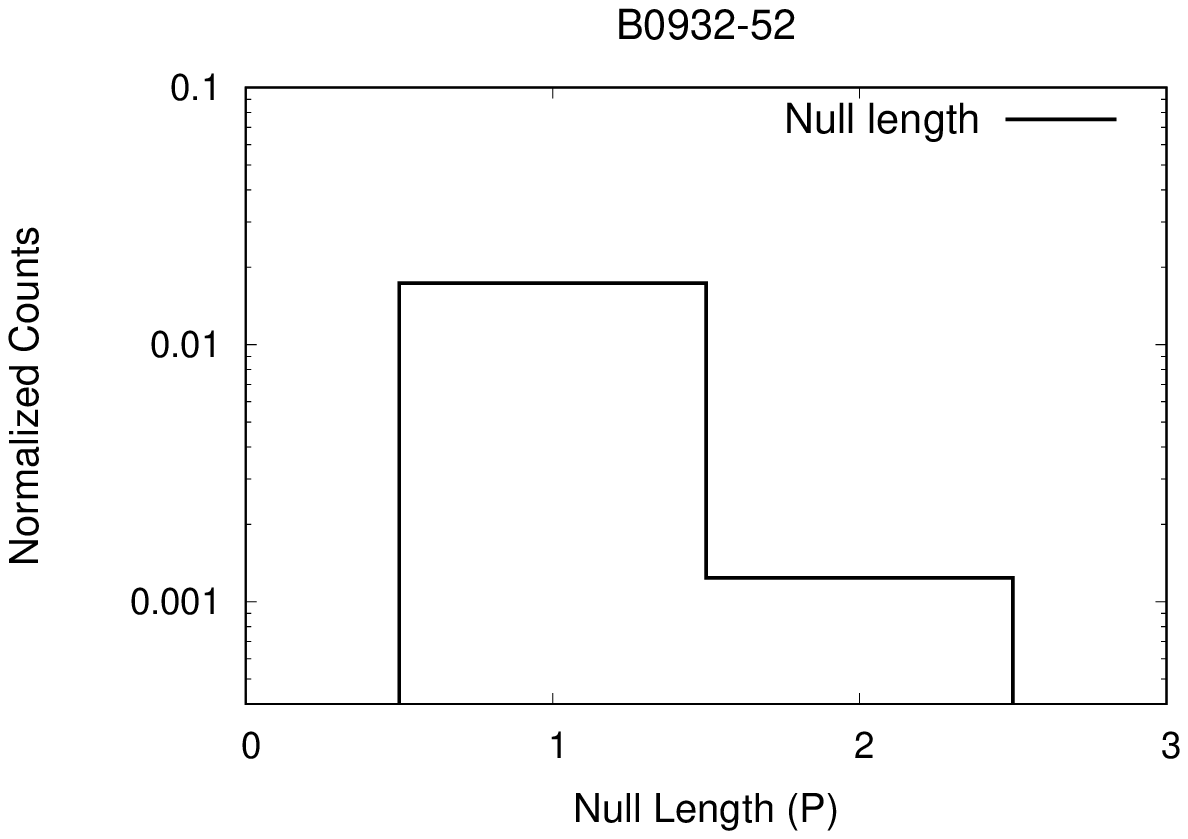}} \\
\mbox{\includegraphics[angle=0,scale=0.57]{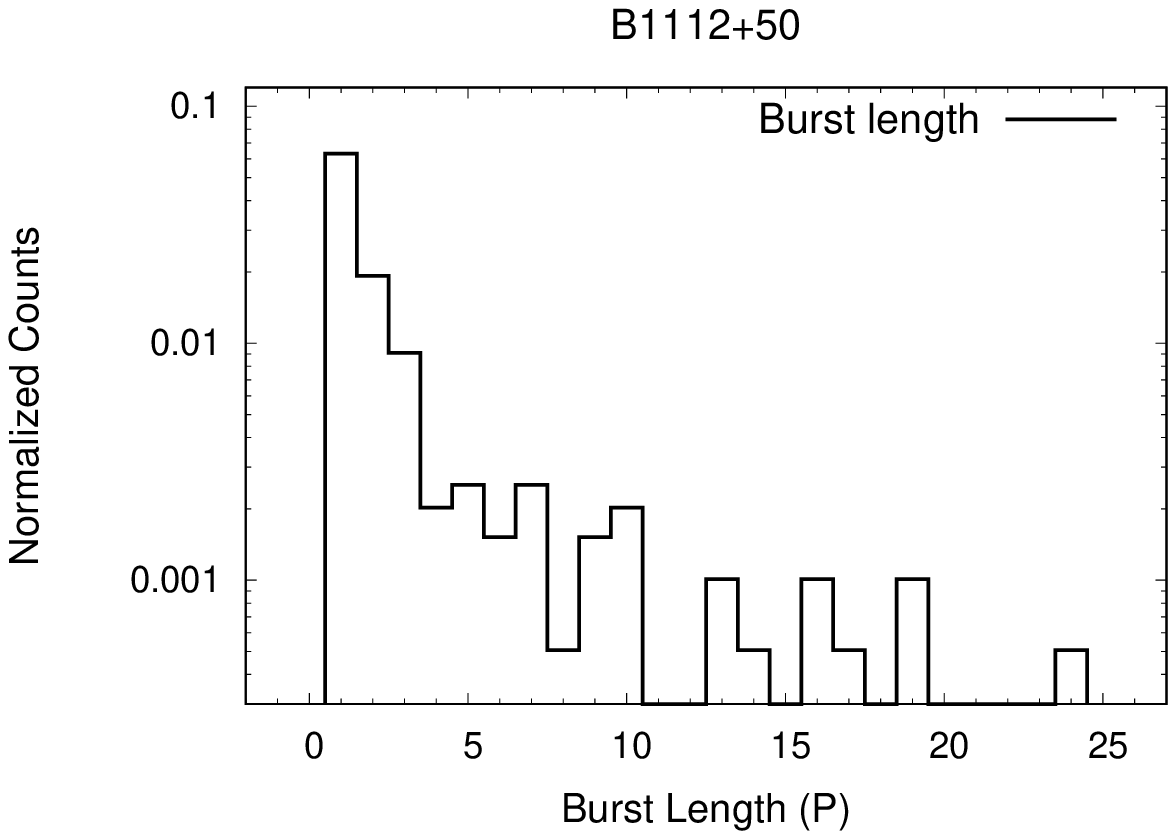}} &
\mbox{\includegraphics[angle=0,scale=0.57]{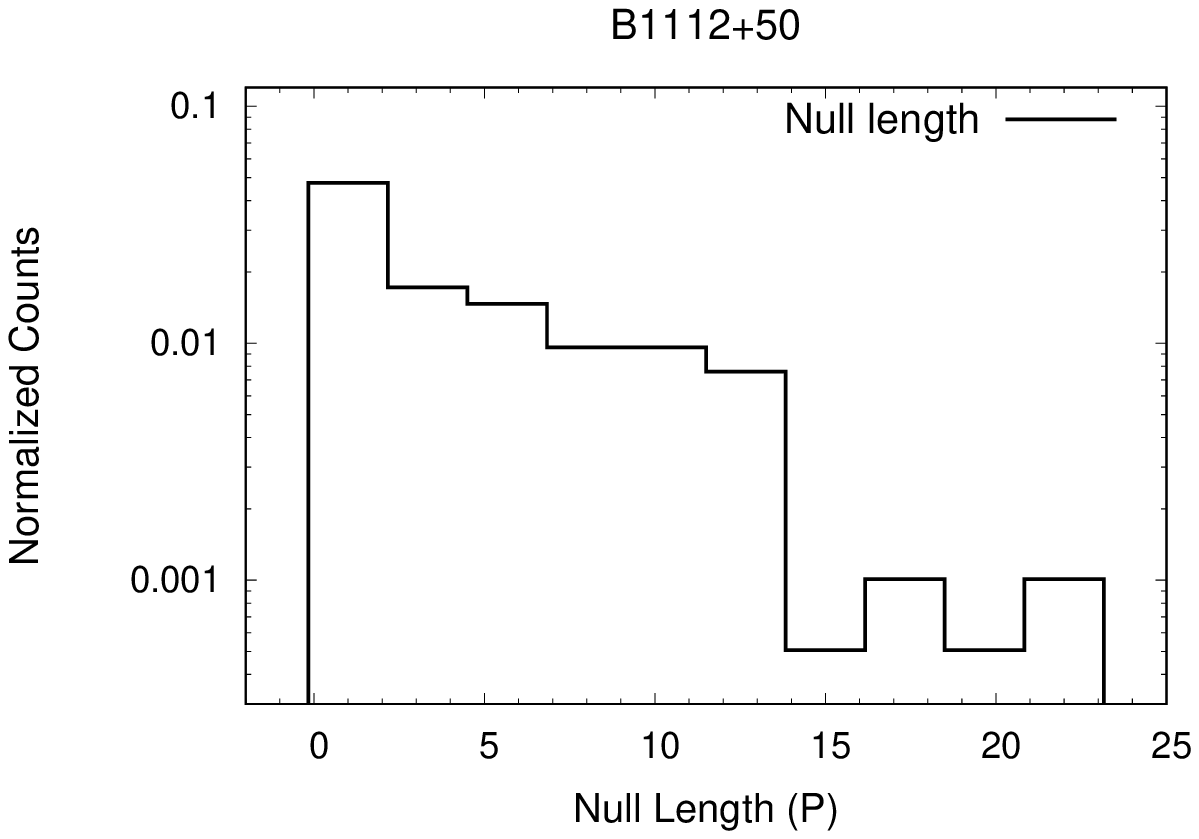}} \\
\mbox{\includegraphics[angle=0,scale=0.57]{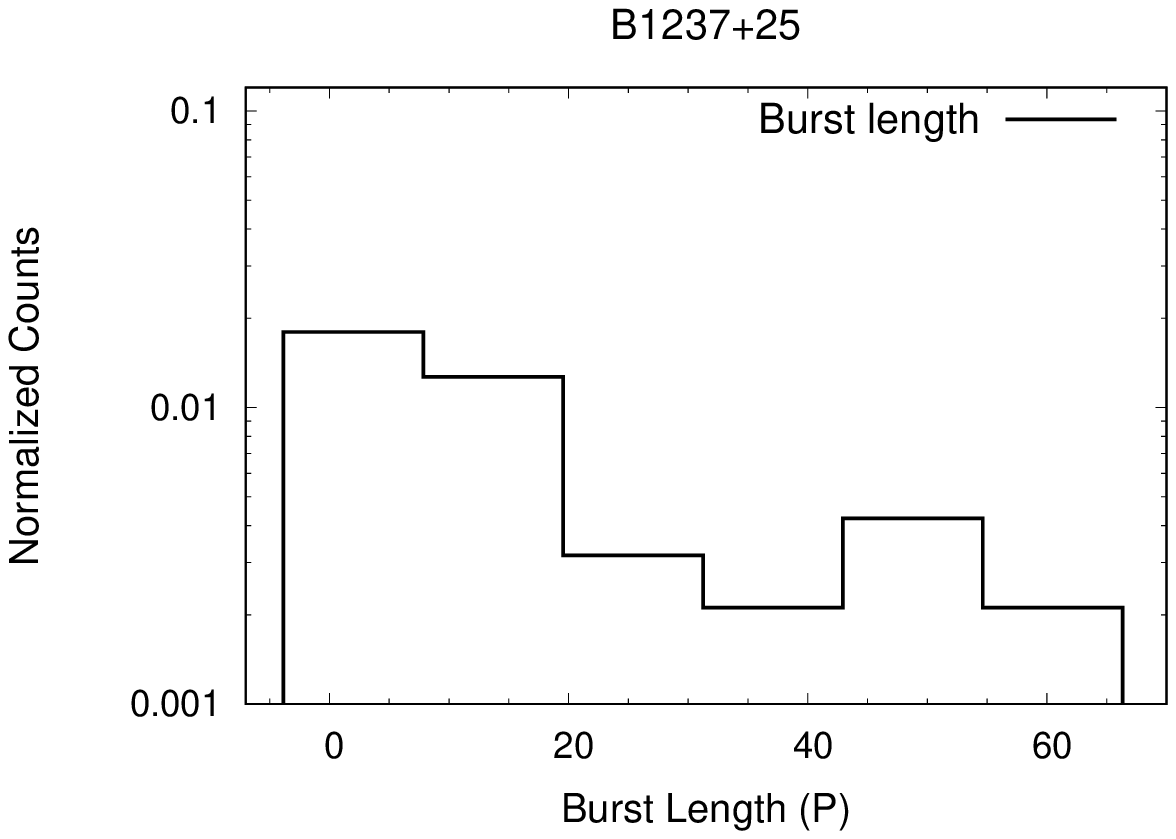}} &
\mbox{\includegraphics[angle=0,scale=0.57]{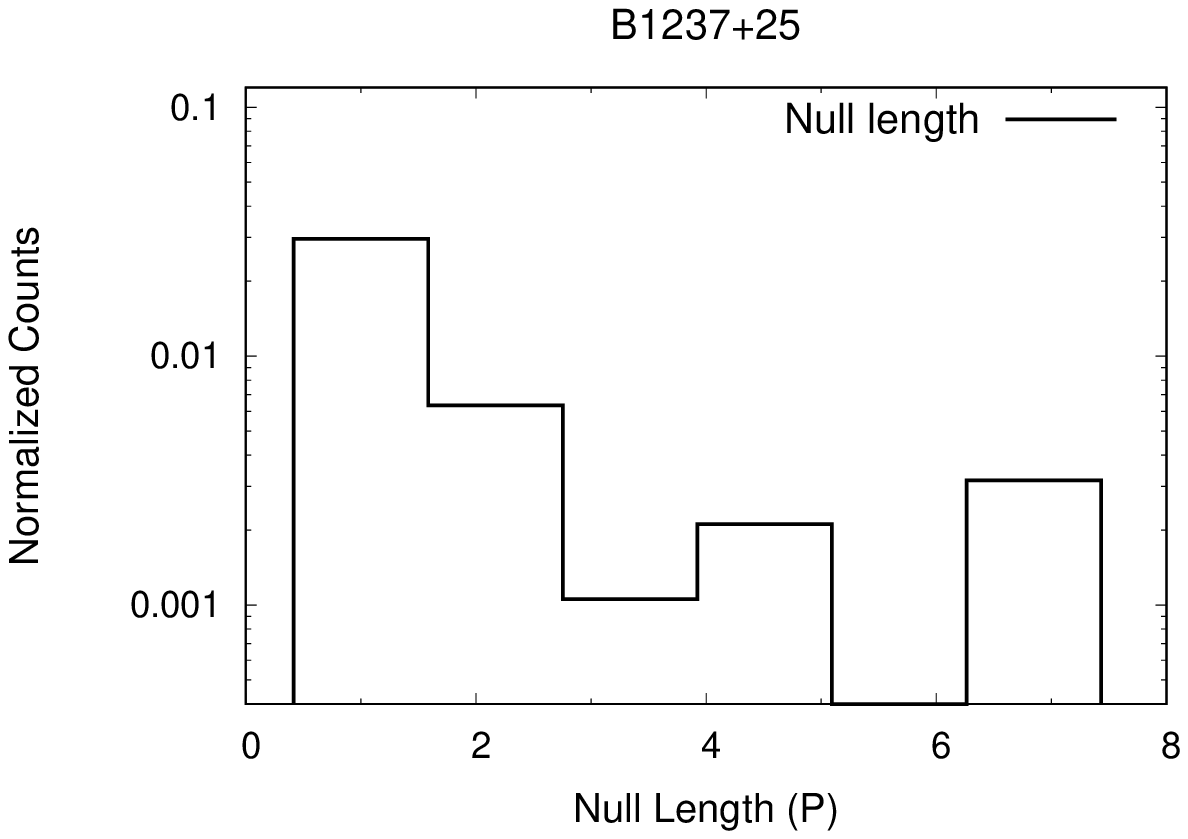}} \\
\end{tabular}
\caption{The Burst length (left panel) and Null length (right panel) distributions.}
\end{center}
\end{figure*}

\clearpage

%3rd set of plots
\begin{figure*}
\begin{center}
\begin{tabular}{@{}cr@{}}
\mbox{\includegraphics[angle=0,scale=0.57]{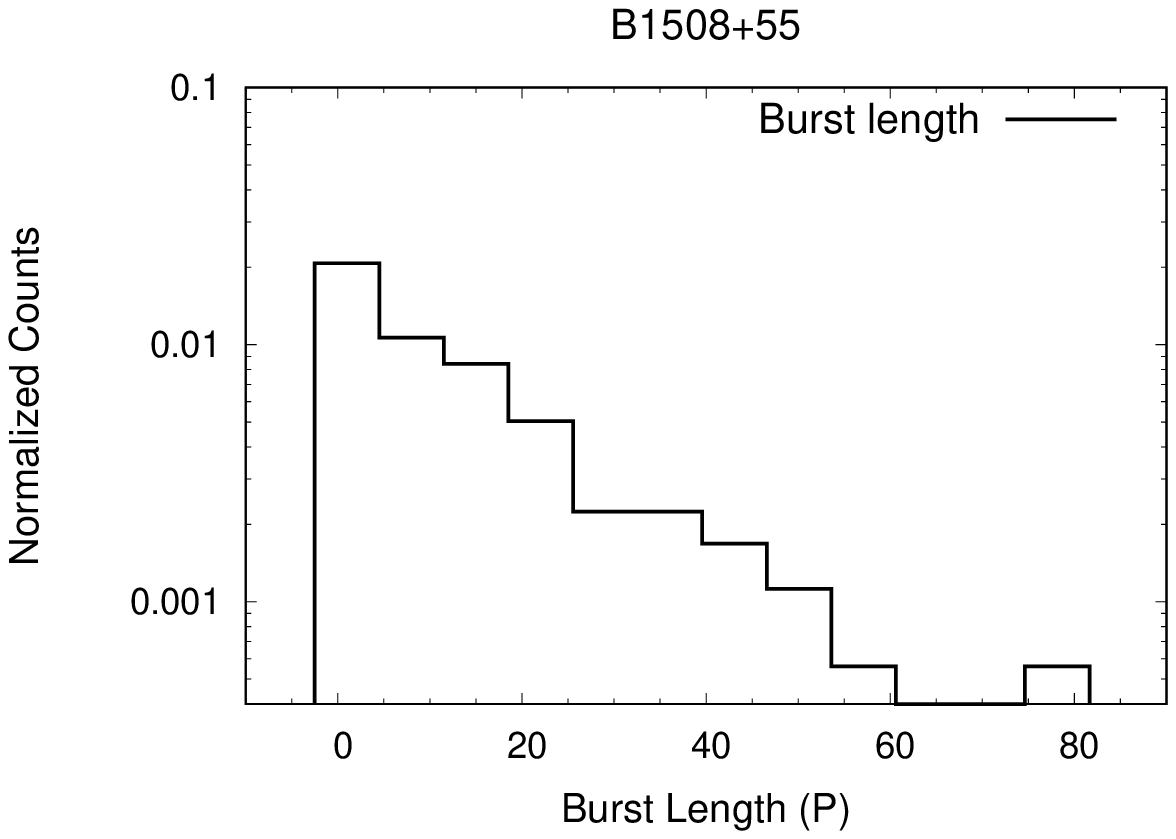}} &
\mbox{\includegraphics[angle=0,scale=0.57]{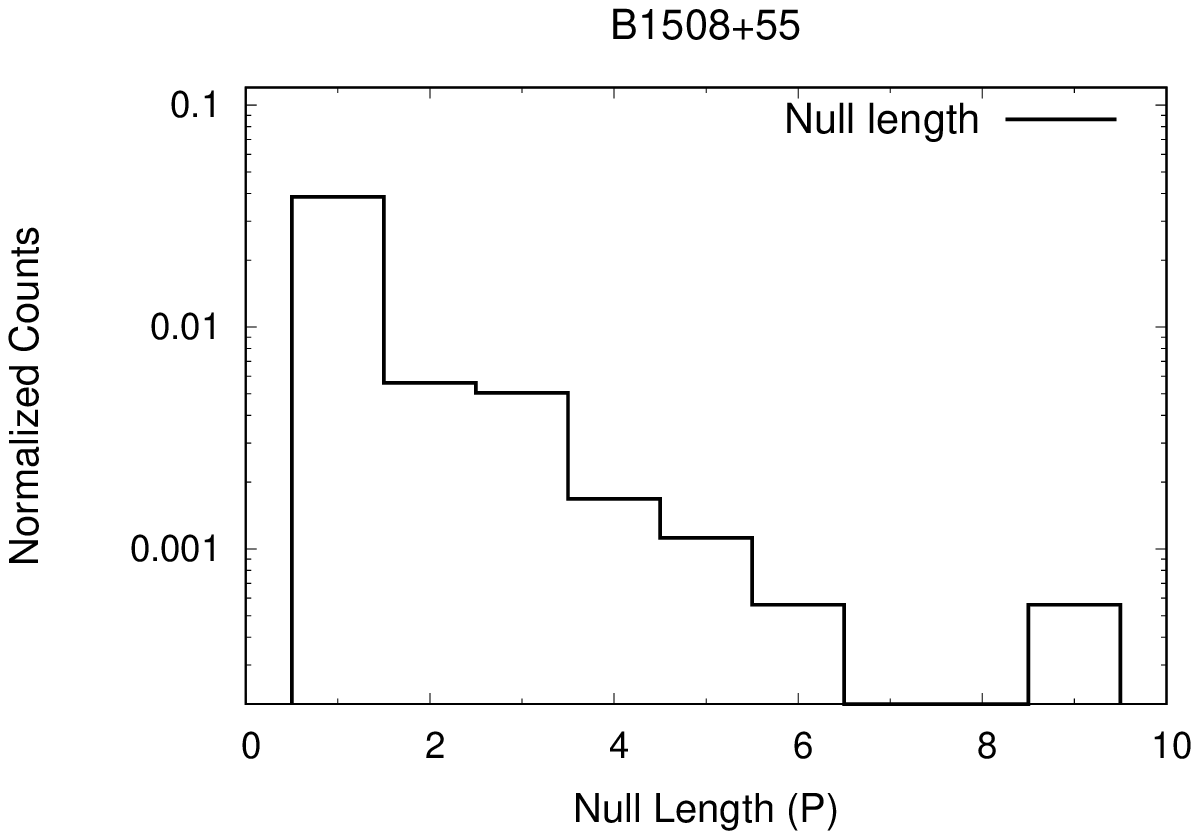}} \\
\mbox{\includegraphics[angle=0,scale=0.57]{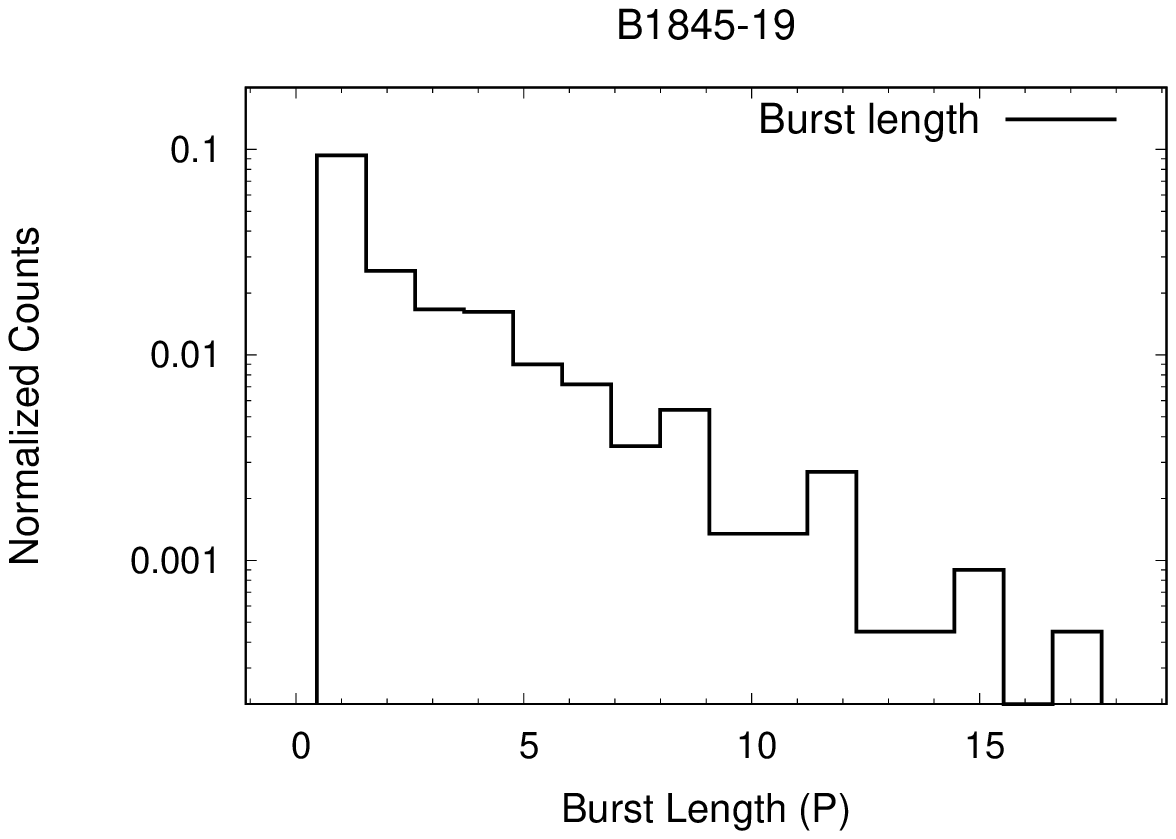}} &
\mbox{\includegraphics[angle=0,scale=0.57]{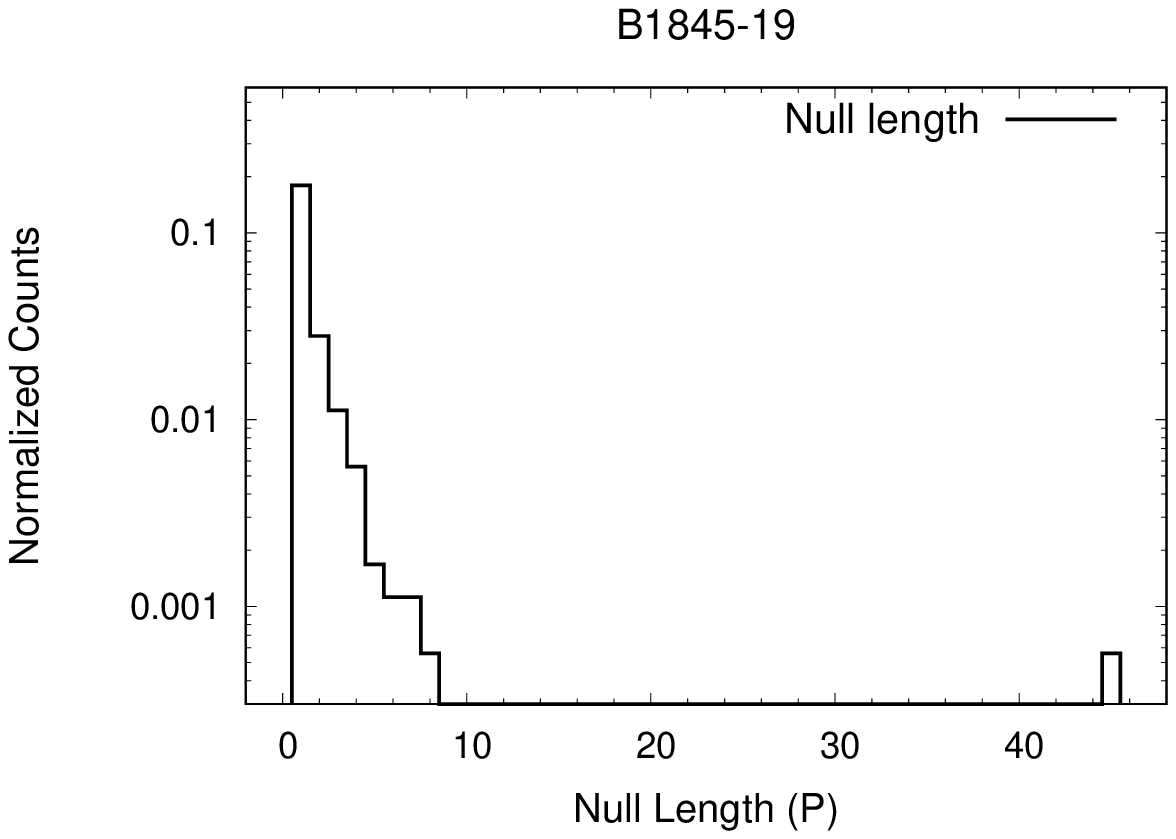}} \\
\mbox{\includegraphics[angle=0,scale=0.57]{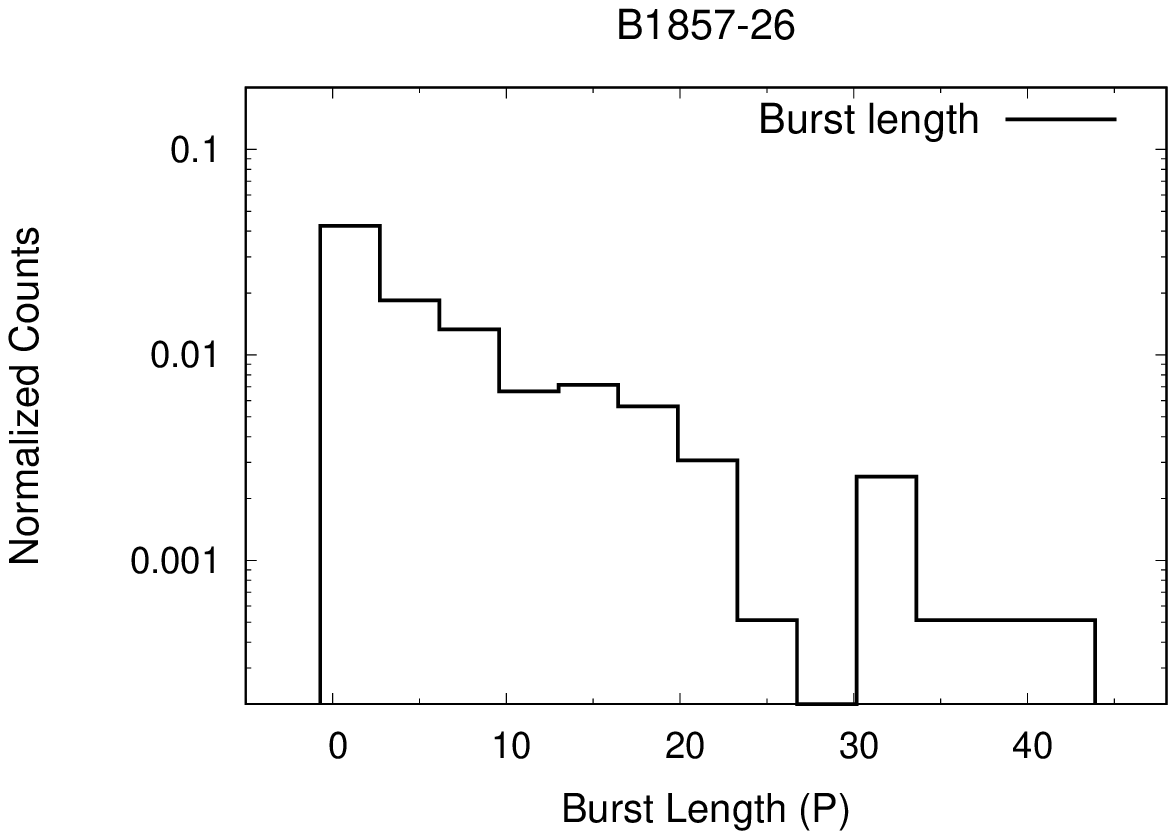}} &
\mbox{\includegraphics[angle=0,scale=0.57]{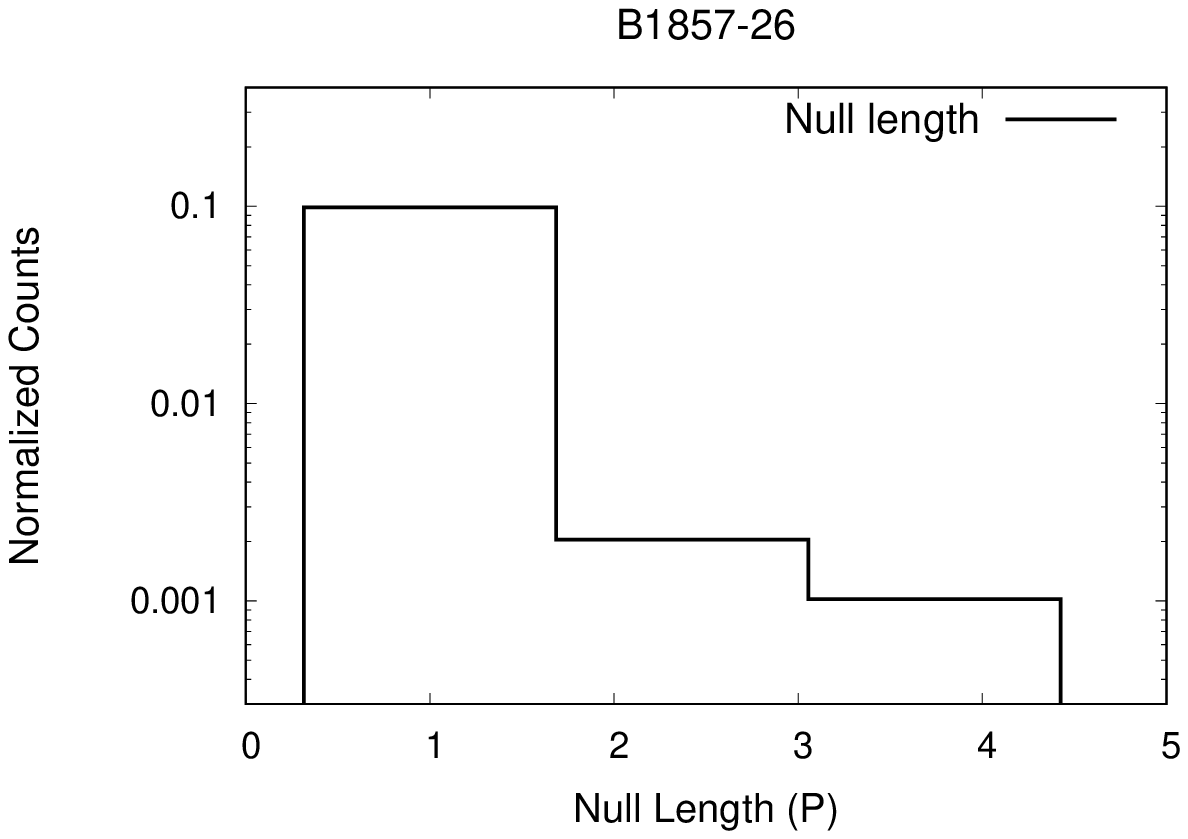}} \\
\mbox{\includegraphics[angle=0,scale=0.57]{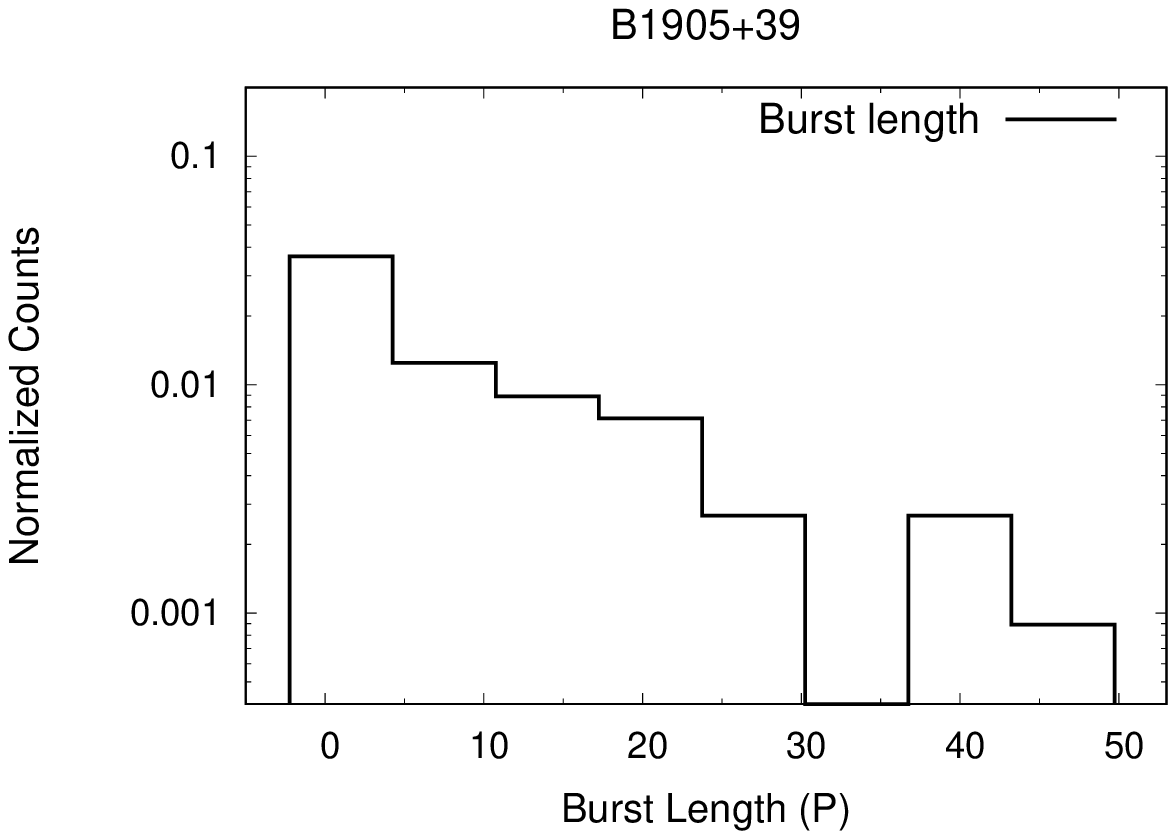}} &
\mbox{\includegraphics[angle=0,scale=0.57]{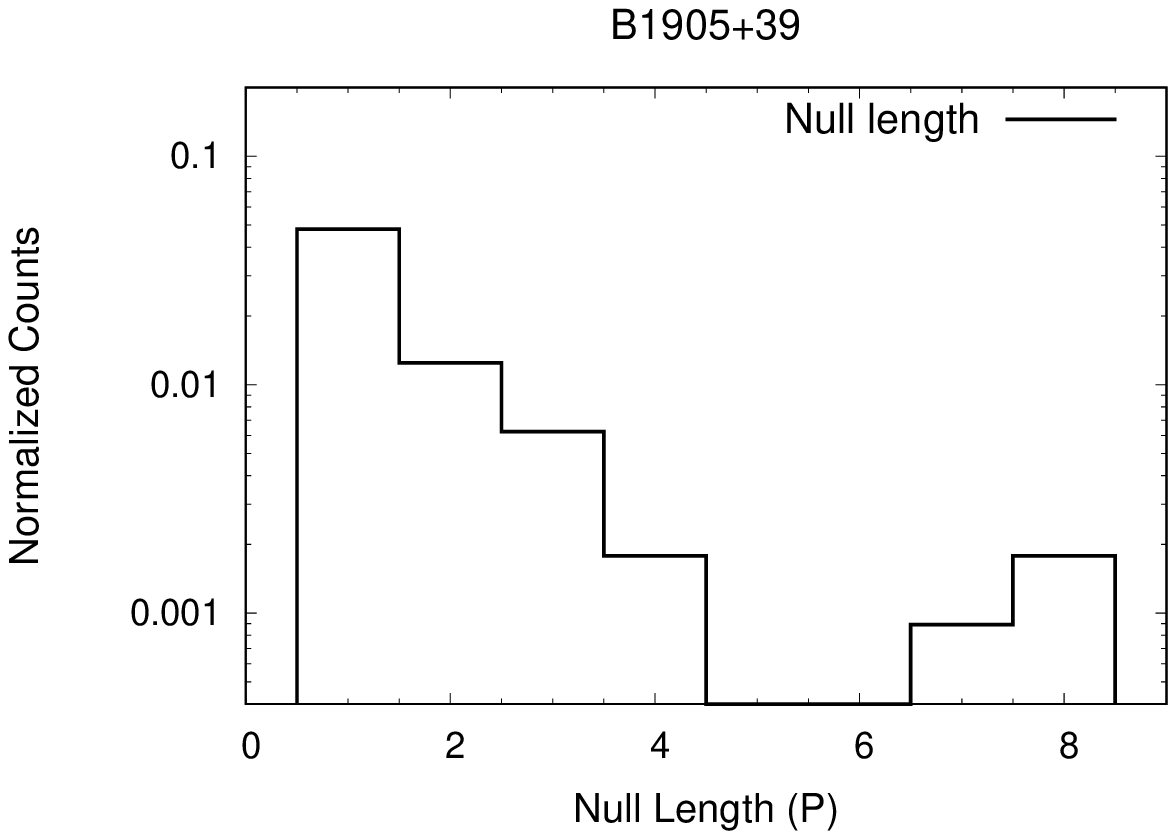}} \\
\end{tabular}
\caption{The Burst length (left panel) and Null length (right panel) distributions.}
\end{center}
\end{figure*}

\clearpage

%4th set of plots
\begin{figure*}
\begin{center}
\begin{tabular}{@{}cr@{}}
\mbox{\includegraphics[angle=0,scale=0.57]{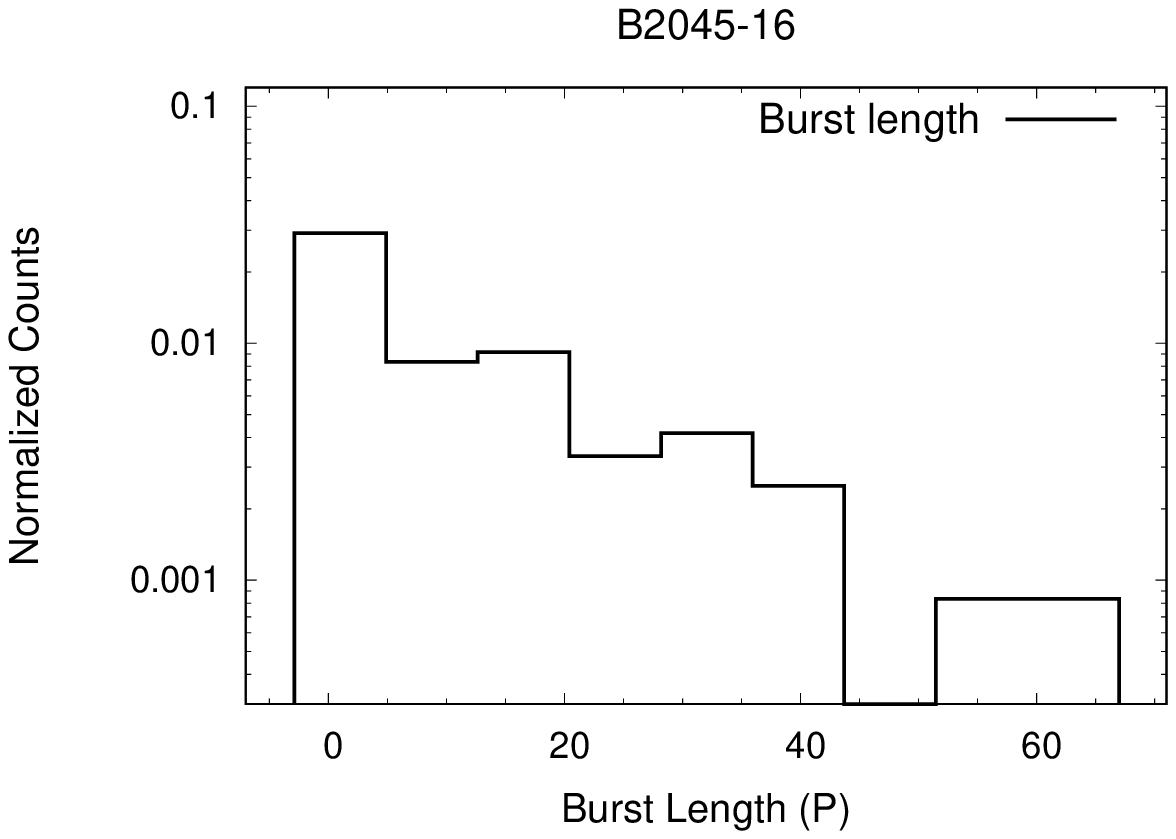}} &
\mbox{\includegraphics[angle=0,scale=0.57]{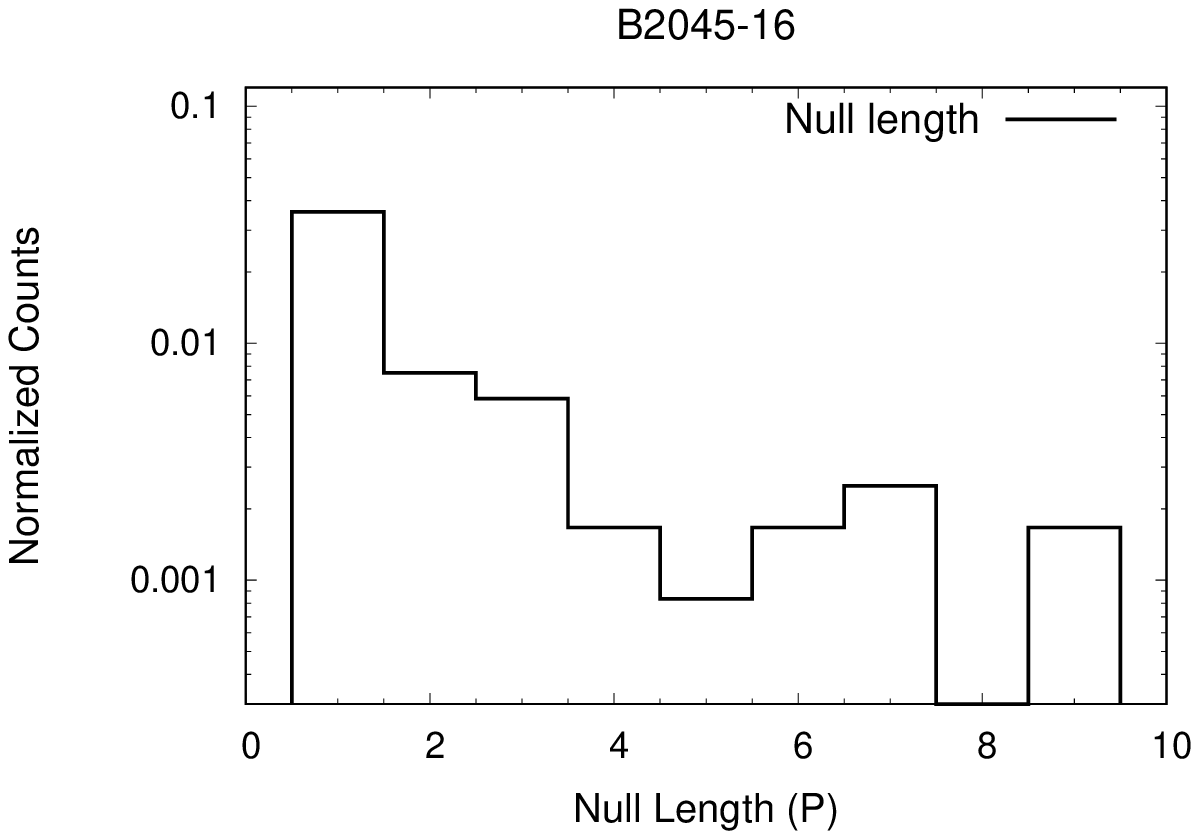}} \\
\mbox{\includegraphics[angle=0,scale=0.57]{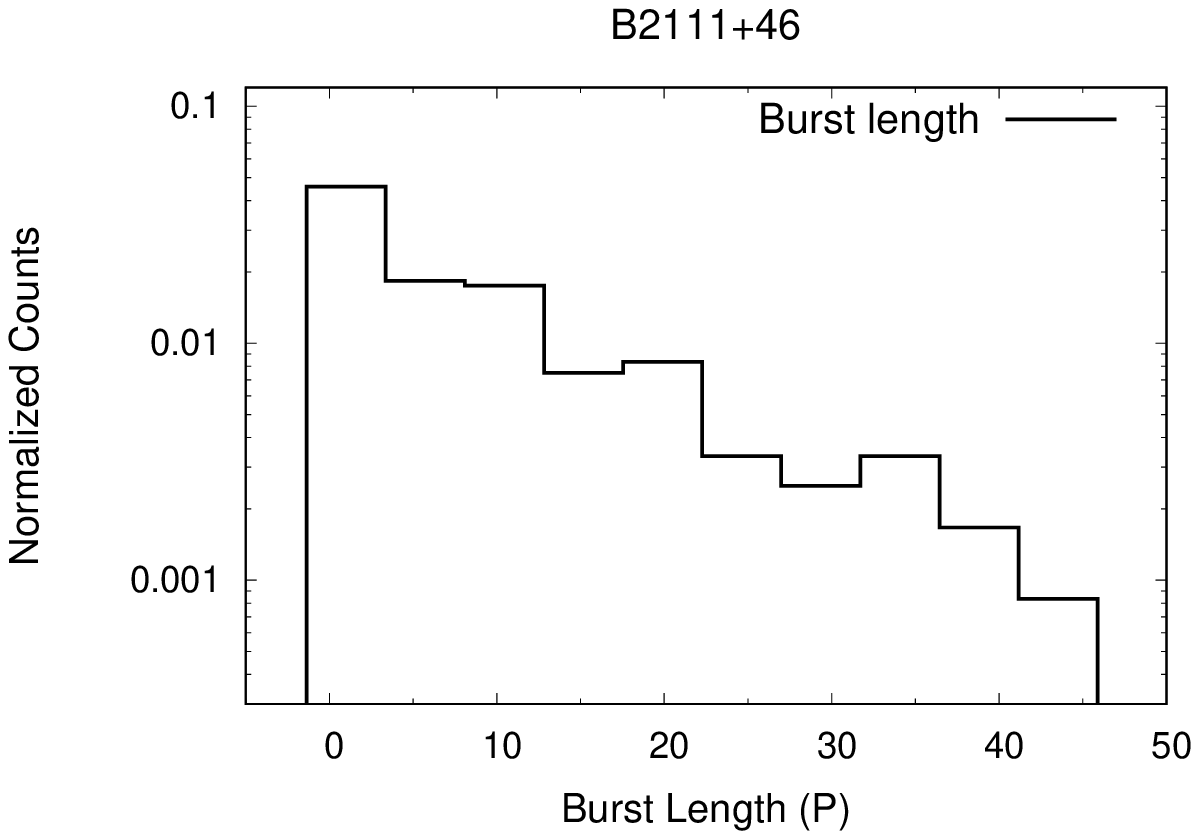}} &
\mbox{\includegraphics[angle=0,scale=0.57]{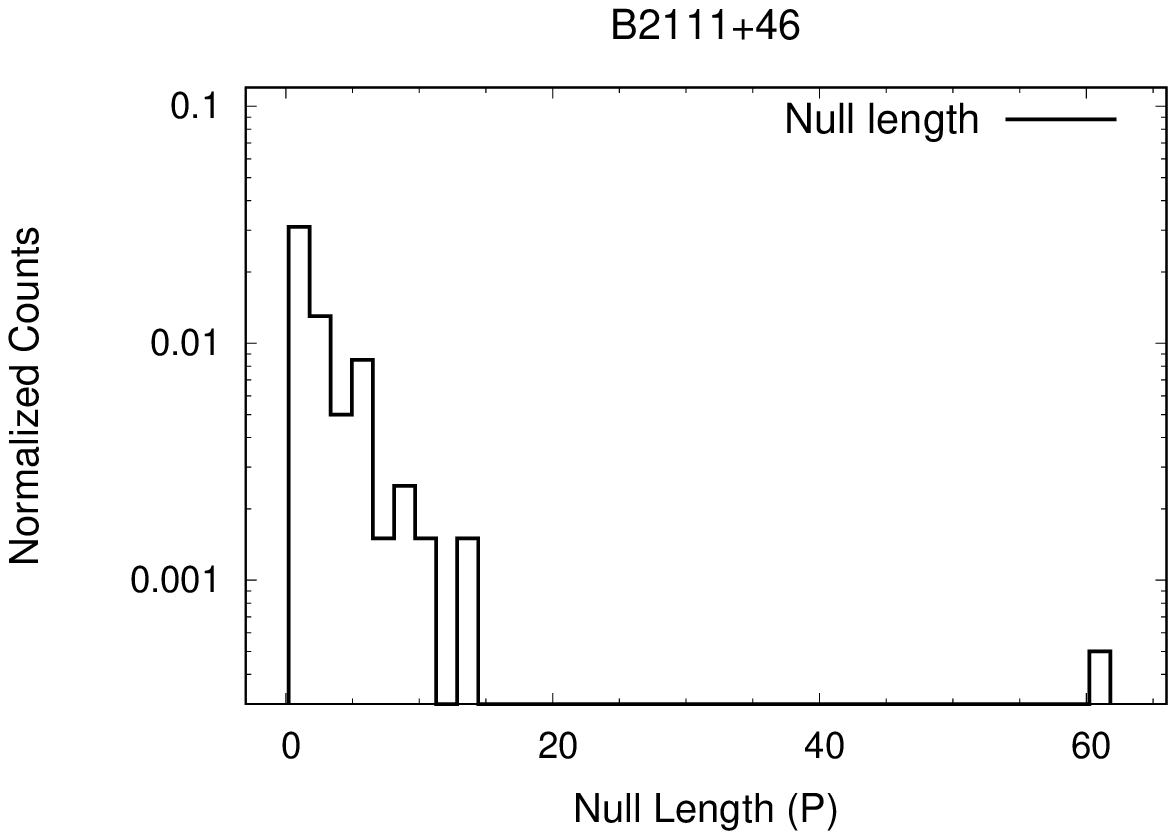}} \\
\mbox{\includegraphics[angle=0,scale=0.57]{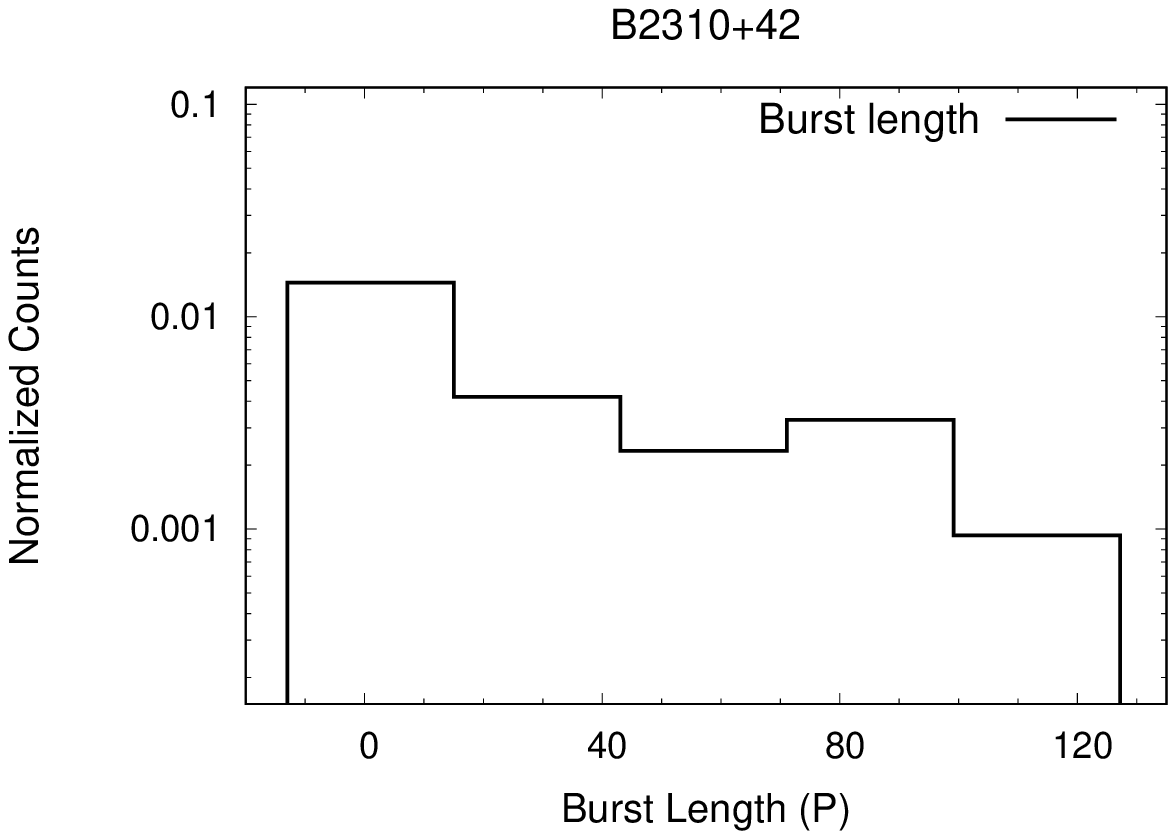}} &
\mbox{\includegraphics[angle=0,scale=0.57]{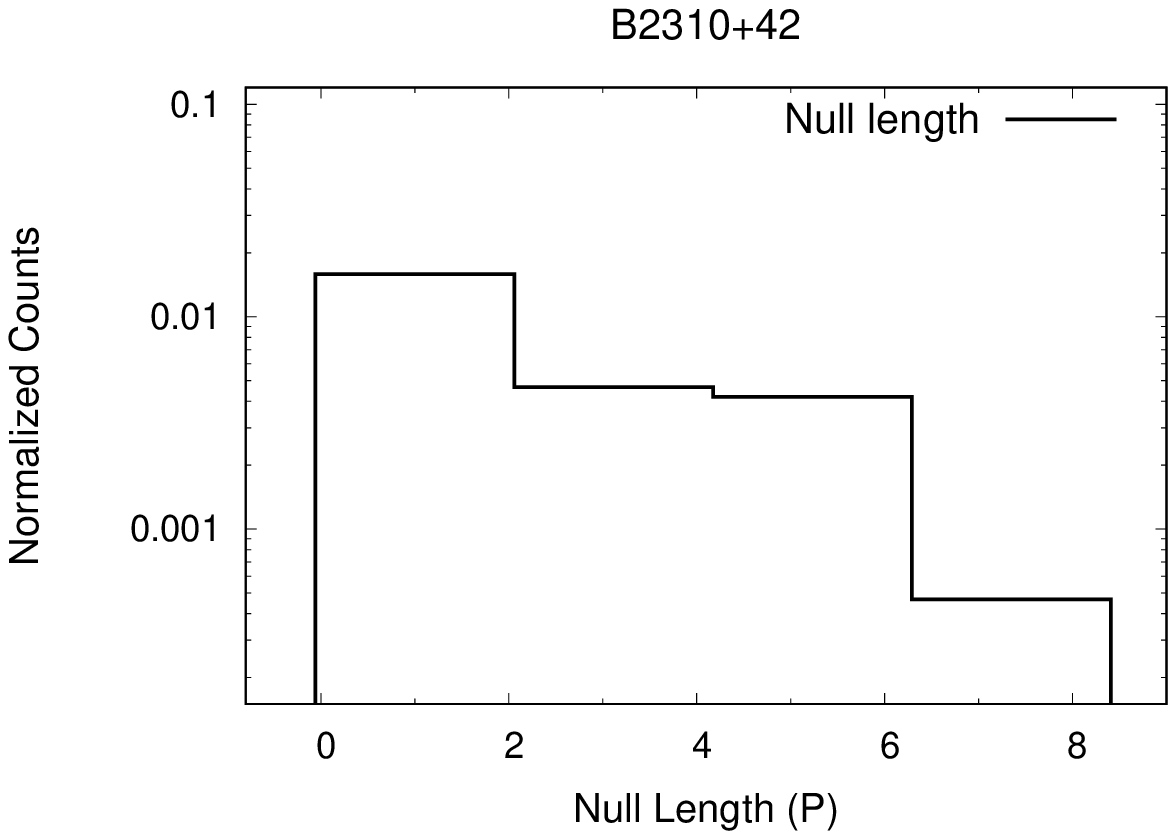}} \\
\mbox{\includegraphics[angle=0,scale=0.57]{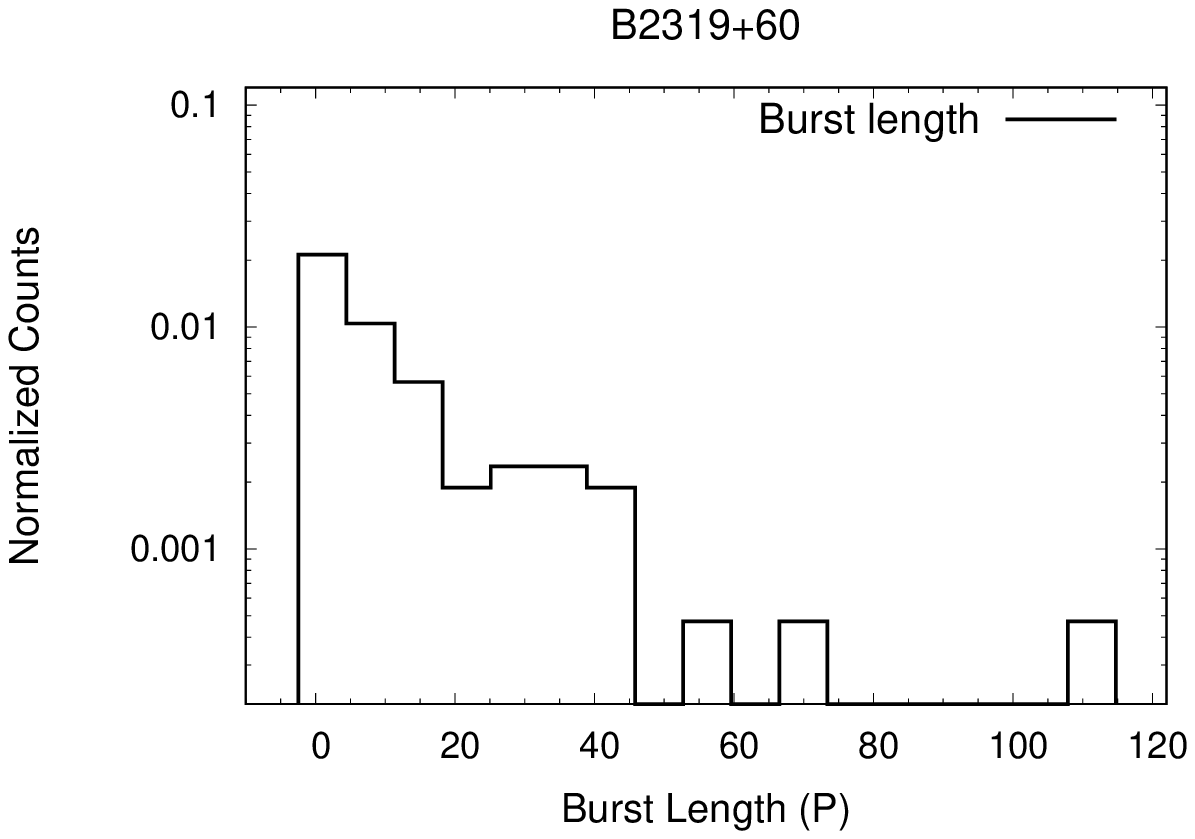}} &
\mbox{\includegraphics[angle=0,scale=0.57]{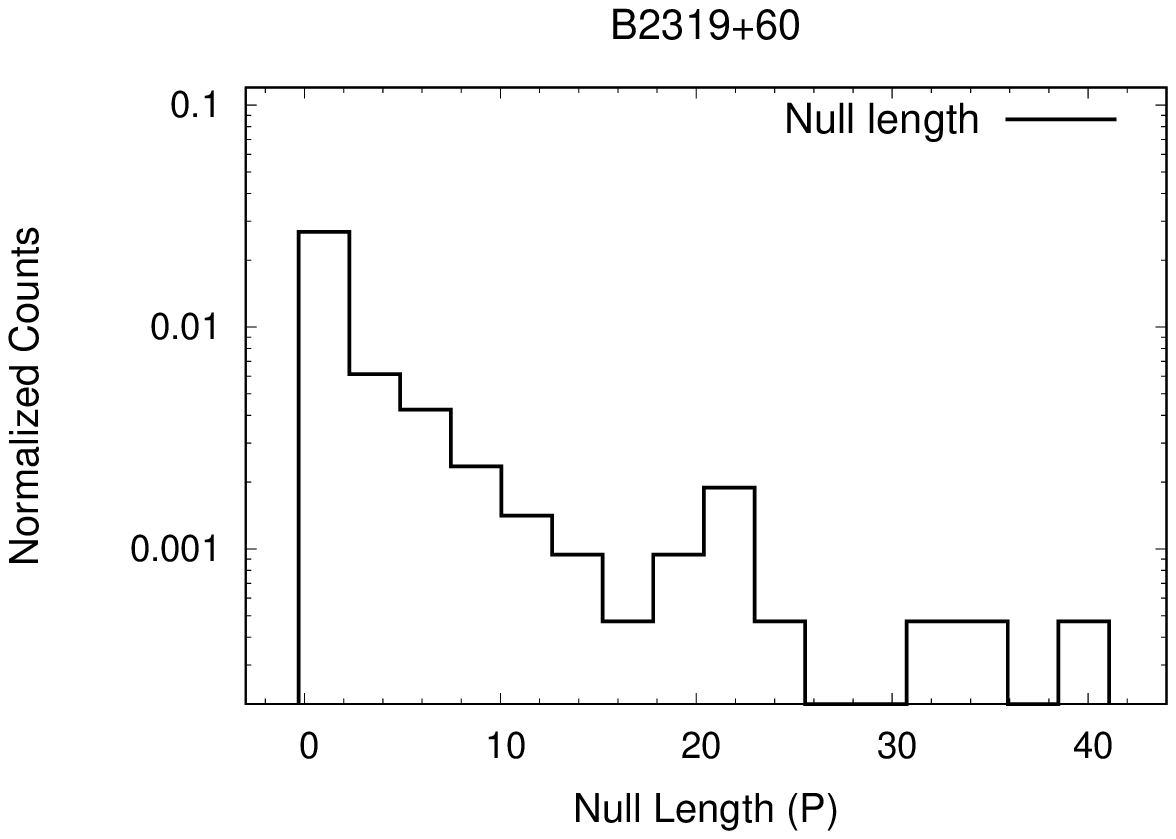}} \\
\end{tabular}
\caption{The Burst length (left panel) and Null length (right panel) distributions.}
\end{center}
\end{figure*}

\clearpage

%5th set of plots
\begin{figure*}
\begin{center}
\begin{tabular}{@{}cr@{}}
\mbox{\includegraphics[angle=0,scale=0.57]{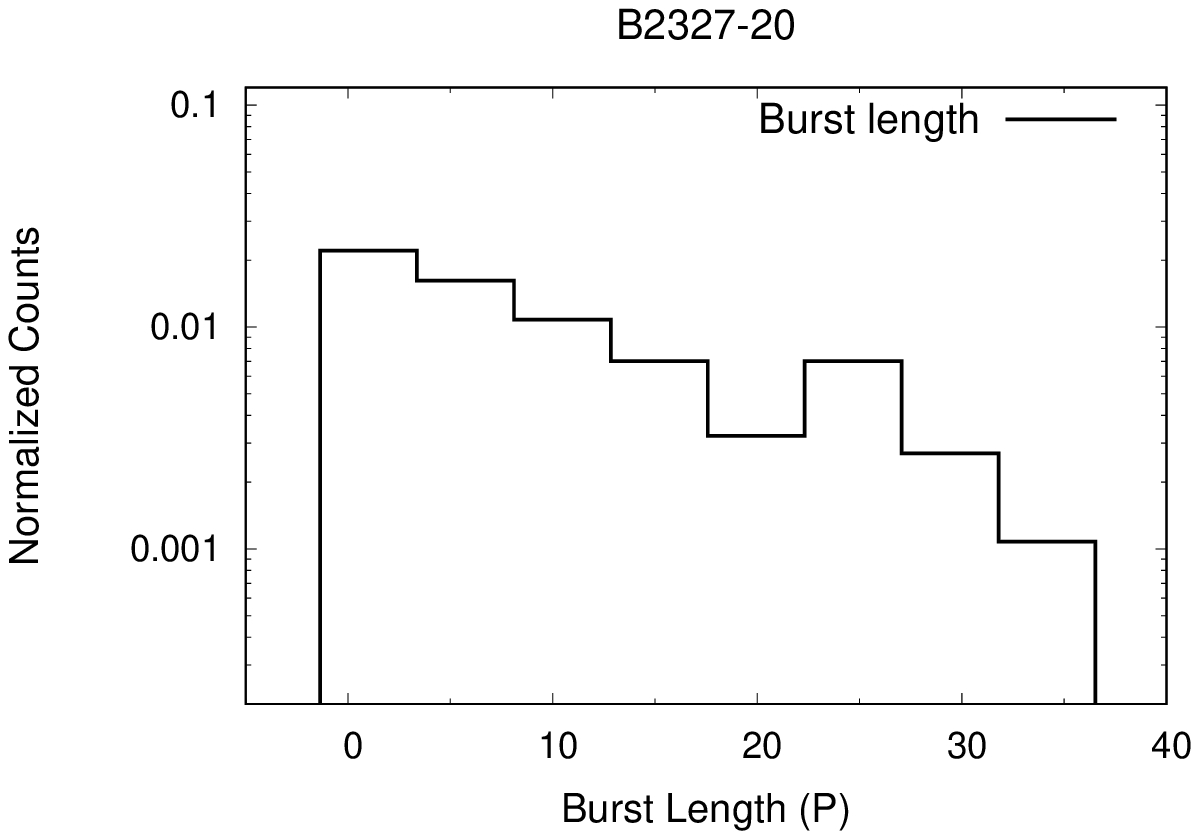}} &
\mbox{\includegraphics[angle=0,scale=0.57]{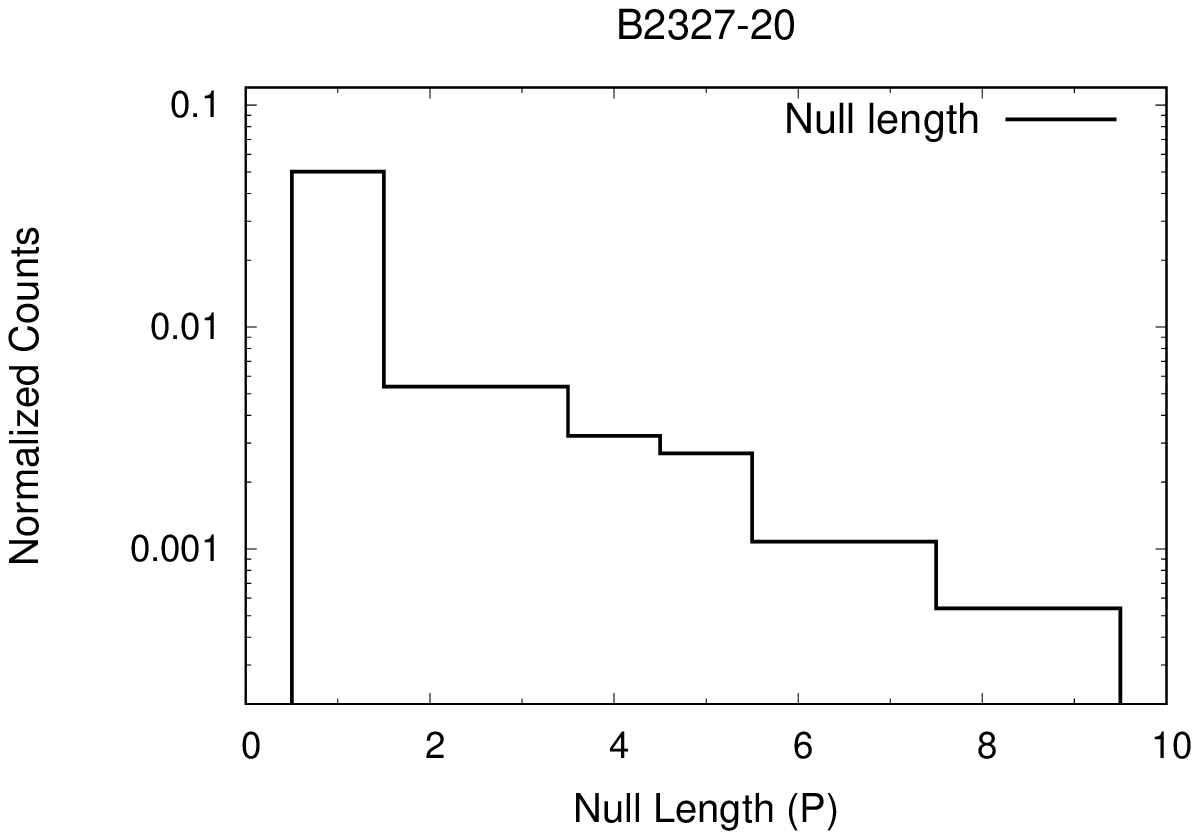}} \\
\end{tabular}
\caption{The Burst length (left panel) and Null length (right panel) distributions.}
\end{center}
\end{figure*}

\clearpage

%% file: appendix3.tex
% J0141+6009
\begin{figure*}
\begin{tabular}{@{}lr@{}}
{\mbox{\includegraphics[scale=0.35,angle=0.]{J0141+6009_LRFSavg_256.ps}}} &
\hspace{50px}
{\mbox{\includegraphics[scale=0.35,angle=0.]{J0141+6009_nullfft_256.ps}}} \\
\end{tabular}
\caption{PSR B0138+59 : The time evolution of the LRFS (left panel) and the 
Null-Burst time series FFT (right panel).}
\label{fig_J0141}
\end{figure*}

%J0452-1759
\begin{figure*}
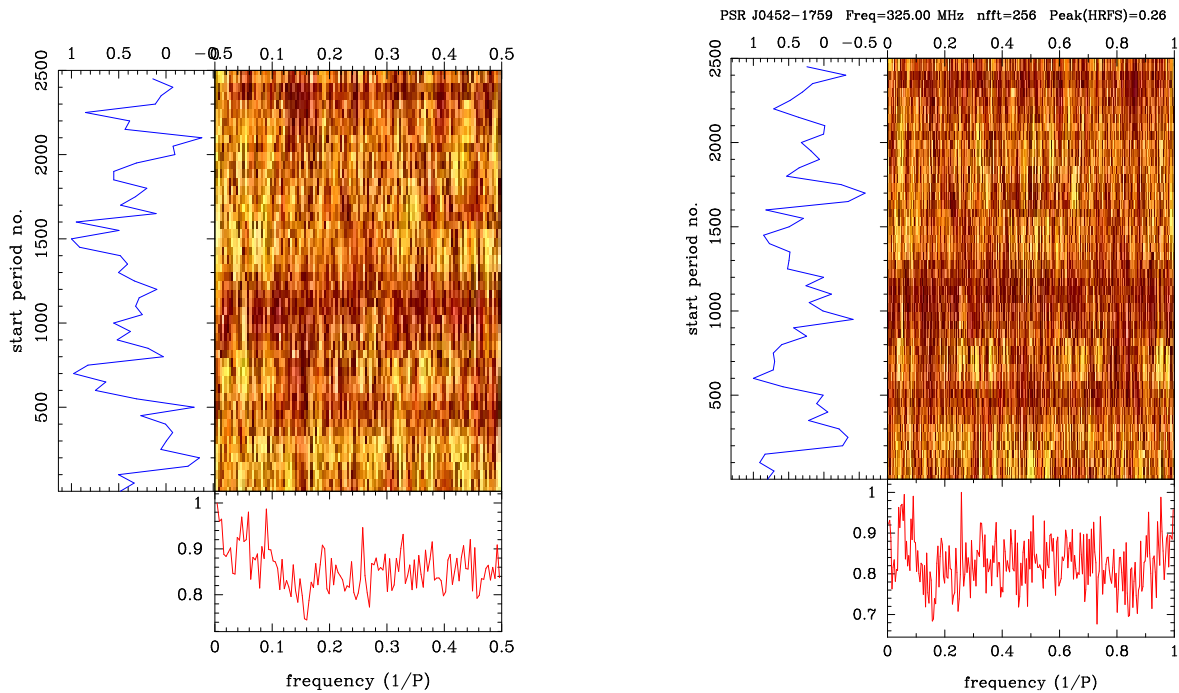

\begin{tabular}{@{}lr@{}}
{\mbox{\includegraphics[scale=0.35,angle=0.]{J0452-1759_LRFSavg_256.ps}}} &
\hspace{50px}
{\mbox{\includegraphics[scale=0.35,angle=0.]{J0452-1759_HRFSavg_256.ps}}} \\
\end{tabular}
\caption{PSR B0447$-$12 : The time evolution of the LRFS (left panel) and the
HRFS (right panel).}
\label{fig_J0452}
\end{figure*}

\clearpage

%J0454+5543
\begin{figure*}
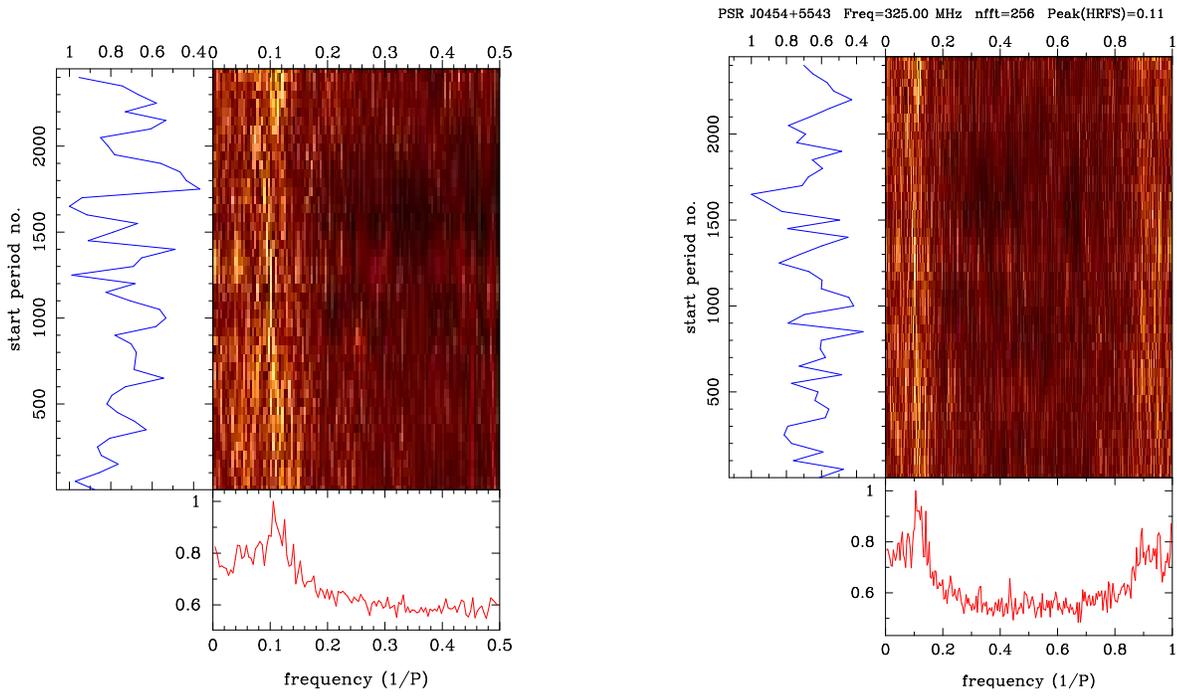

\begin{tabular}{@{}lr@{}}
{\mbox{\includegraphics[scale=0.35,angle=0.]{J0454+5543_LRFSavg_256.ps}}} &
\hspace{50px}
{\mbox{\includegraphics[scale=0.35,angle=0.]{J0454+5543_HRFSavg_256.ps}}} \\
\end{tabular}
\caption{PSR B0450+55 : The time evolution of the LRFS (left panel) and the 
HRFS (right panel).}
\label{fig_J0454}
\end{figure*}

%J0624-0424
\begin{figure*}
\begin{tabular}{@{}lr@{}}
{\mbox{\includegraphics[scale=0.35,angle=0.]{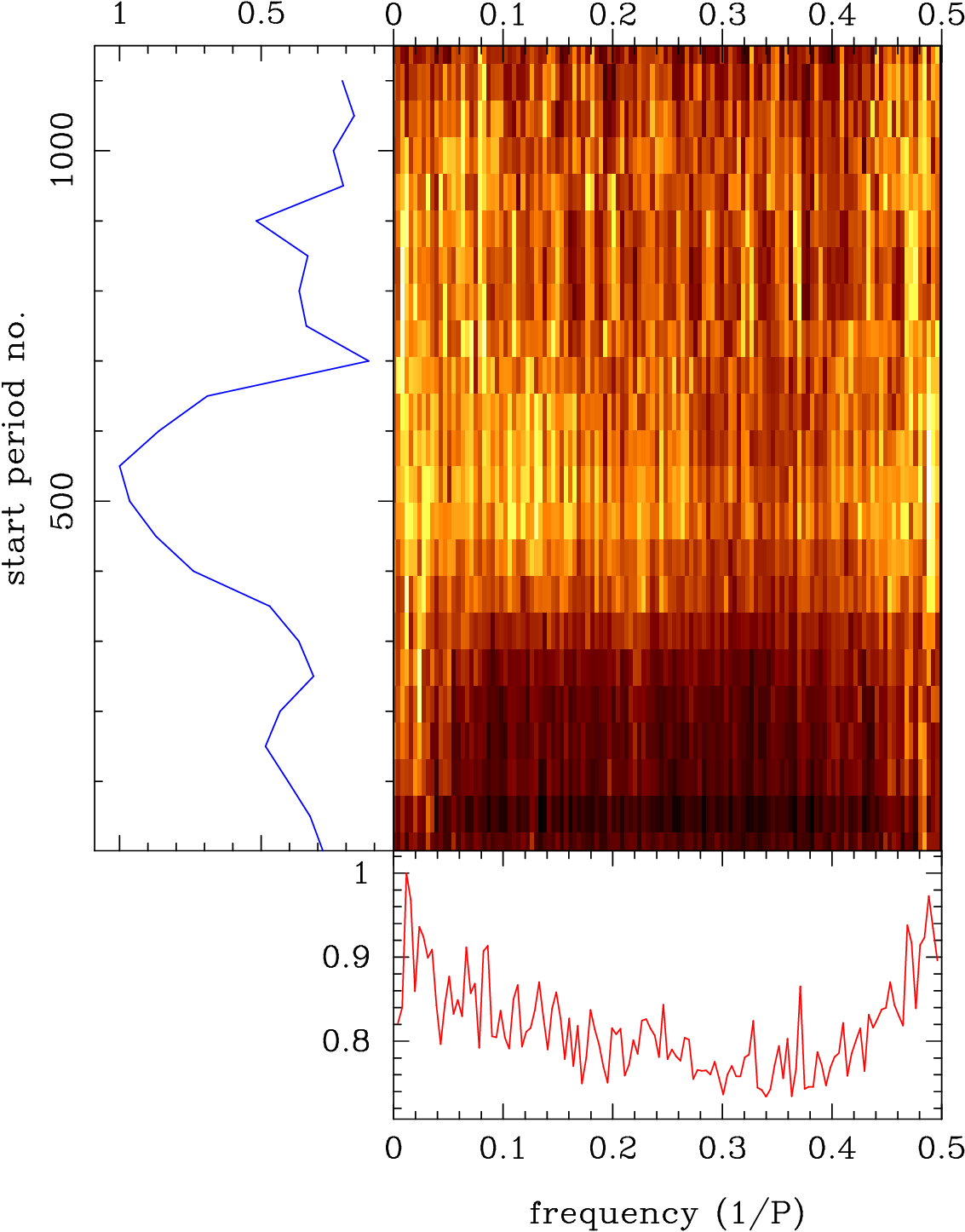}}} &
\hspace{50px}
{\mbox{\includegraphics[scale=0.35,angle=0.]{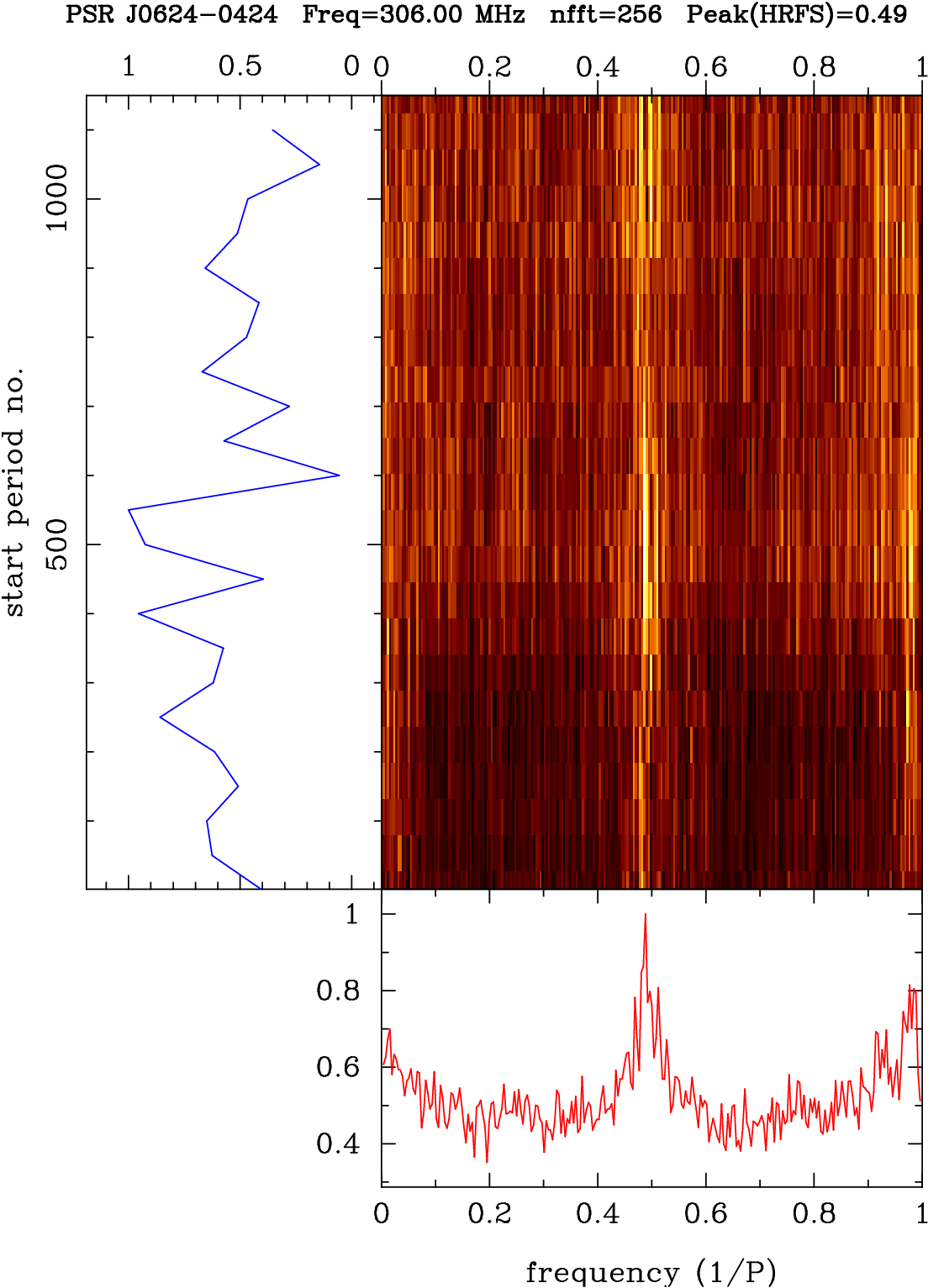}}} \\
\end{tabular}
\caption{PSR B0621$-$04 : The time evolution of the LRFS (left panel) and the 
HRFS (right panel).}
\label{fig_J0624}
\end{figure*}

\clearpage

%J0934-5249
\begin{figure*}
\begin{tabular}{@{}lr@{}}
{\mbox{\includegraphics[scale=0.35,angle=0.]{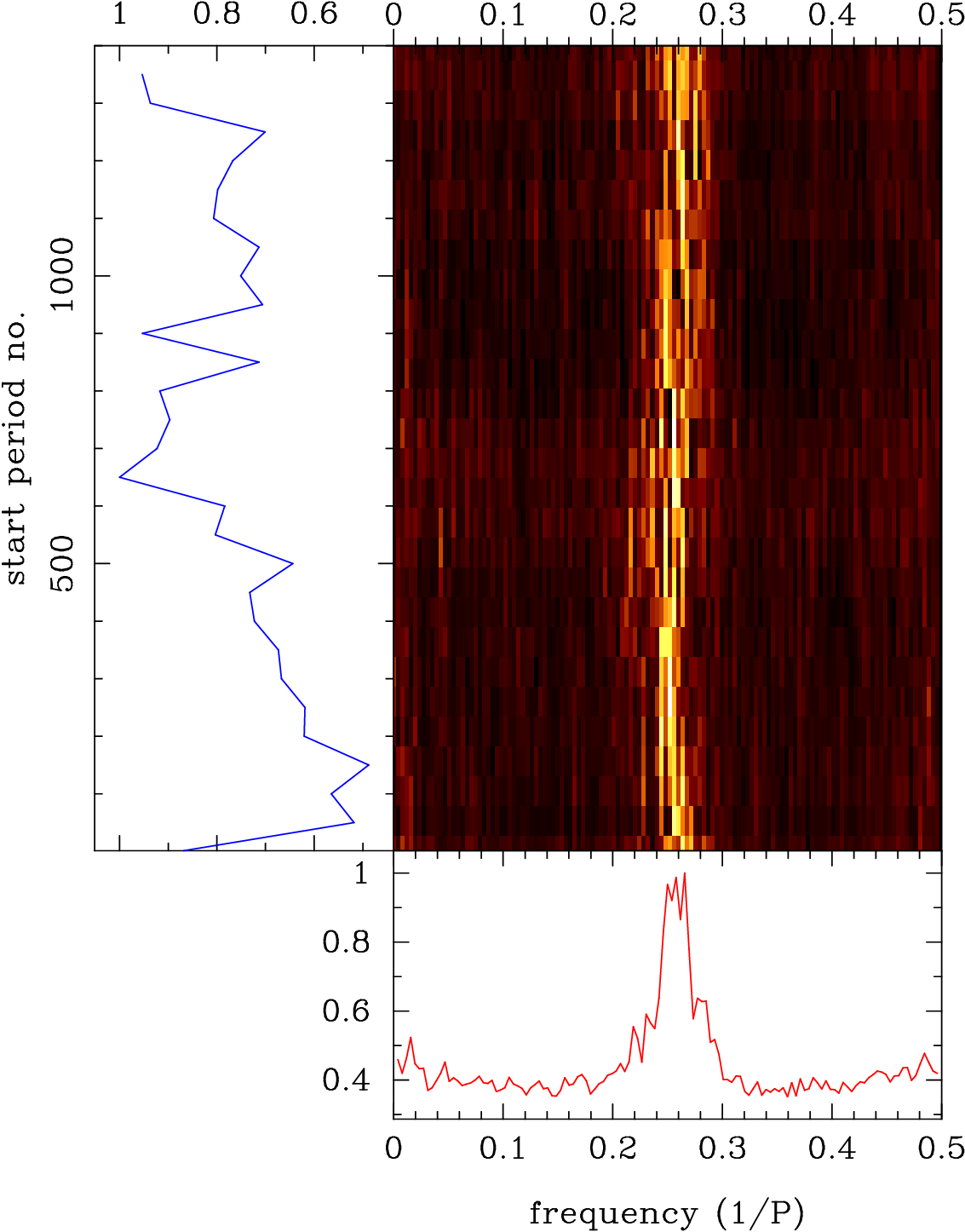}}} &
\hspace{50px}
{\mbox{\includegraphics[scale=0.35,angle=0.]{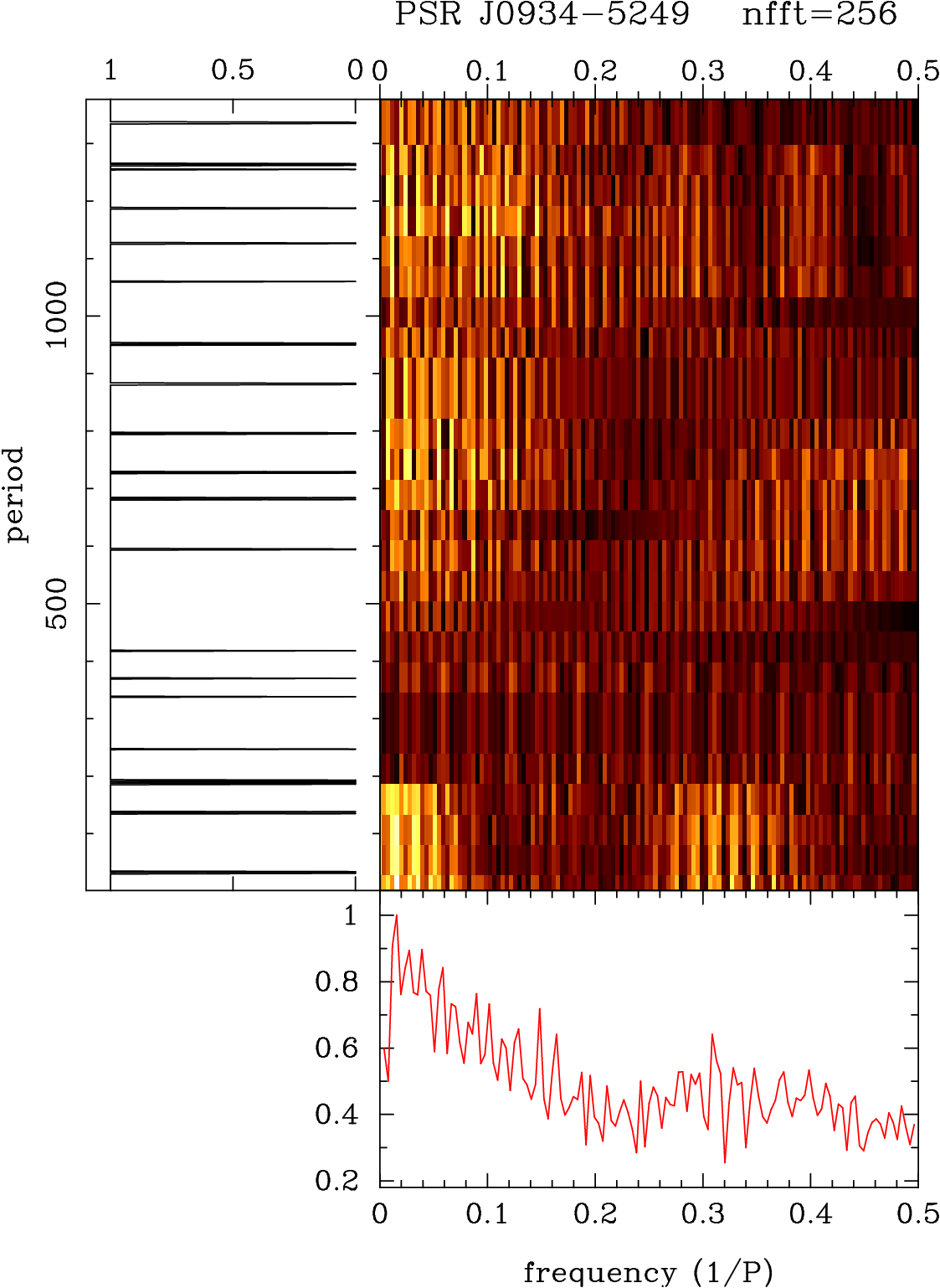}}} \\
\end{tabular}
\caption{PSR B0932$-$52 : The time evolution of the LRFS (left panel) and the
Null-Burst time series FFT (right panel).}
\label{fig_J0934}
\end{figure*}

%J1239+2453
\begin{figure*}
\begin{tabular}{@{}lr@{}}
{\mbox{\includegraphics[scale=0.35,angle=0.]{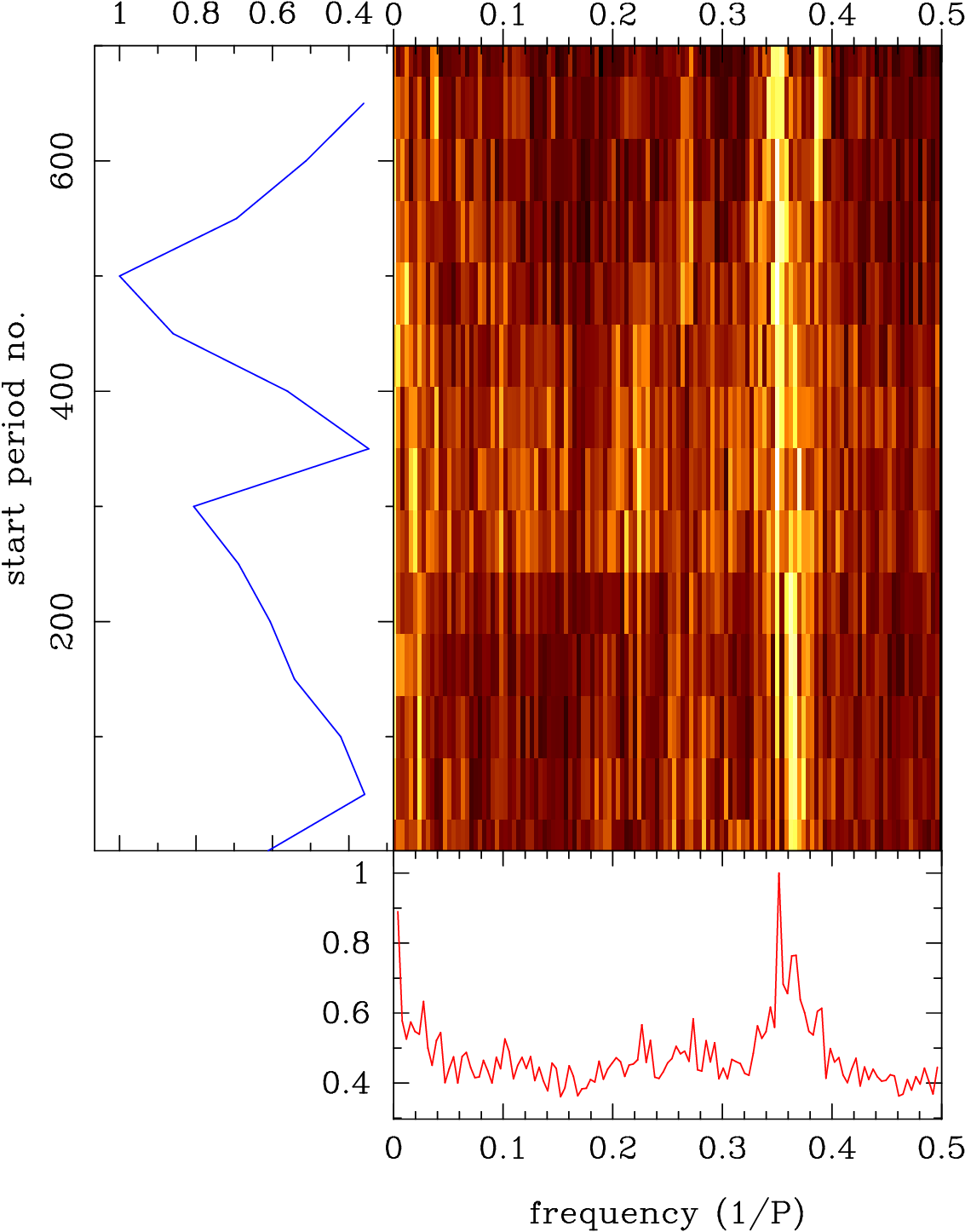}}} &
\hspace{50px}
{\mbox{\includegraphics[scale=0.35,angle=0.]{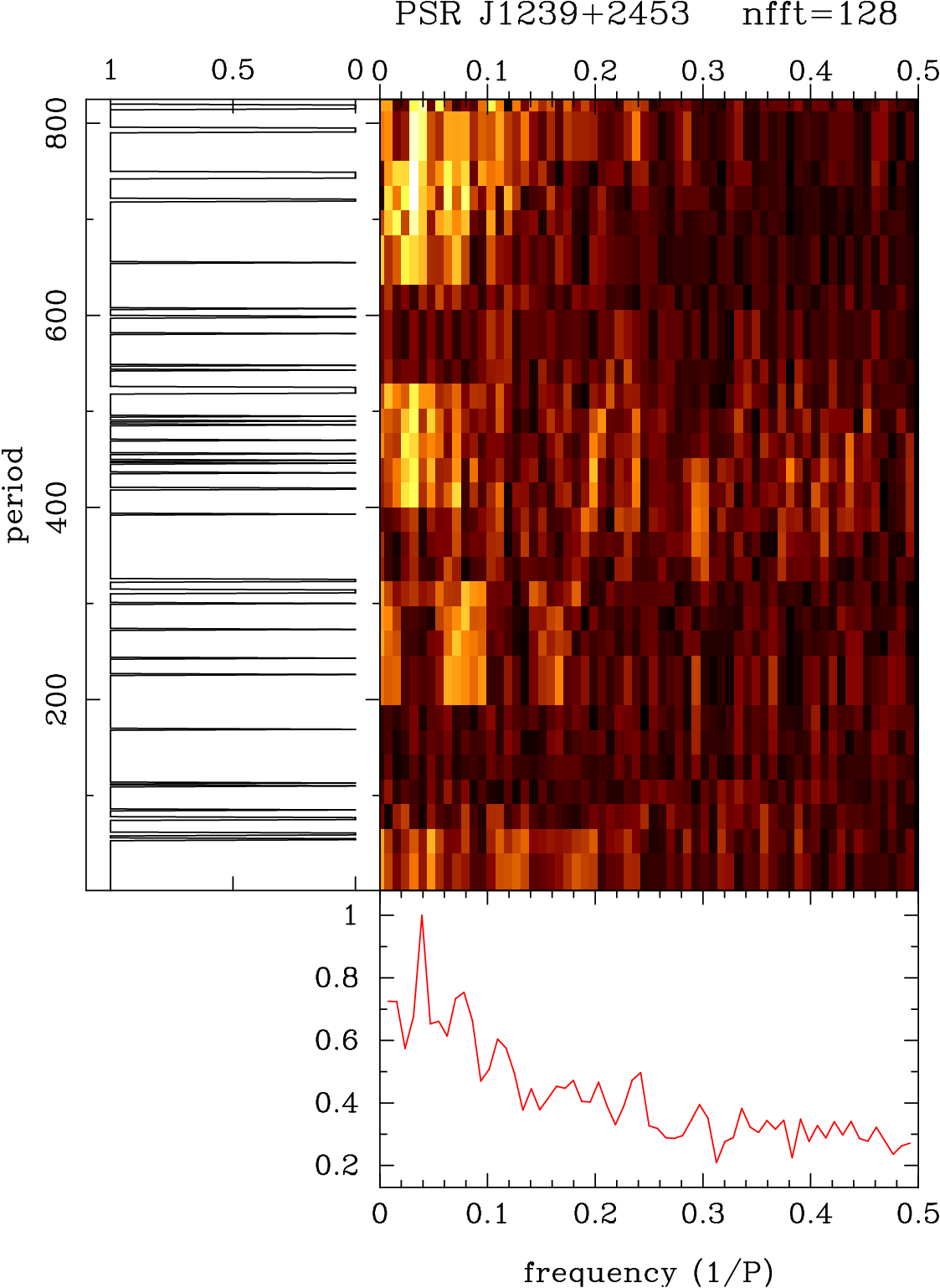}}} \\
\end{tabular}
\caption{PSR B1237+25 : The time evolution of the LRFS (left panel) and the
Null-Burst time series FFT (right panel).}
\label{fig_J1239}
\end{figure*}

\clearpage

%J1509+5531
\begin{figure*}
\begin{tabular}{@{}lr@{}}
{\mbox{\includegraphics[scale=0.35,angle=0.]{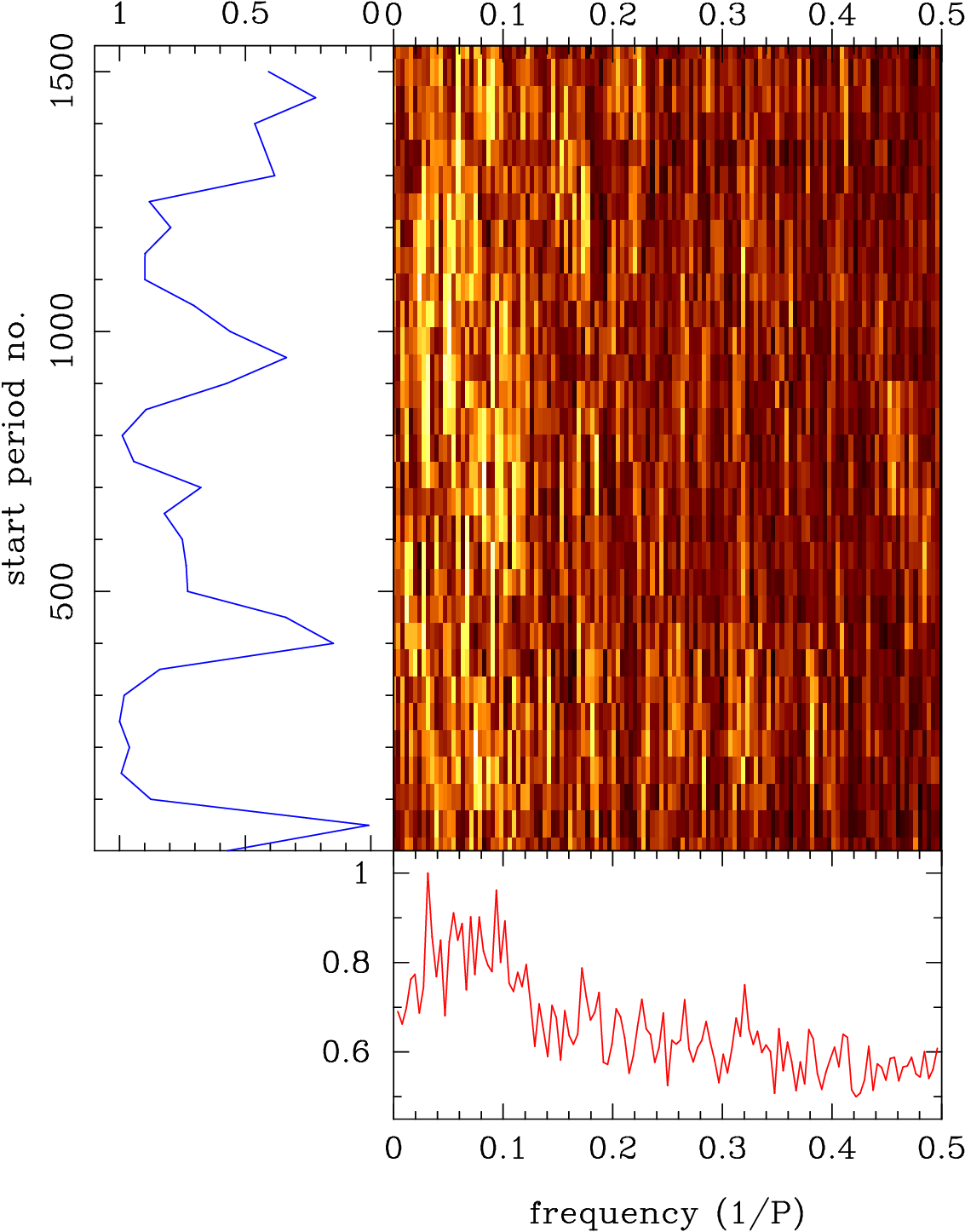}}} &
\hspace{50px}
{\mbox{\includegraphics[scale=0.35,angle=0.]{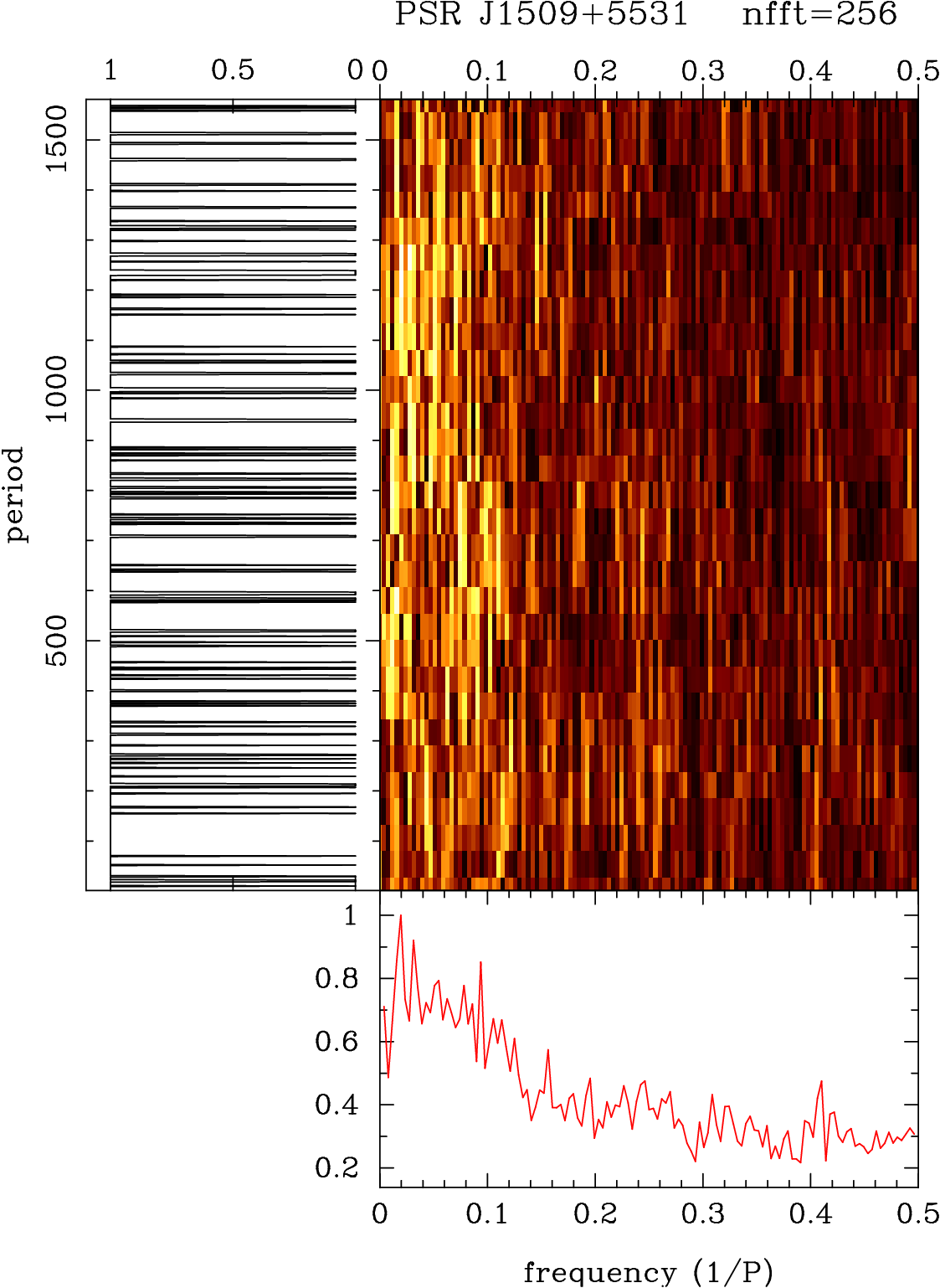}}} \\
\end{tabular}
\caption{PSR B1508+55 : The time evolution of the LRFS (left panel) and the
Null-Burst time series FFT (right panel).}
\label{fig_J1509}
\end{figure*}

%J1514-4834
\begin{figure*}
\begin{tabular}{@{}lr@{}}
{\mbox{\includegraphics[scale=0.35,angle=0.]{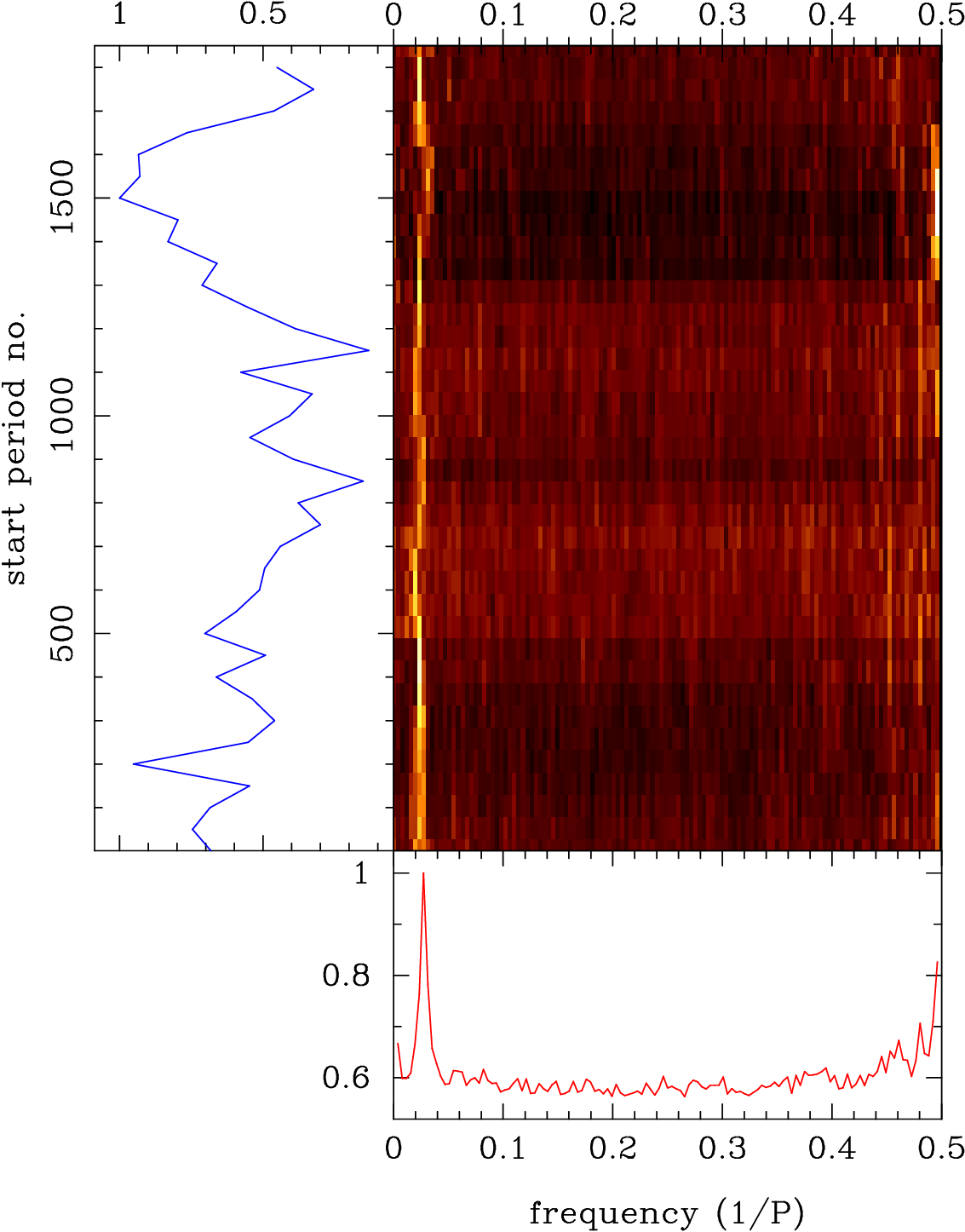}}} &
\hspace{50px}
{\mbox{\includegraphics[scale=0.35,angle=0.]{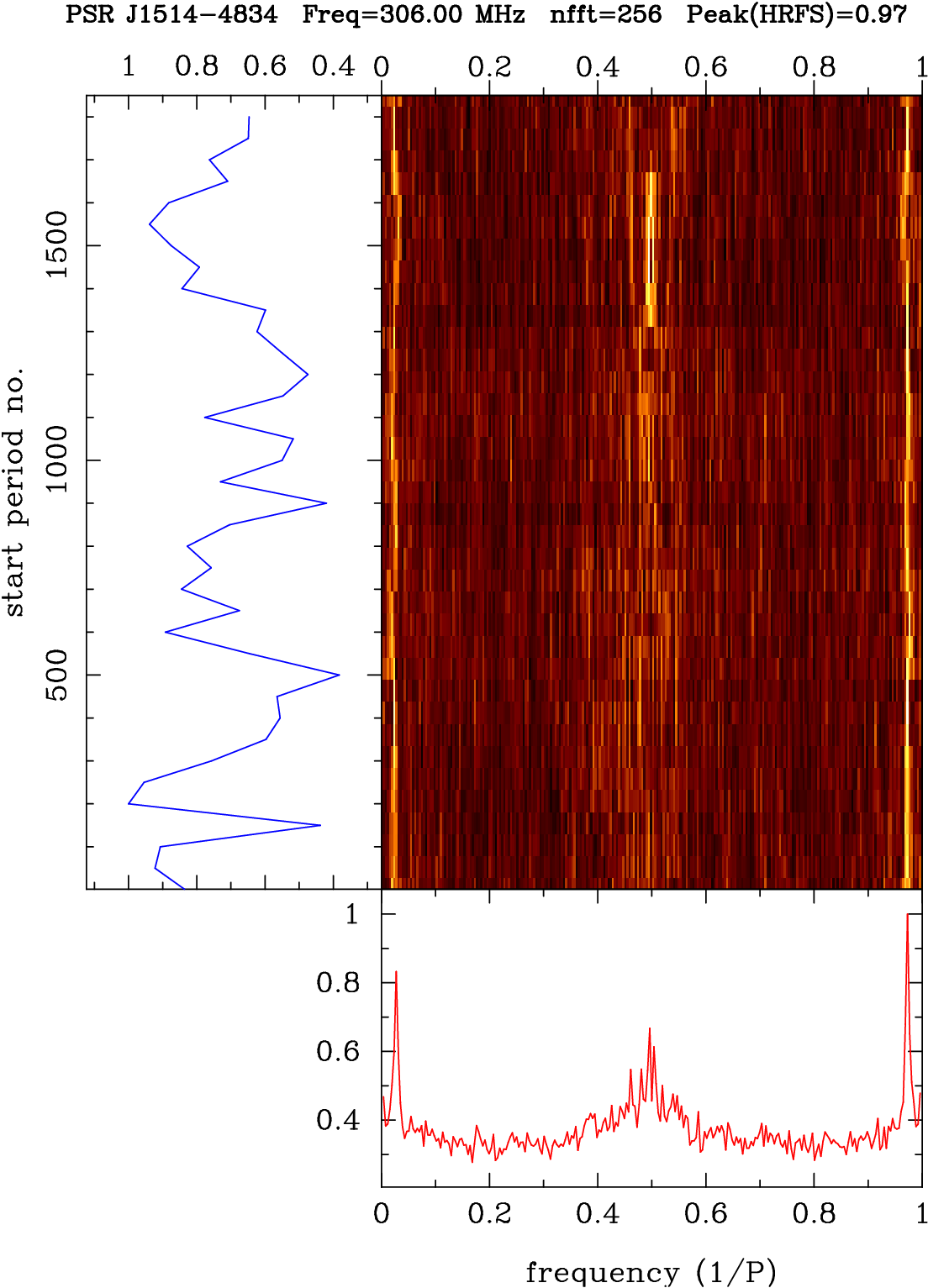}}} \\
\end{tabular}
\caption{PSR B1510$-$48 : The time evolution of the LRFS (left panel) and the
HRFS (right panel).}
\label{fig_J1514}
\end{figure*}

\clearpage

%J1543+0929
\begin{figure*}
\begin{tabular}{@{}lr@{}}
{\mbox{\includegraphics[scale=0.35,angle=0.]{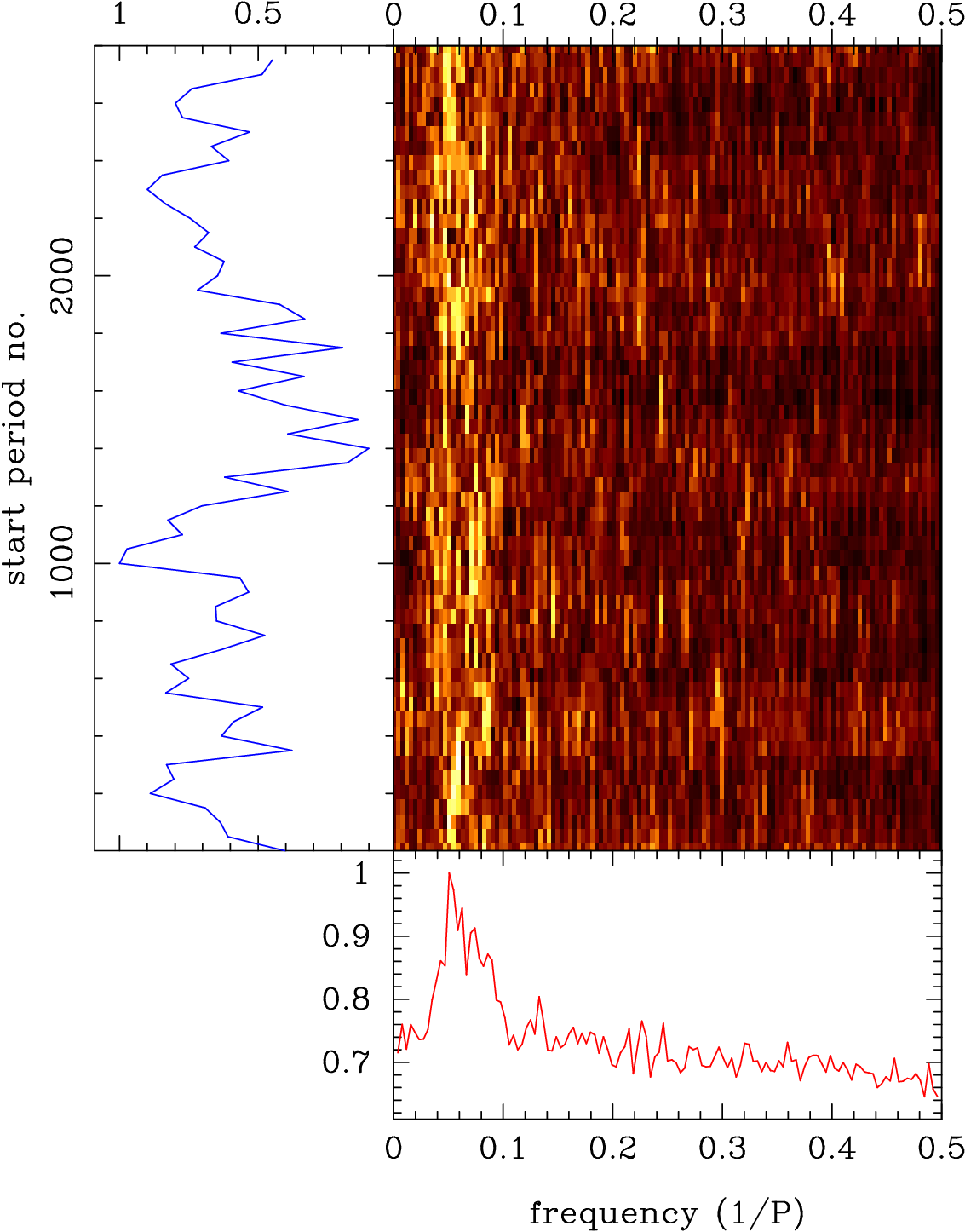}}} &
\hspace{50px}
{\mbox{\includegraphics[scale=0.35,angle=0.]{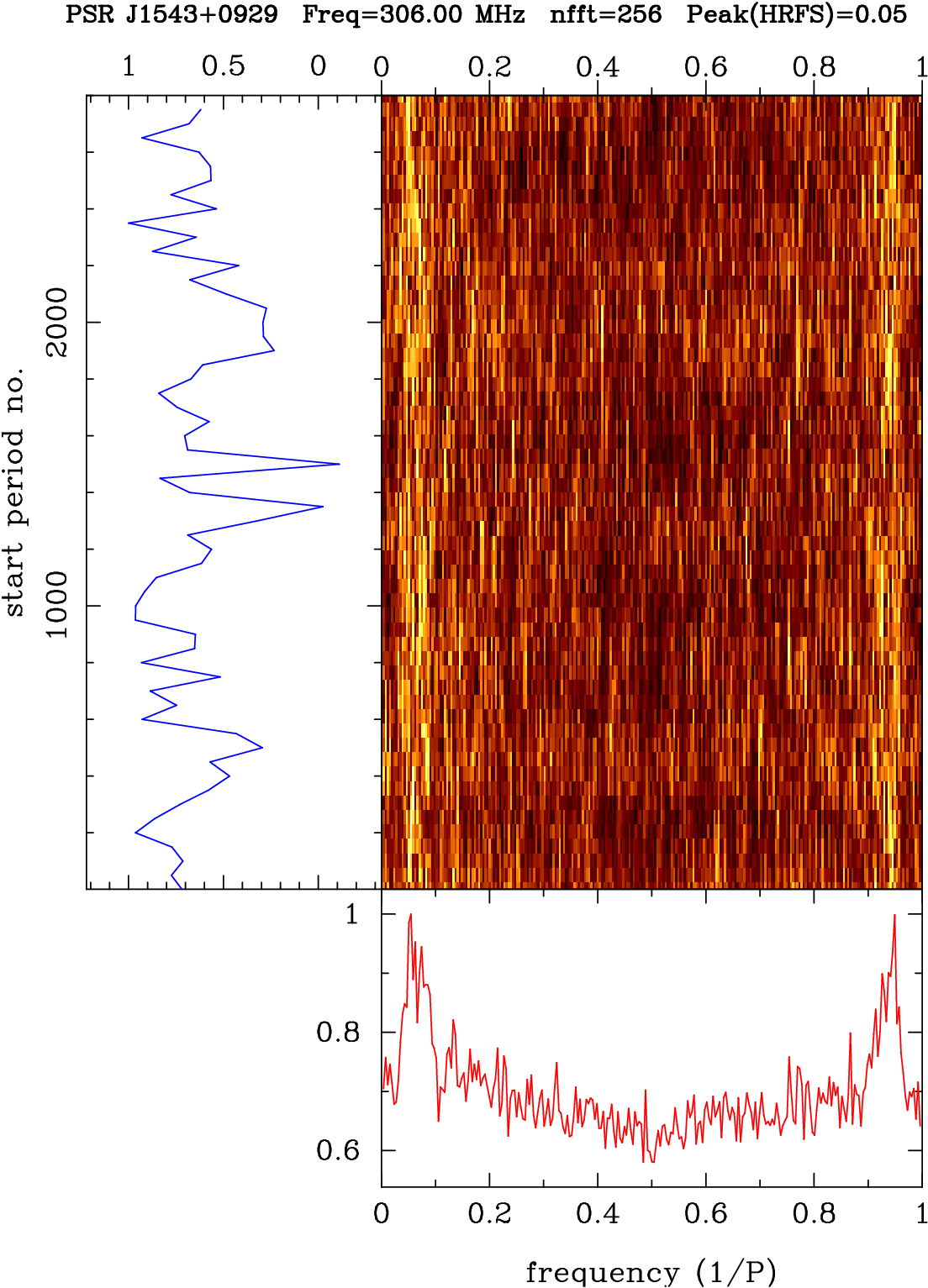}}} \\
\end{tabular}
\caption{PSR B1541+09 : The time evolution of the LRFS (left panel) and the
HRFS (right panel).}
\label{fig_J1543}
\end{figure*}

%J1605-5257
\begin{figure*}
\begin{tabular}{@{}lr@{}}
{\mbox{\includegraphics[scale=0.35,angle=0.]{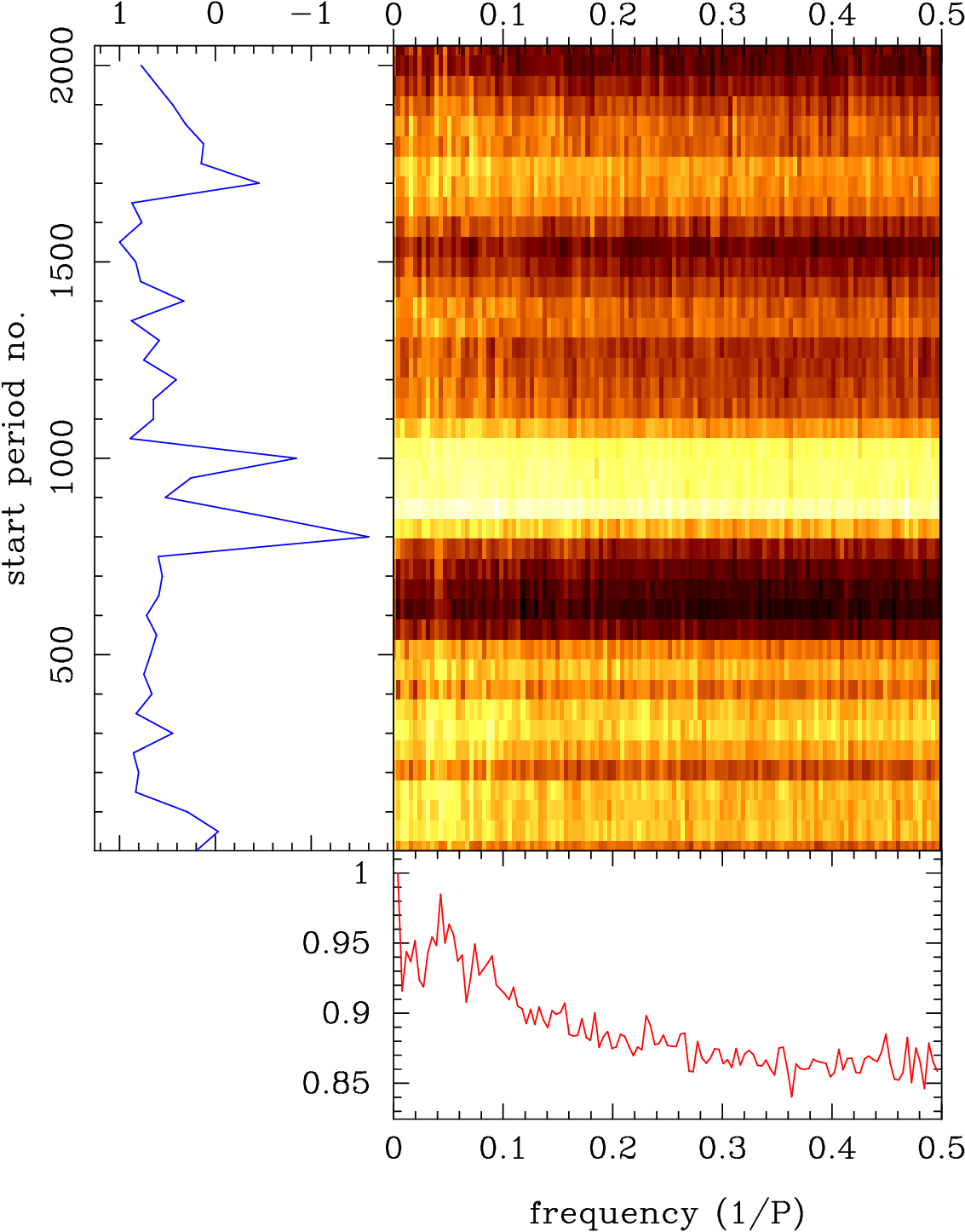}}} &
\hspace{50px}
{\mbox{\includegraphics[scale=0.35,angle=0.]{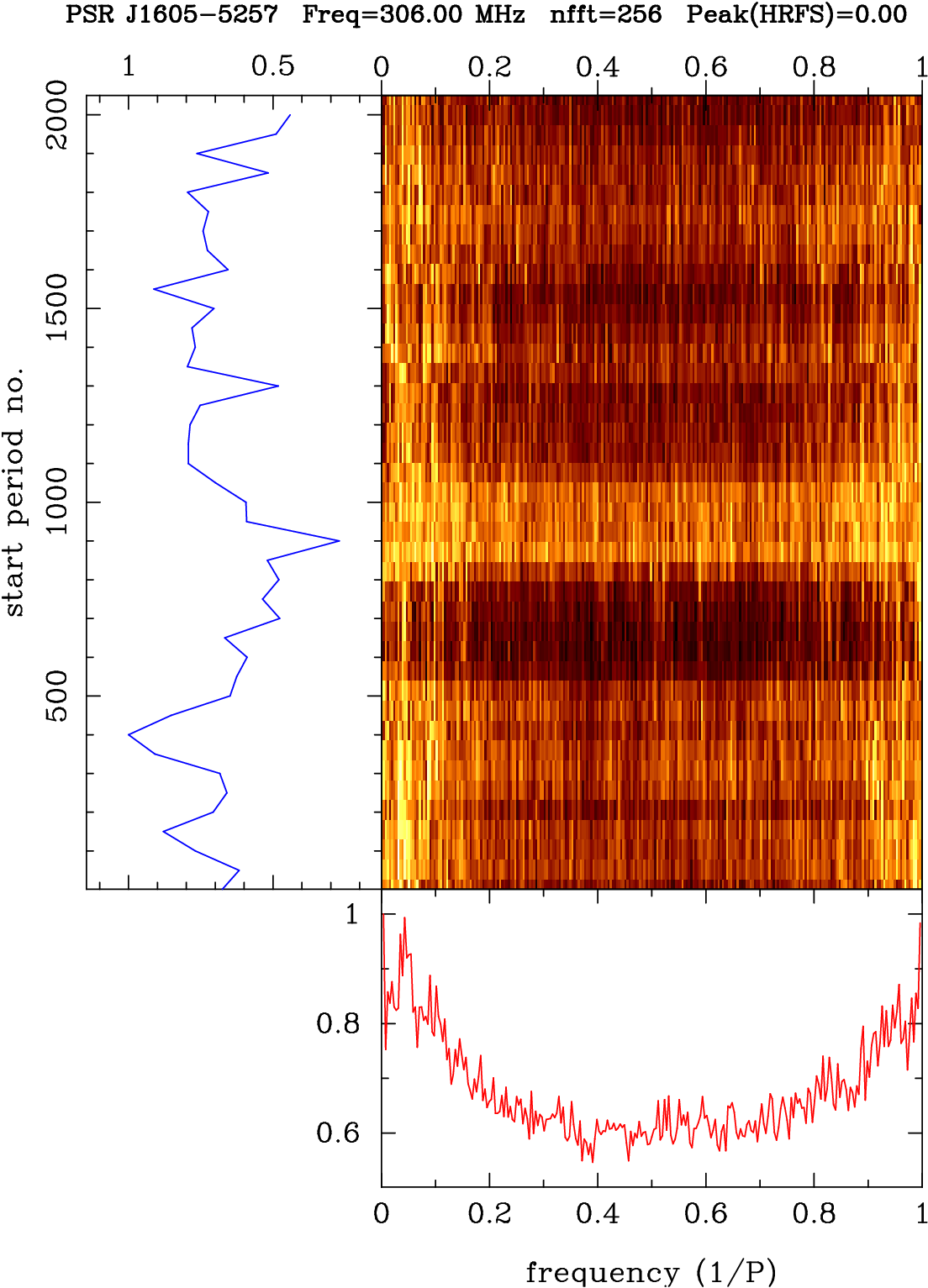}}} \\
\end{tabular}
\caption{PSR B1601$-$52 : The time evolution of the LRFS (left panel) and the
HRFS (right panel).}
\label{fig_J1605}
\end{figure*}

\clearpage

%J1607-0032
\begin{figure*}
\begin{tabular}{@{}lr@{}}
{\mbox{\includegraphics[scale=0.35,angle=0.]{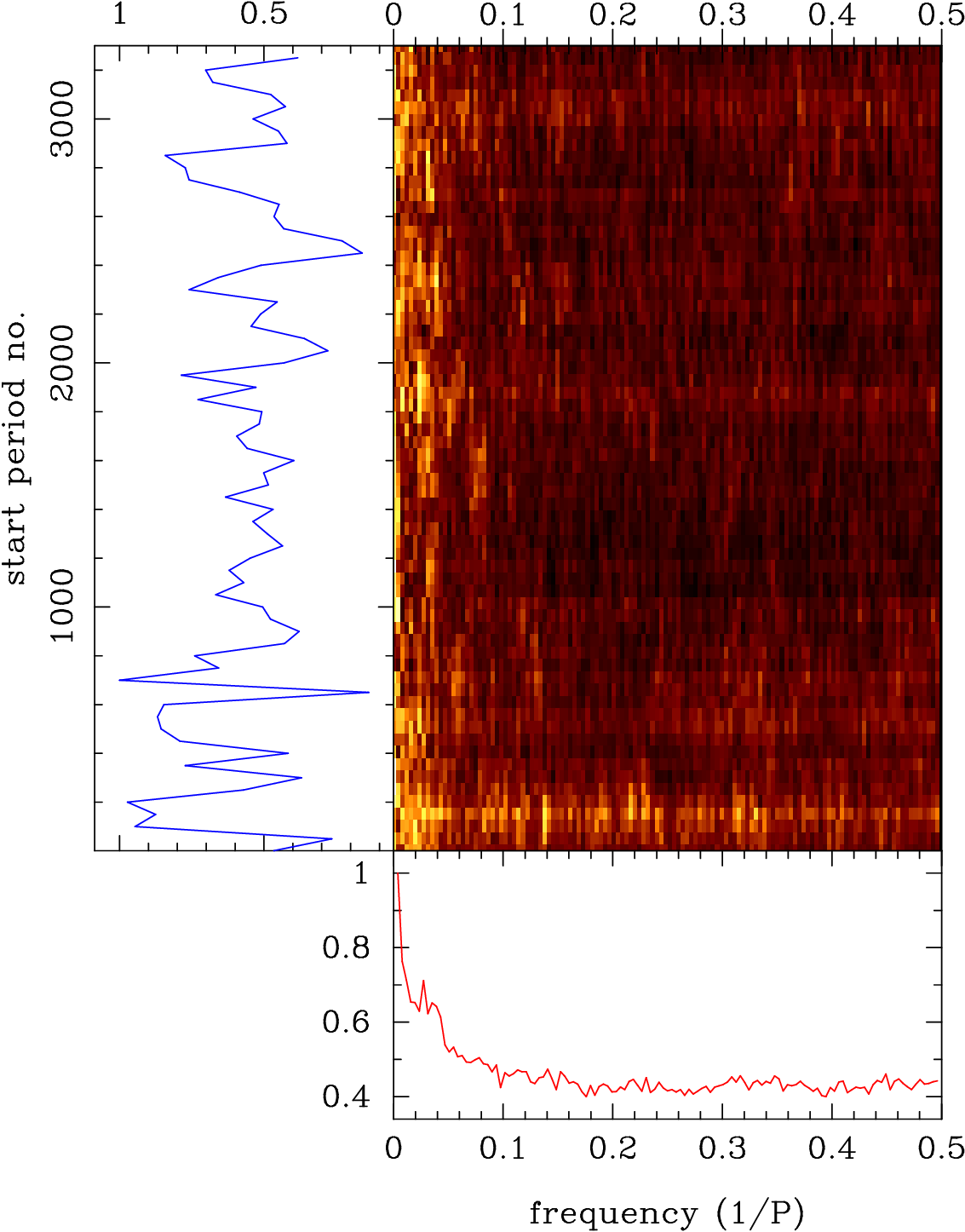}}} &
\hspace{50px}
{\mbox{\includegraphics[scale=0.35,angle=0.]{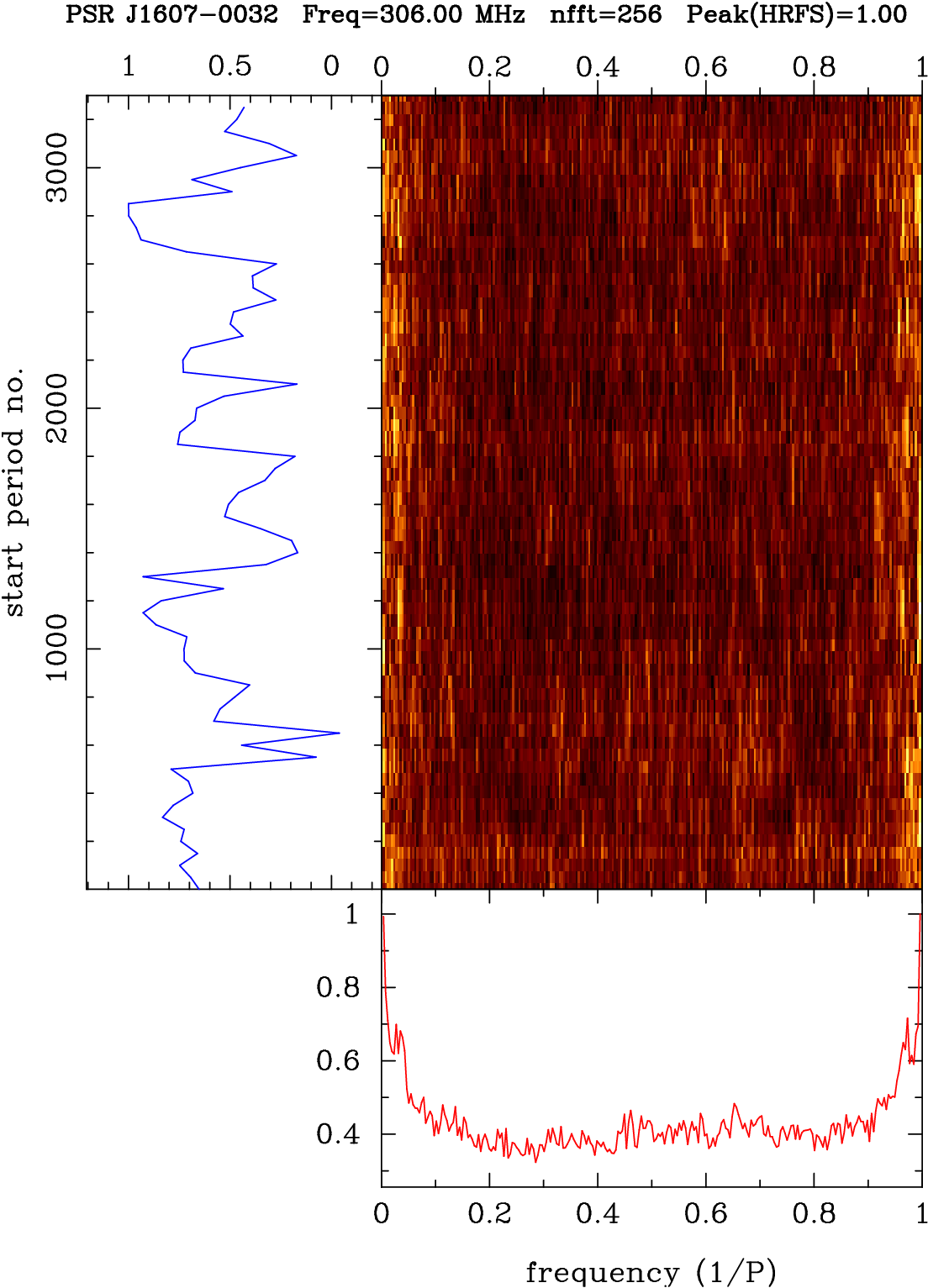}}} \\
\end{tabular}
\caption{PSR B1604$-$00 : The time evolution of the LRFS (left panel) and the
HRFS (right panel).}
\label{fig_J1607}
\end{figure*}

%J1645-0317
\begin{figure*}
\begin{tabular}{@{}lr@{}}
{\mbox{\includegraphics[scale=0.35,angle=0.]{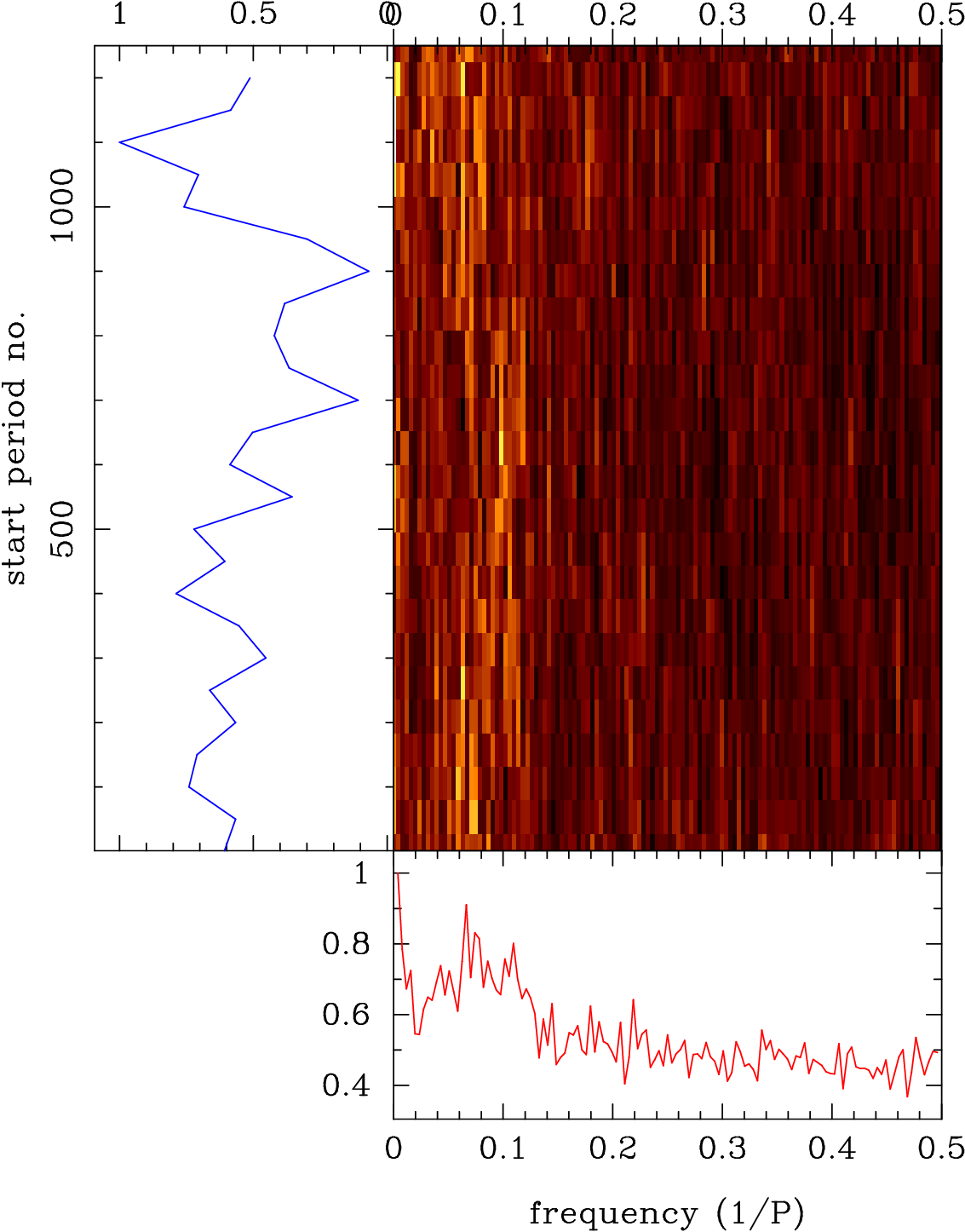}}} &
\hspace{50px}
{\mbox{\includegraphics[scale=0.35,angle=0.]{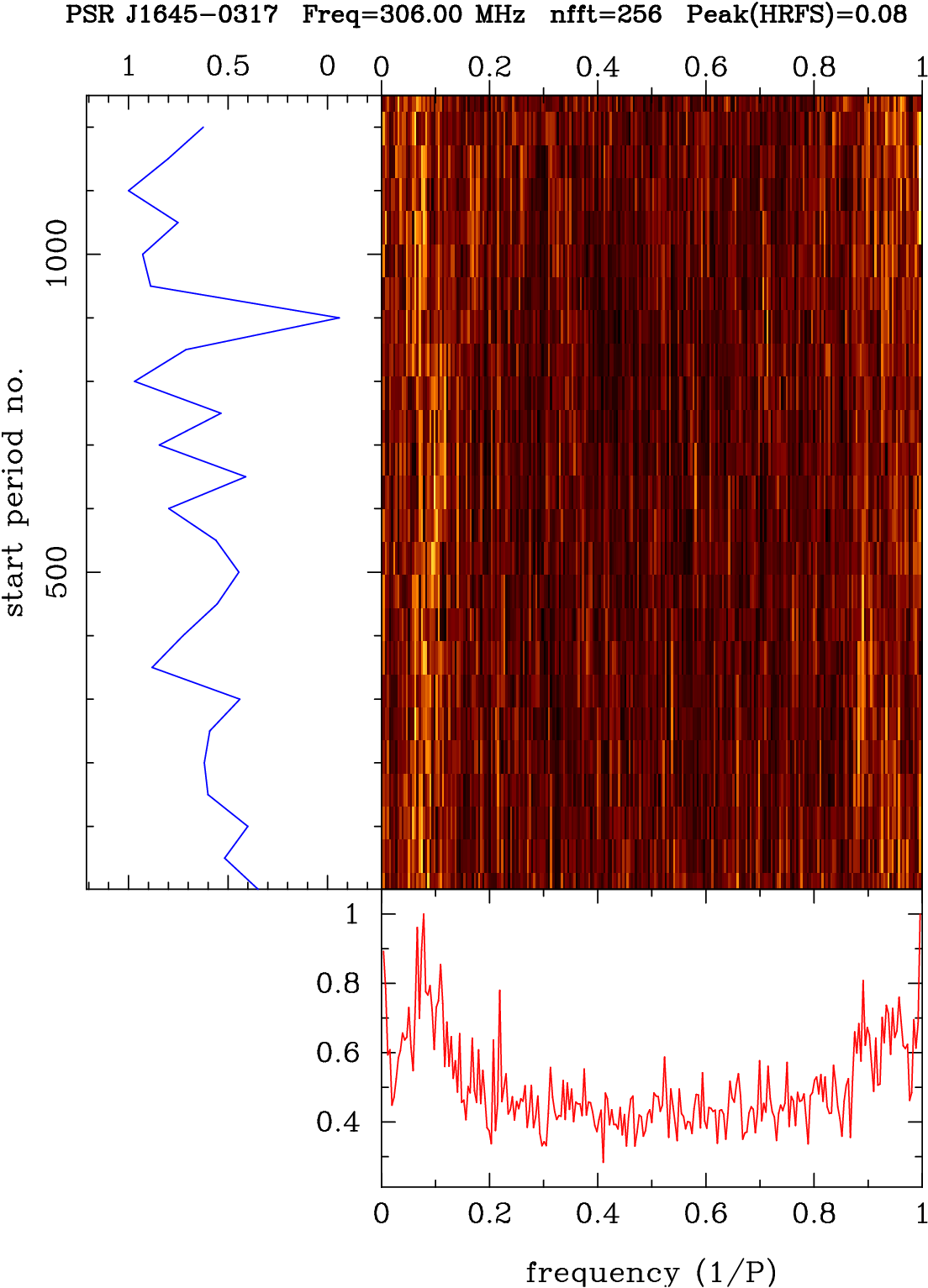}}} \\
\end{tabular}
\caption{PSR B1642$-$03 : The time evolution of the LRFS (left panel) and the
HRFS (right panel).}
\label{fig_J1645}
\end{figure*}

\clearpage

%J1650-1654
\begin{figure*}
\begin{tabular}{@{}lr@{}}
{\mbox{\includegraphics[scale=0.35,angle=0.]{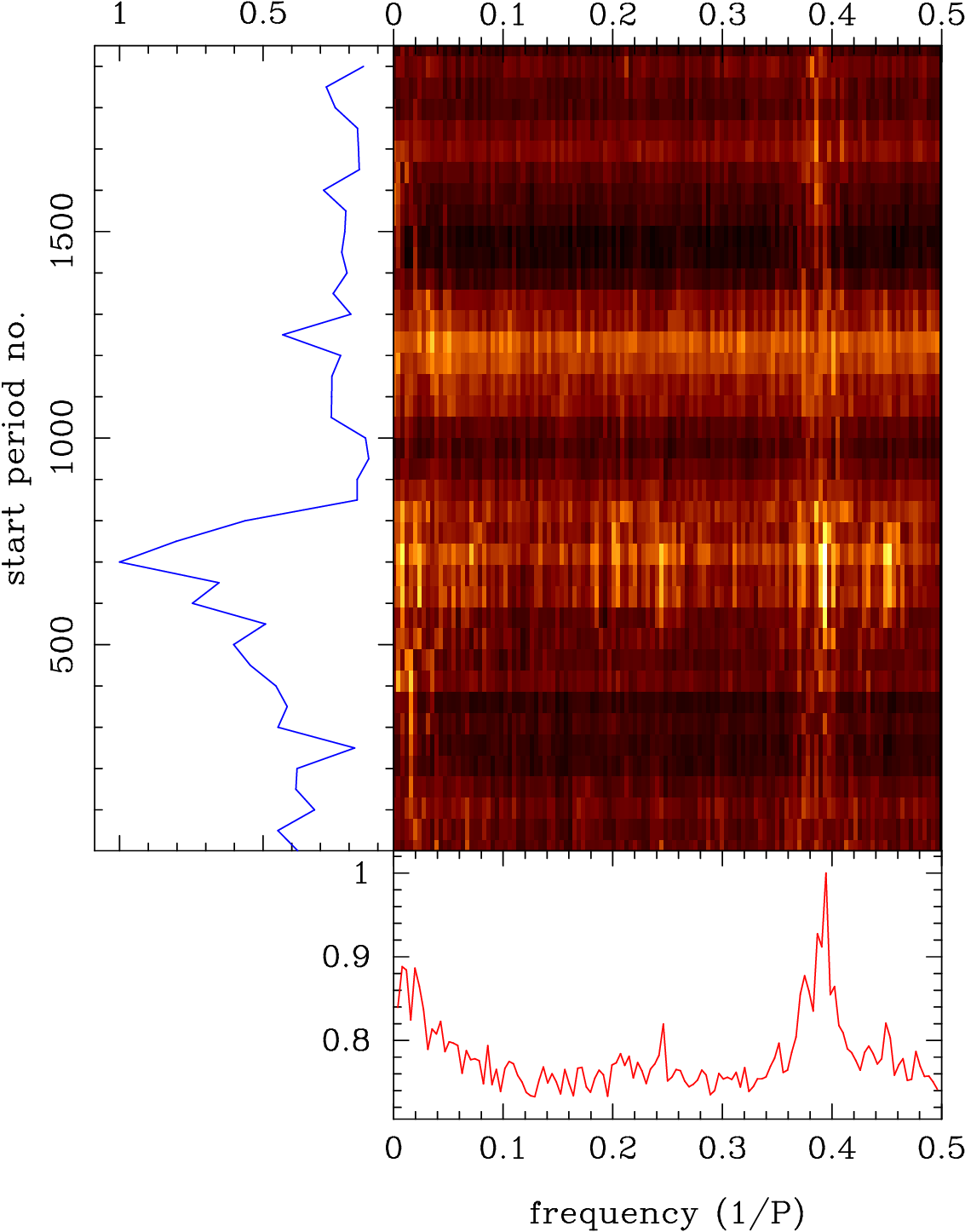}}} &
\hspace{50px}
{\mbox{\includegraphics[scale=0.35,angle=0.]{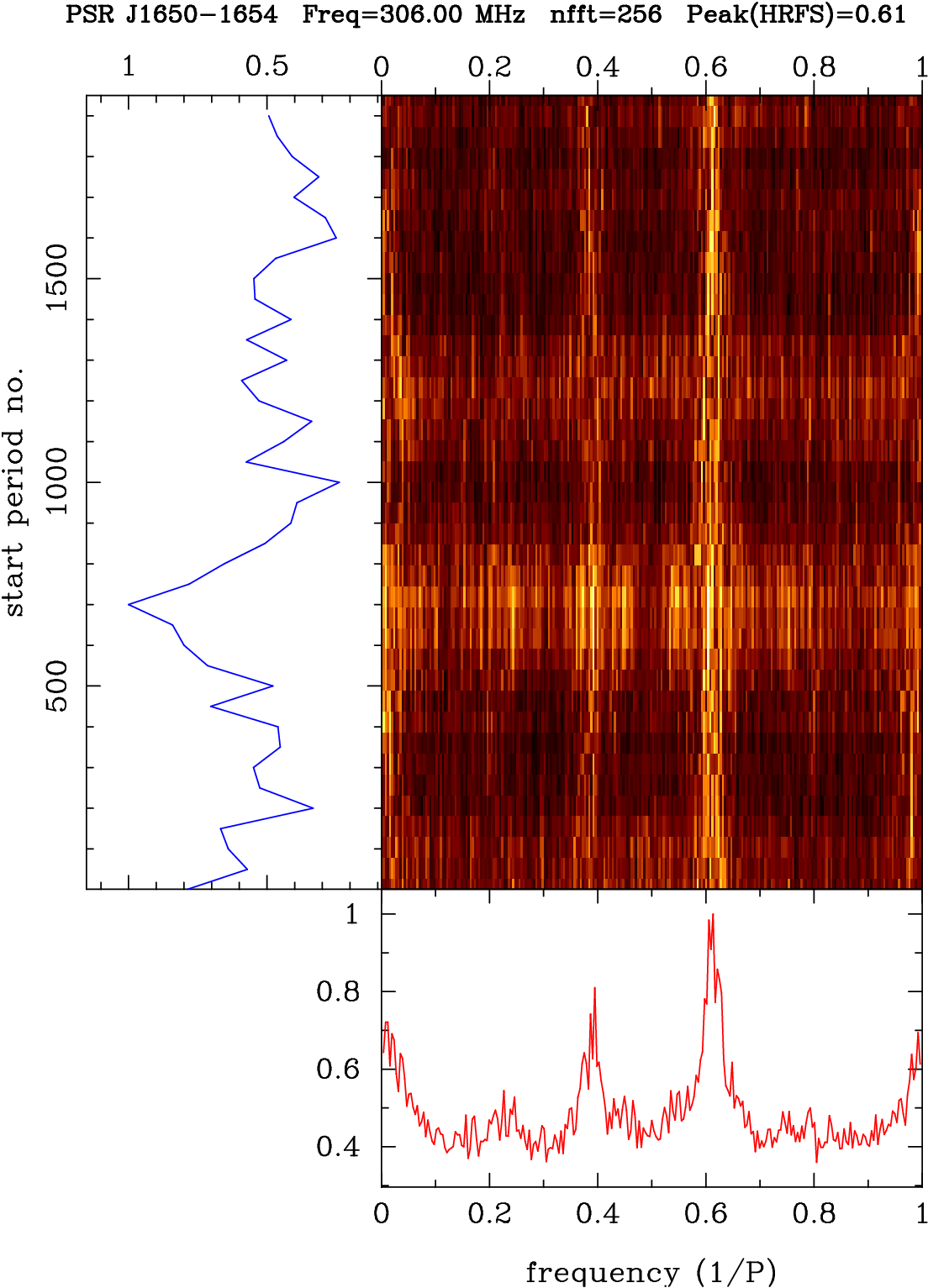}}} \\
\end{tabular}
\caption{PSR J1650$-$1654 : The time evolution of the LRFS (left panel) and the
HRFS (right panel).}
\label{fig_J1645}
\end{figure*}

%J1703-1846
\begin{figure*}
\begin{tabular}{@{}lr@{}}
{\mbox{\includegraphics[scale=0.35,angle=0.]{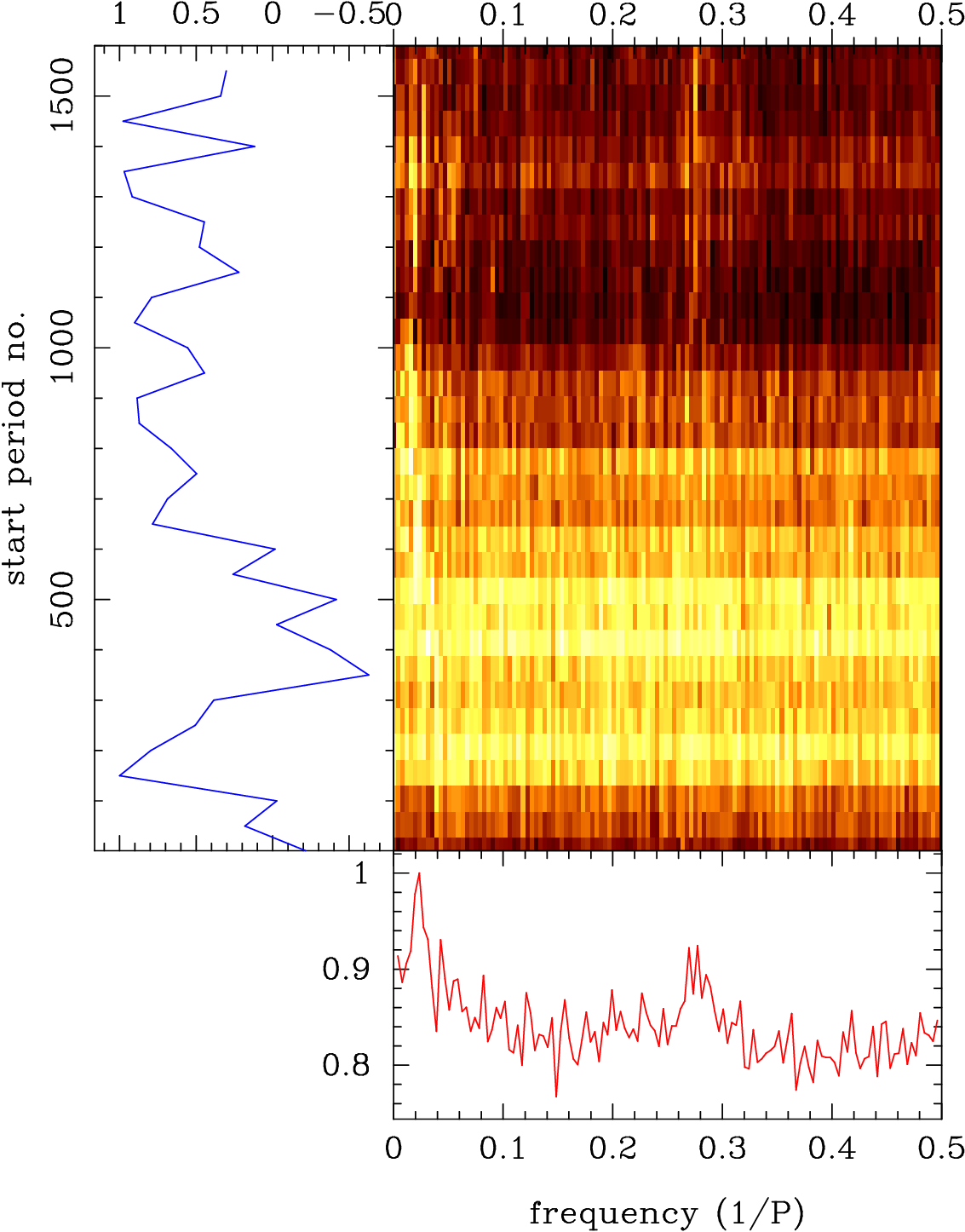}}} &
\hspace{50px}
{\mbox{\includegraphics[scale=0.35,angle=0.]{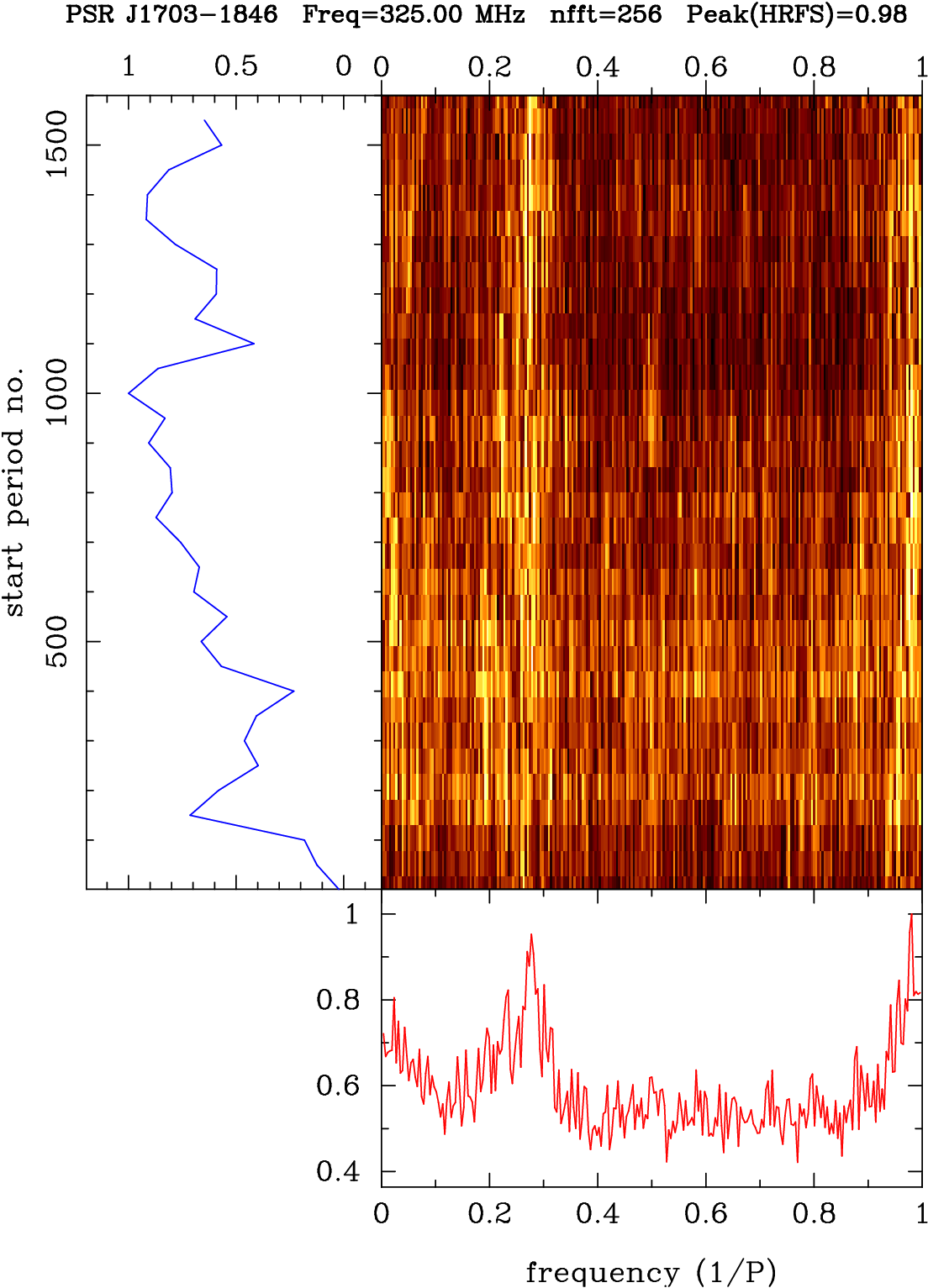}}} \\
\end{tabular}
\caption{PSR B1700$-$18 : The time evolution of the LRFS (left panel) and the
HRFS (right panel).}
\label{fig_J1645}
\end{figure*}

\clearpage

%J1857-1027
\begin{figure*}
\begin{tabular}{@{}lr@{}}
{\mbox{\includegraphics[scale=0.35,angle=0.]{J1857-1027_LRFSavg_256.ps}}} &
\hspace{50px}
{\mbox{\includegraphics[scale=0.35,angle=0.]{J1857-1027_nullfft_256.ps}}} \\
\end{tabular}
\caption{PSR J1857$-$1027 : The time evolution of the LRFS (left panel) and the
Null-Burst time series FFT (right panel).}
\label{fig_J1857}
\end{figure*}

%J1907+4002
\begin{figure*}
\begin{tabular}{@{}lr@{}}
{\mbox{\includegraphics[scale=0.35,angle=0.]{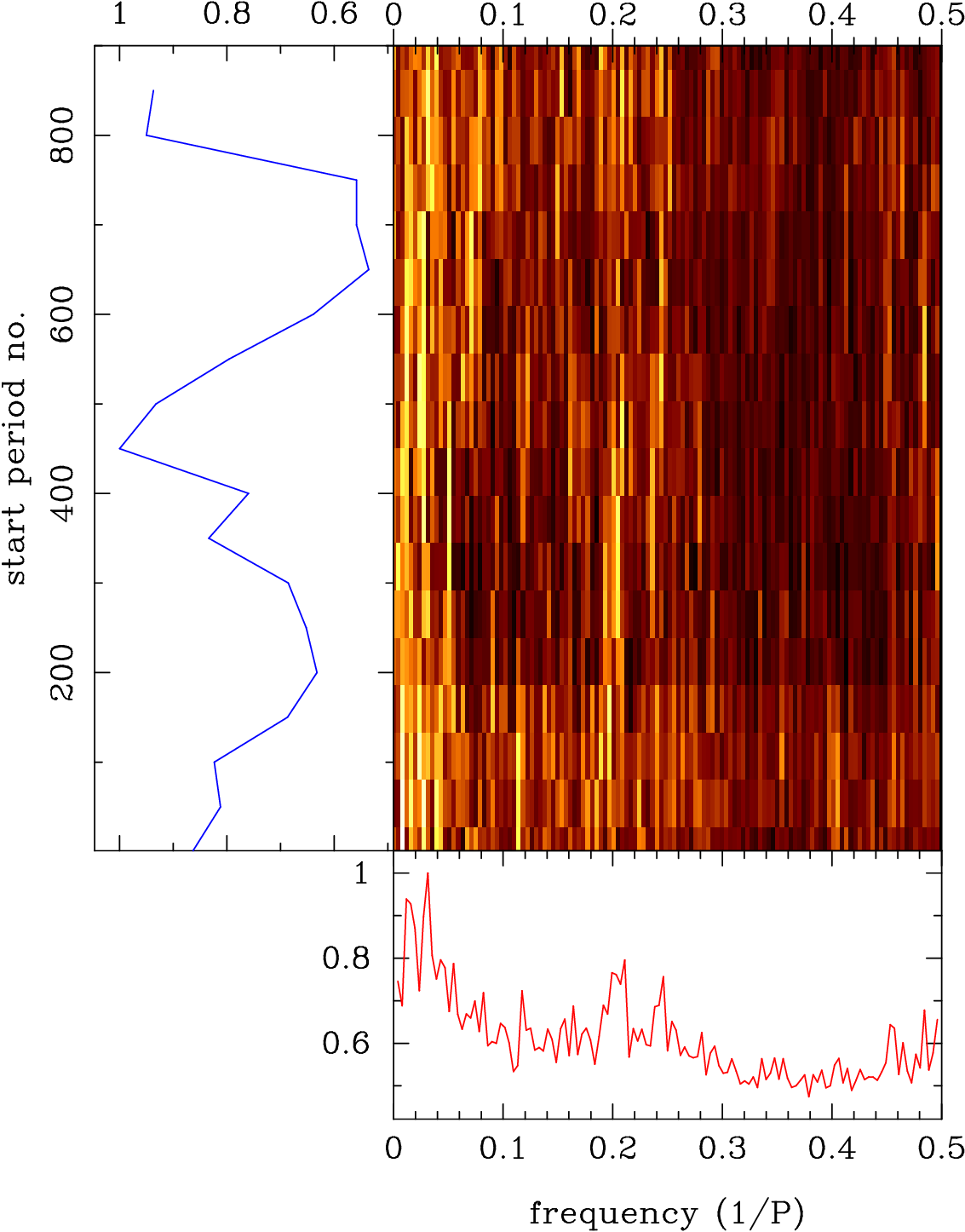}}} &
\hspace{50px}
{\mbox{\includegraphics[scale=0.35,angle=0.]{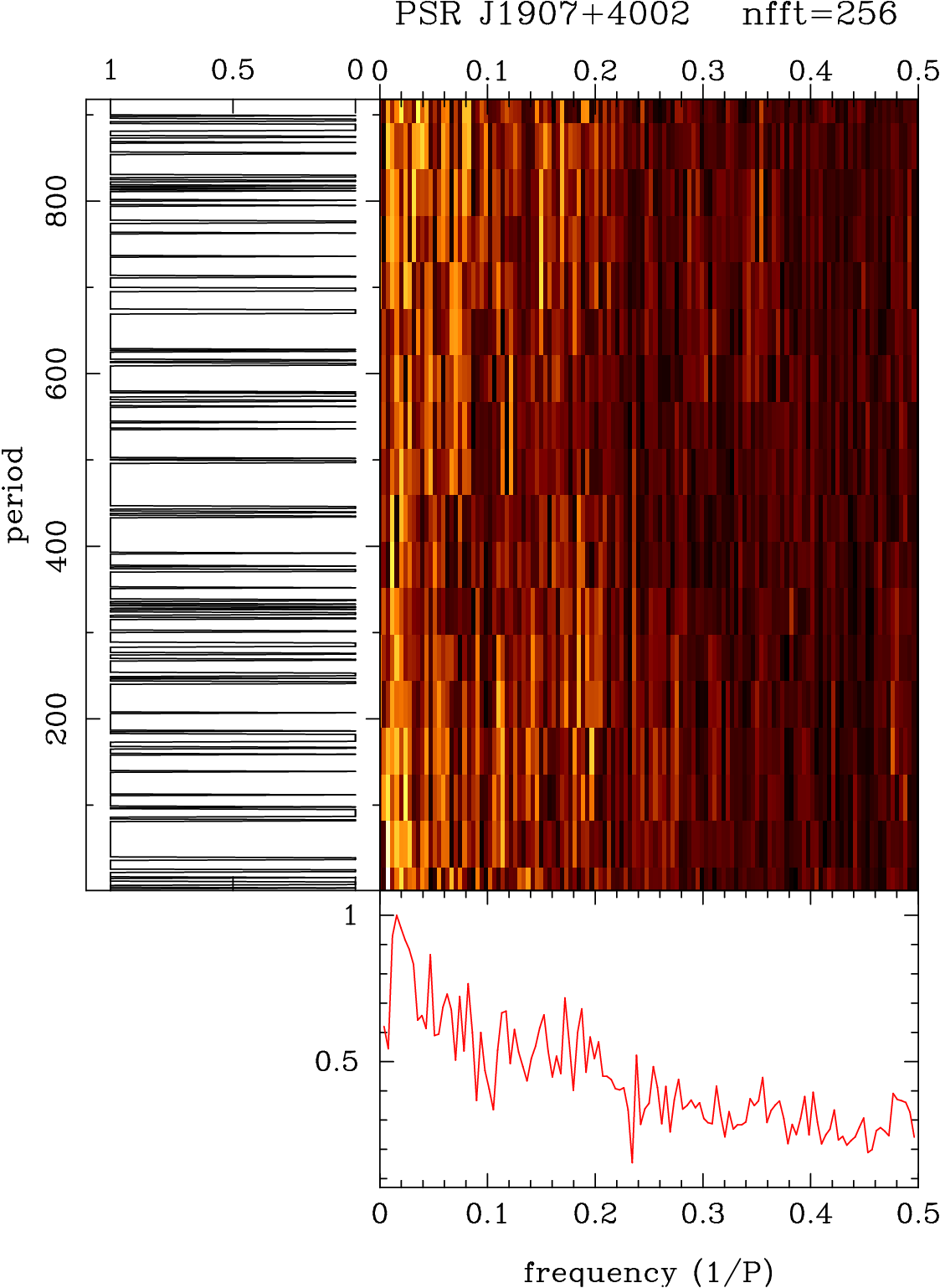}}} \\
\end{tabular}
\caption{PSR B1905+39 : The time evolution of the LRFS (left panel) and the
Null-Burst time series FFT (right panel).}
\label{fig_J1907}
\end{figure*}

\clearpage

%J1932+1059
\begin{figure*}
\begin{tabular}{@{}lr@{}}
{\mbox{\includegraphics[scale=0.35,angle=0.]{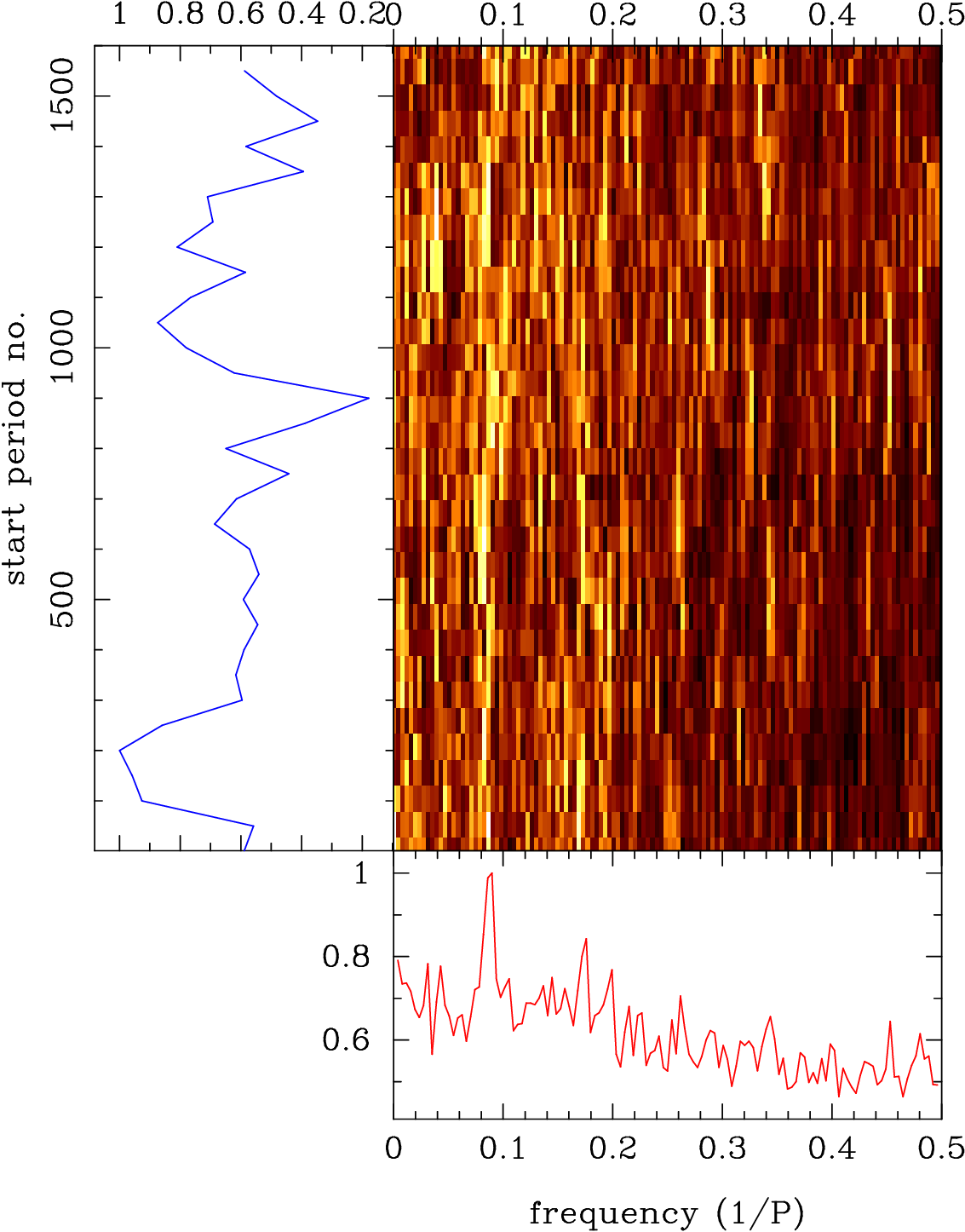}}} &
\hspace{50px}
{\mbox{\includegraphics[scale=0.35,angle=0.]{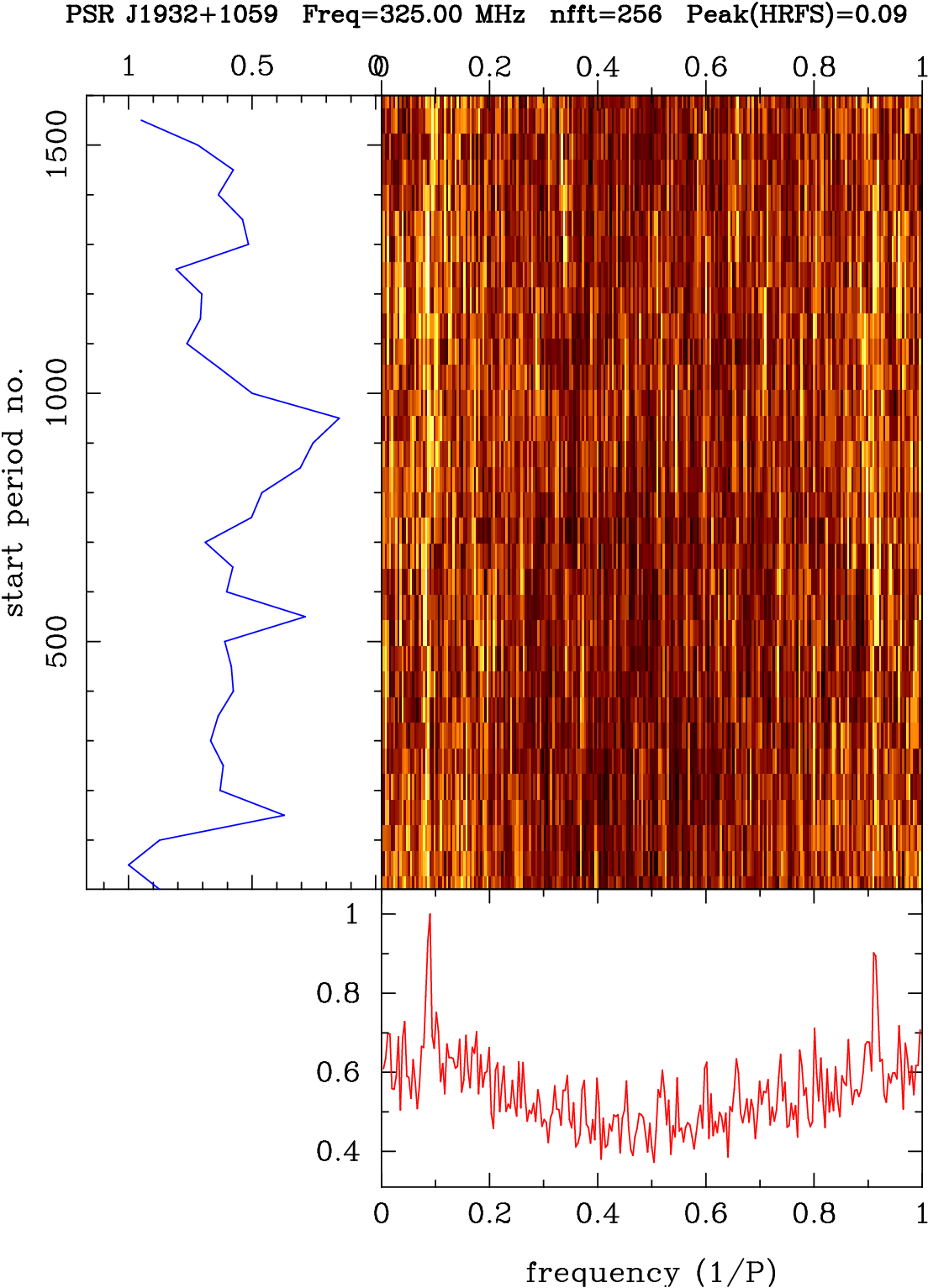}}} \\
\end{tabular}
\caption{PSR B1929+10 : The time evolution of the LRFS (left panel) and the
HRFS (right panel).}
\label{fig_J1932}
\end{figure*}

%J1954+2923
\begin{figure*}
\begin{tabular}{@{}lr@{}}
{\mbox{\includegraphics[scale=0.35,angle=0.]{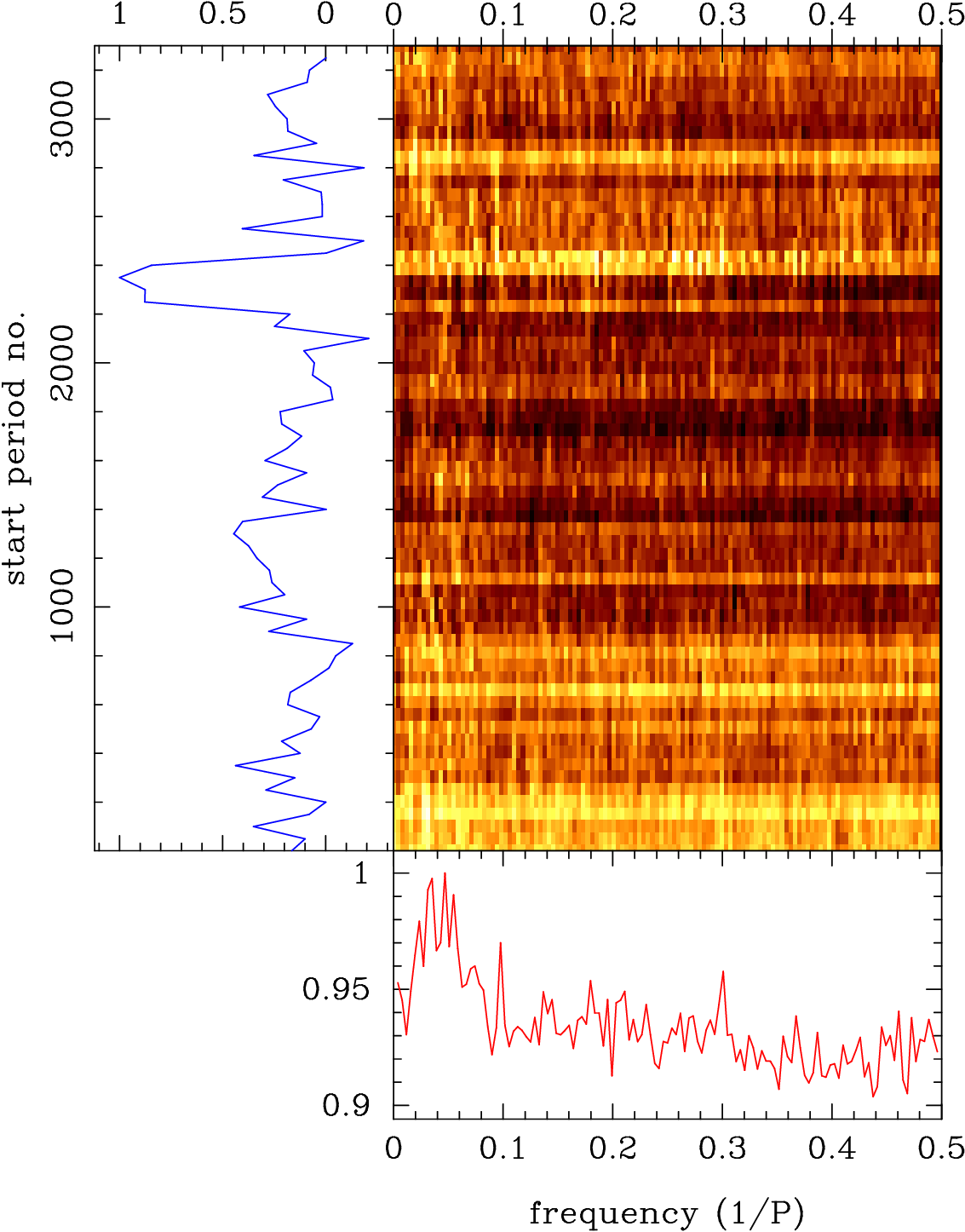}}} &
\hspace{50px}
{\mbox{\includegraphics[scale=0.35,angle=0.]{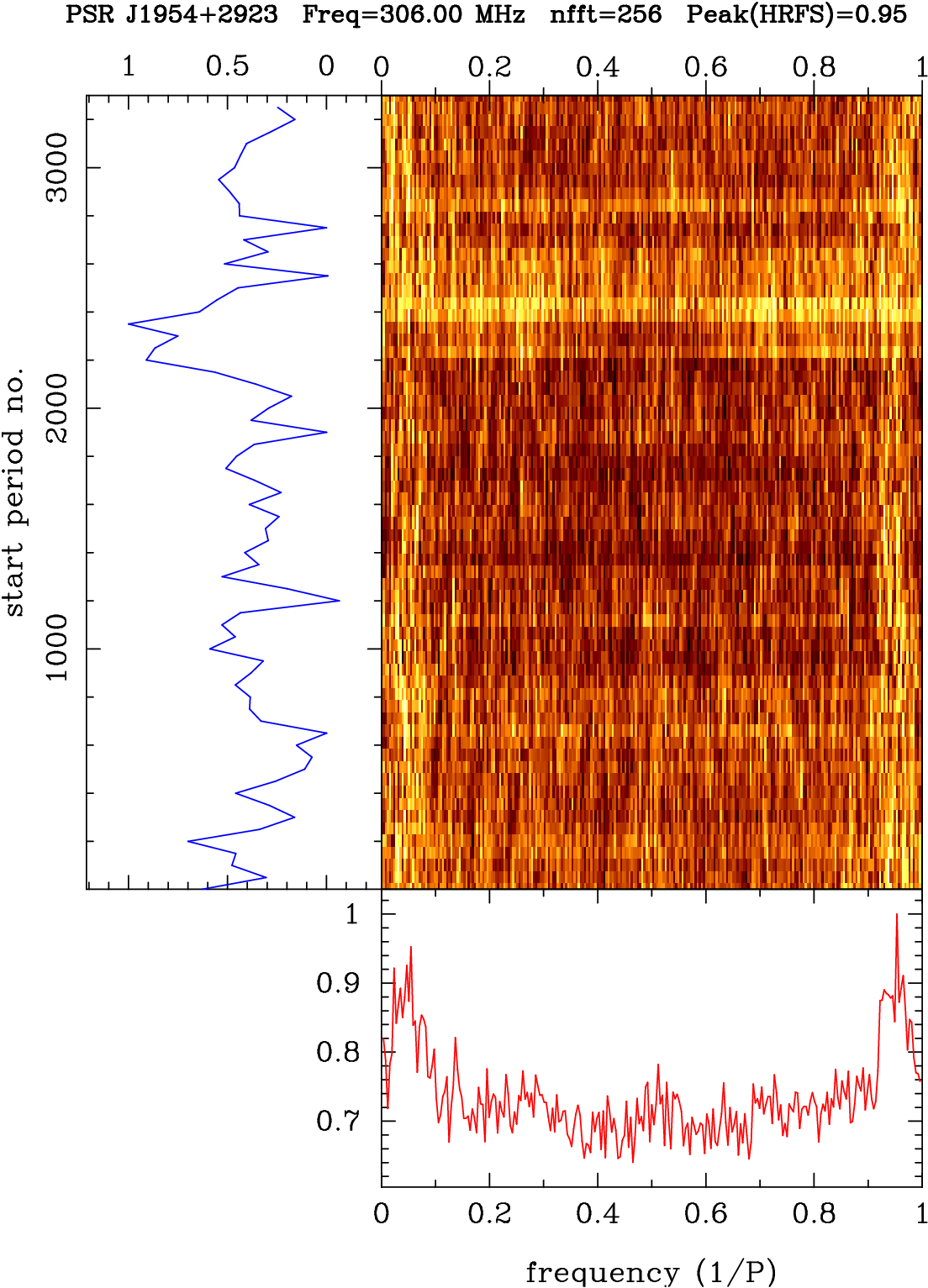}}} \\
\end{tabular}
\caption{PSR B1952+29 : The time evolution of the LRFS (left panel) and the
HRFS (right panel).}
\label{fig_J1954}
\end{figure*}

\clearpage

%J2022+5154
\begin{figure*}
\begin{tabular}{@{}lr@{}}
{\mbox{\includegraphics[scale=0.35,angle=0.]{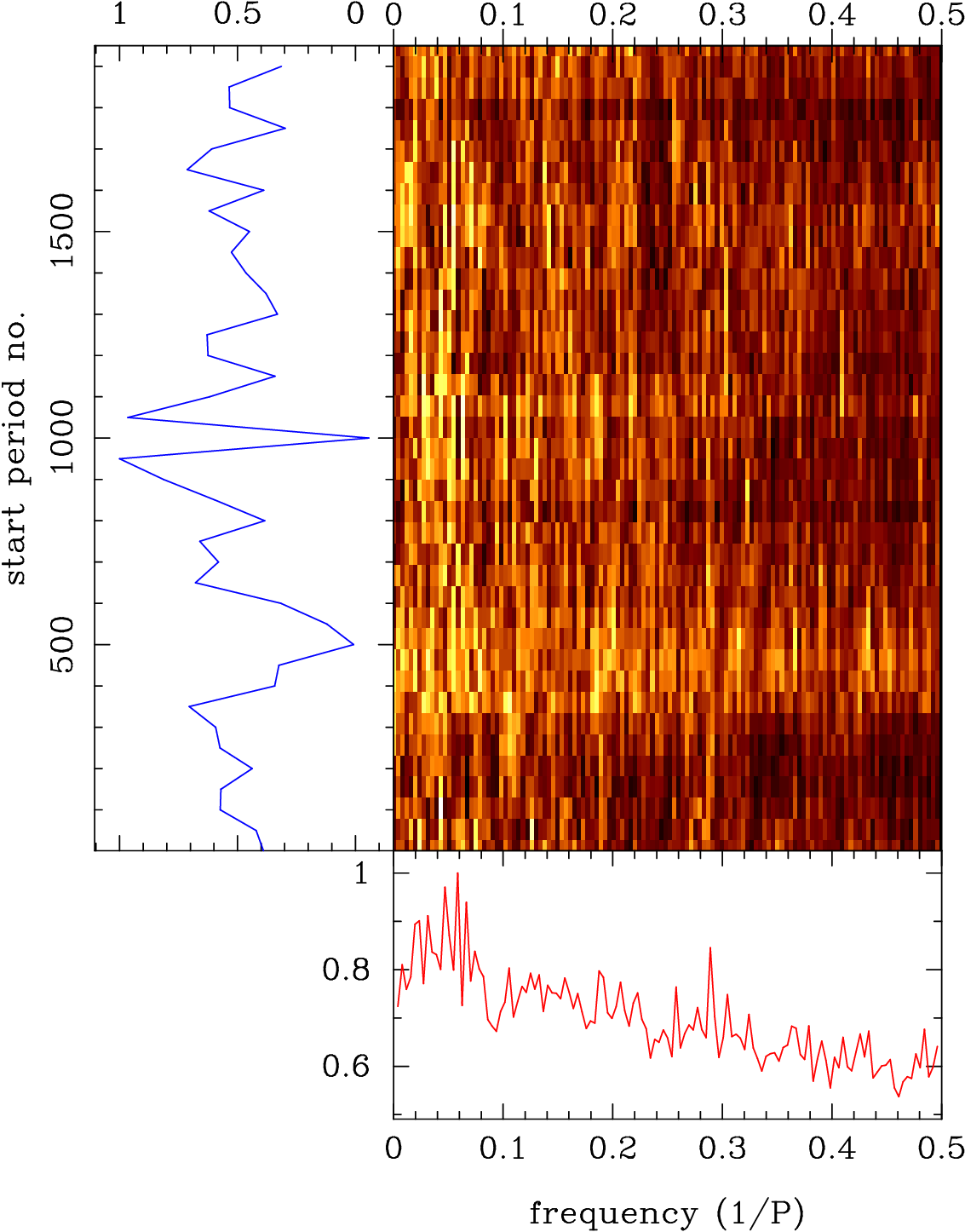}}} &
\hspace{50px}
{\mbox{\includegraphics[scale=0.35,angle=0.]{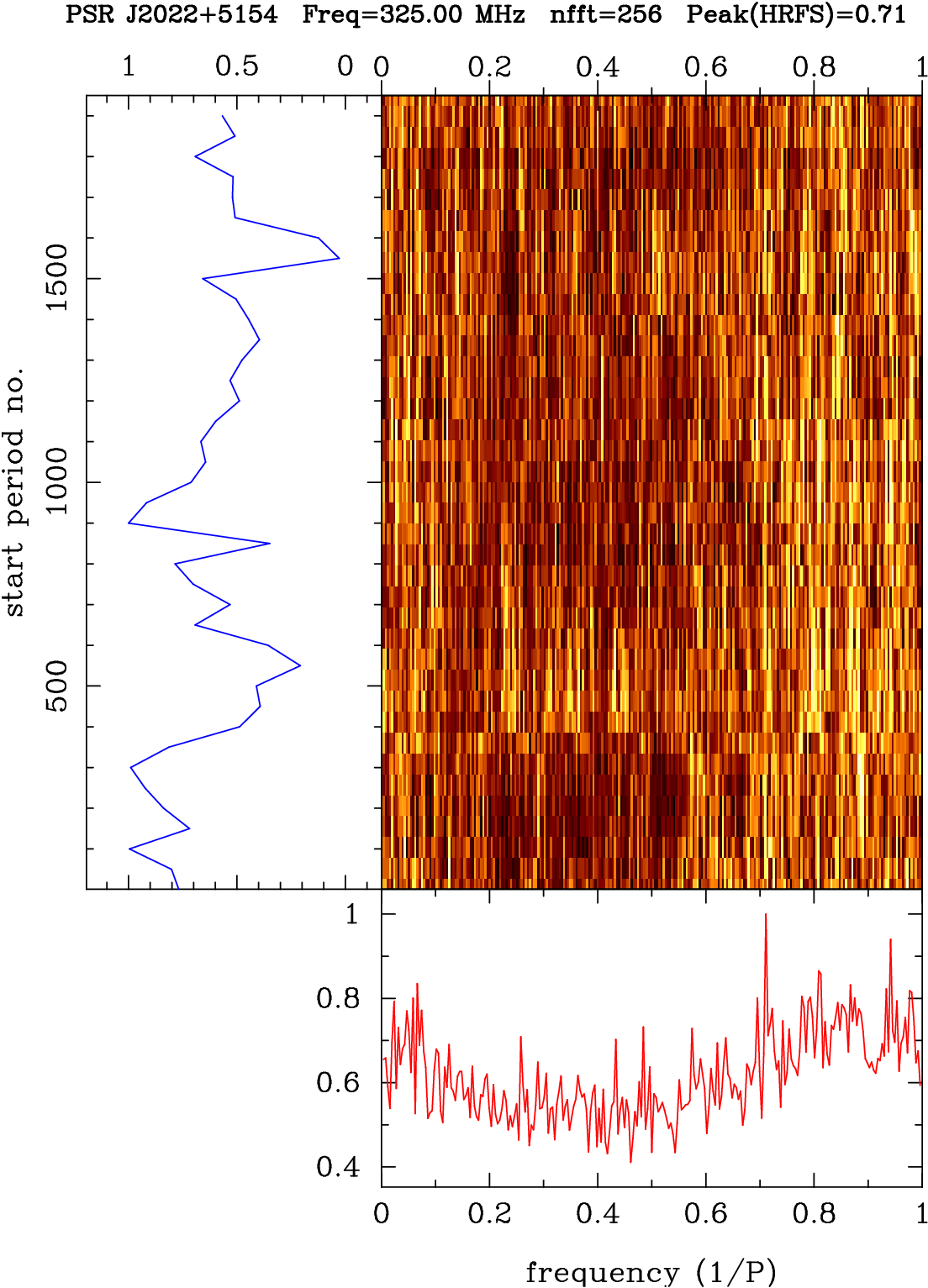}}} \\
\end{tabular}
\caption{PSR B2021+51 : The time evolution of the LRFS (left panel) and the
HRFS (right panel).}
\label{fig_J2022}
\end{figure*}

%J2048-1616
\begin{figure*}
\begin{tabular}{@{}lr@{}}
{\mbox{\includegraphics[scale=0.35,angle=0.]{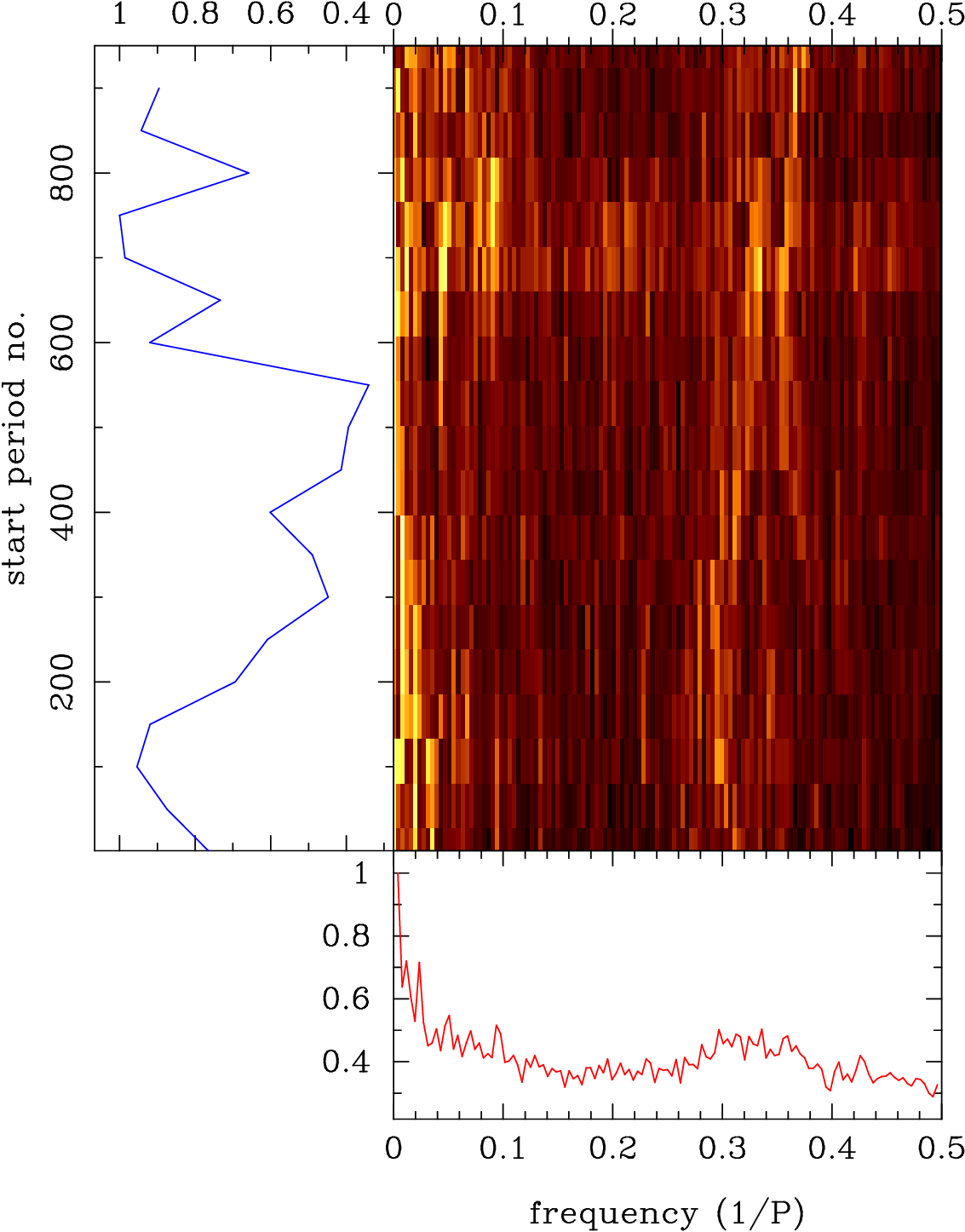}}} &
\hspace{50px}
{\mbox{\includegraphics[scale=0.35,angle=0.]{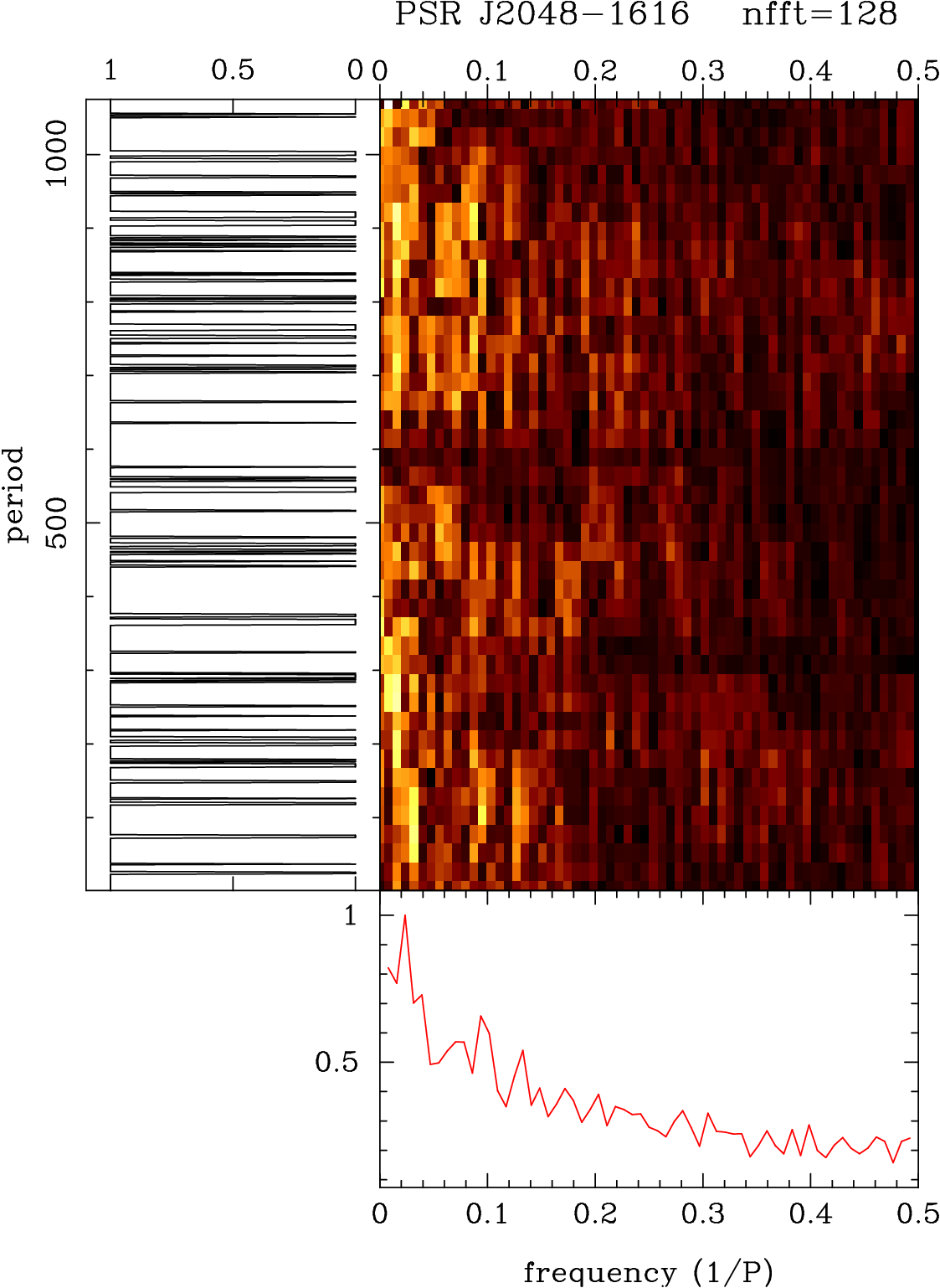}}} \\
\end{tabular}
\caption{PSR B2045$-$16 : The time evolution of the LRFS (left panel) and the
Null-Burst time series FFT (right panel).}
\label{fig_J2048}
\end{figure*}

\clearpage

%J2113+4644
\begin{figure*}
\begin{tabular}{@{}lr@{}}
{\mbox{\includegraphics[scale=0.35,angle=0.]{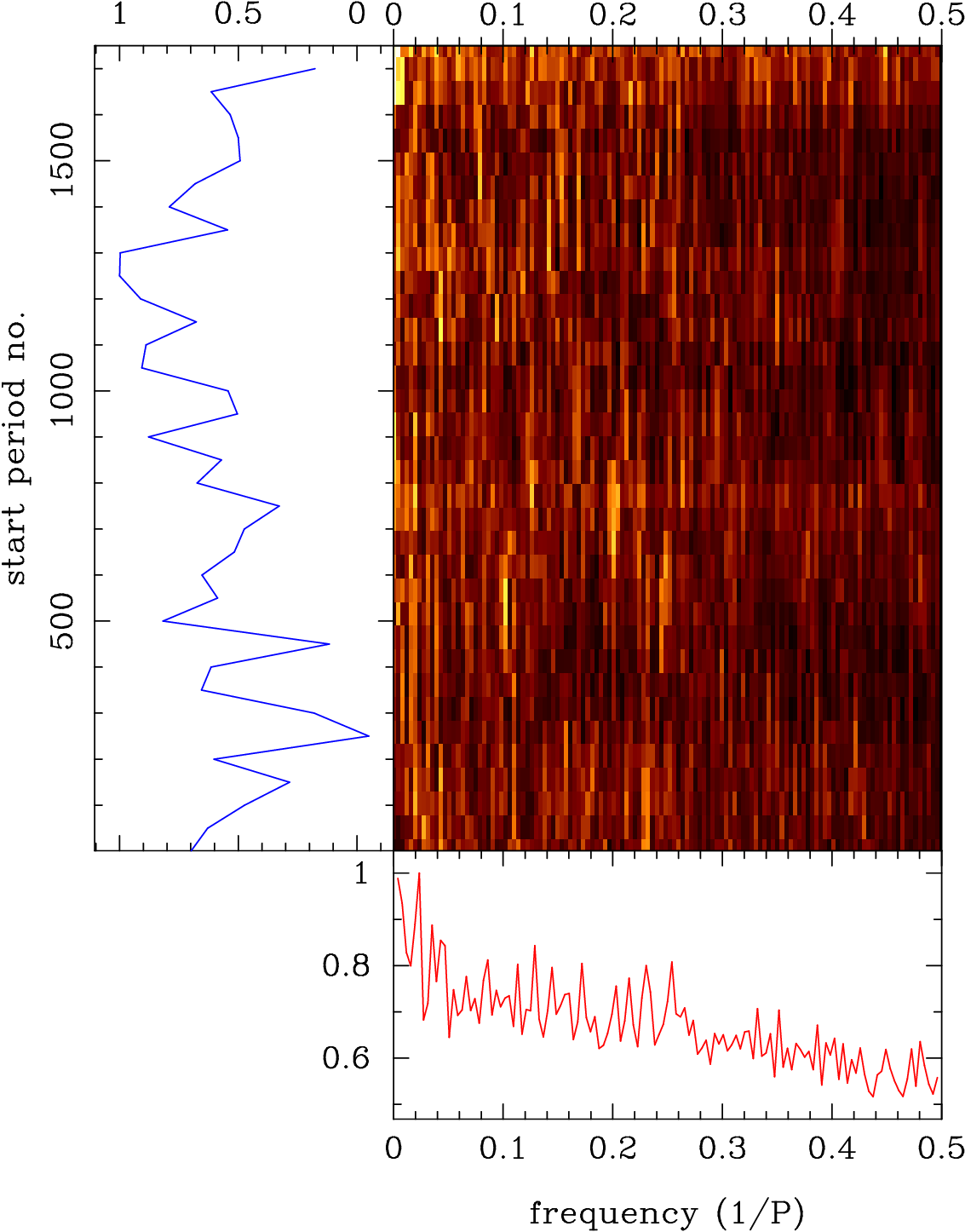}}} &
\hspace{50px}
{\mbox{\includegraphics[scale=0.35,angle=0.]{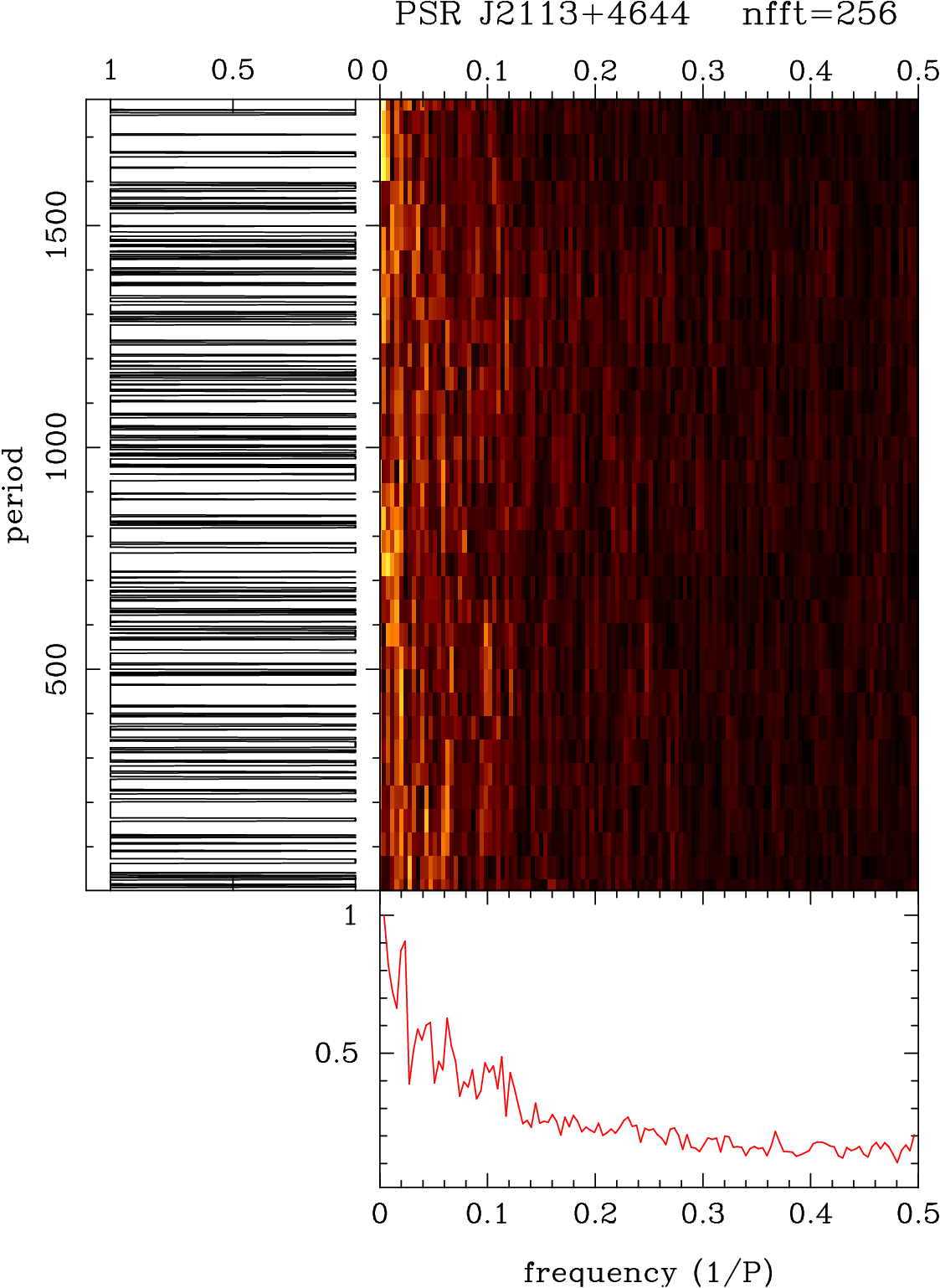}}} \\
\end{tabular}
\caption{PSR B2111+46 : The time evolution of the LRFS (left panel) and the
Null-Burst time series FFT (right panel).}
\label{fig_J2113}
\end{figure*}

%J2313+4253
\begin{figure*}
\begin{tabular}{@{}lr@{}}
{\mbox{\includegraphics[scale=0.35,angle=0.]{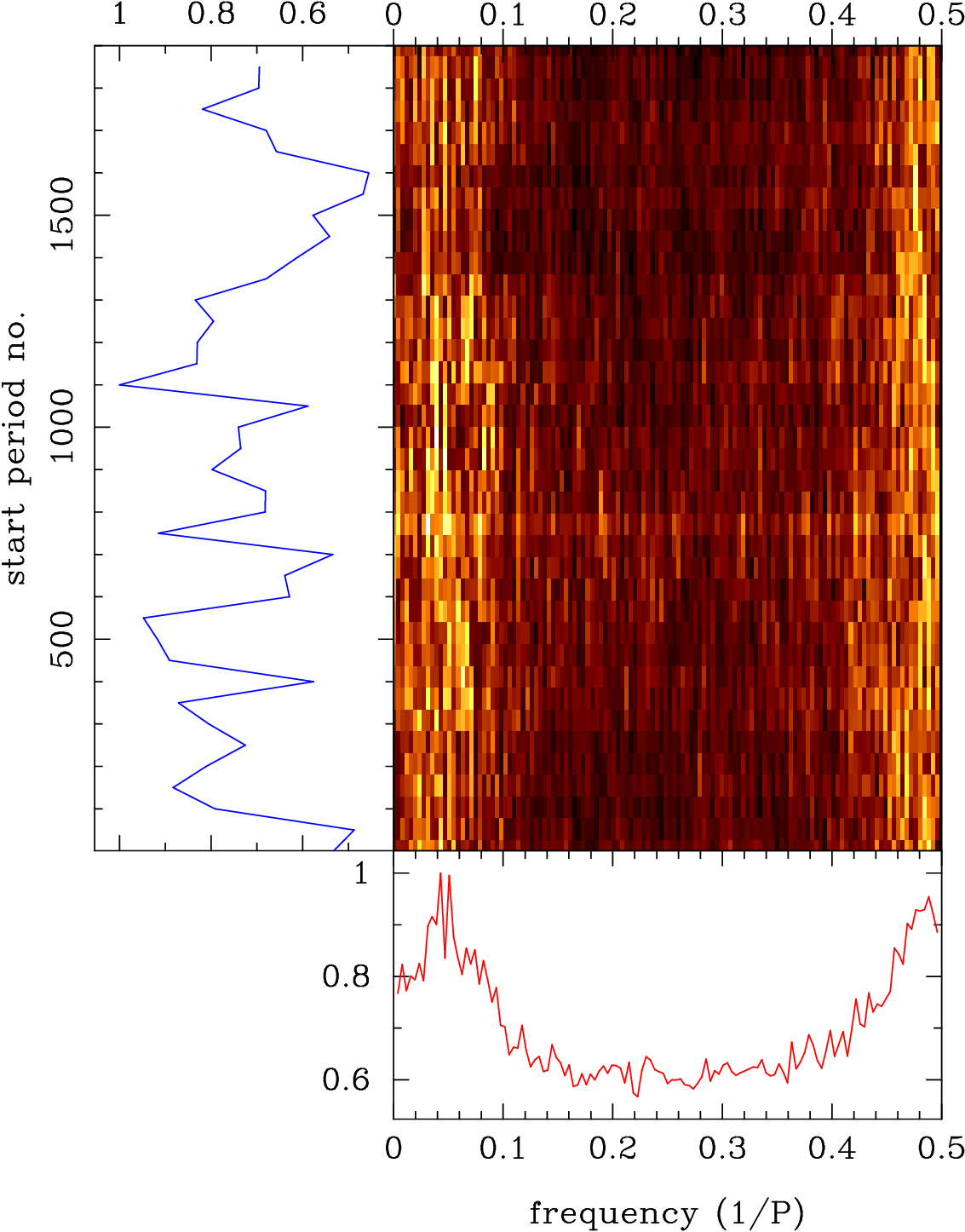}}} &
\hspace{50px}
{\mbox{\includegraphics[scale=0.35,angle=0.]{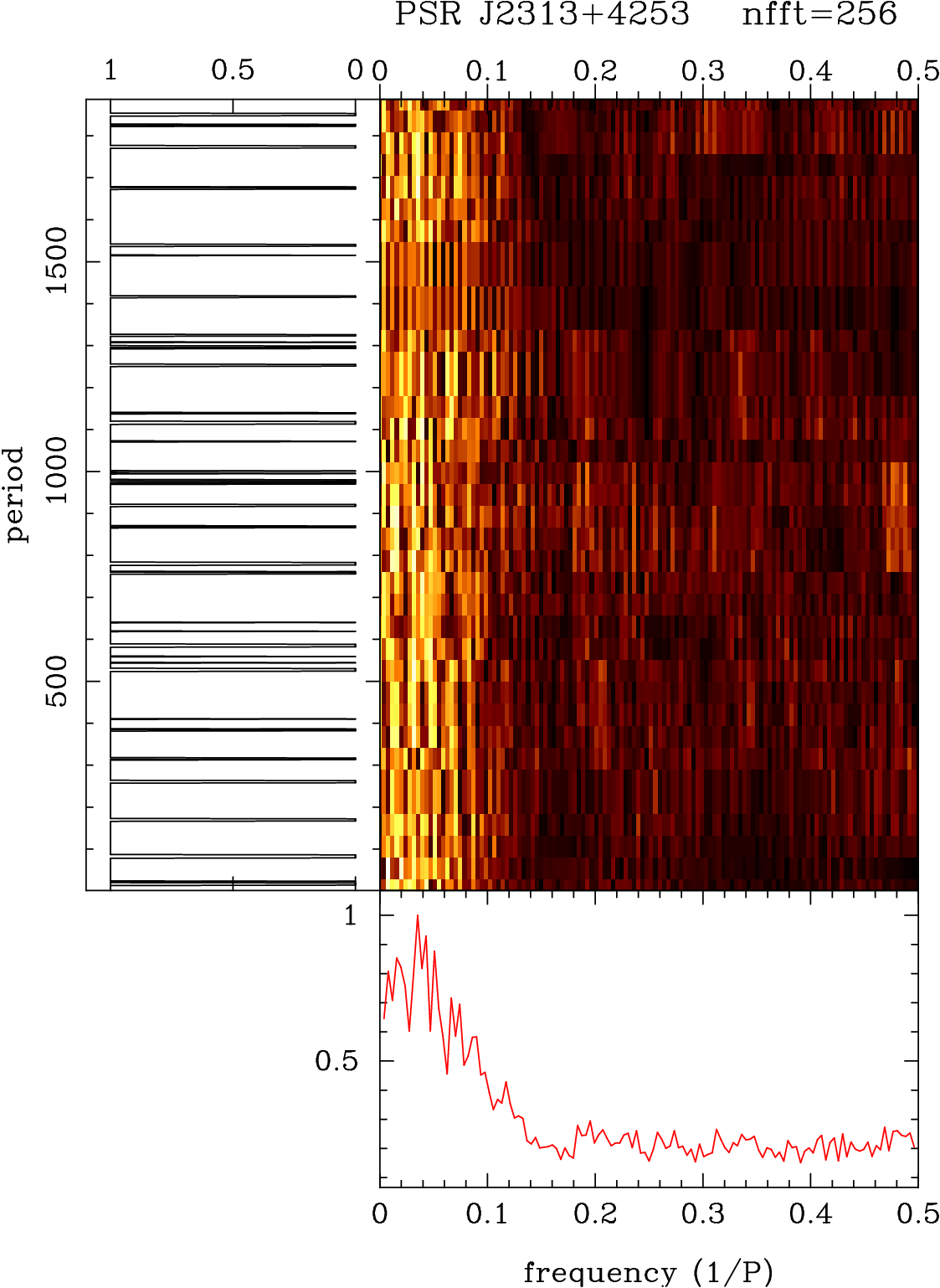}}} \\
\end{tabular}
\caption{PSR B2310+42 : The time evolution of the LRFS (left panel) and the
Null-Burst time series FFT (right panel).}
\label{fig_J2313}
\end{figure*}

\clearpage

%J2321+6024
\begin{figure*}
\begin{tabular}{@{}lr@{}}
{\mbox{\includegraphics[scale=0.35,angle=0.]{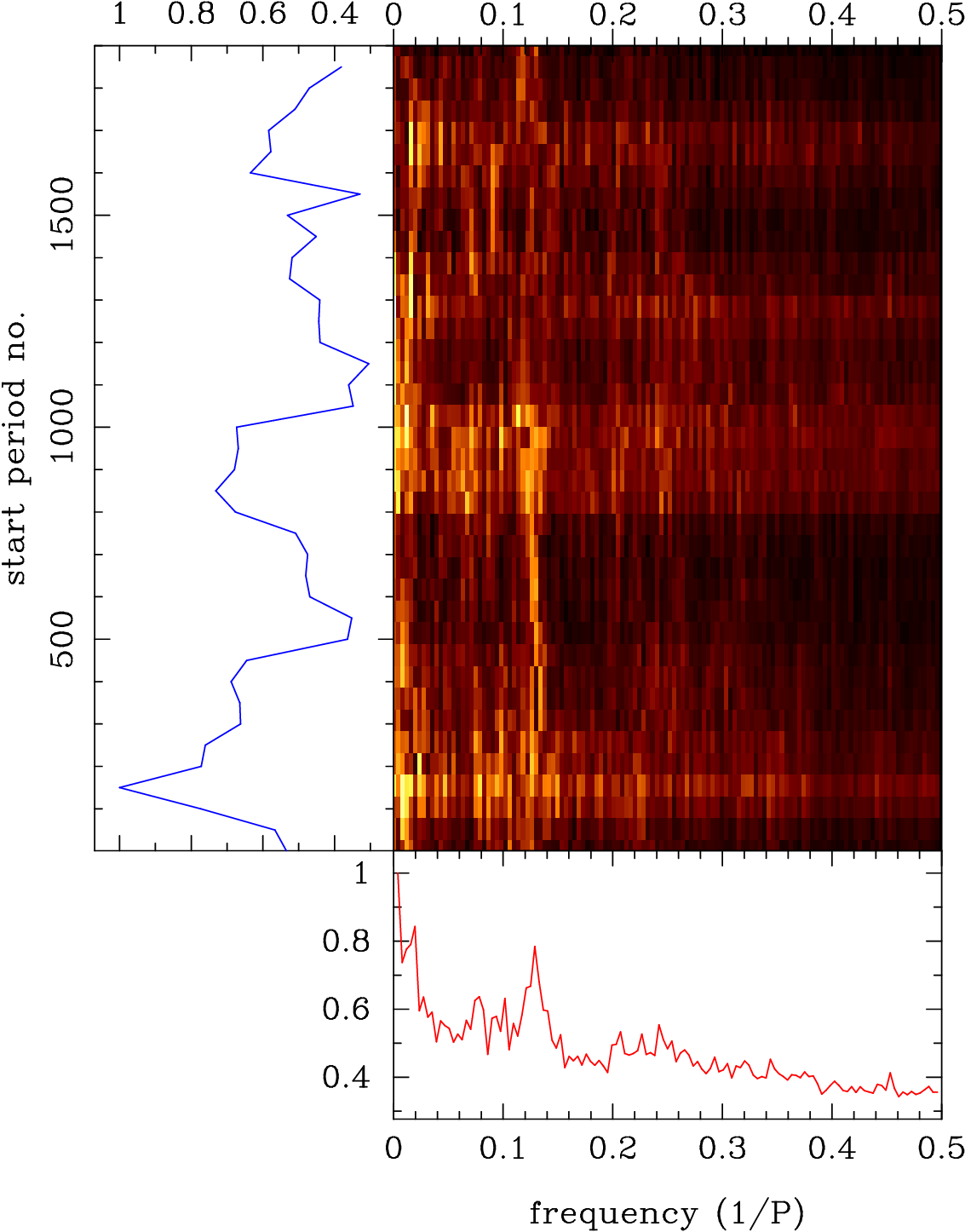}}} &
\hspace{50px}
{\mbox{\includegraphics[scale=0.35,angle=0.]{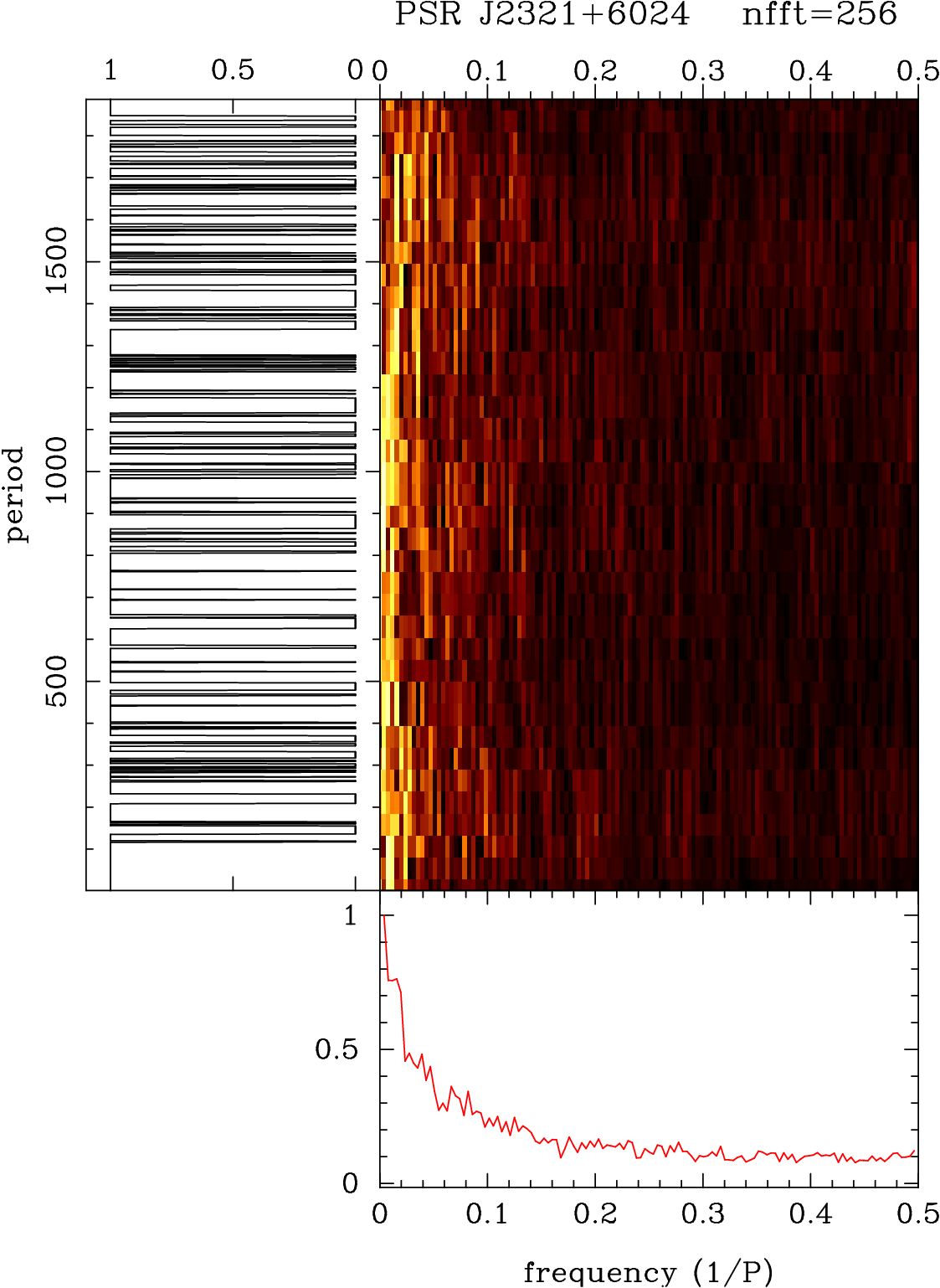}}} \\
\end{tabular}
\caption{PSR B2319+60 : The time evolution of the LRFS (left panel) and the
Null-Burst time series FFT (right panel).}
\label{fig_J2321}
\end{figure*}

%J2330-2005
\begin{figure*}
\begin{tabular}{@{}lr@{}}
{\mbox{\includegraphics[scale=0.35,angle=0.]{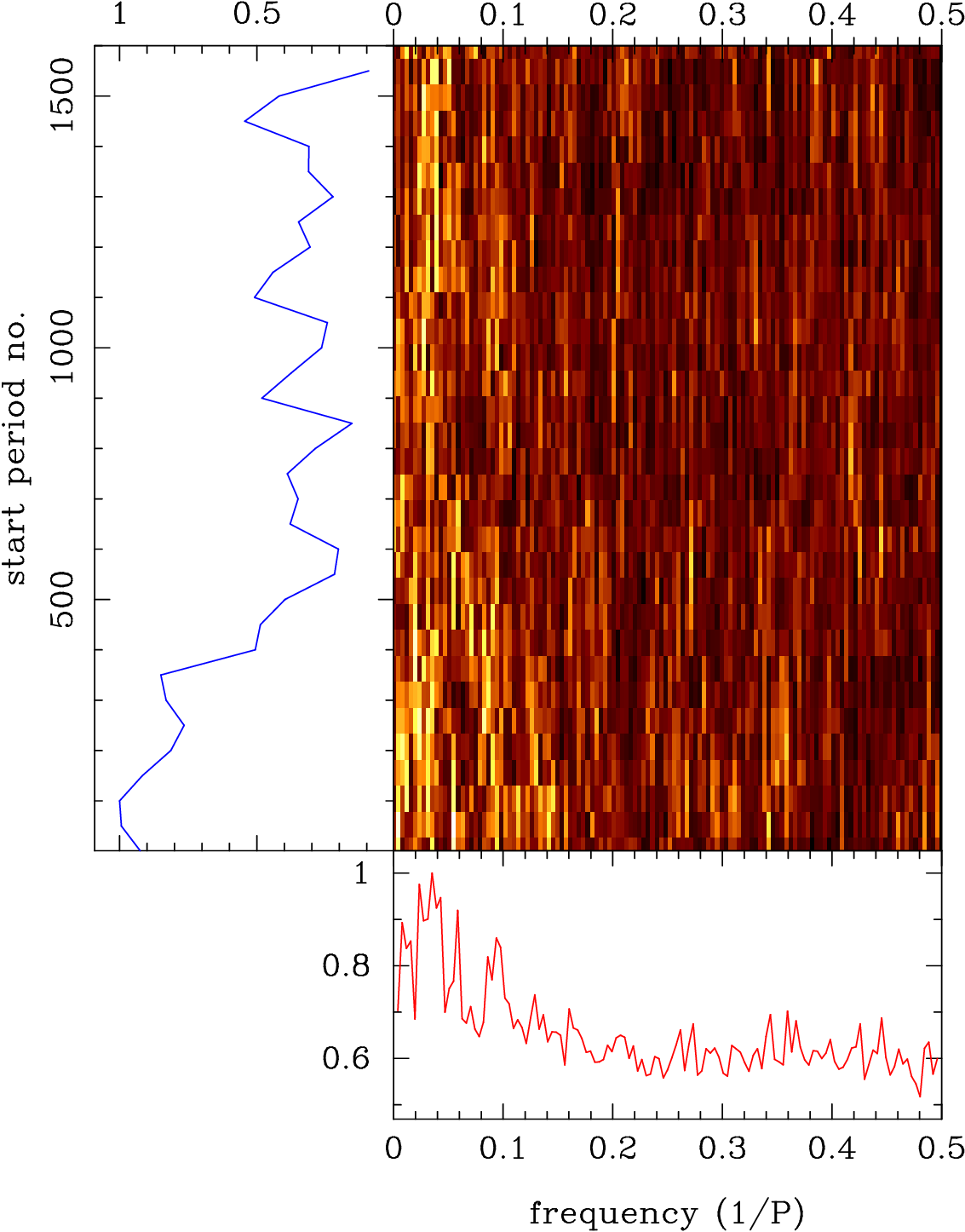}}} &
\hspace{50px}
{\mbox{\includegraphics[scale=0.35,angle=0.]{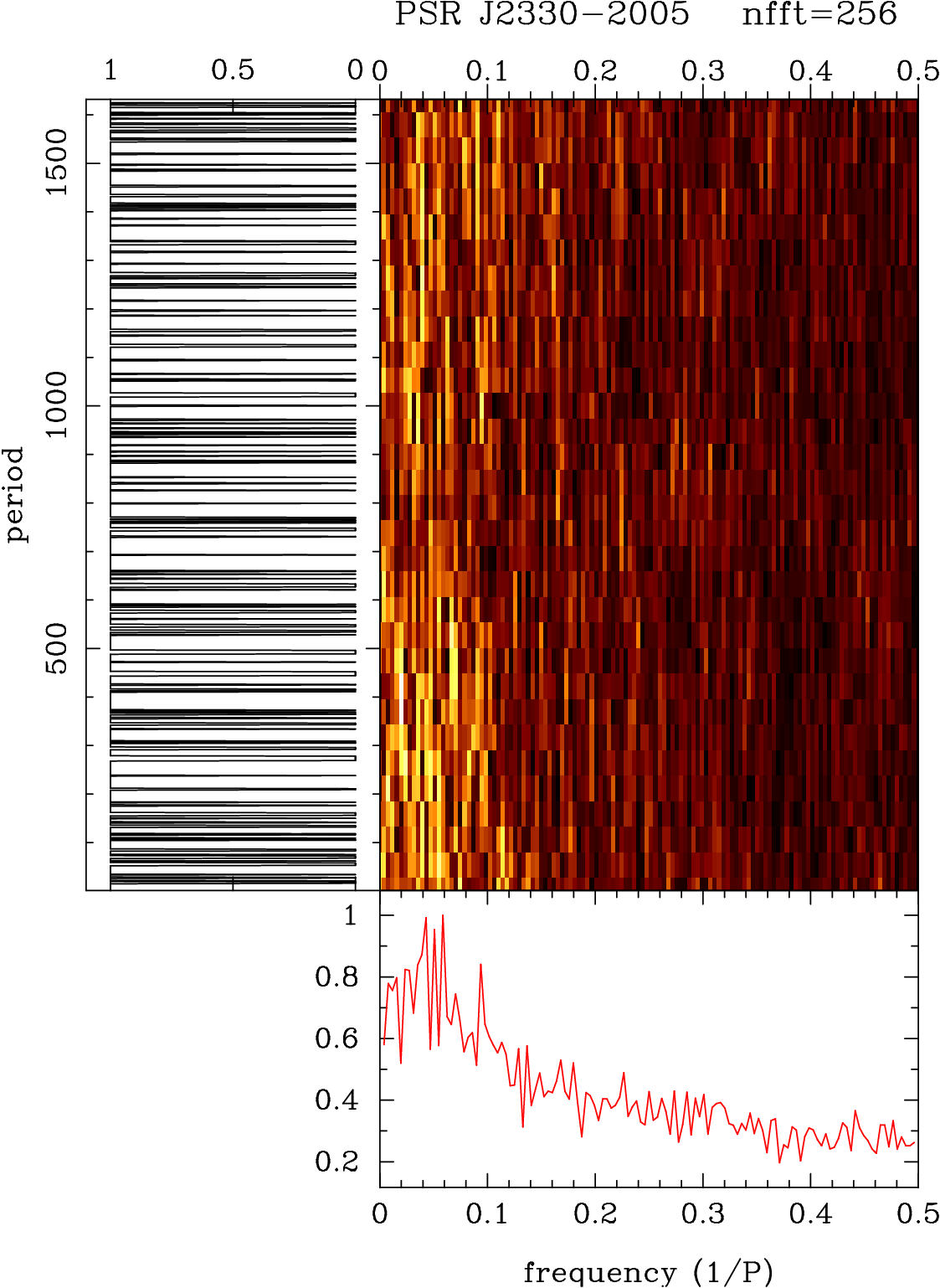}}} \\
\end{tabular}
\caption{PSR B2327$-$20 : The time evolution of the LRFS (left panel) and the
Null-Burst time series FFT (right panel).}
\label{fig_J2330}
\end{figure*}